\documentclass[a4paper,titlepage,11pt]{book}

\usepackage{latexsym}
\usepackage{amsfonts}
\usepackage{amssymb}%
\usepackage[mathscr]{eucal}
\usepackage[tbtags]{amsmath}
\usepackage{amsthmlq}
\usepackage{fancyhdr}
\usepackage{dropping}
\usepackage[sectionbib]{chapterbib} 
\usepackage{sectsty}
\usepackage[T1]{fontenc}
\usepackage{verbatim}
\usepackage{xspace}
\usepackage{fancybox}

\newcommand{\fnref}[1]{~(\ref{#1})}

\renewcommand{\headrulewidth}{.4pt}
\renewcommand{\footrulewidth}{0pt}

\newcommand{\lqheadsec}{\fontencoding{T1}\fontfamily{cmr}\fontseries{m}%
                        \fontshape{sc}\fontsize{9pt}{12}\selectfont}

\newcommand{\lqheadchap}{\fontencoding{T1}\fontfamily{cmr}\fontseries{m}%
                         \fontshape{sl}\fontsize{10pt}{12}\selectfont}

\newcommand{\normalheadfontB}{\fontencoding{T1}\fontfamily{cmr}\fontseries{m}%
                              \fontshape{n}\fontsize{11}{12pt}\selectfont}

\fancypagestyle{plain}{\fancyhf{}\fancyfoot{}%
                       \renewcommand{\headrulewidth}{0pt}%
                       \renewcommand{\footrulewidth}{0pt}}

\subsubsectionfont{\fontfamily{cmdh}\fontseries{m}\large}
\paragraphfont{\normalfont\fontshape{sl}}

\newcommand{\nl}{\par}

   {\begin{Sbox}\begin{minipage}}%
   {\end{minipage}\end{Sbox}\fbox{\TheSbox}}

\makeatletter
\def\cleardoublepage{\clearpage\if@twoside \ifodd\c@page\else
\hbox{}
\thispagestyle{empty}
\newpage
\if@twocolumn\hbox{}\newpage\fi\fi\fi}
\makeatother

\makeatletter
\renewcommand\theequation{\thechapter.\arabic{equation}}
\@addtoreset{equation}{chapter}
\makeatother
\renewcommand{\theequation}{\thechapter.\arabic{equation}}

\theoremstyle{plain}

\newtheorem{thm}{Theorem}[section]
\newtheorem{prop}{Proposition}[section]
\newtheorem{lem}{Lemma}[section]

\newtheorem*{OstReg}{Ostrogradsky's theorem}
\newtheorem*{OstCon}{Generalised Ostrogradsky theorem}
\newtheorem*{OstADM}{Ostrogradsky--\textsc{adm} equivalence theorem}

\theoremstyle{definition}

\newtheorem{exmp}{Example}[section]

\theoremstyle{remark} 

\newtheorem*{rem}{Remark}

\newcommand{\noopsort}[1]{}

\newcommand{\etalchar}[1]{$^{#1}$}

\newcommand{\EuD}{\mathscr{D}}

\newcommand{\EuH}{\mathscr{H}}      
      
\newcommand{\EuM}{\mathscr{M}}      
      
\newcommand{\EuQ}{\mathscr{Q}}      
\newcommand{\EuU}{\mathscr{U}}      

      \newcommand{\cl}{\mathcal{L}}
      
      \newcommand{\cp}{\mathcal{P}}

      \newcommand{\cv}{\mathcal{V}}
\newcommand{\cw}{\mathcal{W}}

	  \newcommand{\Fa}{\mathfrak{A}}
	  \newcommand{\Fb}{\mathfrak{B}}

\newcommand{\fg}{\mathfrak{g}}	  \newcommand{\Fg}{\mathfrak{G}}
	  \newcommand{\Fh}{\mathfrak{H}}

	  \newcommand{\Fl}{\mathfrak{L}}
	  
\newcommand{\fn}{\mathfrak{n}}	  
	  
	  \newcommand{\Fp}{\mathfrak{P}}
	  
	  \newcommand{\Fr}{\mathfrak{R}}
	  
	  \newcommand{\Ft}{\mathfrak{T}}

\newcommand{\rmc}{\mathrm{c}}
\newcommand{\rmd}{\mathrm{d}}

\newcommand{\rmr}{\mathrm{r}}

\newcommand{\rmz}{\mathrm{z}}
\newcommand{\rmA}{\mathrm{A}}
\newcommand{\rmB}{\mathrm{B}}
\newcommand{\rmD}{\mathrm{D}}
\newcommand{\rmE}{\mathrm{E}}

\newcommand{\rmT}{\mathrm{T}}

\newcommand{\rank}{\text{\textrm{rank}}}

\newcommand{\msc}[1]{\text{\normalfont\textsc{#1}}}

\newcommand{\bfg}{\mathbf{g}}

\newcommand{\bfx}{\mathbf{x}}
\newcommand{\bfy}{\mathbf{y}}

\newcommand{\bfL}{\mathbf{L}}

\newcommand{\bfR}{\mathbf{R}}

\newcommand{\bfnab}{\boldsymbol{\nabla}}
\newcommand{\bfGam}{\boldsymbol{\Gamma}}
\newcommand{\bfLam}{\boldsymbol{\Lambda}}

\newcommand{\ttT}{\mathtt{T}}

\newcommand{\sfG}{\mathsf{G}}
\newcommand{\sfH}{\mathsf{H}}

\newcommand{\ie}{i.e.\@\xspace}
\newcommand{\eg}{e.g.,\xspace}
\newcommand{\viz}{viz.\@\xspace}
\newcommand{\cfr}{cf.\@\xspace}

\newcommand{\etal}{et al.\@\xspace}
\newcommand{\dofs}{degrees of freedom\xspace}
\newcommand{\dof}{degree of freedom\xspace}
\newcommand{\gr}{general relativity\xspace}
\newcommand{\einspace}{Einstein space\xspace}
\newcommand{\einspaces}{Einstein spaces\xspace}
\newcommand{\EH}{Einstein--Hilbert\xspace}
\newcommand{\EL}{Euler--Lagrange\xspace}
\newcommand{\EP}{Einstein--Palatini\xspace}
\newcommand{\HP}{Hilbert--Palatini\xspace}
\newcommand{\wrt}{with respect to\xspace}
\newcommand{\adm}{{\normalfont\textsc{adm}}\xspace}
\newcommand{\hog}{higher-order gravity\xspace}
\newcommand{\hotg}{higher-order theories of gravity\xspace}
\newcommand{\flrw}{\textsc{flrw}\xspace}
\newcommand{\ode}{ordinary differential equation\xspace}
\newcommand{\odes}{ordinary differential equations\xspace}
\newcommand{\rhs}{right-hand side\xspace}
\newcommand{\MSc}{MSc\xspace}
\newcommand{\PhD}{PhD\xspace}
\newcommand{\DPh}{D.Ph.\@\xspace}
\newcommand{\Prof}{Prof\xspace}
\newcommand{\Doctor}{Dr\xspace}
\newcommand{\magt}{\textsc{mag}\xspace}

\newcommand{\abs}[1]{\lvert#1\rvert}

\newcommand{\fpR}{f^{\prime} \bigl( R \bigr)}
\newcommand{\fpr}{f^{\prime}}
\newcommand{\fpp}{f^{\prime \prime}}
\newcommand{\fR}{f \bigl( R \bigr)}

\newcommand{\fPhi}{f \bigl( \Phi \bigr)}
\newcommand{\fpPhi}{f^{\prime} \bigl( \Phi \bigr)}

\renewcommand{\qedsymbol}{$\blacksquare$}

\newcommand{\Lagb}{\underline{L}}
\newcommand{\Flb}{\underline{\Fl}}

\newcommand{\piadm}{\pi_{\text{\normalfont{\adm}}}}

\newcommand{\Vo}[1]{#1_{\circ}}
\newcommand{\Levi}{\overset{\circ}{\nabla}}

\DeclareMathOperator{\diff}{\mathrm{Diff}}
\DeclareMathOperator{\diag}{\mathrm{diag}}

\newcommand{\Lie}[1][n]{\cl_{#1}}

\newcommand{\emb}[2][(3)]{{^{#1}} \! #2}

\newcommand{\range}[3]{#1 = #2,\dots, #3}
\newcommand{\Christ}[3]{\genfrac{\{}{\}}{0pt}{1}{\,\!#1}{#2 #3}}
\newcommand{\varD}[2]{\genfrac{}{}{}{}{\delta {#1}}{\delta {#2}}}
\newcommand{\parD}[2]{\genfrac{}{}{}{}{\partial {#1}}{\partial {#2}}}

\newcommand{\tcons}{\; \stackrel{\cdot}{\longrightarrow} \;}

\newcommand{\naught}[2][{(0)}]{{\smash[t]{#2}}^{#1}}

\begin{document}


\allowdisplaybreaks[2]


\frontmatter

\pagenumbering{roman}

\begin{titlepage}
   \noindent\hspace*{6.25mm}
   \begin{minipage}{13.65cm}
      \begin{center}
         \huge{\textbf{Variational Principles and Cosmological Models in
                       Higher-Order Gravity}} 
      \end{center}
   \end{minipage}

\vspace*{\stretch{1}}
\noindent\hspace*{4.45cm}
   \begin{minipage}{6cm}
      \begin{center}
         \Large{\textsc{\textbf{Laurent Querella}}} 
      \end{center}
   \end{minipage}

\vspace*{\stretch{2}}
\noindent\hspace*{2.45cm}
   \begin{minipage}{10cm}
      \begin{center}
         \Large{\textbf{Doctoral dissertation}}   \\
         \vspace*{10mm}  
            \large{\textbf{UNIVERSIT\'E DE LI\`EGE \\
                           Facult' des sciences \\       
                           Institut d'astrophysique et de g'ophysique}} 

         \vspace*{5mm}  
         \large{\textbf{December 18, 1998}}
      \end{center}
   \end{minipage}
   \hspace*{\stretch{1}}
\end{titlepage}

\cleardoublepage


\thispagestyle{empty}

\vspace*{\stretch{1}}

\hfill
\begin{minipage}{10cm}
\large
\begin{quote}
\textit{``O, for a Muse of fire, that would ascend \\
          The brightest heavens of invention; \\
          A kingdom for a stage, princes to act \\
          And monarchs to behold the swelling scene!''} 

\vspace{5mm}

\hfill
--- W. Shakespeare, Henry V, Prologue.
\end{quote}
\normalsize
\end{minipage}

\vspace*{\stretch{2}}

\cleardoublepage


\chapter*{Preface}
\addcontentsline{toc}{chapter}{Preface}

\dropping{2}{T}\textsc{his} doctoral dissertation is the fruit of five years
of full-time research undertaken at the Department of Astrophysics and
Geophysics of the University of Li\`ege on January 1, 1994. Basically, it can 
be thought of as an extension of my \MSc thesis, the scope of which was to 
present in a self-contained fashion the Hamiltonian formulation of a specific 
class of alternative, higher-order theories of gravity, namely those 
relativistic, metric theories based on quadratic curvature Lagrangians. In 
1996 I had the opportunity to stay at the Department of Mathematics of the 
University of the Aegean for two months; the fruitful interplay with my Greek 
collaborators has undoubtedly broadened my research concerns in \gr and 
cosmology and permeates a significant part of this doctoral dissertation. 
\nl
When I started to work on alternative gravity theories, I planned to provide 
the reader of this yet unwritten dissertation with an exhaustive review of 
higher-order theories of gravity. However, I soon realised that the history of
these theories is fairly intricate, hence quite difficult to summarise; an 
exhaustive account with a wide historical perspective would thereby be---very%
---lengthy and not suited to the present work. Still, I am deeply convinced
that understanding the essential motivations of our predecessors greatly helps
in finding one's path in scientific research; but this takes much time, 
thereby implying less published articles: This is not particularly welcomed 
according to modern standards. Fortunately, during the last five years I have 
never been put under pressure to submit a paper every three months or so. 

\vspace*{5mm}
I wish that the interested reader will find this dissertation helpful and 
pleasant to read \ldots

\vfill
\noindent
Cointe, December 1998
\hfill
\textsc{L. Querella}

\newpage

\section*{Acknowledgments} 

\small
First and foremost, I would like to express my deep gratitude to my supervisor
Professor J. Demaret for his tremendous encouragement and reliable guidance, 
without which the bulk of this dissertation would not have come into 
existence, as well as for his open-mindedness about metaphysical questions, 
which inevitably occur in cosmology. \\
\indent
I thank \Prof S. Cotsakis for his enthusiastic attitude, his unalterable
confidence in my ability to succeed, his fruitful ideas, and for inviting me 
at the Department of Mathematics of the University of the Aegean, where some 
part of this research was done. \\
\indent
I am indebted to \Doctor J. Miritzis, a friend and collaborator, for many
interesting discussions, and also for carefully proof-reading Chapter Three.
\\ \indent
I owe a great indebtedness to Academician P. G. L. Leach for carefully proof-%
reading early drafts, for his patience in improving my written English---I 
hope he will not be too deeply depressed when confronted to the actual result%
---but also for challenging me to cycle round my beloved island of Samos. \\
\indent
There are many people whose reliable advices, fruitful ideas, and constructive
criticisms have undoubtedly helped and influenced this research; amongst 
these, I am grateful to \Prof D. Brill, \Doctor A. Burnel, \Doctor H.
Caprasse, \Doctor Y. De Rop, \Prof G. Flessas, \Prof G. T. Horowitz, \Doctor
C. Scheen, and \Prof J. W. York. \\
\indent
I thank the Department of Astrophysics and Geophysics of the University of 
Li\`ege, and in particular Professor J.-P. Swings, Chairman; the faculty;
staff; and students. \\
\indent
I am grateful to \Prof G. Flessas and \Prof N. Hadjissavas for giving me the
opportunity to stay at the Department of Mathematics of the University of the 
Aegean. I also thank the faculty, staff, students, and in particular H. 
Pantazi, for their warm hospitality. \\
\indent
I wish to thank the \emph{English Grammar Clinic}, on the Internet, for 
providing me with valuable insights into numerous subtleties of the English 
language. \\
\indent
More personally, I am indebted to my family for its warm support and I 
dedicate my inmost thoughts to all `Brothers' who share the essential ideals 
of \emph{Philosophia Perennis}. \\
\indent
Last but not least, my special thanks to my cherished wife Pascale, daughter 
Pauline, and son Quentin whose love and generosity proved to be priceless 
antidotes to ``forgetfulness or death.'' \\
\indent 
I also wish to gratefully acknowledge the \textsc{fria} (``Fonds pour la 
formation \`{a} la recherche dans l'industrie et dans l'agriculture'') for 
research fellowships, from Jan 1, 1994 to Dec 31, 1997. I thank the 
\textsc{fnrs} (``Fonds national de la recherche scientifique''), the 
``Patrimoine de l'universit\'e de Li\`ege'' and the Department of Astrophysics
and Geophysics, together with Professor J.-P. Swings, Chairman, for partial 
support by contract No.~\texttt{ARC 94/99-178} (``Action de recherche 
concert\'ee de la Communaut\'e fran\c{c}aise'') and for an assistantship by
contract No.~\texttt{SSTC PAI P4/05}, from Jan 1 to Dec 31, 1998. This
dissertation was typeset using \LaTeXe.
\normalsize

\newpage

\thispagestyle{empty}

\small
\begin{center}
UNIVERSIT\'E DE LI\`EGE \\
Facult\'e des sciences \\
\vspace{3mm}
\textbf{Variational Principles and Cosmological Models in Higher-Order 
        Gravity} \\
\vspace{3mm}
Doctoral dissertation submitted for the degree of \DPh \\
December 18, 1998 \\
\vspace{3mm}
\textsc{Laurent Querella} \\
\vspace{6mm}
\textbf{Abstract} \\
\end{center}

\scriptsize
\noindent
This dissertation investigates three main topics, all of which dealing with 
alternative, higher-order gravity theories in four dimensions. Firstly, we 
study the variational and conformal structure of those theories. Next, we 
analyse their Hamiltonian formulation and in particular its relationship with 
the famous \adm canonical version of \gr. Finally, we study higher-order 
spatially homogeneous cosmologies and exemplify how Hamiltonian methods can be
utilised to simplify the analysis of the associated field equations. \\
\indent
As regards the first topic, we begin by critically reviewing the variational
principle in gravitational theory: We argue that the `\EP', metric-affine 
method of variation, aside from being inherently nonequivalent to the Hilbert,
purely metric variational principle, leads to inconsistencies when applied to 
generalised gravitational actions including higher-order curvature terms. This
conveys us to put forth that one possible scheme that does not exhibit the 
cumbersome features of the `\EP device' is the Lagrange-multiplier version of 
the latter, namely the constrained first-order formalism. Applying this 
constrained method of variation to a general class of nonlinear Lagrangians we
prove that the conformal equivalence theorem of these nonlinear theories with 
\gr and an additional scalar field holds in the extended framework of Weyl 
geometry. As a direct consequence, we demonstrate that the \EP method is a 
degenerate case of the constrained first-order formalism and that it is unable
to deal with Weyl spaces. This investigation sheds another light on what is 
sometimes referred to as the \emph{universality of Einstein's equations}. \\
\indent
Next, we give a detailed account of the Hamiltonian formulation of higher-%
order gravity theories. After a short summary of Dirac's formalism for 
constrained systems, we thoroughly analyse the procedure that enables one to
develop a consistent canonical formulation of any field theory involving
higher derivatives: the generalised Ostrogradsky method. We demonstrate the
effectiveness of this \emph{modus operandi} by expressing nonlinear 
gravitational Lagrangians in canonical form. This conveys us to the main 
result in this part, that is, the equivalence of the Ostrogradsky and \adm 
canonical versions of \gr. We then discuss the issue of boundary terms in the 
light of the Ostrogradsky formalism. We finally obtain the explicit forms of 
the Hamiltonian constraints and canonical equations derived from generic 
quadratic Lagrangians. \\
\indent
The last topic is devoted to the study of spatially homogeneous Bianchi-type
cosmologies in higher-order gravity. We firstly analyse the empty Bianchi-type
IX or \emph{mixmaster} model in the full fourth-order gravity theory---without
resorting to the Ostrogradsky scheme---on approach to the initial singularity;
we prove that the mixmaster chaotic behaviour based on the \textsc{bkl} 
piecewise approximation method is structurally unstable and that there exists 
an isotropic power-law solution reaching the initial singularity in a stable, 
monotonic way. Next, we particularise the aforementioned Ostrogradsky 
canonical formalism to class A Bianchi types in the two distinct variants of 
the generic quadratic theory, namely the pure `$R$-squared' case and the 
conformally invariant, `Weyl-squared' case respectively. In the former we 
reduce the system of canonical equations to a system of autonomous second-%
order coupled differential equations that we solve analytically for Bianchi 
type I. In the latter we prove that the Bianchi-type I system is not 
integrable---in the sense of Painlev\'e---; we determine all particular 
closed-form solutions and discuss their conformal relationship with 
\einspaces.  

\newpage

\thispagestyle{empty}

\small
\begin{center}
UNIVERSIT\'E DE LI\`EGE \\
Facult\'e des sciences \\
\vspace{3mm}
\textbf{Variational Principles and Cosmological Models in Higher-Order 
        Gravity} \\
\vspace{3mm}
Dissertation pr\'esent\'ee par \textsc{Laurent Querella} \\
pour l'obtention du grade de docteur en sciences \\
\vspace{3mm}
18 d\'ecembre 1998 \\
\vspace{6mm}
\textbf{Sommaire} \\
\end{center}

\scriptsize
\noindent
Dans cette th\`ese, nous \'etudions divers aspects li\'es aux th\'eories
alternatives de la gravitation contenant des termes d'ordre sup\'erieur en la
courbure, dans des espaces-temps \`a quatre dimensions. Nous analysons
successivement la structure variationnelle et conforme de ces th\'eories ainsi
que leur formulation hamiltonienne. Ensuite, nous proc\'edons \`a l'\'etude de
mod\`eles spatialement homog\`enes et, en particulier, nous montrons comment
le formalisme canonique peut conduire \`a une simplification dans la recherche
de solutions exactes. \\
\indent
Dans la premi\`ere partie, apr\`es avoir insist\'e sur le fait que le principe
variationel d'\EP n'est, en g\'en\'eral, pas \'equivalent au principe de
Hilbert et qu'il pr\'esente des incoh\'erences manifestes d\`es qu'il est
appliqu\'e \`a des th\'eories autres que la relativit\'e g\'en\'erale dans le
vide, nous avan\c{c}ons qu'un sch\'ema consistant ne souffrant pas de ces
difficult\'es est fourni par la m\'ethode de variation m\'etrique-affine avec
multiplicateurs de Lagrange. Nous d\'emontrons que l'\'equivalence conforme
des th\'eories non lin\'eaires de la gravitation avec la relativit\'e
g\'en\'erale et un champ scalaire additionnel est \'egalement v\'erifi\'ee
dans le cadre g\'eom\'etrique \'etendu des espaces de Weyl. Comme corollaire,
nous prouvons que la m\'ethode d'\EP est un cas d\'eg\'en\'er\'e du formalisme
avec contraintes et qu'elle ne permet pas de travailler en g\'eom\'etrie de
Weyl. Sous cet angle, nous donnons une interpr\'etation diff\'erente de ce qui
est appel\'e, dans la litt\'erature r\'ecente, \emph{universalit\'e des
\'equations d'Einstein}. \\ 
\indent
Nous analysons ensuite de fa\c{c}on d\'etaill\'ee la formulation hamiltonienne
des th\'eories de la gravitation d'ordre sup\'erieur. Apr\`es avoir rappel\'e
les ingr\'edients n\'ecessaires \`a une telle construction, c'est-\`a-dire le
formalisme de Dirac des syst\`emes contraints et la m\'ethode d'Ostrogradsky
g\'en\'eralis\'ee, nous illustrons l'efficacit\'e de cette derni\`ere en
construisant explicitement une formulation canonique des th\'eories \`a
lagran\-giens non lin\'eaires en la courbure. Consid\'erant le lagrangien
d'\EH comme cas particulier, nous d\'emontrons le r\'esultat majeur de cette
partie, \`a savoir l'\'equivalence entre la formulation d'Ostrogradsky de la
relativit\'e g\'en\'erale et le c\'el\`ebre formalisme \adm. Sous ce nouvel
\'eclairage, nous discutons ensuite le probl\`eme des termes de surface pour
les th\'eories d'ordre sup\'erieur. Enfin, nous obtenons la forme explicite
des contraintes et des \'equations canoniques provenant du lagrangien
quadratique le plus g\'en\'eral. \\ 
\indent
La derni\`ere partie du travail est consacr\'ee \`a l'\'etude de mod\`eles
spatialement homog\`enes dans le cadre des th\'eories quadratiques de la
gravitation. Tout d'abord, nous montrons que, au voisinage de la singularit\'e
initiale, le comportement oscillatoire et chaotique du mod\`ele anisotrope
Bianchi IX bas\'e sur l'approximation dite de \textsc{bkl} est
structurellement instable pour la th\'eorie g\'en\'erale purement quadratique.
Ensuite, nous particularisons le formalisme canonique d'Ostrogradsky aux
mod\`eles de Bianchi de classe A, en distinguant les deux variantes
significatives de la th\'eorie quadratique g\'en\'erale, \`a savoir, le cas
o\`u le lagrangien se r\'eduit au carr\'e de la courbure scalaire et le cas
conform\'ement invariant, o\`u le lagrangien est \'egal au produit quadratique
contract\'e du tenseur de Weyl. Dans la premi\`ere variante, nous r\'eduisons
le syst\`eme canonique de d\'epart \`a un syst\`eme d'\'equations
diff\'erentielles autonomes du second-ordre que nous r\'esolvons
analytiquement pour Bianchi I. Dans la seconde, pour ce mˆme mod\`ele, nous
d\'emontrons que le syst\`eme canonique n'est pas int\'egrable, au sens de
Painlev\'e, nous d\'eterminons toutes les solutions analytiques
particuli\`eres et discutons leur relation conforme avec des espaces
d'Einstein. 
 
\normalsize

\tableofcontents

\mainmatter


\chapter{Introduction}
\label{chap:Intro}

\begin{minipage}{11cm}
\begin{quote}
\textit{``By Him who gave to our soul the Tetraktys, 
          which hath the fountain and root of ever-springing nature.''} 
\nl
\hfill
--- Pythagorean's oath.
\end{quote}
\end{minipage}

\vspace{1cm}

\dropping{3}{N}\textsc{otwithstanding} the fact that Einstein's \gr is 
experimentally tested with an overwhelmingly high degree of accuracy---from 
solar system tests to binary pulsars observational data---, it has become a 
peremptory necessity to consider \emph{alternative} theories of gravity. The 
reasons can be summarised as follows. Firstly, although Einstein's theory is 
the simplest---and the most aesthetic---geometrical theory of gravitation 
settled on the basic postulates of the \emph{equivalence principle} and 
\emph{general covariance} of the physical laws, there are no a priori reasons 
whatsoever to restrict the gravitational Lagrangian to be a \emph{linear} 
function of the scalar curvature nor to discard other more complicated 
frameworks obtained upon suitable generalisations of the \EH Lagrangian of 
\gr. Examples of such extensions are: scalar-tensor theories and higher-order 
gravity theories in four dimensions; Kaluza--Klein multidimensional theories; 
gauge theories of gravity, with torsion and `non-metricity'. Secondly, a 
strong research effort has been produced so far in different directions in 
order to formulate a consistent quantum theory of gravity; although the 
electroweak and strong interactions are described by renormalisable quantum 
field theories, Einstein's gravitational theory cannot be quantised according 
to the standard schemes, for the \EH action (with possibly a cosmological 
constant $\Lambda$)
\begin{equation*}
   S_{\msc{eh}} = \int \! \rmd^4 x \sqrt{-g} \, 
                     \frac{c^3}{16\pi G_{\msc{n}}} \bigl( R - 2 \Lambda \bigr)
\end{equation*}
does not define a renormalisable quantum field theory. However, by adding to
this action the most general, covariant action that contains quadratic
curvature terms and dimensionless couplings, namely 
\begin{equation*}
   S_{*} = \int \! \rmd^4 x \sqrt{-g} \, 
              \bigl( 
                 \alpha R^2 - \beta C^2 + 
                 \gamma L_{\msc{gb}} + \delta \square R
              \bigr),
\end{equation*}
where $C^2$ denotes the contracted quadratic product of the Weyl tensor, 
$L_{\msc{gb}}$ the Gauss--Bonnet term (topological invariant), and the 
`box' the d'Alembertian differential operator,%
\footnote{For a list of the conventions and notations used throughout this 
          dissertation, \cfr page \pageref{ConNot}.}
one obtains a power-counting renormalisable theory \cite{utiya=RenCla},%
\footnote{Rigorous renormalisability of the general fourth-order action has 
          been proved by Stelle using \textsc{brs} invariance 
          \cite{stell=RenHig}.}
which is asymptotically free \cite{fradk=RenAsy81}. Secondly, one hopes that 
such alternative theories might provide one with a better approximation, 
semi-classical limit of a yet unknown quantum theory of gravity. Amongst the 
various attempts to understand what a quantum space-time really is, unifying 
schemes such as string theory, supergravity, or more generally M-theory play a 
prominent r\^ole (see, \eg \cite{rovel=StrLoo,gibbo=QuaGra} for very recent 
reviews); and it turns out that higher-derivative terms appear naturally in 
the low-energy effective Lagrangians of some of those theories. Thirdly, the 
standard model of relativistic cosmology suffers from a certain number of 
difficulties that could perhaps be more naturally resolved in the context of 
generalised theories. For instance, whereas the singularity theorems in \gr 
show that the occurrence of space-time singularities is a generic feature of 
any cosmological models (under some reasonable conditions), it might happen 
that in the context of alternative theories those unwanted singularities could 
be avoided; in fact, during the last decade the absence of cosmological 
singularities when higher-order curvature terms are taken into account has 
been pointed out in the literature several times (see, \eg \cite{kanti=SinCos} 
and references therein).
\nl
As stated in the Preface, we do not intend to give an historical perspective
of the development of higher-order theories of gravity. We just would like to
mention that it can be traced back to the early years of \gr, when great
physicists like Weyl, Einstein, Bach, and others were undertaking the first 
investigations aiming at modifying the Hilbert variational principle so as to
unify---on purely geometrical grounds---electromagnetic and gravitational 
phenomena. Although this programme proved to be, so to speak, `chimeric', it
gave a renewed insight into the powerful use of the variational principle in 
gravitational theory and led many researchers to extend its domain of 
applicability to for instance other geometries or greater dimensions. 
Nowadays, the geometrical structure of any theory is intimately connected with
its formulation in terms of an action, in Lagrangian or Hamiltonian form, from 
which are derived the field equations by means of a specific, properly defined
variational principle. Because of the proliferation of various types of 
alternative gravity theories, it is of fundamental importance not just to 
confront their theoretical predictions with observational data but also to 
expressly understand their geometrical structure and the possible 
interconnections between their respective solution spaces. For instance, one
question worth to be addressed in that respect is whether the conformal
equivalence between a certain class of alternative theories and \gr would hold 
in the context of metric-affine variations of generalised actions in Weyl 
geometry; another is to analyse the Hamiltonian formulation of those theories
and possibly their quantisation. As regards higher-order theories though, it 
should be borne in mind that the field equations are much more intricate than 
Einstein's---any method of order reduction is thus most welcomed.
\nl
Whereas plane wave solutions of linearised Einstein gravity propagate two 
physical \dofs, carrying helicity $\pm 2$, the most general quadratic 
gravitational Lagrangian has eight \dofs: a massless spin-$2$ state, a massive 
spin-$0$ state, and a massive spin-$2$ ghost \cite{stell=RenHig,stell=ClaGra}. 
Higher-order gravity theories have not received general acceptance \emph{as 
viable physical theories} because some solutions of the classical theory are 
expected to have no lower energy bound and therefore exhibit instabilities, 
namely `runaway solutions': the linearised field equations propagate ghosts, 
\ie negative-energy modes, and possibly tachyons.%
\footnote{This drawback of higher-order gravity pertains in fact to any theory 
          with higher derivatives; see, \eg \cite{pais=FieThe}.} 
However, crucial \emph{nonperturbative} results have changed the bad 
reputation associated with this nonunitary character. For instance, the 
\emph{zero-energy theorem} states that: Although the theory admits linearised  
solutions with negative energy all exact solutions representing isolated 
(asymptotically flat) systems have precisely zero energy; the solutions to the 
linearised equations with nonzero energy of either sign do not correspond to 
the limit of a one-parameter family of exact solutions \cite{boulw=ZerEne}.
An interesting consequence of this theorem is that one cannot draw conclusions 
from the linearised theory concerning the stability of the full quantum 
theory.  
\nl
One special variant of the generic quadratic theory, which has attracted much 
interest as a promising candidate for quantum gravity, is called 
\emph{conformal gravity};%
\footnote{For a detailed bibliography, the interested reader may consult
          \cite{quere=BibConGra}.}
it is the purely quadratic theory based on an action containing only the 
Weyl-squared, $C^2$ term, which possesses the aesthetically pleasing feature 
of being the local gauge theory of the conformal group and hence locally 
scale invariant. This means that the theory bears no resemblance to Einstein 
gravity classically; it encompasses six \dofs corresponding to massless spin-%
$2$ and spin-$1$ ordinary states and a massless spin-$2$ ghost state 
\cite{lee=CouSta,riege=ParCon}. Conformal gravity satisfies at least two 
remarkable properties \cite{riege=ClaQua}: Birkhoff's theorem holds---in stark 
contrast to the generic quadratic case---and nonperturbative effects can 
confine the ghosts---analogously to the confinement of colour in quantum 
chromodynamics. The conformal fourth-order field equations, first put forth by 
Bach \cite{bach=ZurWey}, are found in differential geometry and in 
mathematical studies of Einstein's field equations of \gr that use conformal 
techniques; they also constitute necessary conditions for a space to be 
conformal to an \einspace. The classical study of the Bach equations is partly
justified by the fact that conformal gravity has been viewed, during the last
decade, as a possible, physically viable alternative theory of gravity that
would be able to resolve some of the problems \gr alone is unable to address 
without ad hoc assumptions such as, for instance, the dark matter hypothesis
\cite{mannh=ExaVac}. However, a deeper investigation reveals other open 
problems that necessitate further consideration (see, \eg \cite{klemm=TopBla}
for the most recent contribution). In particular, little work has been done in 
regard to spatially homogeneous cosmologies.  
\nl
Aside from conformal gravity, it is of great interest to study other variants 
of higher-order theories in the context of cosmology. The Friedmann--%
Lema\^itre--Robertson--Walker (\flrw) spaces, which are isotropic and are 
homogeneous on spacelike sections, constitute the basic pillar of the standard 
model of cosmology, which successfully accounts for many of the observed 
features of our universe. However, this scheme is not free from certain 
riddles: the so-called `horizon', `flatness', and `smoothness' problems. 
These are addressed, more or less adequately, by incorporating the generally
accepted inflationary paradigm into the orthodox isotropic cosmology with 
additional ingredients such as hot and cold dark matter blends---Is anybody 
still laughing at Ptolemaic epicycles? It is known that generalised gravity 
theories such as quadratic or scalar-tensor theories do possess solutions 
exhibiting an inflationary stage; in addition, they could possibly provide us 
with more satisfactory mechanisms to trigger inflation. On the other hand, 
mathematical cosmology focusses more on spatially homogeneous and 
\emph{anisotropic} models since they are fairly more general than the \flrw
universes and because one aims to investigate specific issues, such as for 
instance the genericness of oscillatory, chaotic dynamical regimes on approach
towards the initial singularity, unhindered by the stringent symmetry 
requirement that the universe be isotropic \emph{ab initio}. 
\nl
In this dissertation we examine Lagrangian and Hamiltonian variational methods 
for higher-order gravitational actions and in particular for nonlinear and
conformally invariant theories; we apply these methods to study the classical
solution space of spatially homogeneous cosmological models in higher-order 
gravity. The outline is the following:%
\footnote{A more detailed plan is provided at the beginning of each of those 
          chapters, \ie on pp.~\pageref{chap:VarCon}, \pageref{chap:HamFor}, 
          and \pageref{chap:BiaCos} respectively.}
\addcontentsline{toc}{section}{Outline}
\begin{itemize}
   \item In Chapter Two we begin by critically reviewing the variational
         principle in gravitational theory; we analyse the metric-affine 
         variational method in the context of generalised gravity theories in 
         order to seek in which specific circumstances it can be utilised to 
         deal with extended geometrical settings, such as Weyl geometry. We
         show that the \EP method exhibits inconsistencies and that it is a 
         degenerate case of the constrained first-order formalism, which is 
         the appropriate framework that can incorporate general Weyl spaces.
         We prove the chief result of this chapter, namely that the conformal 
         equivalence theorem of nonlinear theories with \gr and additional 
         scalar fields holds in the extended framework of Weyl geometry. As a  
         direct consequence, we give a different interpretation of a 
         universality property of Einstein's equations found in the context of
         the \EP method.
   \item In Chapter Three we focus on the Hamiltonian formulation of theories
         with higher derivatives. After a short summary of Dirac's formalism 
         for constrained systems and a detailed account of the generalised 
         Ostrogradsky method, we develop a canonical formulation of nonlinear 
         gravitational Lagrangians of the type $\fR$. Next, we prove the most
         important result of this chapter, namely that the Ostrogradsky 
         Hamiltonian formulation of \gr is equivalent to its well known \adm 
         canonical version. We then discuss the issue of boundary terms in the 
         light of the Ostrogradsky formalism and give the explicit forms of 
         the super-Hamiltonian and super-momentum constraints, and of the 
         canonical equations that are derived from generic quadratic 
         Lagrangians.
   \item In Chapter Four we study spatially homogeneous Bianchi-type
         cosmological models in some variants of higher-order gravity. We 
         firstly analyse the empty Bianchi-type IX or mixmaster model in the 
         purely quadratic theory---without resorting to the Ostrogradsky 
         method---on approach to the initial singularity; we prove that the 
         mixmaster chaotic behaviour based on the \textsc{bkl} approximation 
         scheme is structurally unstable and that there exists an isotropic 
         power-law solution reaching the initial singularity in a stable, 
         monotonic way. Next, we particularise the results obtained at the end
         of Chapter Three to class A Bianchi types in the two distinct 
         variants of the generic quadratic theory, namely the pure 
         `$R$-squared' case and the conformally invariant, `Weyl-squared' case 
         respectively. In the former we reduce the system of canonical 
         equations to a system of autonomous second-order coupled differential 
         equations that we solve analytically for Bianchi type I. In the 
         latter we prove that the Bianchi-type I system is not integrable---in 
         the sense of Painlev\'e---; we determine all particular solutions 
         that may be written in closed analytical form and discuss their 
         conformal relationship with \einspaces.
\end{itemize}

\section*{Conventions and notations} \label{ConNot}
\addcontentsline{toc}{section}{Conventions and notations}

In this dissertation we adopt the sign conventions of the gravitation bible
\cite{MISNE=Gra}; in particular we use the metric signature $(- + + +)$.
We also make use of a `customised version' of the \emph{abstract index 
notation} discussed in \cite{WALD=GenRel}: Latin indices of a tensor that 
belong to the beginning of the alphabet $a,b,c,\dots$ denote the type of the 
tensor---not its components---; Latin indices of a tensor that belong to the 
middle of the alphabet $i,j,k,\dots$ refer to spacelike components in the 
Cauchy hypersurfaces $\Sigma_t$ defined by the slicing of space-time; 
however, in Chapter \ref{chap:VarCon} we do not employ Greek indices 
$\alpha,\beta,\gamma,\dots$ to refer to the components of a tensor \wrt a 
specific coordinate space-time basis: we keep Latin indices $a,b,c,\dots$
instead. 
\nl
The symbol $\nabla_a$ usually stands for the covariant derivative operator but
occasionally denotes the associated linear affine connection. In Chapter
\ref{chap:VarCon} we also employ the symbol $\Levi_a$ to refer to the 
covariant derivative operator associated with the Levi-Civita connection, the
components of which in a non-coordinate basis are \label{conv:Cartan}
\begin{equation*}
   \Gamma^{\mu}_{\ \alpha \beta} = 
      \Christ{\mu}{\alpha}{\beta} + 
      \frac12 g^{\mu \nu} 
         \bigl( 
            C_{\beta \nu \alpha} + C_{\alpha \nu \beta} - C_{\nu \alpha\beta}
         \bigr),
\end{equation*}
where $\Christ{\gamma}{\alpha}{\beta}$ are the Christoffel symbols, \ie the
components of the Levi-Civita connection in a natural basis, and the $C$'s are
the structure coefficients of the non-coordinate basis.
\nl
The Riemann curvature tensor is defined by \label{conv:Riem}
\begin{align*} 
   R^a_{\ bcd} \, u^b &= 
      \bigl( \nabla_c \nabla_d - \nabla_d \nabla_c \bigr) u^a, \\
   R_{cdb}^{\ \ \ a} \, v_a &= 
      - R^a_{\ bcd} \, v_a   = 
      \bigl( \nabla_c \nabla_d - \nabla_d \nabla_c \bigr) v_b, 
\end{align*}
for arbitrary vectors $u^b$ and one-forms $v_a$, where $\nabla_a$ is the 
covariant derivative operator. The Ricci tensor is obtained by contraction on 
the first and third indices, that is
\begin{equation*}
   R_{ab} = R^c_{\ acb}.
\end{equation*}
\nl
Gothic characters denote tensor densities; \eg $\Fa^{ab}:=\sqrt{-g}\,A^{ab}$;
round and square brackets around indices denote respectively symmetrisation
and antisymmetrisation (including division by the number of permutations of 
the indices); a tilde denotes conformally transformed quantities. 
\nl
Unless otherwise stated we use \emph{geometrised units}, where the Newton 
gravitational constant $G_{\msc{n}}$ and the speed of light in vacuum $c$ are 
set equal to one: $G_{\msc{n}}=1=c$. 
\nl
We adopt Schouten's nomenclature for the type of spaces considered
\cite{SCHOU=RicCal}. An $L_n$ is a general $n$-dimensional manifold endowed
with a \emph{linear} connection; when the latter is \emph{symmetric} the $L_n$
is called an $A_n$ and is torsion-free. When a metric tensor $g_{ab}$ is 
defined, the compatibility condition does not hold in general, \ie 
$\nabla_c g_{ab} = - Q_{cab} \neq 0$. If $Q_{cab}=0$, the connection is called
\emph{metric} \wrt $g_{ab}$ and the $L_n$ is called a $U_n$; if in addition
the connection is symmetric, one has a $V_n$, \ie \emph{Riemann space}. If 
$Q_{cab}=Q_c \, g_{ab}$, the connection is called \emph{semi-metric}; if in
addition it is symmetric, the $L_n$ is called a $W_n$, \ie \emph{Weyl space}.


\chapter{Variational and conformal structure in \hog}
\label{chap:VarCon}

\begin{minipage}{11cm}
\begin{quote}
\textit{``Sub fide vel spe geometricantis natur{\ae}.''} 
\nl
\hfill
--- Giordano Bruno.
\end{quote}
\end{minipage}

\vspace{1cm}

\dropping{3}{V}\textsc{ariational principles} play a prominent r\^ole in 
theoretical physics; it has become well accepted during this century that any 
fundamental physical theory can be formulated in terms of an action, in 
Lagrangian or Hamiltonian form, from which are derived the equations of motion 
by means of a variational principle. For the last decades this has been raised 
to a metalaw of nature: Nowadays, setting up a (field) theory means that one 
starts to write an action in terms of the fields one considers, even if the 
knowledge of the actual form of their interactions and symmetry properties is 
fragmentary. Specification of the Lagrangian function is determined by 
mathematical and physical requirements like gauge invariance, 
renormalisability, simplicity, and so forth. Yet, certain peculiarities of the 
Lagrangian that arise under symmetry transformations, such as the appearance 
of a total divergence, might indicate ``that one is dealing with some 
approximation or a limiting case of a `better' theory, in which the 
corresponding, possibly modified, symmetries fully preserve the action 
integral'' \cite{traut=MetRem}.
\nl
Even though we do not intend to discuss epistemological or metaphysical issues 
related to the significance of the variational method we would like to 
emphasise that, historically, it has often been the focus of philosophical 
contentions and misconceptions for, in contradiction to the usual 
\emph{causal} description of phenomena, ``the idea of enlarging reality by 
including `tentative' possibilities and then selecting one of these by the 
condition that it minimizes a certain quantity, seems to bring a 
\emph{purpose} to the flow of natural events'' \cite[p.~xxiii]{LANCZ=VarPri}. 
Still, one ought not be disconcerted: For the universal mind of the 
seventeenth and eighteenth centuries, the two ways of thinking did not 
necessarily appear contradictory. Leibniz who had a strong influence on the 
development of the variational method---for example, the present use of the 
word `action' in physics probably originates from Leibniz's expression 
\emph{actio formalis}---had strong teleological propensities, which also 
characterised the ideas of Fermat, Borelli, and Maupertuis. By contrast, the 
sober, matter-of-fact nineteenth century---which still gets hold of numerous 
present-day scientists---looked at the variational principles (of mechanics) 
merely as convenient alternative mathematical formulations of the fundamental 
laws, without any primary importance whatsoever. This pragmatic point of view 
has however changed with the advent of \gr---in the light of which ``the 
application of the calculus of variations to the laws of nature assumes more 
than accidental significance'' \cite{LANCZ=VarPri}---and quantum mechanics---%
especially with regard to Feynman's `sum-over-histories' approach, in relation 
with Dirac's deep intuition \cite{FEYNM=QuaMec,MISNE=GraVarPri}.%
\footnote{In particular, we now know that the \emph{value} of the action 
          integral---and not only its \emph{variation}---is physically 
          relevant.}
\nl
As stated in the Introduction, we aim to analyse the metric-affine variational
method as applied to gravity theories in order to determine in which specific
circumstances it can be regarded as a possible generalisation of the standard 
metric variational principle of \gr. First of all, we make a brief historical 
survey to see how the metric-affine variational principle was introduced in 
\gr.%
\footnote{For a more detailed account, we refer the interested reader to 
          \cite{ferra=VarFor,vizgi=GeoUni,quere=BibEinPal}, and references 
          therein.} 
\nl
The \emph{purely metric} or Hilbert variational principle of \gr was properly 
defined during the years 1914 to 1916 owing to the works of Einstein, Hilbert, 
and Lorentz for any of whom, at that time, the metric tensor was thought of as 
the only fundamental gravitational field. However, from the early works of 
Levi-Civita and Hessenberg in 1917, Weyl and Cartan developed, until 1923, the 
new concept of affine connection on manifolds without a metric structure. For 
many years Weyl's theory of symmetric linear connections \cite{WEYL=SpaTim} 
has been a rich source of inspiration for himself and physicists like 
Eddington and Einstein whose aim was to unify---on purely geometrical grounds%
---gravitational and electromagnetic phenomena by means of an \emph{affine} 
variational principle. Unfortunately their attempts did not meet their hopes 
and Einstein abandoned the purely affine theory on behalf of a \emph{metric-%
affine} variational method in which the metric tensor and the affine 
connection are considered as independent fields through the process of 
variation; Einstein proved exactly, for the first time, what is placed in most 
modern textbooks under the authorship of Palatini---even though Palatini's 
contribution (\cfr equation \eqref{eq:VPPalEqu}) was formulated in a purely 
metric framework \cite{palat=DedInv,PAULI=TheRel}---, namely the equivalence 
of Einstein's field equations of vacuum \gr and the field equations that are 
derived by means of metric-affine independent variation of the \EH 
gravitational action \cite{einst=EinFel}. Once again, Einstein's attempt of a 
geometric unification failed and the metric-affine method lost its interest%
---even if Schr\"odinger revisited the question twenty years later 
\cite{SCHRO=SpaTim}---until the late fifties with the works of Stephenson 
\cite{steph=QuaLag,steph=GenCov} and Higgs \cite{higgs=QuaLag} who analysed 
the field equations obtained from quadratic Lagrangians via a metric-affine 
variational principle; more specifically, they considered those equations as 
an alternative set to Einstein's field equations: The choice of Lagrangians%
---quadratic in the various curvature tensors---was again motivated by Weyl's 
unified theory of gravitation and electromagnetism.%
\footnote{There were numerous attempts towards that goal, which were chiefly
          characterised by a modification of the variational principle through
          different kinds of alteration of its underlined geometric structure: 
          semi-Riemannian manifolds in dimensions greater than four (\eg 
          Kaluza--Klein theories); nonmetric connections (\eg Weyl geometry,
          theories with torsion); purely affine connections (\eg Einstein--%
          Schr\"odinger theory); and so forth. For an exhaustive study on all
          these alternative theories, we refer to the remarkable treatise of
          Tonnelat \cite{TONNE=TheUni}.}
However, Buchdahl raised severe conceptual objections in regard to the self-%
consistency of the method and proposed implicitly to abandon the use of 
`Palatini's device' in gravitational theory \cite{buchd=NonLagPal}. 
\nl
At the same time the so-called `\HP action principle' together with the 
`three-plus-one decomposition' of space-time was invoked successfully by 
Arnowitt, Deser, and Misner (\adm) in order to develop a Hamiltonian 
formulation of \gr \cite[\S 21.2 and \S 21.7]{arnow=DynGen,MISNE=GraVarPri}.%
\footnote{See Subsection \ref{subsec:ConHamIntro} on 
          page \pageref{subsec:ConHamIntro} and Subsection \ref{subsec:CanGR} 
          on page \pageref{subsec:CanGR}.}
However, referring to the `Palatini variational principle' is misleading. As a 
matter of fact, there is no need whatsoever to resort to a metric-affine 
variation in order to rewrite the \EH action in canonical form.%
\footnote{By contrast, the Ashtekar canonical formalism uses explicitly the 
          \HP first-order variational method, with the tetrad field and spin-%
          connection as independent variables, and where only the self-dual 
          part of the curvature is retained in the Lagrangian; see, \eg 
          \cite{pelda=ActGra}.} 
In that respect, the analogy (\cfr \cite[\S 21.2]{MISNE=GraVarPri}) between 
the `Palatini variation' and Hamilton's principle in phase space is 
inaccurate---\emph{stricto sensu} it is wrong.%
\footnote{This was clearly emphasised by El-Kholy, Sexl, and Urbantke who 
          distinguished between what they called the `Palatini principle' and 
          the `formal Palatini method of variation'; the \adm procedure 
          belonging to the second class \cite{elkho=SocPal}.}
In fact, the conjugate momenta are defined in terms of the extrinsic 
curvature, not in terms of the connection; the equivalence of purely metric 
and metric-affine variations in \emph{vacuum} \gr is a mere coincidence (\cfr 
Subsection \ref{subsec:VPEPVar}). 
\nl
In the same spirit as in Stephenson's articles, Yang investigated a theory 
based on a Lagrangian that is quadratic in the Riemann tensor, by analogy with 
the Yang--Mills Lagrangian \cite{yang=IntFor}. Unfortunately, Stephenson's and 
Yang's field equations are tainted by a conceptual mistake occurring in the 
process of variation. Before this error was noticed, several authors proved 
that those equations were leading to generic unphysical solutions and 
consequently they ruled them out.%
\footnote{Notwithstanding this fact, a detailed study of Yang's equations has 
          been published most recently \cite{guilf=YanGra}.} 
\nl
Extending previous results---only valid in vacuum \gr---of Lanczos 
\cite{lancz=EleGen} and Ray \cite{ray=PalVar}, Safko and Elston applied the 
metric-affine variational principle with Lagrange multipliers to quadratic 
Lagrangians \cite{safko=LagMul}.%
\footnote{In spite of several misprints in the resulting formul{\ae} Safko and
          Elston's conclusions are right; \cfr Subsection \ref{subsec:VPCP}.}
Unacquainted with the Ostrogradsky method, they also tried to establish a 
connection between the Lagrange-multiplier version of the variational 
principle and the \adm formalism in order to develop a Hamiltonian formulation 
of quadratic gravity theories (\cfr Subsection \ref{subsec:HamQua}). 
Independently, Kopczy\'nski showed that the introduction of appropriate
constraints into the gravitational action may serve to `unify' the variational 
derivations of distinct theories of gravity such as Einstein's theory and the 
Einstein--Cartan theory \cite{kopcz=PalPri}. The most recent generalisations 
of this constrained method of variation, for manifolds with torsion and 
`non-metricity', lead to the \emph{metric-affine gauge theory of gravity} 
(\magt) \cite{hehl=MetAff}.
\nl 
Following Buchdahl \cite{buchd=QuaLagPal}, we address the questions of the
utility and consistency of this method of variation in the broader context of
generalised gravity theories.%
\footnote{We restrict ourselves to geometries without torsion, \ie symmetric
          connections.}
More specifically, in Section \ref{sec:VarConGR} we briefly present the well-%
known Hilbert method of variation, mainly to settle our notations, and 
critically review the \EP variational principle in \gr. In Section
\ref{sec:VarConHOG} we extend the study of the variational principle to the 
domain of higher-order Lagrangians; the ensuing picture reveals that the 
\EP variational principle is generically unreliable, already at the
classical level. This conclusion conveys us to carefully formulate a 
Lagrange-multiplier version of the metric-affine variational method that we 
call \emph{constrained first-order formalism}. As a first application, we
consider the variation of several higher-order Lagrangians with a Riemannian
constraint and correct Safko and Elston's results \cite{safko=LagMul}. 
Furthermore, we show that the equivalence of the field equations that are 
derived from appropriate actions via this formalism to those produced by 
variation of purely metric Lagrangians is not merely formal but is implied by 
the diffeomorphism covariant property of the associated Lagrangians. In 
Section \ref{sec:VarConStr}, after a brief account of the conformal 
relationship between nonlinear and scalar-tensor gravitational Lagrangians and
\gr with additional scalar fields, we analyse the conformal structure of 
nonlinear gravity theories in the context of the constrained first-order 
formalism; in particular, we prove that the conformal equivalence theorem of 
those theories with \gr plus a scalar field holds in the extended framework of 
Weyl geometry. This investigation enables us to give a different 
interpretation of what has been recently called a `universality property of 
Einstein's equations' \cite{ferra=UniVac,borow=UniEin} and to invalidate
a recent claim on a possible explanation of a---controversial---observed 
anisotropy in the universe \cite{tapia=UniFie,quere=ComUni}. Finally, we point 
out how these results may be further exploited and address a number of new 
issues that arise from this analysis. This work was carried out in 
collaboration with S. Cotsakis and J. Miritzis \cite{cotsa=VarCon} (see also 
\cite{mirit=VarStr,cotsa=MG8,quere=GR15}).

\section{Variational principles in \gr}
\label{sec:VarConGR}

\subsection{Hilbert variation}
\label{subsec:VPEHVar}

Consider a four-dimensional space-time manifold $\EuM$ endowed with a 
Lorentzian metric $g_{ab}$ and assume that the connection $\nabla_c$ be the 
symmetric Levi-Civita connection, \ie $\nabla_c g_{ab}=0$; hence 
$(\EuM,\bfg,\bfnab)$ is a $V_4$, \ie a Riemannian space. In the Lagrangian 
formulation of the theory the Hilbert metric variational principle proceeds 
with the specification of a Lagrangian density $\Fl$, which is assumed to be a 
functional of the metric and its first and possibly higher derivatives, that 
is
\begin{equation} \label{eq:VPEHGenLagDen}
   \Fl = \Fl \bigl( \bfg, \partial \bfg, \partial^2 \bfg, \dots \bigr). 
\end{equation}
In addition, one requires that $\Fl$ be a scalar density of weight $+1$, \ie
$\Fl=\sqrt{-g}\,L$, where $g$ denotes the determinant of the matrix formed 
with the components of $g_{ab}$ and $L$ is the Lagrangian; this enables one to 
form the action integral
\begin{equation} \label{eq:VPEHGenAct}
   S [\bfg] = \int_{\EuU} \! \rmd^4 \Omega \, \Fl, 
\end{equation}
where $\rmd^4\Omega=\rmd x^0\wedge\rmd x^1\wedge\rmd x^2\wedge\rmd x^3$, which 
is taken over a compact region $\EuU$ of the manifold $\EuM$. The field
equations are obtained by requiring that the action \eqref{eq:VPEHGenAct} be 
stationary under arbitrary variations such that the metric and its first
derivatives be held fixed on the boundary $\partial\,\EuU$. This variation 
defines the functional derivative $\Fl_{ab}$ of the Lagrangian density $\Fl$, 
\viz
\begin{equation*}
   \delta S [\bfg] = 
      \int_{\EuU} \! \rmd^4 \Omega \, \Fl_{ab} \, \delta g^{ab},
   \qquad \text{with} \; \Fl_{ab} := \varD{\Fl}{g^{ab}}, 
\end{equation*}
also called the \emph{Euler--Lagrange derivative} of $\Fl$, and the field
equations are 
\begin{equation*}
   \Fl_{ab} = 0.
\end{equation*}
\nl
As is well known, the variational principle implies very important 
differential constraints on the field equations, which hold `off shell', \ie
whether or not the field equations are satisfied; these are the 
\emph{generalised Bianchi identities}, obtained from Noether's second theorem
by taking as a specific class of variations of the metric that induced by 
diffeomorphisms $f:\EuM \rightarrow \EuM$. Since the manifolds $(\EuM,\bfg)$
and $(\EuM,f^*\bfg)$ are physically equivalent, the action functional does not
change under the diffeomorphism $f$; in particular, it remains unaltered under
an infinitesimal coordinate transformation. For such variations, it is not
difficult to see that, at the first order of perturbation, $\delta g^{ab}$ is 
given in terms of the Lie derivative of the metric \wrt the vector field $v^c$ 
that generates the diffeomorphism $f$, that is,%
\footnote{See \cite[Appendix C]{WALD=GenRel}.} 
\begin{equation*}
   \delta g^{ab} = - \Lie[v] g^{ab}
                 = - 2 \nabla^{(a} v^{b)}.
\end{equation*}
Since by definition $\Fl_{ab}$ is a symmetric density of weight $+1$, the 
variational principle yields 
\begin{equation*}
   \delta S [\bfg] = 
      - 2 \int_{\EuU} \! \rmd^4 \Omega \, \Fl_{ab} \bigl( \nabla^a v^b \bigr)
   \equiv 0
\end{equation*}
for all vector fields $v^c$ that vanish on the boundary. Integrating by parts
the last equation and dropping the divergence term one obtains the expected
generalised Bianchi identities, namely
\begin{equation} \label{eq:VPGenBiaIde}
   \nabla^a \Fl_{ab} = 0.
\end{equation}
\nl
The simplest Lagrangian density for gravity is the \EH Lagrangian density 
$\Fl_{\msc{eh}}=\sqrt{-g}\,R$, where $R=g_{ab}R^{ab}$ is the scalar curvature,
to which one may possibly add a cosmological constant term $\Lambda\sqrt{-g}$. 
The corresponding gravitational action is 
\begin{equation} \label{eq:VPEHAction}
   S = \frac{c}{2\kappa^2} 
       \int_{\EuU} \! \rmd^4 \Omega \, \Fl_{\msc{eh}}, 
\end{equation}
where $\kappa^2=8\pi G_{\msc{n}}c^{-2}$ is the Einstein gravitational 
constant.%
\footnote{Hereafter we use `geometrised units', where the Newton gravitational 
          constant $G_{\msc{n}}$ and the speed of light in vacuum $c$ are set 
          equal to one, and we rescale the coordinates to absorb the constant 
          factor $8\pi$.}
The Ricci tensor $R_{ab}$ is expressed in terms of the connection coefficients
$\Gamma^c_{\ ab}$ and their first derivatives, \viz
\begin{equation*} \label{eq:VPRicci}
   R_{ab} := R^c_{\ acb} = \partial_c \Gamma^c_{\ ab} - 
                           \partial_b \Gamma^c_{\ ac} +
                           \Gamma^c_{\ cd} \Gamma^d_{\ ab} - 
                           \Gamma^c_{\ bd} \Gamma^d_{\ ac}.
\end{equation*}
Since the `metricity' or compatibility condition holds, \ie $\bfnab\bfg=0$, 
the $\Gamma$'s are the Christoffel symbols, namely
\begin{equation} \label{eq:VPChrist}
   \Gamma^c_{\ ab} \equiv \Christ{c}{a}{b} = 
      \frac12 g^{cd} 
      \bigl( \partial_b g_{ad} + \partial_a g_{db} - \partial_d g_{ab} \bigr). 
\end{equation}
As the \EH Lagrangian density depends \emph{linearly} on the second-order 
derivatives of the metric, one could discard these higher derivatives through 
a total divergence, the variation of which would not affect the equations of 
motion. Hence, one could start with the so-called `gamma-gamma' first-order 
form of the Lagrangian density for gravity, which is given explicitly by
\begin{equation} \label{eq:VPGamGamLag}
   \Fl = \sqrt{-g} \, g^{ab} \Bigl( 
                                \Gamma^c_{\ ad} \Gamma^d_{\ bc} -
                                \Gamma^c_{\ dc} \Gamma^d_{\ ab}
                             \Bigr).
\end{equation}
This possibility elucidates why Einstein's field equations are second-order
instead of fourth-order differential equations. However, one should bear in
mind that the `gamma-gamma' Lagrangian density is no longer a scalar density.
\nl
Nevertheless, one aims at deriving Einstein's equations in vacuum by requiring 
that the action \eqref{eq:VPEHAction} be stationary under arbitrary variations
of the metric that vanish on the boundary. This is achieved with the help of
the formula $\delta\sqrt{-g}=-\tfrac12\sqrt{-g}\,g_{ab}\delta g^{ab}$ and the 
\emph{Palatini equation} 
\begin{equation} \label{eq:VPPalEqu}
   \delta R^a_{\ bcd} = \nabla_c \bigl( \delta \Gamma^a_{\ bd} \bigr) -
                        \nabla_d \bigl( \delta \Gamma^a_{\ bc} \bigr),  
\end{equation}
which can be easily derived in a locally geodesic coordinate system (see, \eg
\cite{DINVE=IntEin}) and the contraction of which is
\begin{equation} \label{eq:VPPalEquCon}
   \delta R_{ab} = \nabla_c \bigl( \delta \Gamma^c_{\ ab} \bigr) -
                   \nabla_b \bigl( \delta \Gamma^c_{\ ac} \bigr).  
\end{equation}
The Hilbert metric variation of the action \eqref{eq:VPEHAction} is first 
written as%
\footnote{We recall that `Gothicised' quantities denote tensor densities; for 
          instance, $\Fa^{ab}:=\sqrt{-g}\,A^{ab}$, for an arbitrary tensor 
          field $A^{ab}$.} 
\begin{equation*}
   \delta S = \int_{\EuU} \! \rmd^4 \Omega \, 
                 \Bigl( 
                    \delta \fg^{ab} R_{ab} + \fg^{ab} \delta R_{ab} 
                 \Bigr),
\end{equation*}
Making use of equation \eqref{eq:VPPalEquCon} and owing to the compatibility 
condition (in the form $\nabla_c \fg^{ab}=0$) one can transform the second 
term of the integrand as a pure divergence; by Gauss's theorem the 
corresponding integral becomes a surface integral over the boundary 
$\partial\,\EuU$ and vanishes because the variations are assumed to vanish on 
the boundary.%
\footnote{In fact, for general variations such that only the metric be held 
          fixed on the boundary, this surface integral does not vanish; \cfr 
          our subsequent discussion on page \pageref{subsub:CanGRADM} on the 
          r\^ole of boundary terms. \label{foot:VPBouTer}}
Hence the variation of the \EH action reduces to
\begin{equation*}
   \delta S = \int_{\EuU} \! \rmd^4 \Omega \, \Fg_{ab} \delta g^{ab} \equiv 0,
\end{equation*}
where $\Fg_{ab}$ denotes the Einstein tensor density associated with the 
Einstein tensor
\begin{equation*}
   G_{ab} := R_{ab} - \frac12 R g_{ab}.
\end{equation*}
Since the variations $\delta g^{ab}$ and the region of integration $\EuU$ are
arbitrary, one concludes that the variational principle for the action 
\eqref{eq:VPEHAction} implies Einstein's vacuum field equations. 
\nl
In keeping with the above variational principle, when matter comes into play, 
one must add to the gravitational Lagrangian density \eqref{eq:VPEHGenLagDen} 
an appropriate Lagrangian density $\Fl_{\msc{m}}$ for the corresponding 
fields, which assumes a form that is a `generalisation' of its special 
relativistic form---which depends primarily on the field variables, 
collectively called $\psi$---, achieved via the strong principle of 
equivalence according to the `minimal coupling' rule:%
\footnote{This prescription is not free from a certain ambiguity in the sense
          that there are other generalisations---\eg conformal coupling---of 
          the aforementioned special relativistic form that are indeed 
          compatible with the basic postulate of general covariance.} 
\begin{align*}
    &\eta_{ab} \longrightarrow g_{ab},
   &&\partial_a \longrightarrow \nabla_a.
\end{align*}
Observe that the order of the two steps is irrelevant as long as the 
connection is the Levi-Civita connection: For arbitrary connections the 
operation of lowering and raising indices does no longer commute with the 
operation of covariant differentiation. The total action is defined as
$\int (\Fl + \Fl_{\msc{m}})$ and variation of the second term \wrt the metric
defines the stress-energy tensor $T_{ab}$ (\cfr \cite[\S 95]{LANDA=TheCha}) 
so that the full field equations are $G_{ab}=T_{ab}$ supplemented by the
equations of motion for the fields $\psi$. Furthermore, the generalised 
Bianchi identities \eqref{eq:VPGenBiaIde} take the form of the contracted 
Bianchi identities, \ie $\nabla^a G_{ab}=0$, which in turn entail the 
covariant conservation of the stress-energy tensor, as a direct consequence of 
the invariance of the \EH action under diffeomorphisms.

\subsection{Einstein--Palatini variation}
\label{subsec:VPEPVar}

Consider a four-dimensional space-time manifold $\EuM$ endowed with a 
Lorentzian metric $g_{ab}$ but now assume that the connection $\nabla_c$ be an
arbitrary \emph{symmetric} connection, \ie $\nabla_c g_{ab}\neq0$ and zero
torsion; hence $(\EuM,\bfg,\bfnab)$ is an $A_4$, \ie an affine space provided 
with a symmetric linear connection (in which here a metric tensor is also 
defined), which is not necessarily a $V_4$: In particular, no relationship is 
assumed \emph{a priori} between the metric and the connection, \ie they are 
independent from each other. The \EP, metric-affine variational principle
proceeds with the specification of a Lagrangian density $\Fl$ that is 
constructed from the Riemann tensor of the connection and also the metric; 
which is therefore assumed to be a functional of the metric, its covariant 
derivatives up to a certain order, the connection coefficients, and their 
derivatives up to a certain order, that is (formally)
\begin{equation} \label{eq:VPEPGenLagDen}
   \Fl = \Fl \bigl( 
                \bfg, \bfnab \bfg, \bfnab \bfnab \bfg, \dots; 
                \bfGam, \partial \bfGam, \partial^2 \bfGam, \dots
             \bigr). 
\end{equation}
The analogue of the action functional \eqref{eq:VPEHGenAct} is 
\begin{equation} \label{eq:VPEPGenAct}
   S [\bfg, \bfGam] = \int_{\EuU} \! \rmd^4 \Omega \, \Fl. 
\end{equation}
Its variation under arbitrary \emph{independent} variations of the metric and 
the connection that vanish on the boundary $\partial\,\EuU$ is given by
\begin{equation*}
   \delta S [\bfg, \bfGam] = 
      \int_{\EuU} \! \rmd^4 \Omega 
        \Bigl( 
           \Fa_c^{\ ab} \, \delta \Gamma^c_{\ ab} +
           \Fb_{ab} \, \delta g^{ab}
        \Bigr),
\end{equation*}
where $\Fa_c^{\ ab}$ and $\Fb_{ab}$ are the \EL derivatives of $\Fl$ \wrt the
connection and the metric respectively; hence the field equations are
\begin{align*}
    &\Fa_c^{\ ab} = 0,
   &&\Fb_{ab} = 0,
\end{align*}
sometimes called $\bfGam$- and $\bfg$-equations respectively.
\nl
As was first noted by Buchdahl, invoking such a method of variation in a 
gravitational theory is highly objectionable, for the \EP prescriptions 
ascribe to the variational principle the cumbersome task of picking out a 
specific class of spaces amongst all possible $A_4$'s \cite{buchd=NonLagPal}:
It is therefore implicitly assumed that one is dealing with a much broader
geometrical setting than the familiar Riemannian space of \gr; in fact, 
depending on the specific \EP Lagrangian, the general $A_4$ could degenerate
into a more specialised space as for instance a Weyl space $W_4$, or it could
remain totally unspecified, with a completely arbitrary metric tensor. In our
point of view such arbitrariness in the variational principle is obnoxious; 
we recommend that the specific geometry one is dealing with be fixed \emph{ab
initio} even though this is not the usual Riemannian geometry (\cfr 
Subsection \ref{subsec:VPCP}).
\nl
In the case of \gr $\Fl$ can be chosen as the \HP Lagrangian density 
$\Fl_{\msc{hp}}=\fg^{ab} R_{ab} (\bfGam, \partial \bfGam)$, where the Ricci 
tensor depends on the connection coefficients and their first derivatives 
only; hence $\Fl_{\msc{hp}}$ is regarded as a functional of the $10$ metric
components and the $40$ connection coefficients. It turns out that the \EP
method of variation is technically simpler than the Hilbert method of
variation described in the previous subsection. Varying the \HP action \wrt 
the metric $g^{ab}$ one obtains directly 
\begin{equation} \label{eq:VPEPFieEquMet}
   \delta S = \int_{\EuU} \! \rmd^4 \Omega \, R_{ab} \delta \fg^{ab} 
            = \int_{\EuU} \! \rmd^4 \Omega \, \Fg_{ab} \delta g^{ab} 
            \equiv 0,
\end{equation}
whereas variation \wrt the affine connection $\Gamma^c_{\ ab}$ yields, by 
virtue of the contracted Palatini equation \eqref{eq:VPPalEquCon},
\begin{equation*}
   \delta S = \int_{\EuU} \! \rmd^4 \Omega \, \fg^{ab} \delta R_{ab}
            = \int_{\EuU} \! \rmd^4 \Omega \, \fg^{ab} 
                 \Bigl[  
                    \nabla_c \bigl( \delta \Gamma^c_{\ ab} \bigr) -
                    \nabla_b \bigl( \delta \Gamma^c_{\ ac} \bigr)
                 \Bigr].
\end{equation*}
Integrating by parts and discarding the divergence term by the usual argument
one obtains
\begin{equation*}
   \begin{split}
   \delta S &= \int_{\EuU} \! \rmd^4 \Omega 
                  \Bigl[  
                     \nabla_b \fg^{ab} \, \delta \Gamma^c_{\ ac} -
                     \nabla_c \fg^{ab} \, \delta \Gamma^c_{\ ab}
                  \Bigr] \\
            &= \int_{\EuU} \! \rmd^4 \Omega 
                  \Bigl( 
                     \delta^b_c \nabla_d \fg^{ad} - \nabla_c \fg^{ab} 
                  \Bigr) \delta \Gamma^c_{\ ab}
             \equiv 0.
   \end{split}
\end{equation*}
Since the variations $\delta \Gamma^c_{\ ab}$, symmetric in $a$ and $b$, and 
the region of integration $\EuU$ are arbitrary, the symmetric part of the 
expression in round brackets must vanish, \ie
\begin{equation} \label{eq:VPEPFieEquGam}
   \delta^{(b}_c \nabla_d \, \fg^{a)d} - \nabla_c \fg^{ab} = 0.
\end{equation}
This latter equation is equivalent to the metricity condition 
$\nabla_c \fg^{ab}=0=\nabla_c g^{ab}$; hence the connection coefficients 
$\Gamma^c_{\ ab}$ are necessarily the Christoffel symbols $\Christ{c}{a}{b}$.
Therefore one deduces that the field equations obtained from equation 
\eqref{eq:VPEPFieEquMet} coincide exactly with Einstein's vacuum field 
equations. This fact is the source of the commonly accepted belief that the
\EP variational principle is equivalent to the \EH variational principle. 
However, as we shall exemplify below, this is erroneous. As a matter of fact,
for the Lagrangian $L=R$ in vacuum,%
\footnote{Observe that the `gamma-gamma' Lagrangian density is here ruled out 
          as an alternative starting point for the derivation of Einstein's 
          field equations because the difference between 
          \eqref{eq:VPGamGamLag} and the \EH Lagrangian density is a pure 
          divergence only if the compatibility condition is assumed \emph{ab
          initio} \cite{elkho=SocPal}.} 
the equivalence of the field equations turns out to be a mere coincidence. 
Still, in that very case the two variational methods are not equivalent 
because the corresponding boundary conditions are different. As mentioned 
above (\cfr footnote\fnref{foot:VPBouTer}), in the purely metric situation the 
boundary term occurring in the process of variation does not vanish in general 
since the metric only is held fixed on the boundary. By contrast, in the 
metric-affine case the boundary term comes to naught since, in addition to the 
metric, the connection also is held fixed on the boundary. This means that one 
should have to add an \emph{ad hoc} surface term in order to recover the 
Hamiltonian description of the fields at spatial infinity in the case of 
asymptotically flat space-times (\cfr the discussion on boundary terms on 
page \pageref{subsub:CanGRADM}).%
\footnote{This undesirable feature is also present in the Ashtekar formalism,
          as a direct consequence of the use of a \HP Lagrangian; see, \eg 
          \cite{bombe=LagFor}.}
On the other hand, inasmuch as quantum gravity is concerned, the \EH and \HP
Lagrangians---which are only equivalent `on shell'---will most probably give 
different theories since in the path-integral approach for instance one sums 
over all `off-shell' contributions to the action. Furthermore, the choice of 
$\Fl_{\msc{hp}}$ as the starting Lagrangian density for the \EP variational 
principle is very peculiar: The most general \emph{second-order} (torsion-%
free) metric-affine Lagrangian density does in fact involve additional terms 
of the form $(\bfnab \bfg)\cdot(\bfnab \bfg)$ (such as, \eg 
$\nabla_a\sqrt{-g}\,\nabla_b g^{ab}$). In other words there are no selection 
rules that enable to pick out the \HP Lagrangian density, by contrast to what
happens in the metric case, where the requirement of having second-order field 
equations together with the covariance property uniquely determines the \EH
Lagrangian density (up to a cosmological constant). In an interesting paper 
Burton and Mann have recently found, for the aforementioned general second-%
order metric-affine Lagrangian density, the conditions under which the 
connection is in fact the Levi-Civita connection associated with the metric
\cite{burto=PalVarExt}. They have shown that the compatibility condition, 
which arises naturally in the \HP case, is in the more general situation a
constraint that is induced by the breaking of a symmetry of the connection
coefficients under the `deformation transformation' 
$\Gamma^c_{\ ab}\longrightarrow\Gamma^c_{\ ab}+C^c_{\ ab}$ (where $C^c_{\ ab}$
is an arbitrary tensor field symmetric in $a$ and $b$). On the other hand, in
the maximally symmetric case---\ie when no constraints are imposed on the 
tensor field $C^c{}_{ab}$---the connection remains completely undetermined but 
can always be chosen so as to recover Einstein's field equations.%
\footnote{Curiously, the special choice $\Gamma^c_{\ ab}\equiv0$ also turns 
          the $\bfg$-equations into Einstein's field equations.}
\nl
In the presence of matter fields there is an ambiguity in the prescription
of the minimal coupling rule because the compatibility condition between the 
metric and the connection does not hold. Moreover, variation of the total 
action
\begin{equation}
   S = \int_{\EuU} \! \rmd^4 \Omega 
          \Bigl[ 
             \Fr \bigl( \bfg, \bfGam \bigr) +
             \Fl_{\msc{m}} \bigl( \bfg, \psi, \bfnab \psi \bigr) 
          \Bigr],
\end{equation}
gives the following pair of equations: 
\begin{subequations} \label{eq:VPEPfieeq}
   \begin{align}
      &\Fg_{ab} = \Ft_{ab} := - 2 \varD{\Fl_{\msc{m}}}{g^{ab}}, \\
      &\delta^{(b}_c \nabla_d \, \fg^{a)d} - \nabla_c \fg^{ab} = 
          \varD{\Fl_{\msc{m}}}{\Gamma^c_{\ ab}},
   \end{align}
\end{subequations}
which are inconsistent in general owing to the fact that the purely geometric 
parts of these equations are projectively invariant whereas typical sources 
are not \cite{hehl=MetAffI}. The equations \eqref{eq:VPEPfieeq} would be 
equivalent to the full Einstein field equations obtained via the \EH variation 
only if the matter Lagrangian did not depend explicitly on the connection, \ie 
$\delta\Fl_{\msc{m}}/\delta\Gamma^c_{\ ab}\equiv0$. In most circumstances
this will be the case (\eg scalar, Yang--Mills, electromagnetic fields) but
one can pick out examples where this is not. For instance, when Einstein--%
Dirac fields are involved, the connection directly couples to the 
gravitational field thereby breaking the equivalence.%
\footnote{It can however be restored by `healing' the action with 
          \emph{ad hoc} compensating terms \cite{weyl=RemCou}.} 
\begin{rem}
In two dimensions in vacuum the \EP variation is unable to determine the
connection completely \cite{deser=IneFir}. This is can be seen as follows. For 
any dimension $n$, the equation \eqref{eq:VPEPFieEquGam} implies that the 
connection coefficients are given by the formula
\begin{equation} \label{eq:VPEPGamDes}
   \Gamma^c_{ab} = \Christ{c}{a}{b} +    
                   \frac12 \bigl( 
                              \delta_a^c Q_b + \delta_b^c Q_a - g_{ab} Q^c
                           \bigr),
\end{equation}
where we have defined 
$Q_a:=-\nabla_a(\ln\sqrt{-g})=-\partial_a(\ln\sqrt{-g})+\Gamma_a$ with 
$\Gamma_a:=\Gamma^b_{ab}$. The trace of equation \eqref{eq:VPEPGamDes} is
\begin{equation*}
   \Bigl( 1 - \frac{n}2 \Bigr) 
   \bigl[ \partial_a (\ln \sqrt{-g}) - \Gamma_a \bigr] = 0
\end{equation*}
and identically vanishes when $n=2$. Hence the $\Gamma_a$ part of the 
connection is undetermined in two dimensions.
\end{rem}
\nl
For completeness we reiterate that the Ashtekar formulation of canonical 
gravity is grounded on a generalised metric-affine variational principle, 
often referred to as \HP variation \cite{jacob=CovAct,pelda=ActGra}. To be 
more specific, the basic independent (complex) fields are expressed in terms 
of soldering forms and self-dual connections instead of metrics and affine 
connections respectively \cite{ashte=NewVar,ashte=NewHam}. (Those new canonical 
variables can be obtained by a succession of canonical transformations from 
the canonical variables of tetrad gravity \cite{henne=DerAsh}.) The Lagrangian 
for pure (complex) gravity is the so-called \emph{self-dual \HP Lagrangian} 
and the Ashtekar Hamiltonian is obtained upon an appropriate Legendre 
transformation.%
\footnote{This formulation uses explicitly the aforementioned remarkable 
          equivalence property of the \EH and \EP variations.}
In Ashtekar's formalism classical \gr stems from the real sector of the 
complex theory. As a nonperturbative approach to canonical quantum gravity the 
Ashtekar Hamiltonian formulation does not involve generalised---\eg higher-%
order---Lagrangians; therefore the only circumstances where the equivalence 
mentioned above breaks down arises when Dirac spinor fields are considered. In 
that case, because the spinors couple directly to the spin connection, the 
compatibility condition is altered by a term that gives torsion. Nevertheless,
this problem can be avoided if an \emph{ad hoc} term is added to the original 
Lagrangian \cite{tate=NewVar}.

\section{Variational principles in \hog}
\label{sec:VarConHOG}

\subsection{Metric variation}
\label{subsec:VPHOGMet}

In this subsection we write down the \EL derivatives stemming---via purely
metric variation---from higher-order Lagrangians that are quadratic in the 
curvature tensors and from nonlinear gravitational Lagrangians.%
\footnote{For a very detailed derivation of these results in the case of 
          quadratic Lagrangians, we refer to Simon's undergraduate thesis 
          \cite{simon=ModCos}.}
Our main purpose is to provide the subsequent investigation of Chapter
\ref{chap:BiaCos} with the explicit form of the field equations in vacuum and
to enable a direct comparison with the equations that will be obtained in the
next subsection via the metric-affine variational principle. In Subsection
\ref{subsec:VPCP} we will show how the derivation of the fourth-order field
equations can be overwhelmingly simplified by means the constrained first-%
order formalism.
\nl
In a four-dimensional Riemannian space-time $(\EuM,\bfg)$ the general 
gravitational action that contains, besides the \EH term 
$\Fl_{\msc{eh}}=\sqrt{-g}\,R$ (and possibly a cosmological constant) of \gr, 
quadratic invariant combinations of the Riemann curvature tensor, Ricci 
tensor, and scalar curvature is
\begin{equation} \label{eq:VPHOGMetQuaAct}
   S = \int_{\EuM} \! \rmd^4 \Omega \, 
          \Bigl[ 
             \Fl_{\msc{eh}} + \gamma_1 \Fl_1 + \gamma_2 \Fl_2 + \gamma_3 \Fl_3
          \Bigr],
\end{equation}
where $\gamma_i$ for $\range{i}{1}{3}$ are coupling constants and the 
quadratic Lagrangian densities are explicitly given by
\begin{subequations} \label{eq:VPHOGQuaInv}
   \begin{align}
      \Fl_1 &:= \sqrt{-g} \, R^2, \label{QuaInv1} \\
      \Fl_2 &:= \sqrt{-g} \, R^{ab} R_{ab}, \label{QuaInv2} \\
      \Fl_3 &:= \sqrt{-g} \, R^{abcd} R_{abcd}. \label{QuaInv3}
   \end{align}
\end{subequations}
(Note that a fourth possibility exists, namely 
$\Fl_4:=\epsilon^{abcd}R^{ef}_{\ \ ab} R_{efcd}$, where $\epsilon^{abcd}$ is 
the Levi-Civita tensor, but it is of no interest according to parity 
conservation.) 
\nl
The Lanczos action, which is constructed from the Gauss--Bonnet quadratic 
combination $R^{abcd} R_{abcd}-4R^{ab} R_{ab}+R^2$, becomes a topological 
invariant in four dimensions: the Euler--Poincar\'e characteristic of the
manifold (Gauss--Bonnet theorem). The corresponding \EL derivative is the
Bach--Lanczos identity \cite{bach=ZurWey,lancz=RemPro}: 
\begin{equation}
   \Fl_{\msc{bl}}^{ab} = \sqrt{-g} \, \Bigl[ 
                                         C^{acde} C^b_{\ cde} -
                                         \frac14 g^{ab} C_{cdef} C^{cdef}
                                      \Bigr].
\end{equation}
Owing to this remarkable property, only two of the quadratic invariants 
$\Fl_i$ are independent in four dimensions.
\nl
Another interesting quadratic invariant is that constructed from the Weyl
tensor, namely
\begin{equation} \label{eq:VPHOGConf}
   \Fl_{\rmc} = \sqrt{-g} \, C_{abcd} C^{abcd}.
\end{equation}
The associated action is the conformally invariant action in four dimensions, 
which is uniquely determined upon gauging conformal symmetry, \ie upon 
promoting the global conformal group to a local gauge symmetry 
\cite{kaku=GauThe,riege=ClaQua}.
\nl
The fourth-order \EL derivatives corresponding to \eqref{eq:VPHOGQuaInv} are 
respectively:%
\footnote{This can be checked with the \textsc{MathTensor} package for 
          \textsc{Mathematica}; see, \eg \cite{tsant=QuaCur} for the 
          conformally invariant case.}
\begin{subequations} \label{eq:VPHOGELDer}
   \begin{align}
      \Fl_1^{ab} &= 
         \sqrt{-g} \, \Bigl[ 
                         \frac12 g^{ab} R^2 - 2 R R^{ab} + 
                         2 \nabla^a \nabla^b R - 2 g^{ab} \square R
                      \Bigr], \label{ELQuaVar1} \\
      \Fl_2^{ab} &= 
         \sqrt{-g} \, \Bigl[ 
                         \frac12 g^{ab} R_{cd} R^{cd} - 2 R^{bcad} R_{cd} +
                         \nabla^a \nabla^b R - \square R^{ab} - 
                         \frac12 g^{ab} \square R 
                      \Bigr], \label{ELQuaVar2} \\
      \Fl_3^{ab} &=
         \sqrt{-g} \, \Bigl[ 
                         \frac12 g^{ab} R^{cdef} R_{cdef} -
                         2 R^{cdeb} R_{cde}^{\ \ \ \, a} -
                         4 \square R^{ab} + 
                         2 \nabla^a \nabla^b R \notag \\
                 & \qquad \qquad \qquad - 
                         4 R^{bcad} R_{cd} + 
                         4 R^{ca} R^b_{\ c}
                      \Bigr], \label{ELQuaVar3} 
   \end{align}
\end{subequations}
where the box $\square$ is the d'Alembertian second-order differential 
operator. The \EL derivative stemming from \eqref{eq:VPHOGConf} is the tensor 
density associated with the Bach tensor $B_{ab}$ \cite{bach=ZurWey}, which is 
defined by
\begin{equation} \label{eq:BachTens}
   \begin{split}
      B_{ab} &= 2 \nabla^c \nabla^d C_{cabd} + C_{cabd} R^{cd} \\
             &= - \square \Bigl( R_{ab} - \frac{R}6 g_{ab} \Bigr) 
                + \frac13 \nabla_a \nabla_b R
                + \bigl( C_{cabd} + R_{cabd} + R_{cb} g_{ad} \bigr) R^{cd}
   \end{split}
\end{equation}
and is symmetric, trace-free, and conformally invariant of weight $-1$.
\nl
Aside from the quadratic invariants above, we also consider a nonlinear
Lagrangian density that is a smooth arbitrary function of the scalar 
curvature, namely
\begin{equation} \label{eq:VPfrLag}
   \Fl = \sqrt{- g} \, \fR, \qquad  \text{with} \; \fpp \neq 0,  
\end{equation}     
where a prime denotes differentiation \wrt the scalar curvature. The 
corresponding \EL derivative is
\begin{equation} \label{VPfrELdens}
   \fpr R_{ab} - \frac12 f g_{ab} - \nabla_a \nabla_b \fpr + 
   g_{ab} \square \fpr = 0  
\end{equation}   
and was first obtained by Buchdahl \cite{buchd=NonLinCos}.

\subsection{Metric-affine variation}
\label{subsec:VPHOGMetAff}

As already mentioned in the introduction of this chapter, such alternative 
variational principles as the purely affine or the metric-affine methods of
variation were put forth in the early years of the relativistic gravitational 
theory. They were first analysed in the framework of quadratic gravitational 
Lagrangians by Weyl \cite{WEYL=SpaTim} and Eddington \cite{EDDIN=MatThe}.
Later, with the aim of obtaining second-order field equations different from 
Einstein's, Stephenson \cite{steph=QuaLag,steph=GenCov} and Higgs 
\cite{higgs=QuaLag} applied the \EP variational principle to the quadratic 
Lagrangians densities \eqref{eq:VPHOGQuaInv} and Yang investigated a theory 
based on the Lagrangian density \eqref{QuaInv3}, by analogy with the Yang--%
Mills Lagrangian \cite{yang=IntFor}. However, as Buchdahl pointed out 
\cite{buchd=NonLagPal}, there was an error in Stephenson's method of
variation---present in Yang's as well---as he imposed the metricity condition, 
\ie the Levi-Civita connection, \emph{after} the metric-affine variation.
Buchdahl subsequently gave the correct field equations and demonstrated by
means of specific examples that the \EP variational principle is not reliable 
in general \cite{buchd=QuaLagPal}.%
\footnote{In that respect it is strange that none of the most recent works 
          using the \EP method refers to Buchdahl's article.}
\nl
Recent investigations on those field equations derived form higher-order
gravitational Lagrangians via the \EP variational principle may be reviewed
briefly. The nonlinear $\fR$ case has been studied by several authors. In a 
series of papers Shahid-Saless analysed the theory based on a $R+\gamma_1 R^2$ 
Lagrangian with matter \cite{shahi=FirFor,shahi=CurCos,shahi=PalVar}. This 
was generalised to the $\fR$ case by Hamity and Barraco who also studied the 
conservation laws and weak field limit of the resulting equations
\cite{hamit=FirFor}. Ferraris, Francaviglia, and Volovich have recently shown 
that the \EP first-order formalism applied to general nonlinear $\fR$
Lagrangians leads to a series of \einspaces, the cosmological constants of 
which being determined by the specific form of the function $f$
\cite{ferra=UniVac}. Similar results have been obtained in the case of 
$f(R_{ab}R^{ab})$ Lagrangians by Borowiec \etal \cite{borow=UniEin,
borow=ISMC}. First-order variations of a generic set of higher-order curvature
terms appearing in string effective actions have been studied within the
context of the cosmological constant problem by Davis \cite{davis=SymVar}. 
Tapia and Ujevic somehow extend the investigation made by Ferraris \etal on
the universality property of Einstein spaces in the metric-affine setting 
\cite{tapia=UniFie}. As we shall see in Subsection \ref{subsec:VPConStrWeyl},
their claim that ``it is also possible to incorporate a Weyl vector field'' 
and ``therefore the presence of an anisotropy'' proves to be wrong in the 
light of the constrained metric-affine method in Weyl geometry.
\nl
In this subsection we first give the field equations that are derived from the
quadratic Lagrangian densities \eqref{eq:VPHOGQuaInv} via the \EP variational
principle \cite{buchd=QuaLagPal}. Next we analyse in greater detail the \EP 
variation of the nonlinear Lagrangian density \eqref{eq:VPfrLag}. We also 
indicate how Buchdahl's work can be extended to nonlinear Lagrangians of the 
type $f(R_{ab}P^{ab})$ \cite{mirit=VarStr}; in particular, our analysis sheds 
another light on the `universality property' most recently advocated by 
Ferraris \etal \cite{ferra=UniVac,borow=UniEin} and by Tapia and Ujevic 
\cite{tapia=UniFie}. This impels us to present the Lagrange-multiplier version 
of the \EP method, namely the constrained first-order formalism (in 
Subsection \ref{subsec:VPCP}), the consequences of which will be drawn in 
Section \ref{sec:VarConStr}.

\subsubsection{Quadratic case}

Applying the \EP method of variation to the quadratic Lagrangian densities 
\eqref{eq:VPHOGQuaInv} one obtains the following sets of equations:%
\footnote{Only the symmetric part of the Ricci tensor has been retained in
          $\Fl_2$; \cfr the remark on page \pageref{rem:VPRicci}.}
\begin{subequations} \label{eq:VPHOGEPDer}
   \begin{align}
      \delta \Fl_1 \longrightarrow 
         &R \Bigl( R_{(ab)} - \frac14 R g_{ab} \Bigr) = 0, \\
         &\nabla_c \bigl( R \fg^{ab} \bigr) = 0; \\
      \delta \Fl_2 \longrightarrow
         &R_a^{\ c} R_{bc} + R^c_{\ a} R_{cb} -
          \frac12 R_{cd} R^{cd} g_{ab} = 0, \label{eq:VPHOGEPRica} \\ 
         &\nabla_c \Fr^{ab} = 0; \label{eq:VPHOGEPRicb} \\
      \delta \Fl_3 \longrightarrow
         &2 R_{cda}^{\ \ \ \, e} R^{cd}_{\ \ be} - 
          R_{acde} R_b^{\ cde} + 
          R^c_{\ ade} R_{cb}^{\ \ de} -
          \frac12 R_{cdef} R^{cdef} g_{ab} = 0, \\ 
         &\nabla_d \Fr_c^{\ (ab)d} = 0.
   \end{align}
\end{subequations}
For each Lagrangian density the second set of equations has been obtained 
after partial integration, use of the Palatini equation \eqref{eq:VPPalEqu}, 
and elimination of a trace. These equations have been first given by 
Stephenson \cite{steph=QuaLag,steph=GenCov} and Higgs \cite{higgs=QuaLag}. 
Their important features are threefold:
\begin{enumerate} 
   \item \label{en:VPEPConInv}
         They are \emph{conformally invariant}, that is invariant under 
         conformal transformations $\widetilde{g}_{ab}=\Omega^2(\bfx)g_{ab}$, 
         where $\Omega^2(\bfx)$ is an arbitrary, strictly positive, scalar 
         field. This can be easily understood on account of the fact that the 
         affine connection is totally unrelated to the metric and because the 
         Lagrangian densities from which they are derived are quadratic in the 
         curvature. In a purely metric theory this property holds only for the 
         Lagrangian density \eqref{eq:VPHOGConf}. Furthermore, Higgs was the 
         first who showed that the solutions of equations 
         \eqref{eq:VPHOGEPDer} corresponding to $R$-squared and Ricci-squared 
         Lagrangians are conformal to \einspaces with arbitrary cosmological 
         constants \cite{higgs=QuaLag,buchd=QuaLagPal}. 
   \item They are second-order differential equations. This is due to the
         first-order nature of the \EP variation, which crumbles those double 
         covariant derivatives of the curvature tensors that typically occur
         in purely metric variations. In particular, they are obviously 
         \emph{not equivalent} to the fourth-order field equations 
         \eqref{eq:VPHOGELDer} obtained via purely metric variations. 
   \item They do not yield the Levi-Civita connection as is the case for the
         linear \HP Lagrangian and therefore their solutions are \emph{not 
         Riemann spaces}. However, in some specific cases it turns out that
         the underlined manifold is a `disguised' $W_4$, which can be brought
         onto an \einspace upon a suitable conformal transformation (\cfr the 
         discussion on the nonlinear case below). 
\end{enumerate} 
\begin{rem} \label{rem:VPRicci}
The Ricci tensor is not necessarily symmetric as in the Riemannian context; on 
writing $R_{(ab)}:=P_{ab}$ and $R_{[ab]}:=Q_{ab}$ we thereby see that to the 
Lagrangian density $\Fl_2$ in $V_4$ there corresponds a whole family of 
Lagrangian densities in $A_4$, namely 
$\Fl_{2,\alpha}=\sqrt{-g}\,R_{ab}(P^{ab}+\alpha Q^{ab})$, 
where $\alpha$ is an arbitrary constant. In other words there are no selection 
rules that enable to choose $\alpha=0$, \ie $\Fl_2$, in the \EP variational 
principle. On the other hand, there are no a priori reasons whatsoever to 
discard other acceptable Lagrangian densities compatible with the \EP 
prescriptions. For example, one could start with that quadratic combination 
involving only the antisymmetric part of the Ricci tensor, namely
\begin{equation*}
   \Fl_2^* := \sqrt{-g} \, R_{ab} Q^{ab}.
\end{equation*}
However, in that case the ensuing field equations impose only four conditions 
upon the forty $\Gamma$'s and leave the metric tensor totally undetermined
\cite{buchd=QuaLagPal}. This disconcerting situation also happens with
the Ricci-squared Lagrangian: Equations \eqref{eq:VPHOGEPRica} and 
\eqref{eq:VPHOGEPRicb} are fulfilled by any Ricci-flat, \ie $R_{ab}=0$, affine 
space $A_4$ whereas the metric tensor remains quite arbitrary. This sort of 
things never happens with purely metric variations, where the geometry is 
Riemannian from the outset. As Buchdahl, we claim that this degree of 
arbitrariness reflects discredit on the use of \EP variations in a theory of
gravity.
\end{rem}
 
\subsubsection{Nonlinear case}

Applying the \EP method of variation to the metric-affine analogue of the 
nonlinear Lagrangian density \eqref{eq:VPfrLag}, that is
\begin{equation*}
   \Fl = \sqrt{-g} \, f \Bigl( g^{ab} R_{ab}(\bfGam) \Bigr),
\end{equation*}
one obtains the following set of equations:
\begin{subequations} \label{eq:VPfrEPDer}
   \begin{align}
      &\fpr R_{(ab)} - \frac12 f g_{ab} = 0, \label{eq:VPfrEPDera} \\  
      &\nabla_c \bigl( \fpr \fg^{ab} \bigr) = 0. \label{eq:VPfrEPDerb}
   \end{align}
\end{subequations}
Obviously, they differ from the \EL derivative \eqref{VPfrELdens}. To see what 
these equations imply, we first expand equation 
\eqref{eq:VPfrEPDerb}, \viz
\begin{equation*}
   \fpr \nabla_c \fg^{ab} + \fpp \fg^{ab} \nabla_c R = 0.
\end{equation*}
Making use of the formul{\ae} 
$\nabla_a\sqrt{-g}=\partial_a\sqrt{-g}-\sqrt{-g}\,\Gamma_a$ and 
$\nabla_a R=\partial_a R$ and assuming that $\fpr\neq0$ we may write the 
latter equation as 
\begin{equation} \label{eq:VPEPfrint}
   \Bigl[ 
      \partial_c \bigl( \ln \sqrt{-g} \bigr) + 
      \bigl( \ln \fpr \bigr)^{\prime} \partial_c R - 
      \Gamma^d_{\ cd} 
   \Bigr] g^{ab} + 
   \partial_c g^{ab} + \Gamma^a_{\ cd} g^{db} + \Gamma^b_{\ cd} g^{ad} = 0. 
\end{equation}
Contracting with $g_{ab}$ we see that $\Gamma_c$ is a gradient: 
\begin{equation*}
   \Gamma_c = \partial_c \bigl( \ln \sqrt{-g} \bigr) + 
              2 \bigl( \ln \fpr \bigr)^{\prime} \partial_c R.
\end{equation*}
This implies that the Ricci tensor is symmetric since 
$R_{[ab]}=\partial_{[a}\Gamma_{b]}$. Replacing $\Gamma_c$ in equation 
\eqref{eq:VPEPfrint} by the value given above and after little manipulation 
one obtains
\begin{equation}
   \partial_c \bigl( \fpr g_{ab} \bigr) = 
      \fpr \bigl( \Gamma^d_{\ ca} g_{db} + \Gamma^d_{\ cb} g_{ad} \bigr).
\end{equation}
This suggests to define a conformally related metric with conformal factor
$\fpr$; indeed in that case the latter equation becomes
\begin{equation}
   \partial_c \widetilde{g}_{ab} = \Gamma^d_{\ ca} \widetilde{g}_{db} +
                                   \Gamma^d_{\ cb} \widetilde{g}_{ad},  
\end{equation}
thereby implying that the covariant derivative of the new metric 
$\widetilde{g}_{ab}$ with respect to the connection $\nabla_c$ vanishes: hence 
the connection $\Gamma^c_{\ ab}$ is the Levi-Civita connection of the metric 
$\widetilde{g}_{ab}$. On the other hand, equation \eqref{eq:VPEPfrint} is
equivalent to
\begin{equation}
   \nabla_c g^{ab} = \nabla_c \bigl( \ln \fpr \bigr) g^{ab},  
\end{equation}
which seems to show that the \EP variational principle has selected a $W_4$, 
\ie a Weyl space, with $Q_a=\nabla_a\bigl(\ln\fpr\bigr)$ as the Weyl one-form. 
However, this is \emph{not} a $W_4$ but a Riemann space with an `undetermined 
gauge' \cite[p.~134]{SCHOU=RicCal}: \label{fakeweyl} Under a conformal 
transformation $\widetilde{g}_{ab}=\Omega^2 g_{ab}$ in a $W_4$ the Weyl one-%
form transforms as $\widetilde{Q}_a=Q_a-2\nabla_a\bigl(\ln\Omega\bigr)$; if 
$Q_a$ is a \emph{gradient}, it is always possible to choose the conformal 
factor in order to have $\widetilde{Q}_a\equiv0$, which amounts to consider a 
Riemann space---this is the case here since we can `gauge away' the spurious 
Weyl one-form by choosing precisely $\Omega^2:=\fpr$ as the conformal factor. 
On the other hand the analysis of the field equation \eqref{eq:VPfrEPDera} is 
more straightforward. Taking its trace we find
\begin{equation} \label{eq:VPEPfrtr}
   \fpR R = 2 \fR. 
\end{equation}
This latter equation is identically satisfied (up to a constant rescaling 
factor) by the function $\fR=R^2$, as is expected from the fact that any
quadratic invariant metric-affine Lagrangian density is conformally invariant
(\cfr Point \ref{en:VPEPConInv} on page \pageref{en:VPEPConInv}). In that 
specific case the field equations \eqref{eq:VPfrEPDera} reduce to
\begin{equation} \label{eq:VPEPfrR2}
   R_{ab} - \frac14 R g_{ab} = 0,
\end{equation}
provided that $\fpR\neq0$, and the conformal gauge \dof is borne by the scalar 
curvature, which is assumed to be strictly positive. Hence, under an arbitrary 
conformal transformation $\widetilde{g}_{ab}=k^2Rg_{ab}$, with $k^2$ an 
arbitrary positive constant, the equation \eqref{eq:VPEPfrR2} becomes
\begin{equation} \label{eq:VPEPfrR2C}
   \widetilde{R}_{ab} - \frac14 \widetilde{R} \widetilde{g}_{ab} = 0,
\end{equation}
where the conformal scalar curvature is constant,
\viz $\widetilde{R}\equiv k^{-2}$. Indeed, the equation \eqref{eq:VPEPfrR2C}
implies that the underlined manifold is an \einspace with an arbitrary
cosmological constant.
\begin{rem}
It is important to realise that, once the undetermined gauge has been removed,
the conformal invariance is broken: The original `fake' Weyl space has been 
replaced by an \einspace---and \einspaces are of course not conformally 
invariant. In particular, the connection $\Gamma^c_{\ ab}$ being the Levi-%
Civita connection associated with the conformal metric $\widetilde{g}_{ab}$, 
it is totally inaccurate to perform an inverse `conformal transformation' in 
which the connection remains frozen, as was the case in the departing 
counterfeit Weyl space. Strictly speaking it does not make any sense to go 
back to the original---illusory---space. In that respect what is found in the 
literature is either wrong (see \eg Buchdahl's conclusion after his relation 
(3.7) \cite{buchd=QuaLagPal}) or misleading (see \eg Proposition 1 of 
Ferraris \etal \cite{ferra=UniVac}). 
\end{rem}
\nl
If on the other hand one excludes the purely quadratic case and assume that
the function $\fR$ be given prior to the variation, then the trace equation 
\eqref{eq:VPEPfrtr} can be regarded as an algebraic equation to be solved for 
$R$. Denoting the resulting roots $\rho_1,\rho_2,\dots$ one obtains a whole 
series of \einspaces, each characterised by a distinct constant scalar 
curvature. This situation was analysed by Ferraris \etal \cite{ferra=UniVac}
who extended Buchdahl's investigation \cite{buchd=QuaLagPal}. However, if it 
happens for some root $\rho_{i}$ to be such that $\fpr(\rho_i)=0$, then the 
trace equation \eqref{eq:VPEPfrtr} entails that $f(\rho_i)=0$ as well. In that 
instance the field equations \eqref{eq:VPfrEPDer} leave both the metric and 
connection completely undetermined.%
\footnote{This situation is best illustrated with the Lagrangian $L=R^n$ for
          $n>2$. Another interesting example is provided by the Lagrangian 
          $L=aR^2+bR+c$ with $a$, $b$, and $c$ constant: Applying the \EP
          variation one obtains qualitatively different conclusions according 
          to the specific values of the constants $a$, $b$, and $c$; in 
          particular, when $b^2=4ac$ there is only one condition, \ie $R=$ 
          constant, and nothing else is determined.}
\nl
As a special case of the nonlinear theory, one can also apply the \EP 
variation to scalar-tensor Lagrangian densities. Consider for instance the
Brans--Dicke action \cite{brans=MacPri}:
\begin{equation} \label{eq:VPBDAct}
   S = \int \! \rmd^4 \Omega \sqrt{-g} \, 
          \Bigl( 
             \phi^2 g^{ab} R_{ab} - 
             4 \omega \, g^{ab} \nabla_a \phi \nabla_b \phi 
          \Bigr).
\end{equation}
The appropriate conformal factor for which the connection $\Gamma^c_{\ ab}$ is 
the Levi-Civita connection associated with the conformal metric 
$\widetilde{g}_{ab}$ turns out to be the Brans--Dicke scalar field. Rather 
than transforming the field equations one can first transform the Brans--Dicke 
action and thereafter perform the \EP variation. The resulting Lagrangian 
density is
\begin{equation} \label{eq:VPEPBDPlanck}
   \Fl_{\msc{bd}} = \sqrt{-g} \, 
                       \bigl( 
                          \widetilde{R} - 
                          \frac{4\omega}{\phi^2} \, \widetilde{g}^{ab} 
                             \partial_a \phi \, \partial_b \phi 
                       \bigr).
\end{equation}
This corresponds to the `unit-transformed version' of Brans--Dicke's theory in 
which the gravitational constant is effectively constant and rest masses are 
varying (`Einstein frame') \cite{dicke=MacPri}.%
\footnote{The only difference lies in the coupling constant $\omega$, which is
          equal to the \emph{rescaled} Brans--Dicke coupling constant, \ie 
          $\omega\equiv\omega_{\msc{bd}}-\tfrac32$.}
As was first noticed by Van den Bergh, the peculiarity of the Brans--Dicke
action is responsible for a curious fact: Although the purely metric variation 
and the \EP method are not equivalent when they are applied to the action 
\eqref{eq:VPBDAct}, they yield the same dynamics if they are applied to the 
conformally transformed action constructed from \eqref{eq:VPEPBDPlanck}; 
therefore the Einstein frame is singled out as the unit system in which both 
variational principles are equivalent \cite{vande=PalVar}.%
\footnote{See also Burton and Mann's recent investigation 
          \cite{burto=PalVarDil}.}

\subsubsection{Generalisations of the nonlinear case}

The above analysis can be performed on more general Lagrangian densities. By a 
completely analogous procedure one finds Einstein spaces for the following
classes of Lagrangians:
\begin{itemize}
   \item $L=f(R_{ab}P^{ab})$, where $P_{ab}$ is the symmetric part of the 
         Ricci tensor $R_{ab}$ \cite{borow=UniEin};
   \item $L=f(\bfR)$, where $\bfR$ is any higher-order scalar constructed from
         the quantities $g^{ac}P_{cb}(\bfGam)$ \cite{tapia=UniFie};
   \item $L=f(R_{abcd}R^{abcd})$. 
\end{itemize}
\nl
Firstly, we briefly consider the former case.%
\footnote{See Miritzis's Ph.D. dissertation for a thorough investigation
          \cite{mirit=VarStr}.}
Let us define $r:=R_{ab}P^{ab}$. Variation of the corresponding action yields
the following $\bfg$- and $\bfGam$-equations:
\begin{subequations} \label{eq:VPfricEq}
   \begin{align}
      &\fpr P_{ac} P^c_{\ b} - \frac14 f g_{ab} = 0,  \label{eq:VPfricEqa} \\
      &\nabla_c \bigl( \fpr \Fp^{ab} \bigr) = 0. \label{eq:VPfricEqb}
   \end{align}
\end{subequations}
From equation \eqref{eq:VPfricEq}, after some manipulation, we conclude that 
$\Gamma_a$ is a gradient; hence the Ricci tensor is symmetric and $P_{ab}$ can 
be replaced by $R_{ab}$ everywhere. Defining the `reciprocal' tensor 
$\widehat{P}_{ab}$ (\cfr \cite{buchd=QuaLagPal}) by 
$\widehat{P}_{ac}P^{cb}=\delta_a^b$ and setting $p:=\det P_{ab}$ we observe 
that the new metric that is defined by
\begin{equation}  \label{eq:VPfriConfac1}
   \widetilde{g}_{ab} := \frac{\fpr p^{1/2}}{\sqrt{-g}} \widehat{P}_{ab} 
\end{equation}
or by
\begin{equation}  \label{eq:VPfriConfac2}
   \widetilde{g}^{ab} = \frac{\Fr^{ab}}{\fpr p^{1/2}}
\end{equation}
induces the connection $\Gamma^c_{\ ab}$ as its associated Levi-Civita 
connection. Expressing the field equations \eqref{eq:VPfricEqa} in the
conformal frame described by the metric $\widetilde{g}_{ab}$ we find anew an
\einspace, \viz
\begin{equation}
   \widetilde{R}_{ab} = 
      \frac14 \frac{f\sqrt{-g}}{(\fpr)^2 p^{1/2}} \widetilde{g}_{ab}.  
\end{equation}
In the conformally invariant, \ie quadratic ($\fpr\equiv1$) case it is easy to 
show that the conformal factor can once again be chosen to remove the 
undetermined gauge that characterises the illusory Weyl space in the original
frame. This conclusion also holds for the $n$-dimensional conformally 
invariant Lagrangians studied by Tapia and Ujevic \cite{tapia=UniFie}.
\nl
For the latter instance in the list above, varying the corresponding action 
\wrt the metric and connection one obtains 
\begin{subequations} \label{eq:VPEPRiemm}
   \begin{align}
      &- \frac 12 f g_{ab} - \fpr R_a^{\ cde} R_{bcde} 
       + \fpr R^c_{\ ade} R_{cb}^{\ \ de} 
       + 2 \fpr R^c_{\ dae} R_{c \ b}^{\ d \ e} = 0, \\
      &\nabla_d \Bigl( \fpr \Fr_a^{\ (bc)d} \Bigr) = 0, \label{eq:VPEPRiem}
   \end{align}
\end{subequations}
but in contrast to the previous cases there exists no natural way to derive a 
conformal metric $\widetilde{g}_{ab}$ from the field equation 
\eqref{eq:VPEPRiem} unless the Weyl tensor vanishes \cite{davis=SymVar}. 
\nl
For completeness we mention that the inclusion of matter leads in general to
inconsistencies. Moreover, in the nonlinear case the stress-energy tensor no
longer satisfies the conservation equation; still, it is possible to define a
new stress-energy tensor that is conserved but the physical interpretation of 
this generalised conservation law can be put in doubt since in the linearised 
theory for instance the ensuing equation of motion for test particles contains 
an additional term, which disagrees with Newton's law (see 
\cite{hamit=FirFor}). However, most recently, Borowiec and Francaviglia have 
shown that for those classes of nonlinear gravitational Lagrangians that give 
\einspaces it is also possible to derive the Komar expression for the energy-%
momentum complex \cite{borow=ISMC}.
\nl
In order to avoid the difficulties inherent to the \EP variational method, we
now consider its Lagrange-multiplier version. This modification of the 
metric-affine variational principle enables one to choose from the outset the 
type of geometry one wants to deal with. As regards the path-integral approach
to quantum gravity, for instance, this seems more consistent.

\subsection{Constrained first-order formalism}
\label{subsec:VPCP}

As we have indicated in the previous subsections, when one applies the \EP
variational principle one is often confronted with difficulties: Any departure 
from the \HP Lagrangian density 
$\Fl_{\msc{hp}}=\fg^{ab} R_{ab} (\bfGam, \partial \bfGam)$ brings forth 
cumbersome features. Nevertheless, the main shortcoming of the \EP method is 
not that the ensuing field equations are typically not equivalent to those 
that are derived from the `same'---in fact, they are different, since defined
in different functional spaces---Lagrangian densities via the purely metric 
\EH variational method but stems from the fact that it leads to certain 
indeterminacies, thereby vitiating its usefulness in a gravitational theory.
\nl
However that may be, our purpose is not to maintain the status quo; but, if 
one's aim is to examine generalisations of the orthodox variational principle,
one ought to be careful and to proceed gradually, conveyed by the guiding 
principle that one should recover known results as special, limiting, cases of 
the extended framework. We shall prove in this subsection that one consistent 
way of performing this programme is that which proceeds with the Lagrange-%
multiplier version of the \EP variational principle, referred to hereafter as 
the \emph{constrained first-order} method of variation \cite{cotsa=VarCon}.
\nl
The use of Lagrange multipliers in a variational principle is advocated 
whenever one wants to impose some constraint on the independent variables, for
a straight enforcement is in general not consistent with the variation. In the 
context of gravity this was clearly stated for the first time---as far as we
are aware---in Tonnelat's book on Einstein's attempts towards a unified theory 
of the `fundamental fields' \cite{TONNE=TheCha}. In the context of Riemannian
geometry this amounts to add to the Lagrangian density the compatibility
condition as a constraint with Lagrange multiplier; next, one varies the
resulting action \wrt the metric, connection, and Lagrange multiplier, 
considered as independent variables. When the constrained first-order method 
is applied to the \HP Lagrangian, the Lagrange multiplier identically vanishes 
as a consequence of the field equations \cite{ray=PalVar}. This result 
strengthens the interpretation of the equivalence of the \EH and \EP 
variations in vacuum \gr as a pure coincidence. Applied to higher-order
gravitational Lagrangians the constrained method proves to be technically
much simpler than the purely metric equivalent variational method; this was
demonstrated by Safko and Elston \cite{safko=LagMul} and also by Ray
\cite{ray=VarDer}. Nevertheless, the interesting applications of the 
constrained first-order method are those that are defined in the extended 
geometrical framework of \magt theories of gravity (see \cite{hehl=MetAff} and 
references therein). 
\nl
Consider a four-dimensional space-time manifold $\EuM$ endowed with a 
Lorentzian metric $g_{ab}$ and its associated Levi-Civita connection, denoted
hereafter $\Levi_c$ or, indifferently, $\Christ{c}{a}{b}$.%
\footnote{The symbol $\bfnab$ henceforth refers to the independent connection
          that is varied in the constrained metric-affine variational 
          principle.} 
Contradistinctively to the purely metric variation of Subsection
\ref{subsec:VPEHVar} we want to relax the compatibility condition and allow 
for non-metricity during the variation; however, since we aim at recovering a 
specific type of space---\eg Riemannian---after the variation, we must add to 
the original Lagrangian density $\Fl$ (now considered as a metric-affine 
functional of the metric, connection, and possibly matter fields $\psi$) the 
constraint
\begin{equation} \label{eq:VPCPConGen}
   \Fl_{\rmc} \bigl( \bfg, \bfGam, \bfLam \bigr) =
     \sqrt{-g} \, \Lambda_c^{\ ab} 
        \bigl( \Gamma^c_{\ ab} - \Christ{c}{a}{b} - C^c_{\ ab} \bigr),  
\end{equation}
where $\Lambda^c_{\ ab}$ are Lagrange multipliers and $C^c_{\ ab}$ is the
difference tensor between the arbitrary symmetric connection $\Gamma^c_{\ ab}$ 
and the Levi-Civita connection $\Christ{c}{a}{b}$. For instance, in Riemannian 
geometry, \ie $\Levi_c g_{ab}=0$, \eqref{eq:VPCPConGen} takes the form 
\begin{equation}
   L_{\rmc} \bigl( \bfg, \bfGam, \bfLam \bigr) = 
      \Lambda_c^{\ ab} \bigl( \Gamma^c_{\ ab} - \Christ{c}{a}{b} \bigr), 
\end{equation}
whereas in Weyl geometry, \ie $\nabla_c g_{ab}=-Q_c g_{ab}$, it is 
\begin{equation} \label{eq:VPCPWeylCon}
   L_{\rmc} \bigl( \bfg, \bfGam, \bfLam \bigr) =
      \Lambda_c^{\ ab} \Bigl[ 
                          \Gamma^c_{\ ab} - \Christ{c}{a}{b} -
                          \frac12 g^{cd} 
                             \bigl(
                                Q_b g_{ad} + Q_a g_{db} - Q_d g_{ab}
                             \bigr) 
                       \Bigr]. 
\end{equation}
The resulting metric-affine constrained action 
\begin{equation} \label{eq:VPCPConAct}
   S = \int_{\EuM} \! \rmd^4 \Omega \, 
          \Bigl[ 
             \Fl \bigl( \bfg, \bfGam, \psi \bigr) +
             \Fl_{\rmc} \bigl( \bfg, \bfGam, \bfLam \bigr) 
          \Bigr],  
\end{equation}
must be varied \wrt the independent fields $g^{ab}$, $\Gamma^c_{\ ab}$, 
$\Lambda_c^{\ ab}$, and $\psi$. Variation \wrt the metric yields the $\bfg$-%
equations
\begin{equation} \label{eq:VPCPgeq}
   \varD{\Fl}{g^{ab}} \biggr\rvert_{\bfGam} + \Fb_{ab} = 0,  
\end{equation}
where the tensor density $\Fb_{ab}$ is defined by
\begin{equation} \label{eq:VPCPbeq}
   \Fb_{ab} = \sqrt{-g} \, B_{ab}
           := - \frac12 \sqrt{-g} \, \nabla^c 
                   \bigl( 
                      \Lambda_{bac} + \Lambda_{acb} - \Lambda_{cab} 
                   \bigr).  
\end{equation}
Variation \wrt the connection gives the $\bfGam$-equations 
\begin{equation} \label{eq:VPCPgameq}
   \varD{L}{\Gamma^c_{\ ab}} \biggr\rvert_\bfg + \Lambda_c^{\ ab} = 0.  
\end{equation}
Variation \wrt the matter fields $\psi$ yields their respective equations of 
motion. Finally, variation \wrt the Lagrange multipliers restores the 
constraint \eqref{eq:VPCPConGen}. 
\nl
Consider the specific case of Riemannian geometry.%
\footnote{The generalisation to Weyl spaces will be considered in 
          Subsection \ref{subsec:VPConStrWeyl}.}
Solving explicitly the $\bfGam$-equations \eqref{eq:VPCPgameq} for the 
multipliers $\Lambda_c^{\ ab}$ and substituting back the resulting expression 
into the $\bfg$-equations we should obtain field equations equivalent to those 
derived by `Hilbert varying' the metric unconstrained action functional. This 
is expected by construction of the constrained variational method and can be 
checked explicitly on the previous higher-order Lagrangians densities. We 
first write down the resulting $\bfg$- and $\bfGam$-equations together with 
the actual values of the tensor $B^{ab}$ for the Lagrangian densities 
\eqref{eq:VPHOGQuaInv}:%
\footnote{These formul{\ae} correct Safko and Elston's, which contain many
          misprints \cite{safko=LagMul}.} 
\begin{subequations} \label{eq:VPCPQua}
   \begin{align}
      &\frac12 g^{ab} R^2 - 2 R R^{ab} + B^{ab} = 0, \notag \\
      &\Lambda_c^{\ ab} = \bigl( 
                             2 g^{ab} \delta_c^d - 
                               g^{ad} \delta_c^b -
                               g^{db} \delta_c^a
                          \bigr) \nabla_d R, \\
      &B^{ab} = - 2 g^{ab} \square R + 2 \nabla^b \nabla^a R, \notag \\
\intertext{for the Lagrangian density $\Fl_1$;} 
      &\frac12 g^{ab} R_{cd} R^{cd} - R^{ad} R_d^{\ b} - R^{bd} R_d^{\ a} + 
       B^{ab} = 0, \notag \\ 
      &\Lambda_c^{\ ab} = 2 \nabla_c R^{ab} - 
                          \delta_c^a \nabla_d R^{db} - 
                          \delta_c^b \nabla_d R^{ad}, \\ 
      &B^{ab} = - \square R^{ab} +
                  2 \nabla_c \nabla^b R^{ac} - 
                  g^{ab} \nabla_d \nabla_c R^{cd}, \notag \\
\intertext{for the Lagrangian density $\Fl_2$; and} 
      &\frac12 g^{ab} R_{cdef} R^{cdef} -
       2 R^{acde} R^b_{\ cde} + B^{ab} = 0, \notag \\ 
      &\Lambda_c^{\ ab} = 2 \nabla_d R_c^{\ abd} + 2 \nabla_d R_c^{\ bad}, \\
      &B^{ab} = 4 \nabla_d \nabla_c R^{acbd}, \notag  
   \end{align}
\end{subequations}
for the Lagrangian density $\Fl_3$. In the nonlinear case \eqref{eq:VPfrLag} 
we obtain likewise: 
\begin{subequations} \label{eq:VPCPfr}
   \begin{align}
      &\frac12 f g^{ab} - \fpr R^{(ab)} + B^{ab} = 0, \notag \\
      &\Lambda_c^{\ ab} = \frac12 \bigl( 
                                     2 g^{ab} \delta_c^d - 
                                       g^{ad} \delta_c^b -
                                       g^{db} \delta_c^a
                                  \bigr) \nabla_d \fpr, \\
      &B^{ab} = - g^{ab} \square \fpr + \nabla^b \nabla^a \fpr. \notag
   \end{align}
\end{subequations}
It is straightforward to obtain the correspondence with the \EL derivatives
\eqref{eq:VPHOGELDer} and \eqref{VPfrELdens} respectively by taking into 
account the Riemannian constraint and upon substituting in the $\bfg$-%
equations, within each set of \eqref{eq:VPCPQua} and \eqref{eq:VPCPfr}, the 
respective expressions of $B^{ab}$. The outcome is:
\begin{subequations} \label{eq:VPCPFineq}
   \begin{align}
      &\frac12 g^{ab} R^2 - 2 R R^{ab} + 
       2 \nabla^b \nabla^a R - 2 g^{ab} \square R = 0, \\
      &\frac12 g^{ab} R_{cd} R^{cd} - 2 R^{bcad} R_{cd} + 
       \nabla^b \nabla^a R - \square R^{ab} - \frac12 g^{ab} \square R = 0, \\ 
      &\frac12 g^{ab} R_{cdef} R^{cdef} - 2 R^{cdeb} R_{cde}^{\ \ \ \ a} +
       4 \nabla_d \nabla_c R^{acdb} = 0, \\  
      &\fpr R^{ab} - \frac12 f g^{ab} - 
       \nabla^a \nabla^b \fpr + g^{ab} \square \fpr = 0. 
   \end{align}
\end{subequations}
Strictly speaking, in comparison with the usual Hilbert variation, in all 
cases considered so far using the constrained first-order formalism, one
starts from a \emph{different} Lagrangian density, defined in a 
\emph{different} functional space, follows a \emph{different} variational 
method but nevertheless ends up in an \emph{equivalent} set of field
equations. Bearing this in mind, one can trace back this equivalence---which 
is not a mere formal coincidence---from the fact that all our Lagrangian 
densities are \emph{diffeomorphism covariant}. 
\nl
All previous cases can indeed be considered as specialisations of a very 
general Lagrangian $n$-form constructed locally as follows,
\begin{equation} \label{eq:VPCPdiflag}
   \bfL = \bfL \bigl( 
                  g_{ab}, \nabla_{a_1} g_{ab}, \dots,
                  \nabla_{(a_1} \dots \nabla_{a_k)} g_{ab},
                  \psi, \nabla_{a_1} \psi, \dots,
                  \nabla_{(a_1} \dots \nabla_{a_l)} \psi, \gamma
               \bigr).
\end{equation}
More specifically, $\bfL$ is a functional of the dynamical fields $g_{ab}$, 
$\psi$, finitely many of their covariant derivatives \wrt to $\nabla_c$, and 
also other `background fields' collectively referred to as $\gamma$. Referring 
to `$g$ and $\psi$' as `$\phi$', $L$ is called $f$-\emph{covariant}, 
$f\in\diff(\EuM)$, or simply diffeomorphism covariant if 
$\bfL(f^*(\phi))=f^*\bfL(\phi)$, where $f^*$ denotes the induced action of the 
diffeomorphism $f$ on the fields $\phi$. (Note that this definition excludes 
the action of $f^*$ on $\nabla$ or the background fields $\gamma$.) It 
immediately follows that our previous Lagrangians satisfy the above definition 
and, as a result, are diffeomorphism covariant. It is a very interesting 
result, first shown by Iyer and Wald \cite{iyer=SomPro} that if $\bfL$ as 
given in \eqref{eq:VPCPdiflag} is diffeomorphism covariant, then $\bfL$ can be 
reexpressed in the form 
\begin{equation}
   \begin{split}
      \bfL &= \bfL \bigl( 
                      g_{ab}, R_{bcde}, \Levi_{a_1} R_{bcde}, \dots,
                      \Levi_{(a_1} \dots \Levi_{a_m)} R_{bcde}, \\
           &\qquad \qquad \qquad \qquad \qquad
                      \psi, \Levi_{a_1} \psi, \dots, \Levi_{(a_1} \dots 
                      \Levi_{a_l)} \psi 
                   \bigr),
   \end{split}
\end{equation}
where $R_{abcd}$ is the Riemann curvature of $\Levi_c$ and $m=\max(k-2,l-2)$. 
Observe that everything is expressed in terms of the Levi-Civita connection of 
the metric tensor and also that all other fields $\gamma$ are absent. Applying 
Iyer and Wald's theorem to our Lagrangians we immediately see that we could 
have reexpressed them from the outset in a form that involves only the Levi-%
Civita connection and not the original arbitrary connection $\nabla$, and vary 
them to obtain the corresponding `Hilbert' equations. As we showed above, we 
arrived at this result by treating the associated Lagrangians as 
\emph{different} according to whether or not they involved an arbitrary 
symmetric or a Levi-Civita connection. 
\begin{rem}
This does not mean though that any metric-affine Lagrangian could be 
reexpressed as a purely metric Lagrangian. Here, we were allowed to apply Iyer 
and Wald's theorem because we started from a metric Lagrangian and \emph{by 
construction} replaced it with a \emph{constrained} metric-affine Lagrangian,
where the connection is not in fact dynamical.
\end{rem}

\section{Conformal structure of nonlinear gravitational Lagrangians}
\label{sec:VarConStr}

\subsection{Conformal equivalence properties}
\label{subsec:VPConEqu}

There is a huge literature involving conformal equivalence properties in the 
context of gravity theories.%
\footnote{For a very recent and thorough review we refer the reader to the 
          work of Faraoni, Gunzig, and Nardone \cite{farao=ConTra} and 
          references therein.}
In this subsection we are mainly interested in recalling the well-known
conformal equivalence property of nonlinear theories of gravity with 
Einstein's theory and additional scalar fields; in the next subsection we 
shall demonstrate how this classical result can be extended to Weyl geometry. 
We do not intend to discuss the `physicality' issue arising from the 
aforementioned conformal equivalence, that is, the problem of determining
which metric amongst the equivalence class of conformally related metrics is
the physical one. Since the early criticisms of Brans \cite{brans=NonLag}, 
this question has raised a lot of controversies and misinterpretations. In the 
context of stringy gravity it seems to be a crucial---albeit often skirted---%
and fairly intricate issue (see however Dick's constructive criticisms 
\cite{dick=IneJor}): One must indeed decide whether the rank-two tensor 
$g_{ab}$---which defines the `Jordan', or `string frame'---occurring in the 
effective actions of string theory is to be interpreted as the physical metric 
tensor or rather as a unifying object conformally related to the genuine 
metric tensor $\widetilde{g}_{ab}$ of the space-time through additional scalar 
fields. In that respect most theoretical arguments seem to favour the 
(conformal) `Einstein frame' as the relevant set of physical variables; this 
would also be true for a larger class of alternative theories of gravity 
including nonlinear theories of the $\fR$ type, generalised scalar-tensor 
theories, and Kaluza--Klein theories after compactification of the extra 
dimensions: All these theories are dynamically equivalent---in the sense that 
their respective solution spaces are isomorphic to each other---to \gr with 
scalar fields, which are still weird, hypothetical objects 
\cite{brans=GraTen}. 
\nl
The conformal equivalence of \emph{vacuum} nonlinear gravity theories with \gr
plus a scalar field (see, \eg \cite{barro=InfCon}) can be proved within the
extended framework of generalised scalar-tensor theories---which embody all
theories involving scalar fields besides the metric---, described by the 
following generic Lagrangian \cite{hwang=CosPer}:
\begin{equation} \label{eq:VPCSHwang1}
   L = \frac12 f(\phi,R) - 
       \frac12 \omega(\phi) \, g^{ab} \nabla_a \phi \nabla_b \phi - V(\phi).
\end{equation}
The easiest way to proceed is to perform the appropriate conformal 
transformation on the Lagrangian itself, not on the field equations. Indeed,
on defining the conformal factor as 
$\Omega^2 \equiv \fpr \equiv \exp(\sqrt{\tfrac23}\psi)$
it is straightforward to obtain the conformally transformed Lagrangian 
corresponding to \eqref{eq:VPCSHwang1}, namely
\begin{equation} \label{eq:VPCSHwang2}
   \widetilde{L} = \frac12 \widetilde{R} - 
                   \frac12 \frac{\omega}{\fpr} \, \widetilde{g}^{ab} 
                           \widetilde{\nabla}_a \phi 
                           \widetilde{\nabla}_b \phi - 
                   \frac12 \widetilde{g}^{ab} 
                           \widetilde{\nabla}_a \psi 
                           \widetilde{\nabla}_b \psi - 
                   \widetilde{V}(\phi,\psi),
\end{equation}
where the potential term is defined by
\begin{equation} \label{eq:VPCSHwang3}
   \widetilde{V}(\phi,\psi):=\frac1{2(\fpr)^2} \bigl( R \fpr - f + 2V \bigr).
\end{equation}
The original generalised Lagrangian \eqref{eq:VPCSHwang1} has thus been `cast'
into the \EH Lagrangian with an additional scalar field $\psi$ and an 
appropriate potential term $\widetilde{V}(\phi,\psi)$. In most cases,  
introduction of a new scalar field $\widetilde{\phi}$ reduces the expression
\eqref{eq:VPCSHwang2} to the final form
\begin{equation} \label{eq:VPCSHwang4}
   \widetilde{L} = \frac12 \widetilde{R} - 
                   \frac12 \widetilde{g}^{ab} 
                           \widetilde{\nabla}_a \widetilde{\phi} 
                           \widetilde{\nabla}_b \widetilde{\phi} - 
                   \widetilde{V}(\widetilde{\phi}),
\end{equation}
provided that this new scalar field satisfies
\begin{equation*}
   \rmd \widetilde{\phi} = 
      \sqrt{\frac{\omega}{\fpr} \rmd \phi^2 + \rmd \psi^2}.
\end{equation*}
\nl
As a special case of the generic situation above, the conformally transformed nonlinear 
$\fR$ Lagrangian yields the field equations
\begin{equation} \label{eq:VPCSfrconf}
   \widetilde{G}_{ab} = \nabla_a \phi \nabla_b \phi - 
                        \frac12 \widetilde{g}_{ab} 
                           \Bigl[
                              \bigl( \nabla_c \phi \nabla^c \phi \bigr) -
                              2 V(\phi)
                           \Bigr],
\end{equation}
with potential
\begin{equation} \label{eq:VPCSfrpot}
   V (\phi) = \frac12 \bigl( \fpr \bigr)^{-2} \bigl[ R \fpR - \fR \bigr], 
\end{equation}
where it should be understood that $R$ is superseded by $\phi$ through the 
inverse function to $\fpR$.
\nl
The most important consequence of the conformal equivalence property in the
context of higher-order gravity is that the original fourth-order field 
equations can be reduced to second-order equations, thereby simplifying the 
analysis of nonlinear gravity theories (see, \eg \cite{cotsa=CosMod,
mirit=VarStr} and references therein). On the other hand, the dynamical
equivalence of nonlinear theories and \gr plus additional scalar fields can be
demonstrated by performing a Legendre transformation on the original set of
variables (see, \eg \cite{sokol=GR14} and references therein). This procedure%
---which amounts to transform the starting nonlinear theory into a scalar-%
tensor theory---was first used by Teyssandier and Tourrenc to solve the Cauchy 
problem for those theories \cite{teyss=CauPro} and is completely analogous to 
the generalised Ostrogradsky prescriptions studied in Chapter 
\ref{chap:HamFor} (\cfr p.~\pageref{subsub:HamfrDynEquST}); it can be 
generalised to Lagrangians that are functions not only of $R$ but also 
$\square^k R$ for $k\in\mathbb{N}$ \cite{wands=ExtGra}. Yet, it should be
clear that the aforementioned Legendre transformation---sometimes referred to
as a \emph{Helmholtz formalism} \cite{magna=PhyEqu}---does not play the r\^ole 
of the conformal transformation: Nonlinear gravity is \emph{dynamically} 
equivalent to scalar-tensor gravity upon introducing additional scalar fields 
and the latter is then \emph{conformally} equivalent to \gr with one more 
scalar field \cite{wands=ExtGra}.%
\footnote{This slight difference is not merely conventional; it appears more
          clearly when one takes into account the boundary terms occurring in 
          the respective actions.}

\subsection{Generalised conformal structure in Weyl geometry}
\label{subsec:VPConStrWeyl}

For the more general higher-order Lagrangians of the form $f(q)$ where $q=R$, 
$R_{ab}R^{ab}$, or $R_{abcd}R^{abcd}$ and where $f$ is an arbitrary smooth
function, the field equations obtained via the metric-connection formalism are 
of second order (\cfr equations \eqref{eq:VPEPRiemm}) whereas the 
corresponding ones obtained via the usual metric variation are of fourth 
order. At first glance this result sounds very interesting since one could 
foresee that it would perhaps lead to an alternative way to `cast' the field 
equations of these theories in a more tractable, reduced form than the one 
that is usually used for this purpose, namely the conformal equivalence 
theorem (\cfr previous subsection). In this way, certain interpretational 
issues related to the question of the physicality of the two metrics 
associated with the conformal transformation would perhaps be avoided.
Unfortunately, as indicated in Section \ref{sec:VarConHOG}, other difficulties
arise when one uses the \EP method of variation, thereby vitiating its 
reliability as an alternative method to, for instance, reducing the complexity 
of the gravitational field equations. At the end of this subsection, we shall 
see that the origin of these troubles lies in the fact that the metric-affine 
variational principle is in fact a \emph{degenerate} case of the constrained 
first-order formalism, therefrom unable to cope with more general geometrical 
settings such as \emph{`true'} Weyl spaces. 
\nl
We are now interested in investigating the consequences of applying the 
constrained first-order formalism, defined in Subsection \ref{subsec:VPCP}, to
the case of Weyl geometry. We reiterate that a four-dimensional Weyl space 
$W_4$ is an affine space $L_4$ endowed with a linear symmetric and 
\emph{semi-metric} connection, that is
\begin{equation}
   \nabla_c g_{ab} = - Q_c \, g_{ab},
\end{equation}
where $Q_c$ is the Weyl one-form that characterises the geometry. 
The Weyl constraint is \eqref{eq:VPCPWeylCon}; for convenience we recall its 
explicit form:
\begin{equation} \label{eq:VPCSWeylCon}
   L_{\rmc} \bigl( \bfg, \bfGam, \bfLam \bigr) =
      \Lambda_c^{\ ab} \Bigl[ 
                          \Gamma^c_{\ ab} - \Christ{c}{a}{b} -
                          \frac12 g^{cd} 
                             \bigl(
                                Q_b g_{ad} + Q_a g_{db} - Q_d g_{ab}
                             \bigr) 
                       \Bigr]. 
\end{equation}
We apply the constrained first-order formalism to the nonlinear Lagrangian 
$L=\fR$ in vacuum, making use of the contracted Palatini equation 
\eqref{eq:VPPalEquCon} and the following useful formul{\ae}:
\begin{subequations} \label{eq:VPCSFor}
   \begin{align}
      \delta_{\bfg} \Christ{c}{a}{b} 
         &= \frac12 g^{cd} \bigl[ 
                              \nabla_b \bigl( \delta g_{ad} \bigr) + 
                              \nabla_a \bigl( \delta g_{db} \bigr) -
                              \nabla_d \bigl( \delta g_{ab} \bigr) 
                           \bigr] \notag \\
         &\qquad \qquad \qquad \qquad -  
            \frac12 \bigl( Q_b g_{ad} + Q_a g_{db} - Q_d g_{ab} \bigr) 
               \delta g^{cd}, \\
      \delta_{\bfg} C^c_{ab} 
         &= \frac12 \bigl( Q_b g_{ad} + Q_a g_{db} - Q_d g_{ab} \bigr) 
               \delta g^{cd} \notag \\
         &\qquad \qquad \qquad \qquad + 
            \frac12 \bigl( 
                       Q_b \delta g_{ad} + 
                       Q_a \delta g_{db} - 
                       Q_d \delta g_{ab}
                    \bigr) g^{cd}.  
   \end{align}
\end{subequations}
Variation of the nonlinear action
\begin{equation} \label{eq:VPCSAct}
   S = \int_{\EuM} \! \rmd^4 \Omega \, 
          \Bigl[ 
             \sqrt{-g} \, f \bigl( g^{ab} R_{ab}(\bfGam) \bigr) +
             \Fl_{\rmc} \bigl( \bfg, \bfGam, \bfLam \bigr) 
          \Bigr]  
\end{equation}
\wrt the Lagrange multipliers $\Lambda_c^{\ ab}$ is trivial and recovers the 
definition of the Weyl connection in terms of the Levi-Civita connection and 
the Weyl one-form, that is
\begin{equation}
   \Gamma^c_{\ ab} = \Christ{c}{a}{b} +
                     \frac12 g^{cd} \bigl(
                                       Q_b g_{ad} + Q_a g_{db} - Q_d g_{ab} 
                                    \bigr).  
\end{equation}
Taking into account that $\nabla_a\sqrt{-g}=-2Q_a\sqrt{-g}$, variation \wrt 
the metric $g^{ab}$ yields the $\bfg$-equations 
\begin{equation} \label{eq:VPCSgeq}
   \fpr R_{(ab)} - \frac12 f g_{ab} + B_{ab} = 0,  
\end{equation}
where $B_{ab}$ is defined by \eqref{eq:VPCPbeq}. Eventually, variation \wrt 
the connection brings forth the explicit form of the Lagrange multipliers, 
namely 
\begin{equation} \label{eq:VPCSLagMul}
   \Lambda_c^{\ ab} = \delta_c^{(b} \bigl( 
                                       Q^{a)} \fpr - \nabla^{a)} \fpr 
                                    \bigr) -
                      g^{ab} \bigl( Q_c \fpr - \nabla_c \fpr \bigr).
\end{equation}
Substituting back the latter result into equation \eqref{eq:VPCPbeq} we find 
the expression of the tensor $B_{ab}$, that is
\begin{equation} \label{eq:VPCSBTens}
   \begin{split}
      B_{ab} &= 2 Q_{(a} \nabla_{b)} \fpr -
                \nabla_{(a} \nabla_{b)} \fpr + 
                \fpr \nabla_{(a} Q_{b)} - 
                \fpr Q_a Q_b \\
             &\qquad \qquad \qquad - 
                g_{ab} \bigl( 
                          2 Q_c \nabla^c \fpr - Q^2 \fpr - 
                          \square \fpr + \fpr \nabla^c Q_c
                       \bigr).
   \end{split}
\end{equation}
Inserting this result into equation \eqref{eq:VPCSgeq} we obtain the full 
field equations for the nonlinear Lagrangian $L=\fR$ in the framework of Weyl 
geometry, \viz 
\begin{equation} \label{eq:VPCSWfe}
   \fpr R_{(ab)} - \frac12 f g_{ab} - \nabla_a \nabla_b \fpr + 
   g_{ab} \square \fpr = M_{ab},  
\end{equation}
where $M_{ab}$ is defined by 
\begin{equation} \label{eq:VPCSMtens}
   \begin{split}
      M_{ab} &:= - 2 Q_{(a} \nabla_{b)} \fpr 
                 - \fpr \nabla_{(a} Q_{b)} 
                 + \fpr Q_a Q_b \\
             &\qquad \qquad \qquad
                + g_{ab} \bigl( 
                            2 Q_c \nabla^c \fpr - Q^2 \fpr + \fpr \nabla^c Q_c
                         \bigr).  
   \end{split}
\end{equation}
Note that the degenerate case of vanishing Weyl one-form coincides with the 
familiar field equations \eqref{VPfrELdens} obtained in the Riemannian
framework. 
\nl
As discussed in the previous subsection, those equations are conformally 
equivalent to Einstein's equations with a self-interacting scalar field as the 
matter source. We henceforth aim at generalising this important property in 
Weyl geometry. To this end, we define the metric $\widetilde{g}_{ab}$
conformally related to the metric $g_{ab}$ with $\fpr$ as the conformal 
factor. Owing to the fact that the Weyl one-form transforms as 
$\widetilde{Q}_a=Q_a-\nabla_a(\ln\fpr)$ and taking into account the 
formul{\ae} 
$\widetilde{\nabla}_a\equiv\nabla_a$,  
$\widetilde{\square}\equiv
  \widetilde{g}^{ab}\widetilde{\nabla}_a\widetilde{\nabla}_b=
  (\fpr)^{-1}\square$, 
the field equations \eqref{eq:VPCSWfe} become in the conformal frame 
\begin{equation*}
   \fpr \widetilde{R}_{(ab)} - \frac12 \frac{f}{\fpr} \widetilde{g}_{ab} - 
   \widetilde{\nabla}_a \widetilde{\nabla}_b \fpr + 
   \widetilde{g}_{ab} \widetilde{\square} \fpr = \widetilde{M}_{ab}, 
\end{equation*}
where the tensor $\widetilde{M}_{ab}$ is given by 
\begin{equation}
   \begin{split}
      \widetilde{M}_{ab} 
         &= \fpr \widetilde{Q}_a \widetilde{Q}_b -
            \fpr \widetilde{\nabla}_{(a} \widetilde{Q}_{b)} -
            \widetilde{\nabla}_a \widetilde{\nabla}_b \fpr \\
         &\qquad \qquad \qquad +
            \widetilde{g}_{ab} \bigl( 
                                  \fpr \widetilde{\nabla}^c \widetilde{Q}_c - 
                                  \fpr \widetilde{Q}^2 +
                                  \widetilde{\square} \fpr
                               \bigr).
   \end{split}
\end{equation}
Introducing the scalar field $\varphi:=\ln\fpr$ and the potential $V(\varphi)$ 
in the `usual' form, \viz 
\begin{equation} \label{eq:VPCSWpot}
   V \bigl( \varphi \bigr) = \frac12 
                              \bigl( \fpr \bigr)^{-2}
                              \bigl[ R \fpR - \fR \bigr], 
\end{equation}
we can rewrite the field equations in the final form 
\begin{equation} \label{eq:VPCSWfieequ}
   \widetilde{G}_{ab} = \widetilde{M}_{ab}(Q) - 
                        \widetilde{g}_{ab} V \bigl( \varphi \bigr),  
\end{equation}
where we have set 
\begin{subequations}
   \begin{align}
      \widetilde{G}_{ab}    &= \widetilde{R}_{(ab)} - 
                               \frac12 \widetilde{R} \widetilde{g}_{ab} \\ 
\intertext{and} 
      \widetilde{M}_{ab}(Q) &= \widetilde{Q}_a \widetilde{Q}_b - 
                               \widetilde{\nabla}_{(a} \widetilde{Q}_{b)} + 
                               \widetilde{g}_{ab} 
                                  \bigl( 
                                     \widetilde{\nabla}^c \widetilde{Q}_c -
                                     \widetilde{Q}^2
                                  \bigr).
   \end{align}
\end{subequations}
Equivalently we could have obtained the field equations \eqref{eq:VPCSWfieequ}
from the corresponding conformally transformed action, via the constrained
first-order formalism.
\nl
The field equations \eqref{eq:VPCSWfieequ} are Einstein's equations for a
self-interacting scalar field matter source with a potential $V(\varphi)$ and
a source term $\widetilde{M}_{ab}(Q)$ depending on the Weyl one-form
$\widetilde{Q}_a$. If the geometry is Riemannian, \ie $\widetilde{Q}_a=0$, one
recovers the standard conformal result. This will be the case only if the
original Weyl one-form is a gradient, \ie $Q_a\equiv\nabla_a\Phi$, since in
that particular case it can be gauged away by the conformal transformation
$\widetilde{g}_{ab}=(\exp\Phi)g_{ab}$ (\cfr the discussion on page 
\pageref{fakeweyl}). This actually was the case of the \EP variational method 
applied to the Lagrangian $L=\fR$, where the Weyl one-form turned out to be 
$Q_a=\nabla_a(\ln\fpr)$. This fact shows unambiguously that 
\emph{unconstrained} metric-affine variations of the \EP type cannot deal with 
a Weyl geometry and correspond to a degenerate case of the constrained first-%
order method: The field equations obtained from the former can be recovered 
within the constrained setting simply by choosing a very special form of the 
Weyl one-form \cite{quere=GR15} that makes the Weyl space degenerate in a 
Riemann space. As a direct consequence of our investigation, Tapia and 
Ujevic's claim that, upon making use of the \EP variation, it is possible to 
incorporate a Weyl vector field---and therefrom account for the would-be 
observed anisotropy in the universe---is invalidated since this vector field 
describes an undetermined gauge in a Riemann space and can therefore always be 
eliminated \cite{quere=ComUni}. 
\nl
Let us summarise the results obtained in this subsection. Our analysis has 
revealed that a consistent way to investigate generalised theories of gravity, 
without imposing \emph{ab initio} that the geometry be Riemannian, is the 
constrained first-order formalism. Applications to quadratic and $\fR$ 
Lagrangians in the framework of Riemannian and Weyl geometry show that the
unconstrained \EP variational method is a degenerate case corresponding to a 
particular gauge and that the usual conformal structure can be recovered in 
the limit of vanishing Weyl one-form. (The generalisation of the result stated 
above to include arbitrary connections with torsion could be an interesting 
exercise.)
\nl
The physical interpretation of the source term in the field equations 
\eqref{eq:VPCSWfe} is closely related to the choice of the Weyl vector field 
$Q^a$. However, it cannot be interpreted as a genuine stress-energy tensor in 
general since, for instance, choosing $Q^a$ to be a unit timelike, 
hypersurface-orthogonal vector field, the sign of $M_{ab}Q^aQ^b$ depends on 
the respective signs of $\fpR$ and `expansion' $\nabla_a Q^a$.
\nl
The generalisation of the conformal equivalence theorem opens the way to 
analysing cosmology in the framework of these Weyl nonlinear theories by 
methods such as those used in the traditional Riemannian case (\cfr  
\cite{cotsa=MG8}). All the related problems could be tackled by leaving the 
conformal Weyl one-form $\widetilde{Q}_a$ undetermined, while setting it equal
to zero will eventually lead to detailed comparisons with the results already 
known in the Riemannian case.


\chapter{Hamiltonian formulation of \hotg}
\label{chap:HamFor}

\begin{minipage}{11cm}
\begin{quote}
\textit{``And it is right that you should learn all things, both the 
          persuasive, unshaken heart of Objective Truth, and the subjective
          beliefs of mortals, in which there is no true trust.''} 
\nl
\hfill
--- Parmenides of Elea, On Nature (Peri Physis). 
\end{quote}
\end{minipage}

\vspace{1cm}

\dropping{3}{W}\textsc{ithin} the realm of classical mechanics where it sprang 
up the Hamiltonian formalism chiefly served the purpose of a tool for tackling 
dynamical problems. So far Hamiltonian methods have demonstrated their ability 
to prove profound results---the renowned \textsc{kam} theory, for example---%
and to solve otherwise intractable problems in various fields of mathematical 
physics: celestial mechanics, ergodic theory, statistical mechanics, and so 
forth \cite{ARNOL=MetMat}. In addition the conceptual framework provided by 
the canonical formalism has become a convenient starting point for 
quantisation. This is the main reason why Hamiltonian methods have been 
applied to gravity as early as 1930 by Rosenfeld who constructed a quantum-%
mechanical Hamiltonian for the linearised theory of \gr---though he did not 
make any attempt to develop a canonical version of the full theory. This last 
purpose was carried out with the pioneering works of Dirac, Bergmann and many 
others, and ended up as a consistent canonical formalism of Einstein's \gr: 
the celebrated \adm formalism. (For an historical perspective, see 
\cite{dewit=QuaThe1}.) 
\nl
Besides the conceptual interest in the canonical version of \gr, more 
restricted investigations using Hamiltonian methods as technical tools to 
tackle specific problems deserve consideration: Hamiltonian cosmology 
\cite{uggla=HamCos}, relativistic celestial mechanics and black hole physics 
(see references in \cite{beig=ClaThe}), Hamilton--Jacobi theory in 
relativistic \cite{salop=HamJac} and string cosmology \cite{saygi=HamJac}. 
\nl
In this chapter we focus on the Hamiltonian formulation of theories with 
higher derivatives and its application to higher-order gravity theories. We do
not address technical issues related to the quantisation of these theories; 
rather we concentrate on their classical structure, hoping to show that the 
canonical formulation can provide a very powerful method for reducing the 
order of the equations of motion. Applications of the Hamiltonian formalism to 
spatially homogeneous cosmologies is left to Chapter~\ref{chap:BiaCos}. We 
firstly review to some extent in Section~\ref{sec:ConHam} how to deal with 
first-order constrained Hamiltonian systems. Then in Section~\ref{sec:TheHig} 
we give a detailed analysis on the treatment of higher-order field theories by 
means of a generalisation of the so-called \emph{Ostrogradsky construction}. 
Eventually we do attack the most important topic of this thesis in Section~%
\ref{sec:HigOrd}, namely the Hamiltonian formulation of \hotg. In particular, 
we exemplify the effectiveness of the Ostrogradsky method by building up a 
canonical version of gravity theories described by a Lagrangian that is an 
arbitrary function of the scalar curvature. This conveys us to prove the chief
result of this chapter, that is the equivalence of the Ostrogradsky 
Hamiltonian formulation of \gr and the well-known \adm canonical formalism
\cite{quere=Samos}.

\section{Constrained Hamiltonian systems}
\label{sec:ConHam}

\subsection{Introduction}
\label{subsec:ConHamIntro}

Physical theories of fundamental significance are invariant \wrt some group of 
\emph{local} symmetry transformations---gauge transformations for Yang--Mills 
theories; space-time diffeomorphisms for gravity. Such theories are 
generically called \emph{gauge theories} and can be thought of as theories in 
which the physical system under study is described by more variables than the 
number of physically independent \dofs. The physically prevailing variables 
are those that are \emph{gauge invariant} or, in other words, independent of 
the specific local symmetry transformation applied on the system. It is an 
essential characteristic of a gauge theory that the general solution to the 
equations of motion involves arbitrary functions of time: The local symmetry 
relates different solutions stemming from the same initial conditions. In the 
Lagrangian formalism this means that gauge theories are \emph{singular 
systems} (\cfr the remark on page~\pageref{note:DB}). 
\nl
It is unanimously acknowledged that the most exhaustive and reliable treatment 
of gauge systems is that which proceeds through the Hamiltonian formulation.
Nevertheless it is worthwhile to start from the action principle in Lagrangian
form and proceed to the Hamiltonian formulation. Then the very presence of 
arbitrary time-dependent functions in the general solution of the equations of 
motion implies that the canonical variables are not all independent: There 
are conditions on the allowed initial momenta and positions. These relations 
amongst the canonical variables are called \emph{constraints}. As a 
consequence, \emph{all gauge theories are systems with constraints}---the 
converse, however, is not true. This is the reason why most textbooks on the 
quantisation of gauge systems proceed firstly to analysing Hamiltonian 
constrained systems---Henneaux and Teitelboim's book is perhaps the best 
example in that respect \cite{HENNE=QuaGau}. (Historically, the classical 
Hamiltonian formalism has rather exclusively been considered as the 
fundamental setting on which canonical methods of quantisation are rooted.)
\nl
In contrast with the old approaches in theoretical physics, the aim of which 
was to reduce the number of variables entering in the play, the modern way of 
dealing with fundamental systems consists in introducing more powerful%
---gauge---symmetries while increasing the number of variables to make the 
description more transparent. This philosophy has culminated with the 
inception of the elegant and powerful \textsc{brst} formalism; see, \eg 
\cite{HENNE=QuaGau}. 
\nl
There exists a striking difference between gauge theories with \emph{internal 
symmetries} and those which are \emph{generally covariant}, that is, with 
reparameterisation invariance: In the former local symmetry transformations 
are generated by first-class constraints; in the latter the Hamiltonian itself 
is a constraint---the \emph{super-Hamiltonian}. (In most circumstances, 
generally covariant systems do have a zero Hamiltonian; there exist, however, 
counterexamples to this property \cite{HENNE=QuaGau}.) This raises the issue 
of interpreting time in generally covariant theories, for one is led to the 
question whether the Hamiltonian generates the dynamical time evolution or the 
kinematical local symmetries as the other first-class constraints usually do. 
In quantum gravity this issue is particularly puzzling since it is 
intrinsically linked to the various ways of foliating space-time in a one-%
parameter family of spacelike hypersurfaces; see, \eg \cite{ASHTE=ConPro,
isham=CanQua}, and references therein. 
\nl
Circa 1950, the classical treatment of constrained systems was carried out by
Dirac \cite{dirac=GenHam1,dirac=GenHam2,DIRAC=LecQua}, Bergmann and 
collaborators \cite{ander=ConCov,bergm=DirBra}; almost instantaneously, Pirani 
and Schild applied Dirac's methods to the gravitational field 
\cite{piran=QuaEin}. Dirac himself was mainly concerned in the Hamiltonian 
formulation of \gr \cite{dirac=TheGra} and his leading efforts were completed 
in the sixties with the work of Arnowitt, Deser and Misner: the famous \adm 
formalism. Specifically they showed how to use the canonical setting to 
provide a rigorous characterisation of gravitational radiation and energy 
\cite{arnow=CanVar,arnow=DynGen}. (As compared to the gravitational case, the 
application of Dirac's methods to gauge theories---Maxwell electrodynamics and 
Yang--Mills theory---has not been so hastily initiated; see, \eg 
\cite{SUNDE=ConDyn}, and references therein.)
\nl
The geometrisation of the Dirac--Bergmann algorithm was achieved in the late 
seventies by Gotay, Nester, and Hinds \cite{gotay=PreMan,gotay=PreLag1,
gotay=PreLag2}. In the eighties, Batlle \etal obtained general proofs showing 
that the classical Hamiltonian and Lagrangian treatments of gauge theories are 
equivalent \cite{batll=EquLag}. The Dirac--Bergmann theory opened the way up 
to the quantisation of constrained Hamiltonian systems even though canonical 
methods were hard to apply to theories of physical interest. Two distinct 
techniques emerged: the \emph{Dirac method of quantisation}, in which the 
constraints are implemented as operators in Hilbert space, and the 
\emph{reduced quantisation}, where the superfluous phase space \dofs are 
eliminated before quantising, as is the case in the \adm approach. Their 
equivalence is still a matter of controversy; see references in Garc{\'\i}a 
and Pons's article \cite{garci=EquFad}. The development of path-integral 
quantisation methods, incorporating the constraints in the definition of 
Feynman path integrals, and the extension of the formalism to include 
fermionic fields---upon the introduction of Grassmann variables---brought 
forth an appreciable advance which reached its apex with the advent of the 
powerful Hamiltonian \textsc{brst} and Lagrangian \emph{antifield} formalisms. 
\nl
Besides the classical lectures of Dirac \cite{DIRAC=LecQua} there are several 
excellent reviews on the treatment of constrained systems. Some focus more on 
systems with a finite number of \dofs \cite{SUDAR=ClaDyn}, others on field 
theories \cite{HANSO=ConHam}, and some on both \cite{SUNDE=ConDyn,%
GOVAE=HamQua,HENNE=QuaGau,wipf=HamFor,BURNE=IntThe}. For generally covariant 
theories there exists a good monograph \cite{GITMA=QuaFie}. We draw on all 
these valuable sources to briefly summarise the theory of constrained systems, 
with the main purpose of defining the basic concepts and setting the notations 
that will be used throughout this chapter.

\subsection{A short summary of Dirac--Bergmann theory}
\label{subsec:DirBer}

\subsubsection{Action principle in Lagrangian form---Singular Lagrangians}

Consider a conservative holonomic dynamical system the configuration of which
at any instant of time is specified by $K$ independent generalised coordinates 
$q_k$ for $\range{k}{1}{K}$; its time evolution can be derived from an action 
functional,
\begin{equation} \label{DBAction}
   S \bigl[ \gamma \bigr] =
      \int_{t_0}^{t_1} L \bigl( q_k (t), \dot{q}_k (t) \bigr) dt,
\end{equation}
constructed from an appropriate Lagrange function $L(q,\dot{q})$ that depends 
explicitly on the generalised coordinates and their velocities;%
\footnote{At this stage we restrict the discussion to first-order Lagrangians 
          with no explicit time dependence. Systems with higher derivatives 
          are dealt with in Section~\ref{sec:TheHig}. Moreover we assume that 
          there is a finite number of discrete \dofs in order to render this 
          summary as simple as possible.}
it is understood that to any path $\gamma \equiv q_k (t)$ in configuration 
space that takes the initial and final values $q_k (t_0)$ and $q_k (t_1)$ one
associates the value $S[\gamma]$ of the action, which is given by the integral
\eqref{DBAction}. The classical motions of the system are those that make the 
action \eqref{DBAction} stationary under variations $\delta q_k (t)$ of the 
generalised coordinates $q_k$ that vanish at the endpoints $t_0$, $t_1$. The
necessary and sufficient conditions for the action \eqref{DBAction} to be
stationary are the \EL equations 
\begin{equation} \label{DBEL1}
   \frac{\rmd}{\rmd t} \biggl( \parD{L}{\dot{q}_k} \biggr) - 
   \parD{L}{q_k} = 0  \quad \text{for} \; \range{k}{1}{K}.
\end{equation}
We can write equations \eqref{DBEL1} in more detail as
\begin{equation} \label{DBEL2}
   V_k \bigl( q, \dot{q} \bigr) - 
   \sum_{l=1}^K W_{kl} \bigl( q, \dot{q} \bigr) \, \ddot{q}_l = 0
   \quad \text{for} \; \range{k}{1}{K},
\end{equation}
where we have introduced the quantities
\begin{subequations} \label{DBvw}
   \begin{align}
      V_k \bigl( q, \dot{q} \bigr) 
         &:= \parD{L}{q_k} - 
             \sum_{l=1}^K \parD{^2 L}{q_l \partial \dot{q}_k}, \\
      W_{kl} \bigl( q, \dot{q} \bigr)
         &:= \parD{^2 L}{\dot{q}_l \partial \dot{q}_k}.
   \end{align} 
\end{subequations}
We immediately see from expressions \eqref{DBvw} that all equations 
\eqref{DBEL2} are second-order, linearly independent equations, provided the 
Hessian matrix can be inverted, \ie $\det W \neq 0$. Then the accelerations 
$\ddot{q}_k$ at a given time are uniquely determined by the positions and the 
velocities at that time and the Lagrangian is said to be \emph{regular}. The 
general solution to the equations of motion can thus be expressed in terms of 
$2K$ independent constants of integration which are fixed by the initial 
conditions. If, on the other hand, the determinant of the Hessian matrix is 
zero, the accelerations---and thus the dynamics---will not be uniquely 
determined by the positions and the velocities. The general solution to the 
equations of motion will then possibly involve arbitrary functions of time. In 
contradistinction with the former case the Lagrangian is said to be 
\emph{singular}. 
\nl
Henceforth we assume that the Lagrangian $L$ is singular and that the rank of 
its associated Hessian matrix is constant everywhere on the velocity phase 
space and equal to $K-R$ for $R \in \mathbb{N}$, \viz
\begin{equation} \label{DBRank}
   \rank \Biggl( 
            \frac{\partial^2 L}
                 {\partial \dot{q}_k \partial \dot{q}_l} 
         \Biggr) =
   K-R. 
\end{equation}
Therefore, the matrix $W$ has $R$ null-eigenvectors $X^{(i)}$:  
\begin{equation}
   \sum_{l=1}^K X_l^{(i)} \bigl( q, \dot{q} \bigr) \,
                W_{kl} \bigl( q, \dot{q} \bigr) = 0 
   \quad \text{for} \; \range{i}{1}{R}.
\end{equation}
Contracting the \EL equations \eqref{DBEL2} with those eigenvectors we obtain 
necessary and sufficient conditions so that equations \eqref{DBEL2}, 
interpreted as algebraic equations for the unknown $\ddot{q}_k$, have a 
solution; these are
\begin{equation} \label{DBLagCon}
   \phi_i \bigl( q, \dot{q} \bigr) :=
   \sum_{k=1}^K X_k^{(i)} \bigl( q, \dot{q} \bigr) \,
                V_k \bigl( q, \dot{q} \bigr) = 0 
   \quad \text{for} \; \range{i}{1}{R}.
\end{equation}
The independent conditions amongst equations \eqref{DBLagCon} are called 
\emph{Lagrangian constraints}. 
\begin{rem} \label{note:DB}
Some null-eigenvectors can be determined from the generalised Bianchi 
identities, which are obtained through Noether's second theorem; as a direct
consequence, gauge theories are necessarily singular (see, \eg 
\cite{wipf=HamFor}).
\end{rem}

\subsubsection{Hamiltonian formalism}

\paragraph*{Primary constraints.}

The starting point for the Hamiltonian formalism is to define the canonical 
momenta by
\begin{equation} \label{DBCanMom1}
   p_k := \parD{L}{\dot{q}_k} \quad \text{for} \; \range{k}{1}{K}.
\end{equation}
We see that the vanishing of the determinant of the Hessian matrix $W$ is 
precisely the condition that precludes the expression of the velocities as 
functions of the coordinates and momenta. In other words, the momenta 
\eqref{DBCanMom1} are not all independent in this case, and there exist 
amongst equations \eqref{DBCanMom1} some relations
\begin{equation} \label{DBPriCon1}
   \phi_i \bigl( q, p \bigr) = 0 \quad \text{for} \; \range{i}{1}{R},
\end{equation}
which are assumed to be independent. The conditions \eqref{DBPriCon1} are 
called \emph{primary constraints} to emphasise that the equations of motion
were not used to obtain them. Through equations \eqref{DBPriCon1} these 
primary constraints define a $(2K-R)$--dimensional submanifold---the 
\emph{primary constraint surface}---, which we suppose to be smoothly embedded 
in phase space.

\paragraph*{Generalised Legendre transformation---Canonical and Dirac
            Hamiltonians.}

The canonical Hamiltonian is introduced by
\begin{equation} \label{DBCanHam1}
   H_{\rmc} := \sum_{k=1}^K \dot{q}_k \, p_k - L.
\end{equation}
Even though $H_{\rmc}$ as defined by equation \eqref{DBCanHam1} is a function
of the positions and the velocities, its dependence on $(q,\dot{q})$ is quite
specific. Indeed, it is a remarkable property of the Legendre transformation
that the velocities enter $H_{\rmc}$ only through the momenta given by 
definitions \eqref{DBCanMom1}. This follows simply from the evaluation of the
change $\delta H_{\rmc}$ induced by arbitrary independent variations of the
positions and velocities. This means that $H_{\rmc}$ is a function of the
$p$'s and $q$'s. However, it is not uniquely determined as a function of the 
phase space variables since the variations $\delta p_k$ of the momenta are
restricted to preserve the primary constraints \eqref{DBPriCon1}. In other 
words, the canonical Hamiltonian is well defined only on the submanifold
characterised by the primary constraints, and it can be extended arbitrarily
off that manifold. The resulting function, which is not unique, is given by
\begin{equation} \label{DBPriHam1}
   H_{\EuD} := H_{\rmc} + \sum_{i=1}^R \omega^i \bigl( q, p \bigr) \, \phi_i,
\end{equation}
and is called a \emph{Dirac Hamiltonian}. The canonical formalism remains 
unchanged under the replacement $H_{\rmc} \rightarrow H_{\EuD}$. 
\nl
The passage from $q$, $\dot{q}$, $L(q,\dot{q})$ to $q$, $p$, $H_{\EuD}(q,p)$ 
is called a \emph{generalised Legendre transformation}. It enables to cast the 
action principle into Hamiltonian form: The Hamiltonian equations of motion
\begin{align*} 
   \dot{q}_k &= \parD{H_{\rmc}}{p_k} + 
                \sum_{i=1}^R \omega^i \parD{\phi_i}{p_k}, \\
   \dot{p}_k &= - \parD{H_{\rmc}}{q_k} - 
                \sum_{i=1}^R \omega^i \parD{\phi_i}{q_k}, 
\end{align*}
which are equivalent to the original \EL equations \eqref{DBEL1}, can be 
derived from the variational principle
\begin{equation*}
   \delta \, \int_{t_0}^{t_1} 
      \biggl( \sum_{k=1}^K \dot{q}_k \, p_k -
             H_{\rmc} - 
             \sum_{i=1}^R \omega^i \phi_i
      \biggr) \rmd t = 0,
\end{equation*}
for arbitrary variations $\delta q_k$, $\delta p_k$, $\delta \omega^i$ subject
only to the restriction that $\delta q_k$ vanish at the endpoints. The 
variables $\omega^i$, which render the Legendre transformation invertible, 
play here the r\^ole of Lagrange multipliers that are enforcing the primary
constraints \eqref{DBPriCon1}. 

\paragraph*{Poisson bracket---Weak and strong equations.}

Introducing the \emph{Poisson bracket} of two arbitrary functions $f$, $g$, 
defined in phase space, by
\begin{equation}
   \bigl\{ f, g \bigr\} := \sum_{k=1}^K \parD{f}{q_k} \parD{g}{p_k} -
                           \sum_{k=1}^K \parD{f}{p_k} \parD{g}{q_k},
\end{equation}
we can formally write any equation of motion in the canonical formalism as
\begin{equation} \label{DBEqMot}
   \dot{f} = \bigl\{ f, H_{\EuD} \bigr\} = 
             \bigl\{ f, H_{\rmc} \bigr\} + 
             \sum_{i=1}^R \omega^i \bigl\{ f, \phi_i \bigr\}.
\end{equation}
At this stage it is useful to distinguish between \emph{weak} and 
\emph{strong} equations. The primary constraints \eqref{DBPriCon1} do not 
vanish identically throughout phase space and, in particular, they have 
nonzero Poisson brackets with the canonical variables. To take account of this 
property we write the primary constraints \eqref{DBPriCon1} as
\begin{equation} \label{DBWPriCon}
   \phi_i \bigl( q, p \bigr) \approx 0 \quad \text{for} \; \range{i}{1}{R},
\end{equation}
where we have introduced the weak equality symbol ``$\approx$''. More
generally, two functions $f$, $g$ that coincide on the primary constraint
submanifold are said to be \emph{weakly} equal, \viz $f\approx g$. On the 
other hand, an equation that holds throughout phase space is called 
\emph{strong}, and the usual equality symbol ``$=$'' is used in that case.
Accordingly, we may write equations \eqref{DBPriHam1} and \eqref{DBEqMot} 
respectively as
\begin{subequations}
   \begin{align} 
      &H_{\EuD} = H_{\rmc} + 
                  \sum_{i=1}^R \omega^i \bigl( q, p \bigr) \, \phi_i
                \approx H_{\rmc}, \label{DBPriHam2} \\
   \intertext{and}
      &\dot{f} \approx \bigl\{ f, H_{\EuD} \bigr\} 
               \approx \bigl\{ f, H_{\rmc} \bigr\} + 
                       \sum_{i=1}^R \omega^i \bigl\{ f, \phi_i \bigr\}.
      \label{DBEqMot2}
   \end{align}
\end{subequations}

\paragraph*{Dirac--Bergmann algorithm---Total Hamiltonian.}

A basic consistency requirement is that the primary constraints be preserved
when time evolution is considered. This gives rise to the \emph{consistency 
conditions} \cite{dirac=GenHam1,ander=ConCov}
\begin{equation} \label{DBConsCond}
   \dot{\phi}_i \approx \bigl\{ \phi_i, H_{\rmc} \bigr\} + 
      \sum_{j=1}^R \omega^j \bigl\{ \phi_i, \phi_j \bigr\} \approx 0
   \quad \text{for} \; \range{i}{1}{R},
\end{equation}
which, for each value of the index $i$, correspond to one of the three 
possibilities: 
\begin{enumerate}
   \item Equation \eqref{DBConsCond} is trivially satisfied, \ie 
         $0\stackrel{!}{\approx}0$.
   \item Equation \eqref{DBConsCond} is actually an equation to be satisfied 
         by the Lagrange multipliers $\omega$.
   \item Equation \eqref{DBConsCond} reduces to a relation independent of
         $\omega$, that is, a new constraint on the $p$'s and the $q$'s which 
         is said to be \emph{secondary}---in contrast with ``primary''---to 
         emphasise that the equations of motion were used to obtain it. 
\end{enumerate}
For any secondary constraint we must again impose a consistency condition 
similar to \eqref{DBConsCond} and perform the above analysis. This process 
must be repeated until all consistency equations have been exhausted. The 
outcome of this algorithm then consists in:
\begin{enumerate}
   \item a complete set of $R$ primary and $S$ secondary constraints,
         \begin{equation}
            \phi_m \approx 0 \quad \text{for} \; \range{m}{1}{R+S=M};
         \end{equation}
   \item restrictions on the Lagrange multipliers $\omega$,
         \begin{equation} \label{DBResCond}
            \bigl\{ \phi_m, H_{\rmc} \bigr\} + 
            \sum_{i=1}^R \omega^i \bigl\{ \phi_m, \phi_i \bigr\} \approx 0
            \quad \text{for} \; \range{m}{1}{M}.
         \end{equation}
\end{enumerate}
Conditions \eqref{DBResCond} may be thought of as a set of $M$ nonhomogeneous
linear equations in the $R$ unknowns $\omega^i$. (Note that since 
$R\leqslant M$ some compatibility conditions---\ie further secondary 
constraints---could arise to ensure consistency; they must be treated through 
the algorithm as well.) The general solution of equations \eqref{DBResCond} is 
of the form
\begin{equation} \label{DBLagMul}
   \omega^i \approx \mu^i + \sum_{\alpha=1}^A \lambda^{\alpha} \nu_{\alpha}^i,
\end{equation}
where $\mu^i$ is a special solution of equation \eqref{DBResCond} and 
$\nu_{\alpha}^i$ is a complete set of $A$ linearly independent solutions of 
the associated homogeneous system
\begin{equation}
   \sum_{i=1}^R \nu_{\alpha}^i \bigl\{ \phi_m, \phi_i \bigr\} \approx 0.
\end{equation}
Since the coefficients $\lambda^{\alpha}$ are totally arbitrary, equation
\eqref{DBLagMul} means that the multipliers $\omega$ have been resolved into
one arbitrary part and one part that is fixed by the consistency conditions
\eqref{DBResCond}; for more details, see \cite{HENNE=QuaGau}. 
\nl
As a direct consequence of this analysis we may write the equation of motion 
\eqref{DBEqMot2} equivalently as
\begin{equation}  \label{DBEqMot3}
   \dot{f} \approx \bigl\{ f, H_{\rmT} \bigr\}, 
\end{equation}
with the \emph{total Hamiltonian}, 
\begin{equation}  \label{DBTotHam}
   H_{\rmT} = H_{\EuD}^{\prime} + 
              \sum_{\alpha=1}^A \lambda^{\alpha} \, \phi_{\alpha},
\end{equation}
which is defined as a sum of a Dirac Hamiltonian $H_{\EuD}^{\prime}$,
\begin{subequations}
   \begin{align} 
      H^{\prime}_{\EuD} 
         &= H_{\rmc} + \sum_{i=1}^R \mu^i \phi_i, \label{DBPriHam3} \\
\intertext{and a specific linear combination of the primary constraints 
           $\phi_{\alpha}$,}
      \phi_{\alpha} 
         &= \sum_{i=1}^R \nu_{\alpha}^i \, \phi_i. \label{DBPhialp1}
   \end{align}
\end{subequations}
Therefore the general solution to the canonical equations \eqref{DBEqMot3}
will involve $A$ arbitrary functions of time, as expected for a singular
system.

\paragraph*{First-class and second-class constraints.}

The concept of first-class and second-class functions on phase space was
introduced by Dirac \cite{dirac=GenHam1,DIRAC=LecQua}; it plays a central 
r\^ole in the Hamiltonian formalism. 
\nl
A function $f(q,p)$ is said to be \emph{first class} if its Poisson brackets 
with every constraint vanish weakly, that is
\begin{equation} 
   \bigl\{ f, \phi_m \bigr\} \approx 0, \quad \forall \range{m}{1}{R}.
\end{equation}
The set of first-class functions is closed under the Poisson bracket 
\cite{dirac=GenHam2}. Directly available examples of first-class functions are
$H_{\EuD}^{\prime}$ and $\phi_{\alpha}$, defined by \eqref{DBPriHam3} and 
\eqref{DBPhialp1} respectively. Hence the total Hamiltonian \eqref{DBTotHam} 
is the sum of the first-class Dirac Hamiltonian $H_{\EuD}^{\prime}$ and an 
arbitrary linear combination of the primary first-class constraints 
$\phi_{\alpha}$. The number of arbitrary functions $\lambda^{\alpha}$ is thus 
equal to the number of primary first-class constraints. What makes first-class
constraints so important is that they generate gauge transformations.%
\footnote{For a more thorough discussion on the exact meaning of first-class 
          constraints, especially with regard to some controversy found in the 
          literature on the so-called Dirac's conjecture, see 
          \cite{GOVAE=HamQua,HENNE=QuaGau}. \label{foot:DirCon}}   
It indicates there is more than one set of canonical variables that 
corresponds to a given physical state. To overcome this ambiguity further
restrictions may be imposed on the canonical variables: one proceeds to a 
\emph{gauge-fixing} procedure (see, \eg \cite{BURNE=IntThe}), which is not 
unique; its specific implementations actually determine the corresponding 
methods of canonical quantisation---\eg Dirac quantisation, \textsc{brst} 
method, reduced phase space quantisation \cite{burne=ChoGau}.%
\footnote{In the \textsc{brst} approach, the phase space is extended rather
          than reduced.} 
\nl
Conversely, if there exists at least one constraint such that its Poisson
bracket with $f$ does not vanish weakly, then $f$ is said to be \emph{second
class}. If $\chi_{\beta}$ denotes a complete set of second-class constraints,
then the matrix $C$, the elements of which are precisely the Poisson brackets 
between the $\chi$'s,
\begin{equation} 
   C_{\beta \beta^{\prime}} = 
      \bigl\{ \chi_{\beta}, \chi_{\beta^{\prime}} \bigr\},
\end{equation}
is nonsingular and its inverse is denoted as 
$C^{-1}_{\beta \beta^{\prime}} \equiv C^{\beta \beta^{\prime}}$. (This 
property may also serve as an alternative definition of the second-class
constraints \cite{BURNE=IntThe}.)

\paragraph*{Dirac bracket.}

Second-class constraints are associated with redundant \dofs which could be 
solved in terms of the other \dofs provided that a generalised Poisson bracket 
referring to the remaining \dofs only would be defined. This is achieved with 
the \emph{Dirac bracket}, which is defined by
\begin{equation} \label{DBDirBra}
   \bigl\{ f, g \bigr\}_{\EuD} := 
      \bigl\{ f, g \bigr\} -
      \sum_{\beta, \beta^{\prime}} 
         \bigl\{ f, \chi_{\beta} \bigr\} \,
         C^{\beta \beta^{\prime}} \,
         \bigl\{ \chi_{\beta^{\prime}}, g \bigr\}, 
\end{equation}
for any arbitrary functions $f$, $g$ on phase space, and which is always
consistent with the second-class constraints. 
\nl
Owing to definition \eqref{DBDirBra}, second-class constraints become strong 
equations, and we may write the equation of motion \eqref{DBEqMot3} as
\begin{equation}  \label{DBEqMot4}
   \dot{f} \approx \bigl\{ f, H^{\star} \bigr\}_{\EuD}, 
\end{equation}
where $H^{\star}$ stands for any first-class Hamiltonian that generates time
evolution: for instance, the total Hamiltonian $H_{\rmT}$, or a more general 
choice with the \emph{extended Hamiltonian} $H_{\rmE}$ which includes, in the 
linear combination of first-class constraints, \emph{secondary} first-class 
constraints as well (see \cite{HENNE=QuaGau,GOVAE=HamQua} and footnote%
\fnref{foot:DirCon} on page~\pageref{foot:DirCon}). 

\subsubsection{Generally covariant systems}
\label{subsub:GenCov}

Systems that are invariant under arbitrary coordinate transformations are
analogous to the parameterised form of mechanics in which the Hamiltonian and
the time variable are introduced as a conjugate pair of canonical variables 
corresponding to a new \dof. The resulting theory is invariant under an 
arbitrary reparameterisation. In field theory one can also introduce 
arbitrary labels for the spatial coordinates: The theory becomes invariant 
under an arbitrary change of the space-time coordinates, just like generally 
covariant theories. This is the reason why theories of gravity are said to be
\emph{already parameterised} systems. The $3+1$--splitting of space-time, 
which is a crucial step towards their canonical formulation (see Section~%
\ref{sec:HigOrd}), has the virtue of ``de-parameterising'' the theory. 

\paragraph*{Time as canonical variable.}

Consider the action functional \eqref{DBAction} and assume for simplicity that
the Lagrangian $L$ be regular. In order to raise the physical time $t$ as a 
canonical variable we parameterise the theory with respect to a new time 
parameter $\tau$ such that $t=t(\tau)$ and $\rmd t/\rmd\tau\neq 0$.
The original time is then interpreted as an additional generalised coordinate, 
\ie $q_0 := t$. The action functional \eqref{DBAction} thus becomes
\begin{equation} \label{DBAct2}
   S = \int_{\tau_0}^{\tau_1} \rmd \tau \, q_0^{\prime} \,
          L \bigl( q_k, \frac{q_k^{\prime}}{q_0^{\prime}} \bigr) 
     \equiv \int_{\tau_0}^{\tau_1} \rmd \tau \, L_{\tau} 
           \bigl( q_m, q_m^{\prime} \bigr)
   \quad \text{for} \; \range{m}{0}{K},
\end{equation}
where a prime denotes differentiation \wrt the parameter $\tau$. It is 
invariant under any further reparameterisation 
$\tau^{\star}=\tau^{\star}(\tau)$. The conjugate momenta are given by
\begin{subequations}
   \begin{align}
      p_k^{(\tau)} &= \parD{L_{\tau}}{q_k^{\prime}} 
                    = \parD{L}{\dot{q}_k}
                    = p_k \quad \text{for} \; \range{k}{1}{K}, 
      \label{DBConjMoma} \\
      p_0^{(\tau)} &= \parD{L_{\tau}}{q_0^{\prime}} 
                    = L - \sum_{k=1}^K \parD{L}{\dot{q}_k} \dot{q}_k
                    = - H \bigl( q_k, p_k \bigr). \label{DBConjMomb} 
   \end{align}
\end{subequations}
Hence relation \eqref{DBConjMomb} is a primary constraint, 
\begin{equation} \label{DBPriCon2}
   H_0 \bigl( q, p \bigr) := p_0^{(\tau)} + H \bigl( q_k, p_k \bigr)
                          \approx 0,
\end{equation}
and the canonical Hamiltonian
\begin{equation} \label{DBCanHam2}
   H_{\rmc}^{(\tau)} = \sum_{m=0}^K q^{\prime}_m \, p_m^{(\tau)} - L_{\tau}
                     = q^{\prime}_0 \, H_0 \bigl( q, p \bigr)
\end{equation}
identically vanishes---a striking feature due to the parameterisation 
invariance. Since $H_0 \approx 0$ is the only constraint, the total 
Hamiltonian is simply
\begin{equation} \label{DBTotHam2}
   H_{\rmT}^{(\tau)} = N (\tau) H_0 \bigl( q, p \bigr),
\end{equation}
where $N(\tau)$ is an arbitrary function which is interpreted as a Lagrange 
multiplier in the variational principle
\begin{equation} \label{DBParAct}
   \delta \, \int_{\tau_0}^{\tau_1} \rmd \tau
      \biggl( \sum_{m=0}^K q^{\prime}_m \, p_m^{(\tau)} -
             N (\tau) H_0 \bigl( q, p \bigr)
      \biggr) = 0.
\end{equation}
In particular, variation of \eqref{DBParAct} \wrt the momentum $p_0^{(\tau)}$ 
yields $N(\tau)=\rmd t/\rmd \tau$; hence the original canonical theory is 
recovered if we choose the synchronous temporal gauge, \ie $N \equiv 1$. 

\paragraph*{Generalisation to field theory.}

The extension of the above parameterisation procedure to a field theory that
is described by a Lagrangian density 
$\Fl \bigl( \phi, \partial_{\mu} \phi \bigr)$ is achieved by interpreting the 
space-time coordinates as four new field variables, \ie 
$x_{\mu} = x_{\mu} (\underline{x}_{\nu})$. Thus there are four extra conjugate 
momenta; four constraint equations $H_{\mu} \approx 0$ are required to relate 
these momenta to the Hamiltonian density and the field momentum density. It is 
convenient to consider a foliation in terms of hypersurfaces which are 
labelled by constant `times', that is $\underline{x}^0=\text{const.}$ Let 
$n^{\alpha}$ be the unit normal vector field to these hypersurfaces. Then we 
decompose the original set of constraints $H_{\mu}$ into their parts normal 
and tangential to the hypersurfaces, namely $H_{\bot}$ and $H_i$ for $i=1,2,3$ 
respectively. The ensuing total Hamiltonian is
\begin{equation}  \label{DBTotHam3}
   H_{\rmT} = N H_{\bot} + \sum_{i=1}^3 N^i H_i. 
\end{equation}
In the context of general relativity the primary constraints $H_{\bot}$ and 
$H_i$ are called the \emph{super-Hamiltonian} and the \emph{super-momentum} 
respectively; see \cite[page~521]{MISNE=Gra}. The four corresponding Lagrange 
multipliers $N^{\mu}$ specify how to move forward in time from one 
hypersurface to the other as well as onto one and the same hypersurface---they 
are called the \emph{lapse} and \emph{shift} functions respectively; see, \eg 
Kucha{\v r}'s review for a very detailed account \cite{kucha=CanQua}.

\section{Theories with higher derivatives}
\label{sec:TheHig}

\subsection{Introduction}
\label{subsec:TheHigIntro}

The Hamiltonian formulation of theories with higher derivatives was firstly
developed by Ostrogradsky almost one and a half century ago 
\cite{ostro=MemIso}. Basically, the Ostrogradsky method \cite{DOUBR=GeoCon,%
WHITT=TreAna} is adapted only to systems described by a \emph{regular} 
Lagrangian, that is, a Lagrange function the associated Hessian matrix of 
which \wrt the highest-order time derivatives has a nonzero determinant. The 
underlying idea of this method---and of its subsequent generalisations---%
consists in introducing besides the original configuration variables a new set 
of coordinates that encompasses each of the successive time derivatives of the 
original Lagrangian coordinates so that the initial higher-order regular 
system be reduced to a first-order system. In order to recover the standard 
interpretation of the old coordinates when time evolution is considered it is 
essential to insert into the formalism the definition of the new coordinates 
in terms of the old ones by means of a Lagrange-multiplier technique. As a 
direct consequence, the auxiliary \dofs that enable the order-lowering in the 
initial Lagrangian are constrained: One must resort to Dirac's approach for 
building up a consistent Hamiltonian formalism \cite{govae=HamFor}. 
\nl
It seems that the very first use of Ostrogradsky's method was undertaken by 
Kerner in an attempt to develop a Hamiltonian formalism for Wheeler--Feynman 
electrodynamics (cited in \cite{jaen=RedOrd}). It is a remarkable result 
indeed in relativistic particle dynamics that, for conservatively interacting 
particles, there does not exist an ordinary single-time Lagrangian or 
Hamiltonian description if the position coordinates belong to a Lorentz frame. 
Rather, it turns out that in a local-in-time representation all higher time 
derivatives must appear in the Lagrangian, which is therefore an infinite-%
order Lagrange function. Now, because infinite-order Lagrangian systems 
exhibit cumbersome mathematical features, there have been many attempts to 
weed out all higher time derivatives occurring in the higher-order equations 
of motion by utilising the equations of motion of lower orders. In such a 
reduction process the question naturally arises whether the original infinite%
-order Lagrangian can also be reduced to an ordinary Lagrangian. The answer is 
not trivial for, in general, a system of second-order ordinary differential 
equations does not admit an ordinary Lagrangian description. On the contrary, 
the existence of a Hamiltonian formulation is warranted owing to the Lie--%
K\"onig theorem \cite{WHITT=TreAna}. Moreover, as Ja\'en \etal pointed out, a 
straight substitution into the equations of motion implies the intromission of 
some constraints that modify the very nature of the variational principle 
underlying the Lagrangian formalism. It is precisely for this reason that 
Ja\'en \etal tackled the problem of finding the reduced Hamiltonian by means 
of the techniques of constrained Hamiltonian dynamics together with the 
Ostrogradsky method in the case of Wheeler--Feynman electrodynamics 
\cite{jaen=RedOrd} (see also Ellis's canonical formalism for a second-order 
Lagrangian \cite{ellis=CanFor}). 
\nl
More generally, in the problem of reduction of higher-order Lagrangians 
describing the dynamics of systems of point particles, which are given as 
formal power series in some ordering parameter, Damour and Sch\"afer developed 
a new method, called `the method of redefinition of position variables', that 
enabled them to eliminate consistently the higher time derivatives, directly 
at the Lagrangian level \cite{damou=RedPos}. 
\nl
This type of reduction process was also examined by Grosse-Knetter in the 
context of general effective higher-order Lagrangians \cite{gross=EquHam,
gross=EffLag}. However, it must be objected that Grosse-Knetter's treatment is 
not reliable. The major criticism that one could raise has to do with the 
simplifying assumption that $\delta^4(0)$-terms occurring in the path-integral 
formalism may be neglected \emph{on the mere analogy} of what happens in the 
first-order case. As a matter of fact these terms should not be discarded 
unless a suitable regularisation method would be successfully applied. Beyond 
that particular point, the proofs---quite elusive, by the way---make an 
inappropriate use of the results obtained by Damour and Sch\"afer, which are 
valid only for effective Lagrangians that can be written as formal power 
series. Furthermore no scrupulous constraint analysis is performed when 
discussing the equivalence of Lagrangian and Hamiltonian path-integral 
quantisations. In that respect the claim that any higher-order effective 
Lagrangian can be reduced to a first-order one \emph{without} introducing 
extra \dofs---by the way, ``eliminating the unphysical effects associated with 
Ostrogradsky additional \dofs'' \cite{gross=EffLag}---is not proved by any 
rigorous analysis whatsoever.
\nl 
\label{page:BL}
On the other hand, Gitman \etal have worked out a consistent method of 
building up a Hamiltonian formalism for any constrained system with higher 
derivatives (see \cite{GITMA=QuaFie} and references therein). Buchbinder and 
Lyakhovich improved the treatment in a form that is more appropriate for 
theories of gravity (see \cite{BUCHB=EffAct} and references therein); more
specifically, they applied the method to the most general quadratic 
gravitational action in four dimensions \cite{buchb=CanQua}; then, with 
Karataeva, they extended the analysis to the realm of multidimensional 
quadratic gravity \cite{buchb=MulGra}. Within the particular subclass of 
singular second-order Lagrangians Nesterenko \cite{neste=SinLag} and Batlle 
\etal \cite{battl=LagHam} discussed, in a more geometrical setting, the 
relationship between the Lagrangian and Hamiltonian frameworks; Galv\~a and 
Lemos examined the peculiar case of singular second-order Lagrangians that 
differ from a regular first-order Lagrangian by a total time derivative of a 
function of both the coordinates and velocities \cite{galva=QuaCon}; prompted 
by the strange superstition that dealing with Ostrogradsky Lagrangian 
constraints ravels the mind, Schmidt contrived an `alternate Hamiltonian 
formalism' \cite{schmi=ConRel} (see also Kasper's similar approach 
\cite{kaspe=FinHam}). 
\nl
In the general case Pons achieved a rigorous unification of the Dirac 
formalism for constrained systems and the Ostrogradsky method for higher-order 
Lagrangians \cite{pons=OstThe}. Saito \etal developed a similar formalism that 
they applied to the aforementioned case of a Lagrangian describing the 
gravitational interaction of two point particles \cite{saito=DynFor}. 
\nl
Subsequent formal investigations inspired by the Ostrogradsky method may be 
reviewed briefly. In order to study the interplay of higher-order Lagrangian 
and Hamiltonian formalisms, Gr\`acia \etal introduced `partial Legendre--%
Ostrogradsky transformations' and constructed some interesting geometric 
structures \cite{graci=HigLag}; within this geometrisation scheme Gr\`acia and 
Pons extended previous significant results \cite{graci=HamApp} to the study of 
Noether-symmetry transformations for higher-order Lagrangians 
\cite{graci=GauTra}. Chitaya \etal have analysed the constrained Ostrogradsky 
method for constructing the generators of local symmetry transformations 
\cite{chita=OstMet}. Kaminaga has showed that the adjunction of total 
derivative terms to a higher-order Lagrangian does not affect either the 
classical or the quantum canonical structure of the system 
\cite{kamin=QuaMec}.%
\footnote{In the Lagrangian formalism this problem is trivial: The equations 
          of motion are not affected by the addition of a total derivative 
          term to the Lagrangian.}
Nakamura and Hamamoto have analysed the path-integral formalisms associated 
with different variants of the Ostrogradsky construction: the standard, the 
constrained, and the generalised methods respectively \cite{nakam=HigDer}. 
Most recently, Nirov has elaborated a \textsc{brst} formalism for systems that 
are invariant under gauge transformations with higher-order time derivatives 
of gauge parameters \cite{nirov=BRSFor,nirov=OstPre}; Pimentel and Teixeira 
have generalised the Hamilton--Jacobi approach for higher-order singular 
systems \cite{pimen=GenHam}. 
\nl
Very few applications of the Ostrogradsky method to specific higher-order 
field theories have been actually undertaken; for instance, de Urries and 
Julve have demonstrated the physical equivalence of relativistic scalar field 
theories with higher derivatives and their reduced second-order counterpart:
They compare the standard Ostrogradsky procedure to a Lorentz-invariant method 
based on the use of the so-called `Helmholtz Lagrangian' \cite{julve=OstFor}. 
In the context of \hotg, especially in quantum cosmology, the situation is not 
very different. For a quadratic Lagrangian involving $R$ and $R^2$ terms 
Kasper has compared Buchbinder and Lyakhovich's generalised Ostrogradsky 
formalism to another method that is characterised by the introduction of a 
scalar field at the Lagrangian level; he obtains approximate solutions to the 
Wheeler--DeWitt equation, with a closed \textsc{flrw} ansatz 
\cite{kaspe=WheDeW} (Kasper's analysis has been improved by Pimentel and 
Obreg\'on who solved the same equation analytically \cite{pimen=QuaCos}). 
Pimentel \etal have obtained solutions to the Wheeler--DeWitt equation 
corresponding to the Taub model in the case of a pure quadratic $R^2$ 
Lagrangian \cite{pimen=GenQua}. In a very recent work, Ezawa \etal have 
developed a canonical formalism for $f(R)$ theories of gravity 
\cite{ezawa=CanFor}; the formulation presented in Section~\ref{sec:HigOrd} is 
in total agreement with their approach although a slightly different choice 
for the canonical variables does actually simplify the analysis.%
\footnote{The completion of this work was achieved during a visit at the
          Department of Mathematics of the University of the Aegean (Samos, 
          Greece) in May 1996.}   
\begin{rem}
The Ostrogradsky method has also been used to highlight the fundamental
drawbacks inherent in higher-order theories \cite{eliez=ProNon,simon=HigDer,%
schmi=StaHam,schmi=ConRel}. 
\end{rem}

\subsection{Ostrogradsky's method for regular systems}
\label{subsec:OstroReg}

Consider a system with a finite number $K$ of discrete \dofs represented by 
generalised coordinates $x_k$ for $\range{k}{1}{K}$ in configuration space; 
for the sake of clarity we assume that these coordinates are commuting 
variables though it should be clear that exactly the same considerations are 
valid for commuting and anticommuting \dofs, as well as for infinite discrete 
or noncountable set of coordinates---the former case corresponds to the 
treatment of bosonic and fermionic types of \dofs and the latter typically 
that of field theories. Consider now that the system under study be described 
by some time-independent---it is for convenience only that we restrict the 
analysis to time-independent Lagrangians; time-dependent Lagrange functions 
can also be treated along the same lines as developed hereafter---%
\emph{regular} Lagrange function
\begin{equation}
   L \left( x_k, \dot{x}_k, \ddot{x}_k, \dots, x_k^{(\alpha_k)} \right)
   \quad \text{for} \; \alpha_k \geqslant 1 \; \text{and} \; \range{k}{1}{K},
   \label{OstroLag1}    
\end{equation}
where we adopt the convention 
$\bigl( \tfrac{\rmd}{\rmd t} \bigr)^i x_k = x_k^{(i)}$. The index $\alpha_k$ 
denotes the maximal order of all time derivatives of the coordinate $x_k$ 
appearing explicitly in the Lagrangian \eqref{OstroLag1}. (Note that when 
$\alpha_k = 1$ for all \dofs one recovers as a special case of the present 
general treatment the familiar situation encountered in classical mechanics 
for first-order Lagrangians.) In the original formulation of the Ostrogradsky 
method it is also assumed that the Lagrange function does depend on at least 
the first-order time derivative of each \dof in order to prevent the 
occurrence of undesirable constraints. 
\nl
As usual we assume that the dynamical time evolution of the system is obtained 
upon extremising the action functional constructed from the Lagrangian 
\eqref{OstroLag1} under variations $\delta x_k (t)$ of the configuration 
variables $x_k$. The variational principle yields the 
$2 N^{\text{\textrm{th}}}$-order \EL equations 
\begin{equation}
   \sum_{\sigma_k=0}^{\alpha_k} 
      \bigl( -1 \bigr)^{\sigma_k} 
      \left( \frac{\rmd}{\rmd t} \right)^{\sigma_k}
      \parD{L}{x_k^{(\sigma_k)}}
   = 0,
   \quad N = \sup_k \bigl\{ \alpha_k \bigr\}. 
   \label{OstroEL1}    
\end{equation}
\nl
Adopting Ostrogradsky's prescriptions we introduce new quantities
$p_{k,\gamma_k}$ for $\range{\gamma_k}{0}{(\alpha_k-1)}$ (which depend on 
derivatives of the coordinates $x_k$ up to order $2\alpha_k-\gamma_k-1$) that 
are defined by the following recursion relations
\begin{subequations}
   \label{OstroConjMom1}
   \begin{align}
      p_{k,\alpha_k-1} &= \parD{L}{x_k^{(\alpha_k)}},
         \label{OstroConjMom1a} \\
      p_{k,\beta_k-1}  &= \parD{L}{x_k^{(\beta_k)}} -
                          \frac{\rmd}{\rmd t} p_{k,\beta_k}
         \quad \text{for} \; \range{\beta_k}{1}{(\alpha_k-1)}.           
         \label{OstroConjMom1b}
   \end{align}
\end{subequations}
We can now prove the following result (see, \eg \cite[pp.~265]{WHITT=TreAna}).
\begin{OstReg} \label{thm:OstReg}
Let $L$ be a regular Lagrangian depending on generalised coordinates $x_k$ and 
their time derivatives up to order $\alpha_k \geqslant 1$ for 
$\range{k}{1}{K}$. Consider new quantities $p_{k,\gamma_k}$ for 
$\range{\gamma_k}{0}{(\alpha_k-1)}$ that are introduced by the recursion 
relations \eqref{OstroConjMom1}. The \EL equations \eqref{OstroEL1} derived 
from $L$ are equivalent to the system of canonical equations that is obtained 
from the Hamiltonian
\begin{equation*}
   H = \sum_{k=1}^{K} \sum_{\gamma_k=0}^{\alpha_k-1}
          \dot{x}_k^{(\gamma_k)} \, p_{k,\gamma_k} - L. 
\end{equation*}
\end{OstReg}
\begin{proof}
Owing to the recursion relations \eqref{OstroConjMom1} the \EL equations 
\eqref{OstroEL1} take the simple form
\begin{equation}
   \parD{L}{x_k} - \frac{\rmd}{\rmd t} p_{k,0} = 0
   \quad \text{\normalfont{for}} \; \range{k}{1}{\alpha_k},
   \label{OstroEL2}    
\end{equation}
which is reminiscent of Hamilton's equations of motion. To make this statement 
more specific consider the quantity defined by
\begin{equation}
   H = \sum_{k=1}^{K} \sum_{\gamma_k=0}^{\alpha_k-1}
          \dot{x}_k^{(\gamma_k)} \, p_{k,\gamma_k} - L, 
   \label{OstroHam1}    
\end{equation}
where the variables $p_{k,\gamma_k}$ are determined through the recursion 
relations \eqref{OstroConjMom1}. As it stands, the function $H$ should depend explicitly 
on variables $x_k$ and their derivatives up to order $(2 \alpha_k -1)$, on 
account of the definitions \eqref{OstroConjMom1} for the quantities 
$p_{k,\gamma_k}$ the dependence of which on the velocities 
$x_k^{(\sigma_k)}$ for $\range{\sigma_k}{1}{(2 \alpha_k-1)}$ is manifest. 
However, taking the differential of the function $H$ defined above we obtain 
the identity
\begin{equation}
   \rmd H = \sum_{k=1}^{K} \sum_{\gamma_k=0}^{\alpha_k-1}
               \Bigl( 
                  x_k^{(\gamma_k+1)} \rmd p_{k,\gamma_k} -
                  \dot{p}_{k,\gamma_k} \rmd x_k^{(\gamma_k)}
               \Bigr) -
            \sum_{k=1}^{K} 
               \biggl( \parD{L}{x_k} - \dot{p}_{k,0} \biggr) \rmd x_k,
   \label{OstroDHam1}    
\end{equation}
where the last sum turns out to be a combination of the \EL equations. This 
result establishes that the function $H$ defined by \eqref{OstroHam1} depends 
on variables $x_k^{(\sigma_k)}$ for $\range{\sigma_k}{0}{(2 \alpha_k-1)}$ only 
through a dependence of the variables $x_k^{(\gamma_k)}$ and $p_{k,\gamma_k}$ 
themselves, \viz $H=H \bigl( x_k^{(\gamma_k)}, p_{k,\gamma_k} \bigr)$. Note 
that this property holds even in the case of a singular Lagrangian; however, 
here, unhampered by any degeneracies we are able to perform a suitable
Legendre transformation: We regard variables 
$\bigl( x_k^{(\gamma_k)}, p_{k,\gamma_k} \bigr)$ as being canonically 
conjugate pairs and we only solve for the first-order time derivatives of the 
variables $x_k^{(\alpha_k-1)}$ in terms of variables $p_{k,\alpha_k-1}$ and 
variables $x_k^{(\gamma_k)}$; throughout this procedure the latter variables 
must be considered as independent of one another rather than being time 
derivatives of order $\gamma_k$ of the coordinates $x_k$. Finally the identity 
\eqref{OstroDHam1} shows that the \EL equations \eqref{OstroEL1} are 
equivalent to the system of canonical equations
\begin{subequations}
   \label{OstroH1}    
   \begin{align}
      \dot{x}_k^{(\gamma_k)} &= \parD{H}{p_{k,\gamma_k}},
         \label{OstroH1A} \\
      \dot{p}_{k,\gamma_k}  &= - \parD{H}{x_k^{(\gamma_k)}},
         \label{OstroH1B} 
   \end{align}
\end{subequations}
obtained from the Hamiltonian \eqref{OstroHam1}.
\end{proof}
\par%
Let us summarise the whole Ostrogradsky procedure. Firstly, we introduce the
\emph{conjugate momenta} $p_{k,\alpha_k-1}$ with definition 
\eqref{OstroConjMom1a}, where it should be understood that the variables 
$\bigl( x_k, \dot{x}_k, \dots, x_k^{(\alpha_k-1)} \bigr)$ are independent.
Owing to the fact that our Lagrangian is regular, we are able to invert 
relations \eqref{OstroConjMom1a} for the velocities 
$\dot{x}_k^{(\alpha_k-1)}$, \viz
\begin{equation}
   x_k^{(\alpha_k)} = \dot{x}_k^{(\alpha_k-1)}
                    = \dot{x}_k^{(\alpha_k-1)} 
                         \left( 
                            x_k, \dot{x}_k, \dots, x_k^{(\alpha_k-1)},
                            p_{k,\alpha_k-1} 
                         \right).
   \label{OstroVelo1}    
\end{equation}
The remaining conjugate momenta $p_{k,\beta_k-1}$ for
$\range{\beta_k}{1}{(\alpha_k-1)}$ are then determined by the recursion 
relations \eqref{OstroConjMom1b}; it should be stressed, however, that one 
should not solve these recursion relations for the variables 
$\dot{x}_k^{(\beta_k-1)}$ in terms of the variables 
$\bigl( x_k, \dot{x}_k, \dots, x_k^{(\beta_k-1)} \bigr)$, the conjugate 
momenta $\bigl( p_{k,\beta_k}, \dots, p_{k,\alpha_k-2} \bigr)$, and time 
derivatives of the latter, since the variables 
$\bigl( x_k, \dot{x}_k, \dots, x_k^{(\alpha_k-1)} \bigr)$ are regarded as 
independent in the Ostrogradsky method. \\
Secondly, the canonical Hamiltonian is defined by equation \eqref{OstroHam1}, 
where the velocities $\dot{x}_k^{(\alpha_k-1)}$ have been substituted in 
accordance with equation \eqref{OstroVelo1}; explicitly it is
\begin{equation}
   \begin{split}
      H \Bigl( x_k^{(\gamma_k)}, p_{k,\gamma_k} \Bigr) 
         &= \sum_{k=1}^{K} \dot{x}_k^{(\alpha_k-1)} 
               \Bigl( 
                  x_k, \dot{x}_k, \dots, x_k^{(\alpha_k-1)}, p_{k,\alpha_k-1} 
               \Bigr) \, p_{k,\alpha_k-1} \\
         & \qquad 
           - L \Bigl( 
                  x_k, \dot{x}_k, \dots, x_k^{(\alpha_k-1)}, 
                  \dot{x}_k^{(\alpha_k-1)} (\dots)
               \Bigr) \\
         & \qquad 
           + \sum_{k=1}^{K} \sum_{\beta_k=1}^{\alpha_k-1}
                x_k^{(\beta_k)} \, p_{k,\beta_k-1}. 
   \end{split}
   \label{OstroHam2}    
\end{equation}
Eventually it yields Hamilton's equations \eqref{OstroH1}, which can be
written down---upon the introduction of the fundamental Poisson brackets on 
the phase space spanned by the pairs of canonically conjugate variables 
$\bigl( x_k^{(\gamma_k)}, p_{k,\gamma_k} \bigr)$---as the canonical system
\begin{subequations}
   \label{OstroH2}    
   \begin{align}
      \dot{x}_k^{(\gamma_k)} &= \bigl\{ x_k^{(\gamma_k)}, H \bigr\},
         \label{OstroH2A} \\
      \dot{p}_{k,\gamma_k}   &= \bigl\{ p_{k,\gamma_k}, H \bigr\}.
         \label{OstroH2B} 
   \end{align}
\end{subequations}
After the Poisson brackets have been computed, it is necessary to impose the 
condition that the variables $x_k^{(\gamma_k)}$ be time derivatives of order 
$\gamma_k$ of the original coordinates $x_k(t)$. This requirement may be 
implemented \emph{ab initio} by firstly associating with each variable 
$x_k^{(\gamma_k)}$ an auxiliary independent \dof; then replacing the original 
Lagrangian \eqref{OstroLag1} by an \emph{extended} Lagrange function, wherein 
the definitions of the new \dofs in terms of the old ones are inserted as 
constraints with Lagrange multipliers. It is actually the modern way of  
building up an Ostrogradsky formulation and its advantages are twofold. 
Firstly, it shifts the study of higher-order Lagrangians to the analysis of 
the usual type of dynamical systems for which powerful techniques have been 
developed so far (see, for instance \cite{GOVAE=HamQua}); it thus renders the 
necessity of a separate discussion of the quantisation of higher-order systems 
void of any justification whatsoever. Moreover the extension of the formalism 
to embody the case of singular higher-order systems may be done in a natural 
way along the lines of Dirac's approach (\cfr Subsection~%
\ref{subsec:OstroSing}). Secondly, canonical or path-integral quantisations of 
such higher-order systems may be achieved unhindered by the possible ambiguity 
that could arise when dealing with the variables $x_k^{(\gamma_k)}$ and their 
first-order time derivatives $\dot{x}_k^{(\gamma_k)}$: Without introducing 
auxiliary \dofs it could indeed be confusing to perform a Legendre 
transformation on the variables $x_k^{(\alpha_k-1)}$ only, while leaving the 
other derivatives untouched. 
\nl
We illustrate the Ostrogradsky construction with a simple example.
\begin{exmp}
We take the Lagrangian 
$L=\dot{x}^2/2-\omega^2 x^2/2 - \epsilon^2 \ddot{x}^2/2$ corresponding to a
simple (unit-mass) harmonic oscillator with an acceleration-squared term. We
firstly consider $x$ and $y:=\dot{x}$ as independent variables; the Lagrangian
then becomes $L=y^2/2-\omega^2 x^2/2 - \epsilon^2 \dot{y}^2/2$. We define
Ostrogradsky momenta as 
$p_y:=\partial L/\partial \dot{y}=-\epsilon^2 \dot{y}$ and 
$p_x:=\partial L/\partial y - \dot{p}_y=y+\epsilon^2 \ddot{y}$ respectively.
Only the former relation is inverted in terms of $\dot{y}$; we obtain the
Hamiltonian of the system, 
$H(x,p_x,y,p_y)=yp_x-p_y^2/2\epsilon^2-y^2/2+\omega^2x^2/2$, from which the 
canonical equations are readily derived: They are $\dot{x}=y$, 
$\dot{y}=-p_y/\epsilon^2$, $\dot{p}_x=-\omega^2x$, and $\dot{p}_y=-p_x+y$
respectively. We recognise easily the definitions of $y$ and momenta. Finally 
we recover the \EL equations from the canonical equations for the momenta, \ie 
$\epsilon^2 x^{(4)} + \ddot{x} + \omega^2 x = 0$.
\end{exmp}   
\nl
Let us see now how the original Ostrogradsky construction can be treated 
within the framework of Dirac's Hamiltonian formalism for constrained systems. 
A separate analysis of regular and singular higher-order systems is at this
stage unnecessary for, in both cases, the higher-order Lagrangian is reduced 
to a first-order Lagrangian exhibiting primary constraints.

\subsection{Constrained Ostrogradsky construction}
\label{subsec:OstroSing}

\subsubsection{Lagrangian formalism}

Consider a system with $K$ \dofs $x_k$ for $\range{k}{1}{K}$ and assume that
the associated Lagrangian (with higher derivatives) be given by expression 
\eqref{OstroLag1}. Let us introduce new \emph{independent} variables 
$q_{k,\gamma_k}$ for $\range{\gamma_k}{0}{(\alpha_k-1)}$ through the following 
recursion relations
\begin{subequations}
   \label{OstroQ1}    
   \begin{align}
      q_{k,\beta_k} &= \dot{q}_{k,\beta_k-1} 
                       \quad \text{for} \; \range{\beta_k}{1}{(\alpha_k-1)},
         \label{OstroQ1A} \\
      q_{k,0}       &= x_k. 
         \label{OstroQ1B}
   \end{align}
\end{subequations}
Clearly this choice corresponds to assuming that the successive time
derivatives $x_k^{(\gamma_k)}$ are specified as independent variables, for
relations \eqref{OstroQ1} imply
\begin{equation}
   q_{k,\gamma_k} \equiv x_k^{(\gamma_k)} 
      \quad \text{for} \; \range{\gamma_k}{0}{(\alpha_k-1)}.
   \label{OstroQ2}
\end{equation}
Preservation of the standard interpretation of the old coordinates when time
evolution is considered requires the relations \eqref{OstroQ1A} to be brought
into the Lagrangian formalism as primary constraints. The initial higher-%
order Lagrange function $L$ is replaced by an \emph{extended} first-order 
Lagrangian $\Lagb$ (with Lagrange multipliers $\lambda_{k,\beta_k}$ for 
$\range{\beta_k}{1}{(\alpha_k-1)}$ as additional variables) that is given by
\begin{equation}
   \Lagb \bigl( 
            q_{k,\gamma_k}, \dot{q}_{k,\gamma_k}, \lambda_{k,\beta_k} 
         \bigr) := 
       L \bigl( q_{k,\gamma_k}, \dot{q}_{k,\alpha_k-1} \bigr) +
       \sum_{k=1}^K \sum_{\beta_k=1}^{\alpha_k-1}
          \bigl( q_{k,\beta_k} - \dot{q}_{k,\beta_k-1} \bigr)
          \lambda_{k,\beta_k}.
   \label{OstroLb1}
\end{equation}
The variables $(q_{k,\gamma_k}, \lambda_{k,\beta_k})$ thus encompass the 
independent \dofs of the extended Lagrangian system, with 
$(q_{k,\beta_k}, \lambda_{k,\beta_k})$ being \emph{auxiliary} \dofs as 
compared to the original coordinates $x_k (=: q_{k,0})$. 
\nl
Before resorting to Dirac's analysis we must ensure that both Lagrange 
functions $L$ and $\Lagb$ yield equivalent equations of motion 
\cite{govae=HamFor}. 
\begin{prop} \label{prop:Ost1}
Let $L$ be the original Lagrangian \eqref{OstroLag1}. Let $\Lagb$ be the
associated extended Lagrangian \eqref{OstroLb1}. The \EL equations 
corresponding to $L$ and $\Lagb$ respectively are equivalent.
\end{prop}
\begin{proof}
The standard variational principle calls for the action functional constructed 
from the Lagrangian \eqref{OstroLb1} to be stationary under variations of the 
independent \dofs. With respect to the auxiliary \dofs, it yields the 
equations 
\begin{subequations}
   \label{OstroLC1}    
   \begin{align}
      \scriptstyle{\mathtt{[\delta q_{k,\beta_k}]}} 
      &\longrightarrow 
         \parD{L}{q_{k,\rho_k}} + 
         \frac{\rmd}{\rmd t} \lambda_{k,\rho_k+1} +
         \lambda_{k,\rho_k} = 0 
         \quad \text{\normalfont{for}} \; \range{\rho_k}{1}{(\alpha_k-2)}, 
         \label{OstroLC1a} \\
      \scriptstyle{\mathtt{[\delta q_{k,\alpha_k-1}]}} 
      &\longrightarrow 
         \parD{L}{q_{k,\alpha_k-1}} - 
         \frac{\rmd}{\rmd t} 
            \biggl( \parD{L}{\dot{q}_{k,\alpha_k-1}} \biggr) +
         \lambda_{k,\alpha_k-1} = 0,  
         \label{OstroLC1b} \\
      \scriptstyle{\mathtt{[\delta \lambda_{k,\beta_k}]}} 
      &\longrightarrow 
         q_{k,\beta_k} - \dot{q}_{k,\beta_k-1} = 0,  
         \label{OstroLC1c} \\
\intertext{whereas the variation \wrt the original coordinates gives} 
      \scriptstyle{\mathtt{[\delta q_{k,0}]}} 
      &\longrightarrow 
         \parD{L}{q_{k,0}} + \frac{\rmd}{\rmd t} \lambda_{k,1} = 0. 
         \label{OstroLC1d} 
   \end{align}
\end{subequations}
These last equations \eqref{OstroLC1d} are the actual equations of motion of 
the system: The former set of equations \eqref{OstroLC1a}--\eqref{OstroLC1c}
contains in fact constraint equations that determine both the auxiliary \dofs 
$q_{k,\beta_k}$ as the successive time derivatives of the original coordinates 
$q_{k,0}$ (equation \eqref{OstroLC1c}) and the Lagrange multipliers 
$\lambda_{k,\beta_k}$ in terms of successive partial derivatives of the 
original Lagrangian $L$ (equations \eqref{OstroLC1a} and \eqref{OstroLC1b});
we may inject the ensuing multipliers into equations \eqref{OstroLC1d} to 
obtain the equations of motion
\begin{equation}
   \sum_{\gamma_k=0}^{\alpha_k-1} 
      \bigl( -1 \bigr)^{\gamma_k} 
      \biggl( \frac{\rmd}{\rmd t} \biggr)^{\gamma_k} 
      \parD{L}{q_{k,\gamma_k}} + 
   \bigl( -1 \bigr)^{\alpha_k} 
   \biggl( \frac{\rmd}{\rmd t} \biggr)^{\alpha_k} 
   \parD{L}{\dot{q}_{k,\alpha_k-1}} = 0, 
   \label{OstroEL3}
\end{equation}
which in turn reduce to the \EL equations \eqref{OstroEL1} as soon as we
enforce the constraints \eqref{OstroLC1c}. 
\end{proof}
\begin{rem}
Hitherto the procedure has been completely general, that is, irrespective of 
the regular or singular nature of the original Lagrangian: henceforth we focus 
the analysis on singular Lagrangians.
\end{rem} 

\subsubsection{Hamiltonian formalism}

We now assume that the original higher-order Lagrangian $L$ be singular---%
though the present analysis applies equally well to regular systems---and that 
the rank of its associated Hessian matrix be constant everywhere and equal to 
$K-R$ $(R \in \mathbb{N})$, \viz
\begin{equation} \label{OstroRank}
   \rank \Biggl( 
            \frac{\partial^2 L}
                 {\partial x_k^{(\alpha_k)} \partial x_l^{(\alpha_l)}} 
         \Biggr) =
   \rank \Biggl( 
            \frac{\partial^2 \Lagb}
                 {\partial \dot{q}_{k,\alpha_k-1} 
                  \partial \dot{q}_{l,\alpha_l-1}} 
         \Biggr) =
   K-R. 
\end{equation}
With the aim of developing a Hamiltonian formulation on the basis of the 
Lagrangian $\Lagb$ we proceed to the standard Dirac analysis. 
\nl
The configuration space of the extended system is spanned by the set of 
variables $(q_{k,\gamma_k}, \lambda_{k,\beta_k})$ for 
$\range{\gamma_k}{0}{(\alpha_k-1)}$ and $\range{\beta_k}{1}{(\alpha_k-1)}$. We
define the momenta canonically conjugate to the independent \dofs as
\begin{subequations}
   \label{OstroConjMom2}    
   \begin{align}
      p_{k,\beta_k-1}  &:= \parD{\Lagb}{\dot{q}_{k,\beta_k-1}}  
                         = - \lambda_{k,\beta_k}, 
         \label{OstroConjMom2a} \\  
      p_{k,\alpha_k-1} &:= \parD{\Lagb}{\dot{q}_{k,\alpha_k-1}}  
                         = \parD{L}{\dot{q}_{k,\alpha_k-1}}, 
         \label{OstroConjMom2b} \\  
      \pi_{k,\beta_k}  &:= \parD{\Lagb}{\dot{\lambda}_{k,\beta_k}}  
                         = 0,
         \label{OstroConjMom2c}   
   \end{align}
\end{subequations}
where we have used equation \eqref{OstroLb1}. We thus see from equations 
\eqref{OstroConjMom2} that the phase-space \dofs 
$(q_{k,\gamma_k}, p_{k,\gamma_k}; \lambda_{k,\beta_k}, \pi_{k,\beta_k})$
are not all independent; hence we may identify the following set of primary 
constraints,
\begin{subequations}
   \label{OstroHCons1}    
   \begin{align}
      & \varphi_{k,\beta_k} = p_{k,\beta_k-1} + \lambda_{k,\beta_k} \approx 0,
        \label{OstroHCons1a} \\
      & \pi_{k,\beta_k} \approx 0,      
        \label{OstroHCons1b} \\
\intertext{to which we add the constraints stemming from the singular
           character of the original Lagrangian (\cfr equation 
           \eqref{OstroConjMom2b}),}
      & \phi_i \bigl( q_{k,\gamma_k}, p_{k,\alpha_k-1} \bigr) \approx 0
        \quad \text{for} \; \range{i}{1}{R}. 
        \label{OstroHCons1c} 
   \end{align}
\end{subequations}
The first set of primary constraints \eqref{OstroHCons1a}--%
\eqref{OstroHCons1b} originates in the specific way the auxiliary \dofs have 
been introduced into the extended Lagrange function $\Lagb$ (\cfr equation 
\eqref{OstroLb1}); given the fundamental Poisson brackets on the phase space,
\begin{align*} \label{OstroHPB1}    
   \bigl\{ q_{k,\gamma_k}, q_{l,\gamma_l} \bigr\}           &= 0 =
   \bigl\{ p_{k,\gamma_k}, p_{l,\gamma_l} \bigr\},          
   \quad \,
   \bigl\{ q_{k,\gamma_k}, p_{l,\gamma_l} \bigr\} = 
   \delta_{kl} \, \delta_{\gamma_k \gamma_l}, \\
   \bigl\{ \lambda_{k,\beta_k}, \lambda_{l,\beta_l} \bigr\} &= 0 =
   \bigl\{ \pi_{k,\beta_k}, \pi_{l,\beta_l} \bigr\},        
   \quad
   \bigl\{ \lambda_{k,\beta_k}, \pi_{l,\beta_l} \bigr\} = 
   \delta_{kl} \, \delta_{\beta_k \beta_l}, 
\end{align*}
these primary constraints satisfy 
\begin{equation}
   \bigl\{ \varphi_{k,\beta_k}, \pi_{l,\beta_l} \bigr\} = 
      \delta_{kl} \, \delta_{\beta_k \beta_l} 
   \label{OstroHPB2}    
\end{equation}
and hence are second class (they will be removed from the formalism later 
on). 
\nl
Before pursuing Dirac's analysis let us show how it is possible to recover
Ostrogradsky's prescriptions for the definition of the momenta 
$p_{k,\gamma_k}$ \cite{pons=OstThe}. 
\begin{prop} \label{prop:Ost2}
Let $\Lagb$ be the \emph{extended} Lagrangian associated with the original
singular Lagrangian $L$. The \EL equations obtained from $\Lagb$ \wrt the 
auxiliary coordinates $q_{k,\beta_k}$ (for $\range{\beta_k}{1}{\alpha_k-1}$)
are equivalent to the recursion relations \eqref{OstroConjMom1} defining the 
Ostrogradsky momenta $p_{k,\gamma_k}$ (for $\range{\gamma_k}{0}{\alpha_k-1}$).
\end{prop}
\begin{proof} 
It is sufficient to examine the \EL equations obtained from the Lagrangian 
$\Lagb$ \wrt the auxiliary coordinates $q_{k,\beta_k}$, namely
\begin{equation}
   \parD{\Lagb}{q_{k,\beta_k}} - 
   \frac{\rmd}{\rmd t} 
      \biggl( \parD{\Lagb}{\dot{q}_{k,\beta_k}} \biggr) = 0
   \quad \text{\normalfont{for}} \; \range{\beta_k}{1}{(\alpha_k-1)}. 
   \label{OstroEL4} 
\end{equation}
Taking definitions \eqref{OstroLb1} and constraints \eqref{OstroHCons1a} into 
account equations \eqref{OstroEL4} are equivalent to
\begin{equation}
   p_{k,\beta_k-1} = \parD{L}{q_{k,\beta_k}} - 
                     \frac{\rmd}{\rmd t} p_{k,\beta_k}
   \quad \text{\normalfont{for}} \; \range{\beta_k}{1}{\alpha_k-1}, 
   \label{OstroEL5}
\end{equation}
which, owing to the definitions \eqref{OstroQ1A} of the auxiliary \dofs, 
coincide with the recursion relations \eqref{OstroConjMom1b}. On the other 
hand expressions \eqref{OstroConjMom2b} correspond to definition 
\eqref{OstroConjMom1a}. Hence, as announced, we recover the definitions 
\eqref{OstroConjMom1} of the Ostrogradsky momenta.
\end{proof}
\begin{rem}
The remaining \EL equations
\begin{equation}
   \parD{\Lagb}{q_{k,0}} - 
   \frac{\rmd}{\rmd t} \biggl( \parD{\Lagb}{\dot{q}_{k,0}} \biggr) = 0
   \label{OstroEL6}
\end{equation}
are equivalent to the standard \EL equations for the Lagrangian $L$ (\cfr 
equation \eqref{OstroLC1d} and the subsequent discussion). 
\end{rem}
\par%
Of particular concern is the prevailing r\^ole played by the momenta
$p_{k,\alpha_k-1}$: They are indeed the only variables not involved in the
primary constraints \eqref{OstroHCons1a}, \eqref{OstroHCons1b}, and their 
conjugate coordinates $q_{k,\alpha_k-1}$ are the only auxiliary variables 
the velocities of which do appear in the original Lagrangian $L$ (compare 
equations \eqref{OstroConjMom2a} with equations \eqref{OstroConjMom2b}). These 
remarks about the special r\^ole played by the pairs of conjugate variables
$(q_{k,\alpha_k-1}, p_{k,\alpha_k-1})$ motivate the definition of a 
\emph{restricted} canonical Hamiltonian that gives the energy corresponding to 
the Lagrangian $L$ when the variables 
$q_{k,\beta_k-1}$ for $\range{\beta_k}{1}{(\alpha_k-1)}$ have been frozen; it 
is
\begin{equation}
   H_{\rmr} \bigl( q_{k,\gamma_k}, p_{k,\alpha_k-1} \bigr) := 
      \sum_{k=1}^K \dot{q}_{k,\alpha_k-1} \, p_{k,\alpha_k-1} -
      L \bigl( q_{k,\gamma_k}, \dot{q}_{k,\alpha_k-1} \bigr) 
   \label{OstroHam3}    
\end{equation}
and dwells in the restricted phase space---irrespective of whether the
relations \eqref{OstroConjMom2b} are invertible or not \cite{HANSO=ConHam,%
SUNDE=ConDyn}. 
\nl
We define the canonical Hamiltonian of the system in accordance with the 
usual prescription:
\begin{equation}
   H_{\rmc} = \sum_{k=1}^{K} \sum_{\gamma_k=0}^{\alpha_k-1}
                 \dot{q}_{k,\gamma_k} \, p_{k,\gamma_k} +
              \sum_{k=1}^{K} \sum_{\beta_k=1}^{\alpha_k-1}
                 \dot{\lambda}_{k,\beta_k} \, \pi_{k,\beta_k} -
              \Lagb \left( 
                       q_{k,\gamma_k}, \dot{q}_{k,\gamma_k}, 
                       \lambda_{k,\beta_k}
                    \right).
   \label{OstroHam4}    
\end{equation}
Making use of the constraint equations \eqref{OstroHCons1} and definition 
\eqref{OstroHam3} of the restricted Hamiltonian we may write the canonical 
Hamiltonian \eqref{OstroHam4} as
\begin{equation}
   H_{\rmc} \left( q_{k,\gamma_k}, p_{k,\gamma_k} \right) = 
      H_{\rmr} \bigl( q_{k,\gamma_k}, p_{k,\alpha_k-1} \bigr) +
      \sum_{k=1}^{K} \sum_{\beta_k=1}^{\alpha_k-1} 
         p_{k,\beta_k-1} \, q_{k,\beta_k}. 
   \label{OstroHam5}    
\end{equation}
\nl
We now proceed to the Dirac analysis of the system the canonical Hamiltonian 
of which is given by \eqref{OstroHam4}. Firstly, we write down the Dirac 
Hamiltonian of the system
\begin{equation}
   H_{\EuD} := H_{\rmc} +
               \sum_{k=1}^{K} \sum_{\beta_k=1}^{\alpha_k-1}
                  \Bigl( 
                     \mu^{k,\beta_k} \, \varphi_{k,\beta_k} +
                     \nu^{k,\beta_k} \, \pi_{k,\beta_k}
                  \Bigr) + 
               \sum_{i=1}^{R} \omega^i \phi_i,
   \label{OstroHam6}    
\end{equation}
with new Lagrange multipliers $\mu^{k,\beta_k}, \nu^{k,\beta_k}$, and 
$\omega^i$ associated respectively to the primary constraints 
\eqref{OstroHCons1} of the extended system. However, as indicated above, the 
first two constraints \eqref{OstroHCons1a} and \eqref{OstroHCons1b} are second 
class: They can be removed provided the canonical Poisson bracket is replaced 
by the appropriate Dirac bracket. Further we can easily check that the 
consistency algorithm does not generate secondary constraints from these 
primary ones: Their Dirac bracket with the Dirac Hamiltonian \eqref{OstroHam6} 
must vanish on the constraint surface; this requirement yields a unique 
determination of the multipliers $\mu^{k,\beta_k}$ and $\nu^{k,\beta_k}$. We 
thus solve these second-class constraints, \ie 
$\lambda_{k,\beta_k} = - p_{k,\beta_k-1}$ and $\pi_{k,\beta_k} = 0$; hence 
the Dirac Hamiltonian \eqref{OstroHam6} simplifies to
\begin{equation}
   H = H_{\rmc} + \sum_{i=1}^{R} \omega^i \phi_i.
   \label{OstroHam7}    
\end{equation}
The local symplectic structure on the phase space is specified through the
fundamental canonical Dirac brackets
\begin{equation}
   \bigl\{ q_{k,\gamma_k}, q_{l,\gamma_l} \bigr\}_{\EuD} = 0 =
   \bigl\{ p_{k,\gamma_k}, p_{l,\gamma_l} \bigr\}_{\EuD},          
   \quad 
   \bigl\{ q_{k,\gamma_k}, p_{l,\gamma_l} \bigr\}_{\EuD} = 
      \delta_{kl} \, \delta_{\gamma_k \gamma_l}, 
   \label{OstroHDB1}    
\end{equation}
and time evolution results from the knowledge of this symplectic structure and 
the explicit form of the Hamiltonian \eqref{OstroHam7}. 
\nl
Henceforth we have all the prerequisites at our disposal to generalise 
Ostrogradsky's theorem for singular Lagrangians \cite{pons=OstThe}; we proceed 
gradually, establishing partial results which will be collected eventually. 
\begin{prop} \label{prop:OstCon1}
Let $H$ be the Dirac Hamiltonian \eqref{OstroHam7} associated with the 
singular system. The Hamilton--Dirac equations \wrt the variables 
$q_{k,\beta_k-1}$ (for $\range{\beta_k}{1}{\alpha_k-1}$) are equivalent to 
the Lagrangian constraints \eqref{OstroQ1A}. 
\end{prop}
\begin{proof}
We readily obtain the equations
\begin{equation}
   \dot{q}_{k,\beta_k-1} = \bigl\{ q_{k,\beta_k-1}, H \bigr\}_{\EuD}
                         = \parD{H}{p_{k,\beta_k-1}} 
                         = q_{k,\beta_k},
   \label{OstroHDeq1}    
\end{equation}
which, obviously, are equivalent to the Lagrangian constraints 
\eqref{OstroQ1A}.
\end{proof}
\begin{prop} \label{prop:OstCon2}
Let $H$ be the Dirac Hamiltonian \eqref{OstroHam7} associated with the 
singular system. The Hamilton--Dirac equations \wrt the variables 
$q_{k,\alpha_k-1}$ are equivalent to the definition of the momenta
$p_{k,\alpha_k-1}$.
\end{prop}
\begin{proof}
We expand the equations of motion for the variables $q_{k,\alpha_k-1}$, \viz
\begin{equation}
   \begin{split}
      \dot{q}_{k,\alpha_k-1} 
         &= \bigl\{ q_{k,\alpha_k-1}, H \bigr\}_{\EuD} \\
         &= \bigl\{ q_{k,\alpha_k-1}, H_{\rmr} \bigr\}_{\EuD} +
            \sum_{i=1}^{R}
            \omega^i \bigl\{ q_{k,\alpha_k-1}, \phi_i \bigr\}_{\EuD} \\
         &= \parD{H_{\rmr}}{p_{k,\alpha_k-1}} +
            \sum_{i=1}^{R}
            \omega^i \parD{\phi_i}{p_{k,\alpha_k-1}}.
   \end{split}
   \label{OstroHDeq2}    
\end{equation}
These correspond to the first half of Hamilton's equations for the Lagrangian 
$L$, where the variables $q_{k,\beta_k-1}$ have been frozen. Since the 
functions $H_{\rmr}$ and $\phi_i$ depend solely on the momenta 
$p_{k,\alpha_k-1}$, the Dirac brackets in \eqref{OstroHDeq2} may be viewed as 
the canonical brackets defined in the restricted phase space 
$(q_{k,\alpha_k-1}, p_{k,\alpha_k-1})$. We observe that it is always possible, 
in principle, to express the multipliers $\omega^i$ as functions of the 
coordinates and velocities $(q_{k,\gamma_k}, \dot{q}_{k,\alpha_k-1})$ if we 
solve the equations 
\cite{pons=OstThe,HENNE=QuaGau} 
\begin{equation} \label{OstroHDeq3}    
   \begin{split}
      \dot{q}_{k,\alpha_k-1} 
         &= \parD{H_{\rmr}}{p_{k,\alpha_k-1}} 
            \Bigl( 
               q_{k,\gamma_k}, p_{k,\alpha_k-1} 
               \bigl( q_{k,\gamma_k}, \dot{q}_{k,\alpha_k-1} \bigr)
            \Bigr) \\
         & \quad
           + \sum_{i=1}^{R}
             \omega^i \bigl( q_{k,\gamma_k}, \dot{q}_{k,\alpha_k-1} \bigr)
             \parD{\phi_i}{p_{k,\alpha_k-1}}
             \Bigl( 
                 q_{k,\gamma_k}, p_{k,\alpha_k-1} 
                 \bigl( q_{k,\gamma_k}, \dot{q}_{k,\alpha_k-1} \bigr)
             \Bigr),
   \end{split}
\end{equation}
and provided that all the constraints, $\phi_i \approx 0$ for 
$\range{i}{1}{R}$, are independent.%
\footnote{This corresponds to the \emph{irreducible} case that we assume for 
          simplicity. The reducible case could be treated without much 
          difficulty (see \cite{HENNE=QuaGau}).} 
Moreover, the existence of these extra variables $\omega^i$ enables us to
invert the Legendre transformation defined from $(q,\dot{q})$--space to the
constraint surface in $(q,p,\omega)$--space by means of the one-to-one
correspondence
\begin{equation*}
\begin{cases}
   q_{k,\gamma_k} = q_{k,\gamma_k}, \\
   p_{k,\alpha_k-1} = \parD{L}{\dot{q}_{k,\alpha_k-1}} 
                      \bigl( q_{k,\gamma_k}, \dot{q}_{k,\alpha_k-1} \bigr), \\
   \omega^i = \omega^i \bigl( q_{k,\gamma_k}, \dot{q}_{k,\alpha_k-1} \bigr),
\end{cases}
   \! \iff \!
\begin{cases}
   q_{k,\gamma_k} = q_{k,\gamma_k}, \\
   \dot{q}_{k,\alpha_k-1} = \parD{H_{\rmr}}{p_{k,\alpha_k-1}} +
                            \omega^i \parD{\phi_i}{p_{k,\alpha_k-1}}, \\
   \phi_i \bigl( q_{k,\gamma_k}, p_{k,\alpha_k-1} \bigr) = 0. 
\end{cases}
\end{equation*}
Consequently, equations \eqref{OstroHDeq2} and the definition 
\eqref{OstroConjMom2b} of the momenta $p_{k,\alpha_k-1}$ are equivalent.
\end{proof}
\par%
Before considering the Hamilton--Dirac equations \wrt the conjugate momenta we 
can prove the following useful lemma \cite{pons=OstThe}.
\begin{lem} \label{lem:OstCon}
Under the current assumptions the following identity holds:
\begin{equation} \label{OstroDHam3}
   \parD{H_{\rmr}}{q_{k,\gamma_k}} =
      - \sum_{i=1}^{R}
        \omega^i \bigl( q_{k,\gamma_k}, \dot{q}_{k,\alpha_k-1} \bigr)
           \parD{\phi_i}{q_{k,\gamma_k}} 
      - \parD{L}{q_{k,\gamma_k}}.
\end{equation}
\end{lem}
\begin{proof}
From the above remarks on the invertible character of the Legendre 
transformation we may infer the identity
\begin{equation*} \label{OstroHam8}    
   \begin{split}
      &H_{\rmr} \Bigl( 
                   q_{k,\gamma_k}, 
                   p_{k,\alpha_k-1} 
                      \bigl( 
                         q_{k,\gamma_k}, \dot{q}_{k,\alpha_k-1}
                      \bigr) 
                \Bigr) \\
      &\qquad 
       \equiv 
       \sum_{k=1}^{K} 
       p_{k,\alpha_k-1} \bigl( q_{k,\gamma_k}, \dot{q}_{k,\alpha_k-1} \bigr) 
       \, \dot{q}_{k,\alpha_k-1} - 
       L \bigl( q_{k,\gamma_k}, \dot{q}_{k,\alpha_k-1} \bigr), 
   \end{split}
\end{equation*}
whence we obtain
\begin{equation*}
   \parD{H_{\rmr}}{q_{k,\gamma_k}} =
      \sum_{l=1}^{K}
      \biggl( 
         \dot{q}_{l,\alpha_l-1} - 
         \parD{H_{\rmr}}{p_{l,\alpha_l-1}} 
      \biggr) \,
   \parD{p_{l,\alpha_l-1}}{q_{k,\gamma_k}} - \parD{L}{q_{k,\gamma_k}}.
\end{equation*}
After the application of equations \eqref{OstroHDeq3}, this becomes
\begin{equation}
   \parD{H_{\rmr}}{q_{k,\gamma_k}} =
      \sum_{l=1}^{K} \sum_{i=1}^{R}
      \omega^i \bigl( q_{k,\gamma_k}, \dot{q}_{k,\alpha_k-1} \bigr)
         \parD{\phi_i}{p_{l,\alpha_l-1}} 
         \parD{p_{l,\alpha_l-1}}{q_{k,\gamma_k}} -
      \parD{L}{q_{k,\gamma_k}}.
   \label{OstroDHam2}    
\end{equation}
Since the identity 
\begin{equation}
   \phi_i 
      \Bigl( 
         q_{k,\gamma_k}, 
         p_{k,\alpha_k-1} \bigl( 
                             q_{k,\gamma_k}, \dot{q}_{k,\alpha_k-1} 
                          \bigr)
      \Bigr) \equiv 0 
   \label{OstroHCons2}    
\end{equation}
obviously holds, equation \eqref{OstroDHam2} reduces to the expected result, 
namely
\begin{equation}
   \parD{H_{\rmr}}{q_{k,\gamma_k}} =
      - \sum_{i=1}^{R}
        \omega^i \bigl( q_{k,\gamma_k}, \dot{q}_{k,\alpha_k-1} \bigr)
           \parD{\phi_i}{q_{k,\gamma_k}} 
      - \parD{L}{q_{k,\gamma_k}}.
\end{equation}
\end{proof}
\par%
Consider now the second half of Hamilton's equations.
\begin{prop} \label{prop:OstCon3}
Let $H$ be the Dirac Hamiltonian \eqref{OstroHam7} associated with the 
singular system. The Hamilton--Dirac equations \wrt the conjugate momenta 
$p_{k,\beta_k}$ (for $\range{\beta_k}{1}{\alpha_k-1}$) are equivalent to the 
Ostrogradsky recursion relations \eqref{OstroEL5}.
\end{prop}
\begin{proof}
We firstly write down the equations of motion \wrt the conjugate momenta 
$p_{k,\beta_k}$:
\begin{equation}
   \dot{p}_{k,\beta_k} 
      = \bigl\{ p_{k,\beta_k}, H \bigr\}_{\EuD}
      = \bigl\{ p_{k,\beta_k}, H_{\rmr} \bigr\}_{\EuD} +
        \sum_{i=1}^{R}
        \omega^i \bigl\{ p_{k,\beta_k}, \phi_i \bigr\}_{\EuD} -
        p_{k,\beta_k-1}.
\end{equation}
Owing to Lemma~\ref{lem:OstCon}, this reduces to
\begin{equation}
   \dot{p}_{k,\beta_k} = \parD{L}{q_{k,\beta_k}} - p_{k,\beta_k-1} 
   \quad \text{\normalfont{for}} \; \range{\beta_k}{1}{\alpha_k-1},
   \label{OstroEL7}
\end{equation}
which is equivalent to equation \eqref{OstroEL5}. 
\end{proof}
\par%
We are now at the right stage to prove the Ostrogradsky theorem for 
constrained systems \cite{pons=OstThe}.
\begin{OstCon}
Let $L$ be a \emph{singular} La\-grang\-ian depending on generalised 
coordinates $x_k$ and their time derivatives up to order 
$\alpha_k \geqslant 1$ (for $\range{k}{1}{K}$). Let $\Lagb$ be the associated 
\emph{extended} Lagrangian depending on the auxiliary \dofs 
$q_{k,\gamma_k}=x_k^{(\gamma_k)}$ (for $\range{\gamma_k}{0}{\alpha_k-1}$). The 
\EL equations derived from $\Lagb$ are equivalent to the system of canonical 
equations obtained from the Dirac Hamiltonian
\begin{equation*}
   H = H_{\rmc} + \sum_{i=1}^{R} \omega^i \phi_i.
\end{equation*}
\end{OstCon}
\begin{proof}
Owing to Lemma~\ref{lem:OstCon}, Hamilton's equations for the momenta 
$p_{k,0}$,
\begin{equation}
   \dot{p}_{k,0} 
      = \bigl\{ p_{k,0}, H \bigr\}_{\EuD}
      = \bigl\{ p_{k,0}, H_{\rmr} \bigr\}_{\EuD} +
        \sum_{i=1}^{R}
        \omega^i \bigl\{ p_{k,0}, \phi_i \bigr\}_{\EuD} -
        p_{k,\beta_k-1},
   \label{OstroHDeq5}    
\end{equation}
reduce to
\begin{equation}
   \dot{p}_{k,0} = \parD{L}{q_{k,0}},
   \label{OstroEL8}    
\end{equation}
which are identical to the \EL equations \eqref{OstroEL2}. The conclusion is
then readily inferred owing to Proposition \ref{prop:Ost1}, Proposition 
\ref{prop:Ost2}, Proposition \ref{prop:OstCon1}, Proposition 
\ref{prop:OstCon2}, and Proposition \ref{prop:OstCon3}. 
\end{proof}
\par%
The above results establish the equivalence between the original Lagrangian 
equations and Hamilton's equations derived from the Hamiltonian 
\eqref{OstroHam7}. The details of this equivalence are displayed inside the 
box below. \label{box:GenOstThe}
\begin{center}
\begin{tabular}{|lcl|}
\hline
   $\dot{q}_{k,\beta_k-1} = \bigl\{ q_{k,\beta_k-1}, H \bigr\}_{\EuD}$
      &$\iff$ 
      &$q_{k,\beta_k} = \dot{q}_{k,\beta_k-1}$, \\
      &&{\footnotesize{\textsf{[Lagrangian constraints]}}} \\
   $\dot{q}_{k,\alpha_k-1} = \bigl\{ q_{k,\alpha_k-1}, H \bigr\}_{\EuD}$
      &$\iff$
      &$p_{k,\alpha_k-1} = \parD{L}{\dot{q}_{k,\alpha_k-1}}$, \\
      &&{\footnotesize{\textsf{[Definition of momenta}}} 
        ${\scriptstyle{\mathsf{p_{k,\alpha_k-1}}}}$
        {\small{\textsf{]}}} \\
   $\dot{p}_{k,\beta_k} = \bigl\{ p_{k,\beta_k}, H \bigr\}_{\EuD}$
      &$\iff$ 
      &$p_{k,\beta_k-1} = \parD{L}{q_{k,\beta_k}} -
                          \frac{\rmd}{\rmd t} p_{k,\beta_k}$, \\
      &&{\footnotesize{\textsf{[Recursion relations for momenta}}}
        ${\scriptstyle{\mathsf{p_{k,\beta_k-1}}}}$
        {\small{\textsf{]}}} \\
   $\dot{p}_{k,0} = \bigl\{ p_{k,\beta_k}, H \bigr\}_{\EuD}$
      &$\iff$ 
      &$\frac{\rmd}{\rmd t} p_{k,0} = \parD{L}{q_{k,0}}$. \\
      &&{\footnotesize{\textsf{[\EL equations]}}} \\
\hline
\end{tabular}
\end{center}
\par%
The canonical equations thus include: Lagrangian constraints corresponding to 
the auxiliary \dofs; Ostrogradsky's definition of momenta; and the \EL 
equations of motion. 
\nl
\begin{rem}
If we performed the Legendre transformation on the basis of the definition of 
momenta \eqref{OstroConjMom1}---that is, without introducing an extended 
Lagrangian---, then new constraints would arise: Their explicit form would be 
obtained by requiring that the primary constraints \eqref{OstroHCons1c} be 
preserved in time \cite{saito=DynFor}.
\end{rem}
\par
If we write down the Hamilton--Dirac equations in the following 
\emph{nonnormal} form,
\begin{subequations}
   \label{OstHDNnorm}
   \begin{align} 
      \frac{\rmd q_{k,\gamma_k}}{\rmd t} &=  
         \bigl\{ q_{k,\gamma_k}, H_{\rmc} \bigr\}_{\EuD} + 
         \sum_{i=1}^{R}
         \omega^i \bigl( q_{k,\gamma_k}, \dot{q}_{k,\alpha_k-1} \bigr)
         \bigl\{ q_{k,\gamma_k}, \phi_i \bigr\}_{\EuD}, \\
      \frac{\rmd p_{k,\gamma_k}}{\rmd t} &=  
         \bigl\{ p_{k,\gamma_k}, H_{\rmc} \bigr\}_{\EuD} + 
         \sum_{i=1}^{R}
         \omega^i \bigl( q_{k,\gamma_k}, \dot{q}_{k,\alpha_k-1} \bigr)
         \bigl\{ p_{k,\gamma_k}, \phi_i \bigr\}_{\EuD},
   \end{align}
\end{subequations}
then it is clear that we do not impose any constraint on them---constraints
\eqref{OstroHCons1c} are hidden in equations \eqref{OstroEL7}. The ability to
cast the above equations \eqref{OstHDNnorm} into the standard form
\begin{align*} 
   \frac{\rmd q_{k,\gamma_k}}{\rmd t} & \approx  
      \bigl\{ q_{k,\gamma_k}, H_{\rmc} \bigr\}_{\EuD} + 
      \sum_{i=1}^{R}
      \omega^i (t) \bigl\{ q_{k,\gamma_k}, \phi_i \bigr\}_{\EuD}, \\
   \frac{\rmd p_{k,\gamma_k}}{\rmd t} & \approx  
      \bigl\{ p_{k,\gamma_k}, H_{\rmc} \bigr\}_{\EuD} + 
      \sum_{i=1}^{R}
      \omega^i (t) \bigl\{ p_{k,\gamma_k}, \phi_i \bigr\}_{\EuD},
\end{align*}
where $\omega^i(t)$ are arbitrary functions of time, entails the restriction 
of the trajectories in phase space to the surface that is generated by the 
constraints \eqref{OstroHCons1c}. This requirement leads generally to the
determination of some multipliers and to secondary constraints as well.
\nl
The Hamiltonian formulation of higher-order theories obtained by unifying the
Ostrogradsky method with the Dirac formalism for constrained systems is not
satisfactory when having in prospect the development of a canonical formalism 
for gauge theories; the very reason lies in the fact that the usual 
formulation of gauge theories is given in terms of geometrical quantities like
covariant derivatives or curvature tensors: In particular, the prescription
\eqref{OstroQ1} for introducing the auxiliary \dofs does not take into account 
any of those features present in a gauge theory. Therefore, we present 
hereafter in Subsection~\ref{subsec:OstroSingBL} a generalisation of the 
constrained Ostrogradsky construction that will allow us to bring forth a 
consistent Hamiltonian formulation of \hotg. 

\paragraph*{Remark on an alternative formalism.} \label{Rem:AltFor}

A coherent way of building up a Hamiltonian formalism for theories with 
higher derivatives is provided by the constrained Ostrogradsky method, as 
explained above. Curiously enough, some authors devised an alternative 
formalism, in the specific case of second-order gravitational Lagrangians, the
peculiarity of which is to preclude the occurrence of constraints at the 
Lagrangian level \cite{schmi=ConRel,kaspe=FinHam}. The trick is the following: 
add to the original second-order Lagrangian $L_0(x,\dot{x},\ddot{x})$ a total 
time derivative of a second-order arbitrary function $W(x,\dot{x},\ddot{x})$; 
replace \emph{straight} into the action functional---that is without Lagrange
multipliers---the variables $\ddot{x}$ by new independent \dofs $q$; then fix 
the explicit form of the function $W$ by requiring that variation of the 
action \wrt $q$ yield precisely the relation $\ddot{x}=q$. This skirting 
procedure, which takes advantage of the freedom of adding a total derivative 
to the Lagrangian without altering the equations of motion, is misleading: An 
ambiguity arises due to the presence of terms involving $\rmd \dot{x}/\rmd t$. 
For this reason, the would-be advantage of the recipe---the mere absence of 
constraints in the variational principle---is on the contrary a serious 
drawback in comparison with the constrained Ostrogradsky approach. Even worst, 
this contrived formalism comes to naught in the case of \emph{singular} 
second-order Lagrangians: The treatment of the primary constraints is not
compatible with the choice $\ddot{x}=q$ (see the comment on page~%
\pageref{Com:AltFor}).

\subsection{Generalised constrained Ostrogradsky construction}
\label{subsec:OstroSingBL}

The basic idea of the generalised Ostrogradsky method for constrained higher-%
order systems is to allow for a more general definition of the auxiliary 
Ostrogradsky variables $q_{\gamma_k}$ instead of the standard definition 
\eqref{OstroQ2}. At first sight the general formalism that is given in the 
literature provides a satisfactory treatment \cite{buchb=CanQua,BUCHB=EffAct,
GITMA=QuaFie}. However, as explained below, it is tainted with a small 
technical mistake that renders the method ineffective in general; this 
illustrates how pitfalls may arise when one is building up such an abstract 
theoretical setting without testing it on simple examples or toy-models. 
Nevertheless this general formalism can be suitably adapted, taking into 
account the specific features of the system under investigation; we shall come 
back to this point later (\cfr Subsection~\ref{subsec:Hamfr} on $f(R)$ 
theories of gravity); but, in the meantime, we analyse the generalised 
Ostrogradsky method. 
\nl
{\allowdisplaybreaks
Consider a system with $K$ \dofs $x_k$ for $\range{k}{1}{K}$ and assume that
the associated Lagrangian (with higher derivatives) be given by expression 
\eqref{OstroLag1}. Let us introduce new \emph{independent} variables 
$q_{k,\gamma_k}$ for $\range{\gamma_k}{0}{(\alpha_k-1)}$ in accordance with 
the prescriptions
\begin{subequations}
   \label{OstroBLQ1}    
   \begin{align}
      q_{k,\beta_k} &= Q_{k,\beta_k} 
                          \Bigl( 
                             x_l, \dot{x}_l, \dots, x_l^{(\theta_{kl})}
                          \Bigr), 
                       \quad 
                       \theta_{kl} = \min \bigl(\beta_k, \alpha_l-1 \bigr),
         \label{OstroBLQ1A} \\
      q_{k,0}       &= Q_{k,0} = x_k, 
         \label{OstroBLQ1B}
   \end{align}
\end{subequations}
where $Q_{k,\beta_k}$ are arbitrary functions that may depend on generalised
coordinates $x_l$ and their time derivatives up to order $\theta_{kl}$. We
require that relations \eqref{OstroBLQ1A} be invertible in terms of the 
highest-order time derivatives occurring in the explicit form of the 
functions $Q_{k,\beta_k}$; this requisite entails the conditions
\begin{equation} \label{OstroBLDet1}
   \Delta_{\sigma} = \det \biggl( 
                             \parD{Q_{k,\sigma}}{x_l^{(\sigma)}} 
                          \biggr)
                   \neq 0
                   \quad \text{for} \; \range{\sigma}{1}{N-1},
\end{equation}
where the matrices associated with these nonzero determinants 
$\Delta_{\sigma}$ contain only those elements for which the determinant index
$\sigma$ satisfies $\sigma < N \, (= \sup_k \{\alpha_k\})$ (see Example~%
\ref{ex:OstroBL}). 
\nl
In contrast with the standard definition \eqref{OstroQ1} of the auxiliary 
\dofs, definition \eqref{OstroBLQ1} is more general and it enables one to 
choose specific forms for the functions $Q_{k,\beta_k}$ according to the 
actual characteristics of the system under study. Formally, different choices 
of functions $Q_{k,\beta_k}$ will lead to distinct Hamiltonian formulations; 
in that respect, we will address in the sequel the question of how these 
various formulations are connected with one another (see Proposition~%
\ref{prop:OstroBL2}). 
\nl
We consider now a specific example in order to clarify the practical use of 
prescriptions \eqref{OstroBLQ1} and \eqref{OstroBLDet1}.

\begin{exmp} \label{ex:OstroBL}
Let 
$L = L \bigl( x, \dot{x}, \ddot{x}, x^{(3)},
              y, \dot{y}, \ddot{y}, y^{(3)}, y^{(4)} \bigr)$.
We may infer, from the highest-order time derivatives corresponding to $x$ and 
$y$ occurring in $L$, the values of $\alpha_k$, namely $\alpha_x=3$ and 
$\alpha_y=4$ respectively. We introduce new coordinates according to the rules 
\eqref{OstroBLQ1}, that is
\begin{align*}
   q_{x,0} &= x                        
  &q_{y,0} &= y \\
   q_{x,1} &= Q_{x,1} \bigl( x,\dot{x},y,\dot{y} \bigr)                   
  &q_{y,1} &= Q_{y,1} \bigl( x,\dot{x},y,\dot{y} \bigr) \\
   q_{x,2} &= Q_{x,2} \bigl( x,\dot{x},y,\dot{y},\ddot{y} \bigr)  
  &q_{y,2} &= Q_{y,2} \bigl( x,\dot{x},y,\dot{y},\ddot{y} \bigr) \\ 
 &&q_{y,3} &= Q_{y,3} \bigl( x,\dot{x},y,\dot{y},\ddot{y},y^{(3)} \bigr).
\end{align*}
We then construct the determinants $\Delta_{\sigma}$, fulfilling conditions 
\eqref{OstroBLDet1}: 
\begin{equation*}
   \Delta_1 = \begin{vmatrix}
                 \partial Q_{x,1}/\partial \dot{x}
                &\partial Q_{x,1}/\partial \dot{y} \\
                 \partial Q_{y,1}/\partial \dot{x}
                &\partial Q_{y,1}/\partial \dot{y} 
              \end{vmatrix}, \quad
   \Delta_2 = \begin{vmatrix}
                 \partial Q_{x,2}/\partial \ddot{x}
                &\partial Q_{x,2}/\partial \ddot{y} \\
                 \partial Q_{y,2}/\partial \ddot{x}
                &\partial Q_{y,2}/\partial \ddot{y} 
              \end{vmatrix}, \quad
   \Delta_3 = \parD{Q_{y,3}}{y^{(3)}}.
\end{equation*}
\end{exmp}
\nl%
Owing to conditions \eqref{OstroBLDet1} we may solve equations 
\eqref{OstroBLQ1A} in terms of time derivatives of order $\beta_k$ (for
$\range{\beta_k}{1}{\alpha_k-1}$) of the generalised coordinates; the 
resulting functions are
\begin{equation} \label{OstroBLX}
   x_k^{(\beta_k)} = X_k^{\beta_k} \bigl( 
                                      q_{l,0}, \dots, q_{l,\theta_{kl}}
                                   \bigr), \quad
   \det \biggl( \parD{X_k^{\sigma}}{q_{l,\sigma}} \biggr)
   \neq 0.
\end{equation}
Henceforth we slightly depart from what is found in the literature (see, \eg
\cite{buchb=CanQua}), wherein an erroneous result renders the subsequent 
analysis ineffective in general. Nevertheless let us firstly demonstrate the 
mistake. The authors claim that the first-order time derivatives 
$\dot{q}_{k,\beta_k-1}$ could be expressed in terms of the auxiliary \dofs 
$\dot{q}_{l,\rho_l}$ (for $\range{\rho_l}{0}{\theta_{kl}}$) only. However, in 
general this is not true, for differentiating \wrt time the quantities 
$x_k^{(\sigma_k)}$ for $\range{\sigma_k}{1}{(\alpha_k-2)}$, which are given in 
equation \eqref{OstroBLX}, we obtain the series expansion
\begin{equation}
   X_k^{\sigma_k+1} = \sum_{l=1}^K \sum_{\rho_l=0}^{\gamma_{kl}}
          \parD{X_k^{\sigma_k}}{q_{l,\rho_l}} 
          \dot{q}_{l,\rho_l}, \quad \gamma_{kl} = \min (\sigma_k, \alpha_l-1),
\end{equation}
which contains explicitly first-derivative terms of the form 
$\dot{q}_{l,\alpha_l-1}$ if it ever happens that 
$\sigma_k \geqslant (\alpha_l-1)$. Unfortunately this last inequality 
corresponds to the generic case, as it can more easily be checked on Example~%
\ref{ex:OstroBL}. The only means to remove these first-derivative terms is to 
assume that each coordinate $x_k$ occurring in the Lagrange function $L$ have 
one and the same maximal order $N$ (\ie $\alpha_k=N$ $\forall k$). This 
requirement is not a severe restriction though for we could always add a total 
derivative term to the original Lagrangian \eqref{OstroLag1} without modifying 
the Lagrangian dynamics and such that the above assumption would be satisfied. 
However, we must ensure that distinct Hamiltonian formulations stemming from 
Lagrangians that differ by a total derivative only be equivalent; this is 
achieved with the following result.%
\footnote{For our purpose it is sufficient to consider only one \dof.}
\begin{prop} \label{prop:OstroBL0}
Consider two Lagrangians, $L$ (of order $N$) and $L_0$ (of order $N-1$) that 
differ by a total time derivative, \viz
\begin{equation} 
   L \bigl( x, \dot{x}, \dots, x^{(N)} \bigr) =
      L_0 \bigl( x, \dot{x}, \dots, x^{(N-1)} \bigr) +
      \frac{\rmd}{\rmd t} W \bigl( x, \dot{x}, \dots, x^{(N-1)} \bigr).
\end{equation}
The two distinct Hamiltonian formulations that are constructed on $L$ and 
$L_0$ respectively are canonically equivalent; the appropriate canonical 
transformation is defined by 
\begin{subequations}
   \label{OstroBLCanTra}
   \begin{align} 
      q_{\beta-1} &= q^{(0)}_{\beta-1}, \\
      p_{\beta-1} &= p^{(0)}_{\beta-1} + \parD{W}{q^{(0)}_{\beta-1}},
   \end{align}
\end{subequations}
where $(q_{\beta-1}, p_{\beta-1})$ and 
$\bigl( q^{(0)}_{\beta-1}, p^{(0)}_{\beta-1} \bigr)$ (for 
$\range{\beta}{1}{N-1}$) denote the canonical variables associated with the 
Lagrangians $L$ and $L_0$ respectively. 
\end{prop}
\begin{proof}
We firstly examine the Ostrogradsky formalism for the Lagrangian $L_0$. \\
We introduce auxiliary \dofs and their conjugate momenta through the standard 
recursion relations
\begin{align*} 
   q^{(0)}_0 &= x, &
   p^{(0)}_{N-2} &= \parD{L_0}{\dot{q}^{(0)}_{N-2}},\\
   q^{(0)}_{\sigma} &= \dot{q}^{(0)}_{\sigma-1}, &
   p^{(0)}_{\sigma-1} &= \parD{L_0}{q^{(0)}_{\sigma}} -
                         \frac{\rmd}{\rmd t} p^{(0)}_{\sigma}
      \quad \text{\normalfont{for}} \; \range{\sigma}{1}{N-2}.
\end{align*}
We then obtain the canonical Hamiltonian corresponding to $L_0$, namely
\begin{equation*}
   H^{(0)}_{\rmc} \Bigl( q^{(0)}_{\beta-1}, p^{(0)}_{\beta-1} \Bigr) =
      \sum_{\sigma=1}^{N-2} q^{(0)}_{\sigma} \, \naught{p}_{\sigma-1} +
      \dot{q}^{(0)}_{N-2} \, p^{(0)}_{N-2} -
      L_0 \Bigl( q^{(0)}_{\beta-1}, \dot{q}^{(0)}_{N-2} \Bigr).
\end{equation*}
Now we turn to the Ostrogradsky formalism for the Lagrangian $L$. \\
Once again we introduce auxiliary \dofs and their conjugate momenta through 
the standard recursion relations
\begin{align*} 
   q_0 &= x, &
   p_{N-1} &= \parD{W}{q_{N-1}},\\
   q_{\beta} &= \dot{q}_{\beta-1}, &
   p_{\beta-1} &= \parD{L}{q_{\beta}} - \frac{\rmd}{\rmd t} p_{\beta}
      \quad \text{\normalfont{for}} \; \range{\beta}{1}{N-1}.
\end{align*}
We then obtain the canonical Hamiltonian corresponding to $L$, namely
\begin{equation*}
   H_{\rmc} \bigl( q_{\gamma}, p_{\gamma} \bigr) =
      \sum_{\beta=1}^{N-1}  
         \biggl( p_{\beta-1} -  \parD{W}{q_{\beta-1}} \biggr) q_{\beta} - 
      L_0 (q_{\gamma}) \quad \text{\normalfont{for}} \; 
         \range{\gamma}{0}{N-1}.
\end{equation*}
Note that the restricted Hamiltonian \eqref{OstroHam3} does not depend here on 
the momentum $p_{N-1}$; hence the pair of canonical variables 
$(q_{N-1}, p_{N-1})$ are associated with spurious physical \dofs. It is then 
straightforward to see that the canonical transformation \eqref{OstroBLCanTra} 
turns the Hamiltonian $H_{\rmc}$ into the Hamiltonian $H^{(0)}_{\rmc}$.
\end{proof}
\nl
For completeness we write down the explicit form of the nontrivial primary 
constraints of the $N^{\text{\textrm{th}}}$-order theory,  
\begin{subequations}
   \label{OstroBLPriCons}
   \begin{align} 
      \phi_{N-1} &= p_{N-1} - \parD{W}{q_{N-1}} \approx 0, \\
      \phi_{N-2} &= p_{N-2} - \parD{L_0}{q_{N-1}} - \parD{W}{q_{N-2}} 
                    \approx 0.
   \end{align}
\end{subequations}
The former constraint is first class and does not generate any secondary 
constraint; it corresponds to the gauge freedom associated with the choice of
the function $W$, especially with respect to its dependence on the variable
$q_{N-1}$. The last constraint arises from the singular character of the
original Lagrangian $L_0$; its preservation in time generates secondary
constraints which are classified as usual into first-class and second-class
constraints. Furthermore it can be easily shown that the Dirac bracket
structure is preserved under the canonical transformation 
\eqref{OstroBLCanTra} and that both Hamiltonian formulations lead to 
equivalent quantum counterparts via canonical \cite{kamin=QuaMec} or path-%
integral quantisation methods \cite{gross=EquHam}. 

\paragraph*{Further remark on an alternative formalism.} \label{Com:AltFor}

Proposition~\ref{prop:OstroBL0} sheds a new light upon the alternative 
formalism briefly discussed on page~\pageref{Rem:AltFor}. Kasper shows that 
this formalism is canonically equivalent to the Ostrogradsky construction 
performed on the original second-order Lagrangian $L_0$ \cite{kaspe=FinHam}. 
This relationship is even more manifest when one applies the Ostrogradsky 
method to the third-order Lagrangian involving the total time derivative of a 
specific second-order function, the explicit form of which is chosen in 
accordance with the aforementioned recipe. Indeed, the relation 
\eqref{OstroBLCanTra} enables us to write down the corresponding Ostrogradsky 
momenta, $p_0$ and $p_1$, in terms of the original momenta, $p_0^{(0)}$ and 
$p_1^{(0)}$, and compare them with the `alternative' momenta, $\pi_0$ and 
$\pi_1$. We find the relations
\begin{align*}
   p_0 &= p_0^{(0)} + \parD{W}{q^{(0)}_0} \equiv \pi_0, \\
   p_1 &= p_1^{(0)} + \parD{W}{q^{(0)}_1} 
        = \parD{L_0}{\ddot{x}} + \parD{W}{\dot{x}} \equiv 0, \\
\intertext{as well as the definition of momentum $p_2$,}
   p_2 &= \parD{W}{\ddot{x}} \equiv \pi_1.
\end{align*}
The vanishing of the momentum $p_1$ is a direct consequence of the hypothesis 
that the original Lagrangian be regular and of the condition that has to be 
fulfilled by the function $W$ to render the alternative formalism meaningful. 
In other words, the alternative formalism has been defined to preclude the 
occurrence of the primary constraints \eqref{OstroBLPriCons}. If the standard 
constraint analysis is performed within the Ostrogradsky formalism, then the 
above relations for $p_1$ and $p_2$ are second-class constraints. 
\nl
In agreement with the previous discussion we thus assume that the maximal 
order of variables $x_k$ is one and the same, that is $\alpha_k=N$ 
$\forall k$. The auxiliary variables are introduced with the definition 
\eqref{OstroBLQ1}---with $\theta_{kl} \equiv \beta_k$---such that conditions 
\eqref{OstroBLDet1} be fulfilled. Thus, differentiating equations 
\eqref{OstroBLX} \wrt time we can express the time derivative of variables 
$q_{k,\beta_k-1}$ as
\begin{equation} \label{OstroBLQ2}
   \dot{q}_{k,\beta_k-1} = \EuQ_{k,\beta_k} 
                           \bigl( 
                              q_{l,0}, \dots, q_{l,\beta_k}  
                           \bigr), 
\end{equation}
where the functions $\EuQ_{k,\beta_k}$ can be determined with the recursion
relations
\begin{subequations} 
   \label{OstroBLEuQ}
   \begin{align} 
      X_k^1            &= \EuQ_{k,1}, \\
      X_k^{\sigma_k+1} &= \sum_{l=1}^K \sum_{\rho_l=0}^{\sigma_k}
                          \parD{X_k^{\sigma_k}}{q_{l,\rho_l}} 
                          \EuQ_{l,\rho_l+1}  
                          \quad \text{for} \; \range{\sigma_k}{1}{N-2}.
   \end{align}
\end{subequations}
We introduce \emph{velocities} $v_{k,N-1}$ with the definition
\begin{equation} \label{OstroBLVelo}
   v_{k,N-1} := \dot{q}_{k,N-1}.
\end{equation}
\begin{rem} \label{remvelo}
At first sight, it could seem unnecessary to add more \dofs to the theory:  
Indeed, we did not need to introduce such velocities through the constrained 
Ostrogradsky construction (\cfr Subsection~\ref{subsec:OstroSing}). In the 
literature the consideration of these new variables is merely a loophole: When 
one adopts definition \eqref{OstroBLVelo} the equations of motion for the
variables $q_{k,N-1}$ simply become Lagrangian constraints; the equivalence 
with the definition of momenta $p_{k,N-1}$ is lost and primary constraints 
$\phi_i \approx 0$ do not occur at this stage. So to speak, one eludes the 
discussion given in the proof of Proposition~\ref{prop:OstCon2}, that is the 
discussion on the invertibility of the Legendre transformation associated with 
the presence of primary constraints $\phi_i \approx 0$. The resulting 
formalisms are hybrid-like: At one time velocities are regarded as independent 
\dofs; at another time they are viewed as the first-time derivatives of 
variables $q_{k,N-1}$. This is confusing and this could lead to mistaken 
interpretations (see \cite{buchb=CanQua,BUCHB=EffAct,GITMA=QuaFie}). For that 
very reason we give a more throughout analysis, hereafter.
\end{rem}
\nl
Assuming that we adopt the definition \eqref{OstroBLVelo} for the velocities 
$v_{k,N-1}$ we can prove the following result \cite{buchb=CanQua}.
\begin{prop} \label{prop:OstroBL1}
Under the current assumptions the functional form of the highest-order time 
derivatives of the generalised coordinates $x_k$ is given by
\begin{equation} \label{OstroBLXkn}
   x_k^{(N)} = X_k^N \bigl( q_{l,0}, \dots, q_{l,N-1}, v_{l,N-1} \bigr), 
   \quad 
   \det \biggl( \parD{X_k^N}{v_{l,N-1}} \biggr) \neq 0.
\end{equation}
\end{prop}
\begin{proof}
Taking into account equations \eqref{OstroBLQ1}, \eqref{OstroBLX}, and
\eqref{OstroBLQ2} we obtain successively 
\begin{equation*}
   \begin{split} 
      x_k^{(N)} &= \dot{x}_k^{(N-1)} 
                 = \sum_{l=1}^K \sum_{\rho_l=0}^{N-1}
                   \parD{X_k^{N-1}}{q_{l,\rho_l}} 
                   \dot{q}_{l,\rho_l}, \\
                &= \sum_{l=1}^K \sum_{\rho_l=0}^{N-2}
                   \parD{X_k^{N-1}}{q_{l,\rho_l}} 
                   \EuQ_{l,\rho_l+1} +
                   \sum_{l=1}^K \parD{X_k^{N-1}}{q_{l,N-1}} \dot{q}_{l,N-1},
   \end{split}
\end{equation*}
where the last right-hand side may be written as a function $X_k^N$, which 
could be formally given by equation \eqref{OstroBLXkn}.
\end{proof}
We now interpret equations \eqref{OstroBLQ2} and \eqref{OstroBLVelo} as 
Lagrangian constraints and we replace the original (higher-order) Lagrange 
function $L$ by the extended (first-order) Lagrangian $\Lagb$,
\begin{equation} \label{OstroBLLagb}
   \begin{split}
      \Lagb \bigl( q, \dot{q}, \lambda, v, \mu \bigr) 
         = L \bigl( q, v \bigr) 
         &+ \sum_{k=1}^K \sum_{\beta_k=1}^{N-1}
               \bigl( \dot{q}_{k,\beta_k-1} - \EuQ_{k,\beta_k} \bigr)
               \lambda_{k,\beta_k} \\
         &+ \sum_{k=1}^K 
               \bigl( \dot{q}_{k,N-1} - v_{k,N-1} \bigr)
               \mu_{k,N-1},
   \end{split}
\end{equation}
with Lagrange multipliers $\lambda_{k,\beta_k}$ and $\mu_{k,N-1}$, so as to 
recover the interpretation in terms of the original set of coordinates. In 
close analogy with Proposition~\ref{prop:Ost1} the \EL equations derived from 
the extended Lagrangian $\Lagb$ are equivalent to the \EL equations obtained 
from the original Lagrangian $L$. (The proof sheds no new light upon the 
subsequent analysis.) 
\nl
The momenta canonically conjugate to the independent \dofs are defined
respectively by
\begin{subequations}
   \label{OstroBLConjMom1}    
   \begin{align}
      p_{k,\beta_k-1}             &:= \parD{\Lagb}{\dot{q}_{k,\beta_k-1}}  
                                    = \lambda_{k,\beta_k}, 
                                      \label{OstroBLConjMom1a} \\  
      p_{k,N-1}                   &:= \parD{\Lagb}{\dot{q}_{k,N-1}}  
                                    = \mu_{k,N-1}, 
                                      \label{OstroBLConjMom1b} \\  
      \pi^{(\lambda)}_{k,\beta_k} &:= \parD{\Lagb}{\dot{\lambda}_{k,\beta_k}}  
                                    = 0, \label{OstroBLConjMom1c} \\  
      \pi^{(\mu)}_{k,N-1}         &:= \parD{\Lagb}{\dot{\mu}_{k,N-1}}  
                                    = 0, \label{OstroBLConjMom1d} \\   
      \pi^{(v)}_{k,N-1}           &:= \parD{\Lagb}{\dot{v}_{k,N-1}}  
                                    = 0. \label{OstroBLConjMom1e} 
   \end{align}
\end{subequations}
We thus have five sets of Lagrangian constraints that we denote as 
$\varphi_{k,\beta_k-1} \approx 0$, $\varphi_{k,N-1} \approx 0$, 
$\pi^{(\lambda)}_{k,\beta_k} \approx 0$, $\pi^{(\mu)}_{k,N-1} \approx 0$, and 
$\pi^{(v)}_{k,N-1} \approx 0$ respectively. 
\nl
The canonical Hamiltonian of the system is readily obtained, \viz
\begin{equation} \label{OstroBLHam1}    
   H_{\rmc} \bigl( q, p, v \bigr) = 
      \sum_{k=1}^K \sum_{\beta_k=1}^{N-1}
         p_{k,\beta_k-1} \, \EuQ_{k,\beta_k} +
      \sum_{k=1}^K p_{k,N-1} \, v_{k,N-1} -
      L \bigl( q, v \bigr),
\end{equation}
where the constraints in the extended Lagrangian \eqref{OstroBLLagb} have been
used. 
\nl
Before resorting to the Dirac analysis of the system with canonical 
Hamiltonian \eqref{OstroBLHam1} we must firstly identify the primary 
constraints that characterise the singular theory under study. In contrast 
with the standard treatment---the formalism without the extra velocities 
$v_{k,N-1}$---, these primary constraints do not occur in definition 
\eqref{OstroBLConjMom1b} of the momenta $p_{k,N-1}$. Instead, they occur in 
the canonical equations of motion for the momenta $\pi^{(v)}$, derived from 
the canonical Hamiltonian \eqref{OstroBLHam1}, \viz
\begin{equation} \label{OstroBLCanEqPi}   
   \dot{\pi}^{(v)}_{k,N-1} = - \parD{H_{\rmc}}{v_{k,N-1}}  
                           = \parD{L}{v_{k,N-1}} - p_{k,N-1} \approx 0,
\end{equation}
where we have enforced the preservation in time of the constraint 
\eqref{OstroBLConjMom1e}. In the aforementioned hybrid formalism no rigorous 
constraint analysis is performed and equation \eqref{OstroBLCanEqPi} is used 
to define the primary constraints $\phi_i \bigl( q,p \bigr) \approx 0$. 
However, if the velocities are considered as independent variables---not 
merely a handy notation---, then the above interpretation of equation 
\eqref{OstroBLCanEqPi} is misleading. In other words, to ensure consistency of 
the formalism one must either refrain oneself from introducing velocities $v$ 
as additional---though spurious---\dofs or one has to perform the standard 
constraint analysis, assuming that these velocities are independent---though 
constrained---\dofs. Choosing the first option we are brought back to the 
analysis made in Subsection~\ref{subsec:OstroSing}; adopting the second we 
must consider the Dirac Hamiltonian to be given by
\begin{equation} \label{OstroBLHam2}    
   \begin{split}
      H_{\EuD} := H_{\rmc} 
               &+ \sum_{k=1}^{K} \sum_{\beta_k=1}^{N-1}
                     \Bigl(
                        \eta^{k,\beta_k-1} \, \varphi_{k,\beta_k-1} +
                        \xi^{k,\beta_k}_{(\lambda)} \, 
                           \pi^{(\lambda)}_{k,\beta_k} 
                     \Bigr) \\
               &+ \sum_{k=1}^{K}
                     \Bigl(
                        \eta^{k,N-1} \, \varphi_{k,N-1} +
                        \xi^{k,N-1}_{(\mu)} \, \pi^{(\mu)}_{k,N-1} +
                        \xi^{k,N-1}_{(v)} \, \pi^{(v)}_{k,N-1} 
                     \Bigr),
   \end{split}
\end{equation}
with Lagrange multipliers $\eta$ and $\xi$ associated with the primary 
constraints. 
\nl
Enforcing preservation in time of the primary constraints we obtain
\begin{subequations} \label{OstroBLMult1}    
   \begin{align}
      \varphi_{k,\beta_k-1} \approx 0 
         &\tcons
      \xi^{k,\beta_k}_{(\lambda)} = 
         \parD{L}{q_{k,\beta_k-1}} - 
         \sum_{l=1}^K \sum_{\beta_l=1}^{N-1}
            \biggl( \parD{\EuQ_{l,\beta_l}}{q_{k,\beta_k-1}} \biggr) 
            \lambda_{l,\beta_l},  
      \label{OstroBLMult1a} \\  
      \varphi_{k,N-1} \approx 0 
         &\tcons
      \xi^{k,N-1}_{(\mu)} = 
         \parD{L}{q_{k,N-1}} - 
         \sum_{l=1}^K \sum_{\beta_l=1}^{N-1}
            \biggl( \parD{\EuQ_{l,\beta_l}}{q_{k,N-1}} \biggr) 
            \lambda_{l,\beta_l},  
      \label{OstroBLMult1b} \\  
\intertext{and}
      \pi^{(\lambda)}_{k,\beta_k} \approx 0 
         &\tcons
      \eta^{k,\beta_k-1} = 0, 
      \label{OstroBLMult1c} \\ 
      \pi^{(\mu)}_{k,N-1}  \approx 0
         &\tcons
      \eta^{k,N-1} = 0, 
      \label{OstroBLMult1d} \\ 
      \pi^{(v)}_{k,N-1} \approx 0
         & \tcons
      \chi^{(v)}_{k,N-1} := \parD{L}{v_{k,N-1}} - p_{k,N-1} \approx 0.
      \label{OstroBLMult1e}   
   \end{align}
\end{subequations}
Some multipliers are thus determined through equations 
\eqref{OstroBLMult1a}--\eqref{OstroBLMult1d} while the secondary constraints 
$\chi^{(v)}_{k,N-1}$ do arise from equation \eqref{OstroBLMult1e}. Note that 
all constraints but those associated with the velocities are second class; we 
can thus coherently remove them from the formalism provided that we define the
appropriate Dirac bracket. On the other hand, time-evolution of the secondary 
constraints $\chi^{(v)}_{k,N-1}$ yields the equation
\begin{equation} \label{OstroBLMult2}
   \begin{split}
      \chi^{(v)}_{k,N-1} \approx 0 \tcons 
         &\sum_{l=1}^K  
             \biggl( 
                \frac{\partial^2 L}{\partial v_{l,N-1} \partial v_{k,N-1}}
             \biggr) \xi^{l,N-1}_{(v)} +
          \sum_{l=1}^K \sum_{\beta_l=1}^{N-1}  
             \biggl( 
                \frac{\partial^2 L}
                     {\partial q_{l,\beta_l-1} \partial v_{k,N-1}}
             \biggr) \EuQ_{l,\beta_l} \\
         &\quad +
          \sum_{l=1}^K  
             \biggl( 
                \frac{\partial^2 L}{\partial q_{l,N-1} \partial q_{k,N-1}}
             \biggr) v_{l,N-1} -
          \xi^{k,N-1}_{(\mu)} = 0,
   \end{split} \raisetag{26pt}
\end{equation}
which enables to fix $K-r$ multipliers $\xi_{(v)}$ only, on account of the 
fact that the rank of the Hessian matrix, as given by equation 
\eqref{OstroRank}, is expressed in terms of the velocities $v_{k,N-1}$ by
\begin{equation} 
   \rank \Biggl( 
            \frac{\partial^2 L \bigl( q, v \bigr)}
                 {\partial v_{k,N-1} \partial v_{l,N-1}} 
         \Biggr) =
   K-r. 
\end{equation}
\nl
At this stage one may work out the standard consistency algorithm of the Dirac 
method, ending up eventually with a complete set of primary and secondary 
constraints, which can be classified into first-class and second-class 
constraints depending on the specific features of the theory.
\begin{rem}
In contradistinction to the hybrid approach \cite{buchb=CanQua} equation 
\eqref{OstroBLMult1e} defines constraints, be the system regular or not.
\end{rem}
\nl
This is now the right stage to address the aforementioned issue on the 
possible connection between distinct Hamiltonian formulations stemming from 
different choices of the arbitrary functions $Q_{k,\beta_k}$; the following 
proposition provides us with the appropriate answer \cite{GITMA=QuaFie}.
\begin{prop} \label{prop:OstroBL2}
The Hamiltonian formulations developed from one and the same Lagrangian $L$ by
different means of introducing the auxiliary \dofs in the generalised
Ostrogradsky method are canonically equivalent.
\end{prop}
\begin{proof}
It is sufficient to prove that an arbitrary choice of the functions 
$Q_{k,\beta_k}$ leads to a Hamiltonian formulation that is canonically 
equivalent to the constrained Ostrogradsky Hamiltonian formulation of 
Subsection~\ref{subsec:OstroSing}. \\
We consider the standard Ostrogradsky variables 
$(q_{k,\gamma_k}, p_{k,\gamma_k})$ defined as previously by 
\begin{align*}
   q_{k,0}         &= x_k, &
   q_{k,\beta_k}   &= x_k^{(\beta_k)}, \\
   p_{k,N-1}       &= \parD{L}{\dot{q}_{k,N-1}}, &
   p_{k,\beta_k-1} &= \parD{L}{q_{k,\beta_k}} - \dot{p}_{k,\beta_k}. 
\end{align*}
We introduce the generalised Ostrogradsky conjugate pairs 
$(Q_{k,\gamma_k},P_{k,\gamma_k})$ with the following prescriptions: Firstly, 
the auxiliary \dofs are defined by
\begin{align*}
   x_k             &= Q_{k,0}, \\
   x_k^{(\beta_k)} &= X_k^{\beta_k} 
                      \bigl( Q_{l,0}, \dots, Q_{l,\beta_k} \bigr);
\end{align*}
then, their conjugate momenta are determined through the recursion relations
\begin{align*}
   &P_{k,N-1} = \parD{L}{\dot{Q}_{k,N-1}}, \\
   &\sum_{l=1}^{K} \sum_{\beta_l=1}^{N-1}
       \parD{\EuQ_{l,\beta_l}}{Q_{k,\beta_k}} P_{l,\beta_l-1} = 
    \parD{L}{Q_{k,\beta_k}} - \dot{P}_{k,\beta_k}, 
\end{align*}
which generalise relations \eqref{OstroEL5}. (Note that we do not introduce 
the velocities $v_{k,N-1}$ to keep the comparison between both formalisms more 
transparent.) \\
We try to find the generating function of the canonical transformation that
maps variables $(Q,P)$ onto variables $(q,p)$ \cite{SUDAR=ClaDyn}. On account 
of the above definitions of variables $q_{k,\gamma_k}$ the appropriate choice 
consists in taking the type--2 generating function $F_2 \bigl( Q, p \bigr)$ 
that is defined by
\begin{equation*}
   F_2 \bigl( Q, p \bigr) = \sum_{k=1}^{K} p_{k,0} \, Q_{k,0} +
                            \sum_{k=1}^{K} \sum_{\beta_k=1}^{N-1}
                               p_{k,\beta_k} \, X_k^{\beta_k} \bigl( Q \bigr). 
\end{equation*}
Indeed, from that specific form we obtain the relationship between variables
$q_{k,\gamma_k}$ and $Q_{k,\gamma_k}$, \viz
\begin{align*}
   q_{k,0}       &= \parD{F_2}{p_{k,0}} = Q_{k,0}, \\
   q_{k,\beta_k} &= \parD{F_2}{p_{k,\beta_k}} = X_k^{\beta_k}
                                                \bigl( Q \bigr).
\end{align*}
On the other hand, the conjugate momenta are transformed as
\begin{subequations}
   \begin{align}
      P_{k,0}       &= \parD{F_2}{Q_{k,0}} 
                     = p_{k,0} + 
                       \sum_{l=1}^K \sum_{\beta_l=1}^{N-1}
                          \parD{X_l^{\beta_l}}{Q_{k,0}} \, p_{l,\beta_l}, 
         \label{OstroBLMomPa} \\
      P_{k,\beta_k} &= \parD{F_2}{Q_{k,\beta_k}} 
                     = \sum_{l=1}^K \sum_{\beta_l=1}^{N-1}
                          \parD{X_l^{\beta_l}}{Q_{k,\beta_k}} \, 
                          p_{l,\beta_l}.
         \label{OstroBLMomPb}
   \end{align}
\end{subequations}
Now we must check that the canonical transformation generated by 
$F_2 \bigl( Q, p \bigr)$ be preserving the canonical Hamiltonian, \viz
\begin{equation*}
   H_{\rmc} \bigl( q, p \bigr) \equiv H_{\rmc} \bigl( Q, P \bigr) 
      \Bigr\rvert_{Q,P \rightarrow q,p}
\end{equation*}
We start with the canonical Hamiltonian \eqref{OstroBLHam1} obtained in the 
generalised Ostrogradsky method, namely
\begin{equation*}
   \begin{split}     
      H_{\rmc} \bigl( Q, P \bigr) 
         &= \sum_{k=1}^K 
               P_{k,0} \, \EuQ_{k,1} +
            \sum_{k=1}^K \sum_{\sigma_k=1}^{N-2} 
               P_{k,\sigma_k} \, \EuQ_{k,\sigma_k+1} \\
         &\quad \qquad + 
          \sum_{k=1}^K 
            P_{k,N-1} \, \dot{Q}_{k,N-1} -
          L \bigl( Q, \dot{Q} \bigr).
   \end{split}
\end{equation*}
Utilising relation \eqref{OstroBLMomPa} we expand the first term of 
$H_{\rmc}$, which is 
\begin{equation*}     
   \sum_{k=1}^K P_{k,0} \, \EuQ_{k,1} =
      \sum_{k=1}^K p_{k,0} \, \EuQ_{k,1} +
      \sum_{k,l=1}^K \sum_{\beta_l=1}^{N-1} 
         \parD{X_l^{\beta_l}}{Q_{k,0}} p_{l,\beta_l} \, \EuQ_{k,1}. 
\end{equation*}
Making use of relation \eqref{OstroBLMomPb} we find for the second term of
$H_{\rmc}$ the series expansion
\begin{equation*}     
   \sum_{k=1}^K \sum_{\sigma_k=1}^{N-2} 
      P_{k,\sigma_k} \, \EuQ_{k,\sigma_k+1} =
   \sum_{k,l=1}^K \sum_{\beta_l=1}^{N-1} \sum_{\sigma_k=1}^{N-2}
      \parD{X_l^{\beta_l}}{Q_{k,\sigma_k}} \,
      p_{l,\beta_l} \, \EuQ_{k,\sigma_k+1}. 
\end{equation*}
Repeating once again this procedure for the third term of $H_{\rmc}$ we obtain
\begin{equation*}     
   \begin{split}
      \sum_{k=1}^K P_{k,N-1} \, \dot{Q}_{k,N-1} 
         &= \sum_{k,l=1}^K \sum_{\beta_l=1}^{N-1} 
            \parD{X_l^{\beta_l}}{Q_{k,N-1}} \,
            p_{l,\beta_l} \, \dot{Q}_{k,N-1} \\
         &= \sum_{k,l=1}^K  
            \parD{X_l^{N-1}}{Q_{k,N-1}} \, p_{l,N-1} \, \dot{Q}_{k,N-1}.
   \end{split}
\end{equation*}
In agreement with the proof of Proposition~\ref{prop:OstroBL1} we derive 
\begin{equation*}
   \dot{q}_k^{(N-1)} = \sum_{l=1}^K \sum_{\rho_l=0}^{N-2}
                          \parD{X_k^{N-1}}{Q_{l,\rho_l}} \,
                          \EuQ_{l,\rho_l+1} +
                       \sum_{l=1}^K 
                          \parD{X_k^{N-1}}{Q_{l,N-1}} \, \dot{Q}_{l,N-1},
\end{equation*}
which enables us to rewrite the third term of $H_{\rmc}$ as
\begin{equation*}     
   \sum_{k=1}^K P_{k,N-1} \, \dot{Q}_{k,N-1} =
   \sum_{k=1}^K p_{k,N-1} \, \dot{q}_{k,N-1} -
   \sum_{k,l=1}^K \sum_{\rho_l=0}^{N-2} 
      \parD{X_k^{N-1}}{Q_{l,\rho_l}} \, p_{k,N-1} \, \EuQ_{l,\rho_l+1}.
\end{equation*}
Summing up all these intermediate results we obtain for the canonical 
Hamiltonian $H_{\rmc}$ the expression
\begin{equation*}     
   \begin{split}
      H_{\rmc} \bigl( Q, P \bigr) 
         &= \sum_{k=1}^K p_{k,0} \, \EuQ_{k,1} +
            \sum_{k=1}^K p_{k,N-1} \, \dot{q}_{k,N-1} -
            L \bigl( q, \dot{q} \bigr) \\
         &\quad +
          \sum_{k,l=1}^K \sum_{\beta_l=1}^{N-1} \sum_{\rho_k=0}^{N-2}
          \parD{X_l^{\beta_l}}{Q_{k,\rho_k}} \,
          p_{l,\beta_l} \, \EuQ_{k,\rho_k+1} -
          \sum_{k,l=1}^K \sum_{\rho_l=0}^{N-2} 
          \parD{X_k^{N-1}}{Q_{l,\rho_l}} \, p_{k,N-1} \, \EuQ_{l,\rho_l+1}.
   \end{split}
\end{equation*}
On account of equations \eqref{OstroBLEuQ} it is not difficult to show that
the last two terms of the above expression simplify to
\begin{equation*}
   \sum_{k=1}^K \sum_{\sigma_k=1}^{N-2} p_{k,\sigma_k} \, X_k^{\sigma_k+1}.
\end{equation*}
Hence the canonical Hamiltonian $H_{\rmc}$ reduces to 
\begin{equation*}     
   H_{\rmc} \bigl( q, p \bigr) = 
     \sum_{k=1}^K \sum_{\beta_k=1}^{N-1} p_{k,\beta_k-1} \, q_{\beta_k} +
     \sum_{k=1}^K p_{k,N-1} \, \dot{q}_{k,N-1} -
     L \bigl( q, \dot{q} \bigr),
\end{equation*}
which coincides with the canonical Hamiltonian \eqref{OstroHam5} of the 
standard Ostrogradsky construction.    
\end{proof}
}

\section{Higher-order theories of gravity}
\label{sec:HigOrd}

\subsection{A short summary of canonical general relativity}
\label{subsec:CanGR}

\subsubsection{Three-plus-one splitting of space-time}

Because \gr is an already parameterised theory it seems natural to `de-%
parameterise' it so as to make the constraints manifest and proceed to its 
canonical formulation (\cfr the discussion on page~\pageref{subsub:GenCov}). 
The celebrated path to achieve this programme consists in foliating space-time 
into three-dimensional spacelike hypersurfaces parameterised by a global time 
function.%
\footnote{There are numerous reviews on the $3+1$--splitting of space-time, 
          which was firstly used in the development of the \adm formalism; one 
          can pick out, \eg \cite{arnow=DynGen,brill=QuaGen,MISNE=Gra,%
          macca=QuaCos,kucha=CanQua,quere=ForHam}.}
There are, so to speak, two opposite ways of looking at the reliability of the 
$3+1$--splitting. The first, which is the most convenient to adopt, advocates
its use for any field theory because it yields naturally the determination of
the \dofs, the constraints, the gauge freedom, and the field evolution 
equations \cite{isenb=CanGra}; in \gr, in particular, it provides a good 
insight into the nature of constraints and it simplifies the action principle 
and the initial-value problem \cite{brill=QuaGen}. The second point of view 
stems from the idea that the $3+1$--splitting seems to be contrary to the 
whole spirit of \gr. Furthermore this splitting restricts the topology of 
space-time to be the product of the real line with some three-dimensional 
manifold. In regard to quantum gravity this inhibition is not welcomed for one 
would like to allow all possible topologies of space-time 
\cite{hawki=PatInt1}. Henceforth we adopt the first, quite conservative, point 
of view. 
\nl
The $3+1$--splitting of space-time relies on the following important result 
(see \cite[Chapter~8]{WALD=GenRel}).
\begin{thm}
Let $\EuM$ be a time-orientable globally hyperbolic space-time endowed with a
Lorentzian metric $g_{ab}$ of signature $(- + + +)$. Then 
$\bigl( \EuM, g_{ab} \bigr)$ is stably causal. Furthermore a global time 
function $t$ can be chosen such that each surface of constant $t$ is a Cauchy
surface. Thus $\EuM$ can be foliated by Cauchy surfaces $\Sigma_t$ and the
topology of $\EuM$ is $\mathbb{R} \times \Sigma$, where $\Sigma$ denotes any
Cauchy surface.
\end{thm}
\paragraph*{\adm variables.}

The basic geometric data of this decomposition are:
\begin{itemize} 
   \item[(\texttt{I})]
      the \emph{intrinsic} geometry of the hypersurfaces $\Sigma_t$, described 
      by the induced three-dimensional Riemannian metric $h_{ab}$ on each 
      $\Sigma$, which is defined by the formula
      \begin{equation}
         h_{ab} = g_{ab} + n_a n_b,
      \end{equation}
      where $n^a$ denotes the unit normal vector field to the hypersurfaces 
      $\Sigma_t$ (note that $h_a^{\ b}$ plays the r\^ole of a projection 
      operator from the tangent space to $\EuM$ onto the tangent space to
      $\Sigma$; $h_a^{\ b}: T_P \EuM \mapsto T_P \Sigma$); 
   \item[(\texttt{II})]
      the way one goes from one hypersurface to the other, determined by the 
      \emph{lapse function} $N$ and the \emph{shift vector} $N^a$ which
      enable to decompose a `time-flow' vector field $t^a$ on $\EuM$ 
      satisfying $t^a \nabla_a t = 1$ into its parts normal and tangential to 
      $\Sigma$, \viz
      \begin{subequations}
         \begin{align}
            N   &= - t^a n_a = \Bigl( n^a \nabla_a t \Bigr)^{-1}, \\ 
            N_a &= h_{ab} t^b
         \end{align}
      \end{subequations}
      ($N$ measures the rate of flow of proper time \wrt coordinate time as 
      one moves normally to $\Sigma$, whereas $N^a$ measures the `shift' 
      tangential to $\Sigma$ contained in the vector field $t^a$); 
   \item[(\texttt{III})]
      the way each $\Sigma$ is embedded in $\bigl( \EuM, g_{ab} \bigr)$, 
      provided by the \emph{extrinsic curvature} tensor $K_{ab}$ which is 
      defined by
      \begin{equation} \label{CanGRExtCur1}
         K_{ab} = - h_a^{\ c} h_b^{\ d} \, \nabla_{(c} n_{d)} 
                = - h_b^{\ c} \, \nabla_c n_a 
                = - \frac12 \Lie h_{ab},
      \end{equation}
      where $\Lie$ denotes the Lie derivative along the normal to $\Sigma$ 
      (one may think of $K_{ab}$ as a generalised notion of time derivative 
       that describes the `bending' of $\Sigma$ in space-time). 
\end{itemize} 
Prescriptions (\texttt{I--III}) above imply that, in terms of $N$, $N^a$, and 
$t^a$, we have $n^a = (t^a - N^a)/N$ and hence the inverse space-time metric
can be written as
\begin{equation}
   g^{ab} = h^{ab} - N^{-2} \bigl( t^a - N^a \bigr) \bigl( t^b - N^b \bigr).
\end{equation}
This suggests to choose, as the field variables, the spatial metric $h_{ab}$, 
the lapse function $N$, and the covariant shift vector $N_a$ rather than the 
inverse metric $g^{ab}$, which is usually used in the variational principle. 
This equivalent set of variables is usually referred to as \adm 
\emph{variables}. Furthermore the three-metric $h_{ab}$ uniquely determines 
the derivative operator on $\Sigma$ compatible with $h_{ab}$, which we denote 
as $\rmD_a$. Regarding tensor fields on $\Sigma$ as fields on $\EuM$ with all 
their indices orthogonal to $n^a$ we can re-express the action of $\rmD_a$ in 
terms of that of $\nabla_a$, \ie the derivative operator associated with 
$g_{ab}$, \viz 
\begin{equation} \label{CanGRCovDer1}
   \rmD_c \, T^{a_1 \dots a_k}_{\quad \quad \ b_1 \dots b_l} = 
      h^{a_1}_{\ \ d_1} \dots h_{b_l}^{\ \ e_l} h_c^{\ f} 
      \nabla_f T^{d_1 \dots d_k}_{\quad \quad \ e_1 \dots e_l}.
\end{equation}
The derivative operator $\rmD_a$ brings forth a curvature tensor 
$\emb{R}^a_{\ bcd}$ on $\Sigma_t$. It is easy to show that, in terms of 
$\rmD_a$, the extrinsic curvature tensor in \eqref{CanGRExtCur1} takes the
form
\begin{equation} \label{CanGRExtCur2}
   K_{ab} = \frac1{2N} \bigl( 
                          - \Lie[t]{h_{ab}} + \rmD_a N_b + \rmD_b N_a
                       \bigr).
\end{equation}
\nl%
With respect to a coordinate basis the normal to $\Sigma_t$ has the components 
$n^{\alpha} \equiv (1, - N^i)/N$ and $n_{\alpha} \equiv (-N, 0)$ respectively 
and the metric $g_{ab}$ can be cast into the form
\begin{equation}
   \rmd s^2 = h_{ij} \bigl( \rmd x^i + N^i \rmd t \bigr) \otimes
              \bigl( \rmd x^j + N^j \rmd t \bigr) -
              N^2 \rmd t \otimes \rmd t,
\end{equation}
which enables one to identify its components, $g_{\alpha \beta}$, as well as 
those corresponding to its inverse, $g^{\alpha \beta}$, \viz
\begin{equation}
   g_{\alpha \beta} \equiv 
   \begin{pmatrix}
      N_k N^k - N^2 & N_j \\
      N_i           & h_{ij}
   \end{pmatrix};
   \quad 
   g^{\alpha \beta} \equiv \frac1{N^2}
   \begin{pmatrix}
      -1  & N_j \\
      N_i & N^2 h^{ij} - N^i N^j
   \end{pmatrix}.
\end{equation}
We thus see that the covariant components $g_{ij}$ and $h_{ij}$ coincide, 
whereas the contravariant components $g^{ij}$ and $h^{ij}$ do not. Moreover,
with regard to volume elements, we obtain 
$\sqrt{\smash[b]{- g}} \equiv N \sqrt{h}$ by application of the Frobenius--%
Schur formula. In terms of $N$, $N_i$, and $h_{ij}$, the components of the 
extrinsic curvature tensor defined in \eqref{CanGRExtCur1} and 
\eqref{CanGRExtCur2} are given by
\begin{equation} \label{CanGRExtCur3}
   K_{ij} = \frac1{2N} \bigl( 
                          - \parD{h_{ij}}{t} + N_{i \mid j} + N_{j \mid i}
                       \bigr),
\end{equation}
where a vertical stroke denotes covariant differentiation on $\Sigma_t$ (a 
semi-colon denotes covariant differentiation in space-time). 

\paragraph*{The Gauss, Codazzi, and York equations.}

The space-time curvature tensor $R^a_{\ bcd}$ is connected with the curvature 
tensors on $\Sigma$, \ie $\emb{R}^a_{\ bcd}$ and $K_{ab}$, through the 
\emph{Gauss equation}
\begin{equation} \label{CanGRGaussEq}
   \emb{R}^a_{\ bcd} = h^a_{\ m} h_b^{\ f} h_c^{\ g} h_d^{\ e} R^m_{\ fge} + 
                       2 K^a_{\ [d} K_{c]b}.
\end{equation}
Suitable contractions of this equation with the three-metric $h_{ab}$ yield
\begin{equation} \label{CanGRContGaussEq}
   \emb{R}_{ab} = h_a^{\ c} h_b^{\ d} R_{cd} + 
                  n^e n^f h_a^{\ c} h_b^{\ d} R_{ecfd} + 2 K^c_{\ [b} K_{c]a}.
\end{equation}
In a similar way we derive the \emph{Codazzi equation}
\begin{equation} \label{CanGRCodazziEq}
   h_e^{\ d} h_f^{\ c} h_g^{\ b} n_a R^a_{\ bcd} = 2 \rmD_{[f} K_{e]g},
\end{equation}
and its contraction,
\begin{equation} \label{CanGRContCodazziEq}
   h_c^{\ b} n^a R_{ab} = \rmD_c K - \rmD_a K^a_{\ c}.
\end{equation}
In addition to the Gauss equation \eqref{CanGRGaussEq} and the Codazzi 
equation \eqref{CanGRCodazziEq} we derive, after tedious calculations, the 
\emph{York equation}%
\footnote{It is actually an exercise proposed by York in \cite{MISNE=Gra}; a 
          detailed proof can be found in \cite{quere=ForHam}.}
\begin{equation} \label{CanGRYorkEq}
   n^b n^d h_e^{\ a} h_f^{\ c} R_{abcd} = 
      \Lie K_{ef} + \rmD_{(e} a_{f)} + K_{eg} K^g_{\ f} + a_e a_f,
\end{equation}
where $a_c$ stands for the four-acceleration of an observer moving along the 
normal to $\Sigma$, \viz $a_c := n^b \nabla_b n_c$. Note that in the 
synchronous Gauss system ($N=1$ , $N^i=0$) equation \eqref{CanGRYorkEq} 
reduces to the most commonly found expression
\begin{equation} 
   R_{0i0j} = \parD{K_{ij}}{t} +  K_{ik} K^k_{\ j}, 
\end{equation}
on account of the following formula expressing the Lie derivative of $K_{ij}$
\begin{equation} \label{CanGRLieK}
   \Lie K_{ij} = \frac1{N} \Bigl( 
                              \parD{K_{ij}}{t} - N^k K_{ij \mid k}
                                               - N^k_{\ \, \mid i} K_{kj}   
                                               - N^k_{\ \, \mid j} K_{ik}
                           \Bigr). 
\end{equation}
Contraction of the York equation \eqref{CanGRYorkEq} readily yields 
\begin{equation} \label{CanGRContYorkEq}
   n^a n^b R_{ab} = 
      h^{ab} \Lie K_{ab} + h^{ab} \rmD_{(a} a_{b)} + K_{ab} K^{ab} + a_c a^c.
\end{equation}
Making use of the contracted Gauss equation \eqref{CanGRContGaussEq} and the 
contracted York equation \eqref{CanGRContYorkEq} we obtain a general 
expression for the scalar curvature, \viz
\begin{equation} \label{CanGRScaCur1} 
   R = \emb{R} + K^2 - 3 K_{ab} K^{ab} - 2 h^{ab} \Lie K_{ab} 
               - 2 h^{ab} \rmD_{(a} a_{b)} - 2 a_c a^c. 
\end{equation}
From equations \eqref{CanGRContYorkEq} and \eqref{CanGRScaCur1} we obtain
\begin{equation} \label{CanGREinsTensEq}
   2 n^a n^b G_{ab} = \emb{R} + K^2 - K_{ab} K^{ab}.
\end{equation}
Hence setting the left-hand side in equations \eqref{CanGREinsTensEq} and
\eqref{CanGRContCodazziEq} to zero we find the initial-value constraint
equations of general relativity in vacuum.  
\nl
For completeness we may write the Codazzi equation \eqref{CanGRCodazziEq} and 
the York equation \eqref{CanGRYorkEq} in terms of the Weyl tensor, \viz
\begin{subequations} \label{CanGRCodYorWeyl}
   \begin{align} 
      h_e^{\ d} h_f^{\ c} h_g^{\ b} n_a C^a_{\ bcd} 
         &= 2 \Bigl[ 
                 h_e^{\ d} h_f^{\ c} h_g^{\ b} - 
                 \frac12 h^{bd} \bigl( 
                                   h_{eg} h_f^{\ c} - h_{fg} h_e^{\ c} 
                                \bigr)
              \Bigr] \rmD_{[c} K_{d]b}, \label{CanGRCodYorWeyla} \\
      n^b n^d h_e^{\ a} h_f^{\ c} C_{abcd} 
         &= \frac12 \bigl( h_e^{\ a} h_f^{\ c} - \frac13 h_{ef} h^{ac} \bigr) 
            \notag \\ 
         &  \qquad \times \bigl( 
                             \Lie K_{ac} + \emb{R}_{ac} + K K_{ac} + 
                             \rmD_{(a} a_{c)} + a_a a_c
                          \bigr). \label{CanGRCodYorWeylb}
   \end{align}
\end{subequations}       
\nl
We can specialise the Gauss, Codazzi, and York equations \eqref{CanGRGaussEq}, 
\eqref{CanGRCodazziEq}, and \eqref{CanGRYorkEq}, which are pure tensorial 
expressions, to the \adm basis, which is defined by 
$\{ \vec{e}_{\fn} = (\partial_t - N^i \partial_i)/N, 
    \vec{e}_i = \partial_i \}$.
We obtain respectively
\begin{subequations} \label{CanGRGauCodYorADM}
   \begin{align}
      R^l_{\ ijk}        &= {\emb{R}^l_{\ ijk}} + K_{ik} K^l_{\ j} 
                                                - K_{ij} K^l_{\ k}, \\
      R^{\fn}_{\ \, ijk} &= K_{ij \mid k} - K_{ik \mid j}, \\
      R_{\fn i \fn j}    &= \Lie K_{ij} + \frac{N_{\mid ij}}{N} 
                                        + K_{ik} K^k_{\ j}. 
   \end{align}
\end{subequations}       
The set of equations \eqref{CanGRGauCodYorADM} does provide all the 
information that is necessary to express the components of any four-%
dimensional curvature tensor in terms of the intrinsic and extrinsic three-%
dimensional tensors and of the lapse function and shift vector. For instance, 
the components of the Einstein tensor are given by
\begin{align*}
   G_{\fn \fn} &= \frac12 \bigl( {\emb{R}} + K^2 - K_{ij} K^{ij} \bigr), \\
   G_{\fn k}   &= K_{\mid k} - K^j_{\ k \mid j}, \\
   G_{kl}      &= \emb{G}_{kl} + K K_{kl} - 2 K_{ki} K^i_{\ l} -
                  \frac12 h_{kl} \bigl( K^2 - 3 K_{ij} K^{ij} \bigr) 
                  \notag \\
               &  \quad \qquad 
                + \bigl( 
                     h_{kl} h^{ij} - \delta_k^{\ i} \delta_l^{\ j} 
                  \bigr) \Lie K_{ij}
                + \frac1{N} 
                  \bigl( h_{kl} h^{ij} N_{\mid ij} - N_{\mid kl} \bigr).
\end{align*}
On the other hand, equation \eqref{CanGRScaCur1} for the scalar curvature  
reduces to
\begin{equation} \label{CanGRScaCur2} 
   R = \emb{R} + K^2 - 3 K_{ij} K^{ij} - 2 N^{-1} h^{ij} N_{\mid ij} 
               - 2 h^{ij} \Lie K_{ij}. 
\end{equation}
 
\subsubsection{{\normalfont{\textsc{Adm}}} gravitational Hamiltonian---Surface 
               terms}
\label{subsub:CanGRADM}

In most field theories the Hamiltonian can be derived from the covariant
action in a systematic way. In \gr the situation is more intricate due to the 
presence of a surface term in the Einstein--Hilbert action. This difficulty
was not addressed by Arnowitt, Deser, and Misner (\adm) who conducted the 
first major investigation into \gr---conceived as a dynamical system 
\cite{arnow=QuaThe,arnow=DynStr,arnow=CanVar,arnow=DynGen}. Although they 
highlighted the importance of the constraints and showed that the Hamiltonian 
is precisely the spatial integral of the constraints, they inadvertently 
discarded a total derivative term inherently present in the gravitational 
action. By contrast, DeWitt was the first to realise that the identification 
of the Hamiltonian with the constraints is only valid in the case when the 
three-manifold is compact, without boundaries \cite{dewit=QuaThe1}; in the 
case where the three-manifold is open, boundary conditions become crucial. 
Consider for instance the asymptotically flat situation: DeWitt put the total 
derivative term---which becomes a surface term at spatial infinity---back into 
the Hamiltonian and recovered the fact that the value of the Hamiltonian at a 
solution coincides with the \adm energy. Further investigation by Regge and 
Teitelboim demonstrated that the `correct' boundary conditions, together with 
the requirement that the Hamiltonian and its variations be well defined, lead 
to the existence of ten surface integrals---the Poincar\'e charges---in the 
Hamiltonian of the theory, which are connected with the Poincar\'e group of 
transformations acting at spatial infinity \cite{regge=RolSur,HANSO=ConHam}. 
Beig and \'o Murchadha further re-examined this formulation in the language of 
symplectic geometry (they also corrected a mistake in Regge--Teitelboim's work 
having to do with the generator of boosts) \cite{beig=PoiGro}. 
\nl
On the other hand, Gibbons and Hawking have advocated the addition \emph{ad 
hoc} to the Einstein--Hilbert action of the surface integral 
$2\oint \! K \rmd \Sigma$, the explicit form of which was firstly determined 
by York in a different context \cite{york=RolCon}, where $K$ is the trace of 
the extrinsic curvature of the boundary \cite{gibbo=ActInt}. There are two 
separate arguments requiring that this boundary term be added to the
Einstein--Hilbert action. In the first argument, one demands that the 
solutions of the classical field equations be extrema of the action under all 
perturbations that vanish on the boundary, \ie $\delta h_{ab}=0$ on 
$\partial \EuM$. This means that, on the boundary, only the normal derivatives 
of the induced metric are allowed to vary. In order to satisfy this condition 
one has to add a compensating term to the action the virtue of which is to 
cancel the second-order derivatives of the metric present in the scalar 
curvature. This compensating term is precisely the surface integral of the 
trace of the extrinsic curvature. The second argument, which entails the same 
boundary term, is formulated in the context of the path-integral formulation 
of the theory \cite{hawki=PatInt1,hawki=PatInt2}: \label{path-int} The 
gravitational action must be such that the amplitude to go from an initial to 
a final hypersurface must be independent of any arbitrary intermediate 
hypersurfaces and their associated induced three-metrics.
\nl
Hawking and Horowitz have recently re-examined this long-standing discussion
on surface integrals for manifolds with boundaries \cite{hawki=GraHam1}.%
\footnote{See also Hawking and Hunter's generalisation in the presence of 
          nonorthogonal boundaries \cite{hawki=GraHam2}.}
Starting from the Einstein--Hilbert action they derive the gravitational 
Hamiltonian without discarding any surface term---in contrast with the 
prevalent procedure. In particular, they show that the boundary terms in the 
Hamiltonian come directly from the boundary terms in the action and do not 
need to be added `by hand'. They generalise the definition of the \adm energy 
for space-times that are not asymptotically flat though asymptotically 
approaching a static background solution. (They also discuss the effect of 
horizons and the relation between their area and the total entropy of black 
holes.) We follow their approach. 
\nl
We start with the gravitational action in vacuum,
\begin{equation} \label{CanGREH}
   S [g] = \frac1{16 \pi} \int_{\EuM} \! \sqrt{- g} \, R - 
           \frac1{8 \pi} \oint_{\partial \EuM} \! \rmd \Sigma \, K, 
\end{equation}
where $K$ is the trace of the extrinsic curvature of the boundary and 
$\rmd \Sigma$ is the surface element. (We assume that the boundary 
$\partial \EuM$ consists of an initial and final surface with unit normal 
$n^a$, and a surface near infinity $\Sigma^{\infty}$, with unit normal $r^a$, 
on which $n^a$ is tangent.) As indicated above, the surface term in the action 
\eqref{CanGREH} is required so that the variational principle yield the 
correct equations of motion subject only to the condition that the induced 
three-metric on the boundary be held fixed---see, \eg 
\cite[Appendix E]{WALD=GenRel}, the specific works of Charap and Nelson 
\cite{chara=SurInt}, and York's \cite{york=BouTer} (and references therein). 
Note that the action \eqref{CanGREH} is well defined for spatially compact 
geometries only---\eg closed cosmological models.%
\footnote{For noncompact geometries one must choose a \emph{reference 
          background} which is required to be a static solution to the field 
          equations. The physical action is then the difference between the
          original action and the action that is evaluated on this background
          (see \cite{gibbo=EucQua,hawki=GraHam1} for a foolproof analysis).} 
\nl
Pure divergences are hidden in equation \eqref{CanGRScaCur1}: We derive an 
expression of the scalar curvature where they appear explicitly. We firstly
have
\begin{equation} \label{CanGRScaCur3} 
   R = 2 \bigl( G_{ab} - R_{ab} \bigr) n^a n^b.
\end{equation}
The first term in equation \eqref{CanGRScaCur3} is given by equation 
\eqref{CanGREinsTensEq} whereas the second term can be evaluated from the 
definition of the Riemann tensor given on page \pageref{conv:Riem} with 
$u^a \equiv n^a$, \viz 
\begin{equation} \label{CanGRtotder1}
   n^a n^b R_{ab} = K^2 - K_{ab} K^{ab} - 
                    \nabla_a \bigl( n^a \nabla_b n^b \bigr) +
                    \nabla_b \bigl( n^a \nabla_a n^b \bigr).
\end{equation}
Hence the scalar curvature is given by the general expression
\begin{align} \label{CanGRScaCur4} 
      R &= \emb{R} - K^2 + K_{ab} K^{ab} + 
           2 \Bigl[ \nabla_a \bigl( n^a \nabla_b n^b \bigr) -
                    \nabla_b \bigl( n^a \nabla_a n^b \bigr) \Bigr] \notag \\
        &= \emb{R} - K^2 + K_{ab} K^{ab} - 
           2 \nabla_c \Bigl( K n^c + a^c \Bigr),
\end{align}
which is equivalent to equation \eqref{CanGRScaCur1}. Observe that in the \adm 
basis the volume integral of the four-divergence $\nabla_c (K n^c)$ 
contributes the term $\partial_{\fn} (\sqrt{h} \, K)$ to the action whereas in 
a coordinate basis it yields 
$\partial_t (\sqrt{h} \, K) - \sqrt{h} \, (K N^k)_{\mid k}$; in the latter 
case, adding the four-divergence $\nabla_c a^c$ we recover the expression 
first given by DeWitt \cite{dewit=QuaThe1}, namely 
$-2\partial_t (\sqrt{h} \, K) + 2 \sqrt{h} \, (K N^k -  N^{\mid k})_{\mid k}$.
\nl
When substituted into the action \eqref{CanGREH} the two total derivative 
terms in equation \eqref{CanGRScaCur4} give rise to boundary contributions 
according to the formula 
\begin{equation}
   \int_{\EuM} \! \rmd^4 x \sqrt{-g} \, \nabla_a A^a 
      = \int_{\EuM} \! \rmd^4 x \, \partial_a \Bigl[ \sqrt{-g} \, A^a \Bigr]
      = \oint_{\partial \EuM} \! \rmd \sigma_a A^a,
\end{equation}
where the surface element $\rmd \sigma_a$ is defined by 
$\rmd \sigma_a = \nu_a \, \rmd \Sigma 
               = \epsilon \, \nu_a \sqrt{\abs{h}} \, \rmd^3 x$, 
where $\nu^a$ is the unit normal to the boundary and $\epsilon = \nu^a \nu_a$ 
($\epsilon = -1$ or $+1$ whether $\nu^a$ is timelike or spacelike 
respectively). As indicated above, the explicit form of the surface integral 
was first written down by York in his analysis of canonical gravity based on 
the \adm decomposition of space-time \cite{york=RolCon}. The first total 
divergence in \eqref{CanGRScaCur4} neatly cancels the $\oint K$ surface term 
on the initial and final boundaries: We have indeed 
$- 2 \int_{\EuM} \rmd^4 x \sqrt{- g} \, \nabla_c \bigl( K n^c \bigr) =
- 2 \oint \rmd^3 x \sqrt{h} \, K$. The second, which is orthogonal to the 
normal $n^a$, only contributes to the surface integral near infinity%
\footnote{The boundary near infinity, denoted by $\Sigma^{\infty}$, is 
          foliated by a family of two-surfaces $S^{\infty}_t$ coming from its 
          intersection with $\Sigma_t$.}
and is therefore only relevant for noncompact geometries. This surface 
integral also contains the contribution from the $\oint \! K$ surface term in 
the action \eqref{CanGREH}; explicitly it is
\begin{equation} \label{CanGRSurInt}
   \frac1{8 \pi} \oint_{\Sigma^{\infty}} \! \sqrt{\abs{h}} \,
                 \Bigl(
                    \nabla_c r^c - r_c a^c
                 \Bigr) =
   \frac1{8 \pi} \oint_{\Sigma^{\infty}} \! \sqrt{\abs{h}} \,
                 \Bigl(
                    g^{ab} + n^a n^b
                 \Bigr) 
                 \nabla_a r_b,
\end{equation}
where $r^c$ denotes the unit normal to $\Sigma^{\infty}$. Hence the action 
\eqref{CanGREH} takes the form
\begin{equation} \label{CanGREH2}
   S [h] = 
      \int \! N \rmd t \biggl[ \frac1{16 \pi} \int_{\Sigma_t} \! \sqrt{h} \,
                           \Bigl( \emb{R} + K_{ab} K^{ab} - K^2 \Bigr) -
                           \frac1{8 \pi} \oint_{S^{\infty}_t} \!
                           \sqrt{\abs{h}} \, \emb[(2)]{K}
                    \biggr], 
\end{equation}
where $\emb[(2)]{K}$ is the two-dimensional extrinsic curvature of 
$S^{\infty}_t$ in $\Sigma_t$. 
\nl
We thus take as the gravitational Lagrangian density
\begin{equation} 
   \Fl \bigl[ N, N^i, h_{ij} \bigr] = 
       \frac{N \sqrt{h}}{16 \pi} \Bigl( \emb{R} + K_{ij} K^{ij} - K^2 \Bigr)
\end{equation}
and introduce the canonical momenta $\Pi$, $\Pi_i$, and $\piadm^{ij}$ 
conjugate to $N$, $N^i$, and $h_{ij}$ respectively, namely%
\footnote{Since $N$ and $N^i$ are cyclic variables, they have no dynamics---%
          their conjugate momenta vanish identically. They can be considered 
          just as Lagrange multipliers rather than phase-space variables. 
          (This feature of the gravitational Lagrangian can be traced back to 
          the diffeomorphism-invariance of general relativity.)}
\begin{subequations} \label{CanGRmom1}
   \begin{align}
      \Pi          &= \varD{\Fl}{(\partial_t N)} = 0, \\
      \Pi_i        &= \varD{\Fl}{(\partial_t N^i)} = 0, \\
      \piadm^{ij}  &= \varD{\Fl}{(\partial_t h_{ij})} 
                    = - \frac{\sqrt{h}}{16 \pi} 
                        \bigl( K^{ij} - K h^{ij} \bigr).
   \end{align}
\end{subequations}       
The action \eqref{CanGREH2} can be brought to canonical form by means of the
usual procedure, \viz
\begin{equation} \label{CanGREH3}
   \begin{split}
      S &= \int \! \rmd t 
              \biggl[ 
                 \int_{\Sigma_t}  
                    \Bigl( 
                       \piadm^{ij} \, \partial_t h_{ij} - N \EuH - N^i \EuH_i 
                    \Bigr) \\
        & \qquad \qquad \qquad \qquad -
                 \frac1{8 \pi} \oint_{S^{\infty}_t} 
                    \Bigl( 
                       N \sqrt{\abs{h}} \, \emb[(2)]{K} + 
                       N_i \, \piadm^{ij} \, r_j 
                    \Bigr)
              \biggr], 
   \end{split}
\end{equation}
where we have introduced the following quantities
\begin{subequations} \label{CanGRSuper}
   \begin{align}
      \EuH     &= 16 \pi \, G_{ijkl} \, \piadm^{ij} \piadm^{kl} - 
                 \frac{\sqrt{h}}{16 \pi} \emb{R}, \label{CanGRSupera} \\
      \EuH^i   &= - 2 \pi^{ij}_{\text{\adm} \ \mid j}, \label{CanGRSuperb} \\
\intertext{and the so-called \emph{DeWitt metric},}
      G_{ijkl} &= \frac1{2 \sqrt{h}} 
                  \bigl( h_{ik} h_{jl} + h_{il} h_{jk} - h_{ij} h_{kl} \bigr),
   \end{align}
\end{subequations}       
and where an additional surface term stemming from the Legendre transformation
has been taken into account. The canonical Hamiltonian is thus 
\begin{equation} \label{CanGRCanHam}
   H_{\rmc} = \int_{\Sigma_t}  
              \Bigl( N \EuH + N^i \EuH_i \Bigr) +
              \frac1{8 \pi} \oint_{S^{\infty}_t} 
              \Bigl( 
                 N \sqrt{\abs{h}} \, \emb[(2)]{K} + 
                 N_i \, \piadm^{ij} r_j 
              \Bigr).
\end{equation}
This expression diverges in general. However, once the Hamiltonian for the
reference background has been obtained, one can define the total energy of a
given stationary solution of the field equations to be simply the value of the
\emph{physical} Hamiltonian,
\begin{equation} \label{CanGRTotEne}
   E = \frac1{8 \pi} \oint_{S^{\infty}_t} 
       \biggl[ N \sqrt{\abs{h}} \Bigl( \emb[(2)]{K} - \emb[(2)]{K}_0 \Bigr) + 
               N_i \, \piadm^{ij} r_j 
       \biggr],
\end{equation}
where the subscript `0' refers to the background solution. This expression,
firstly obtained by Hawking and Horowitz, generalises the famous \adm energy 
for asymptotically flat space-times \cite{hawki=GraHam1}. 
\begin{rem} \label{rem:SurTer}
As indicated above, surface terms play a crucial r\^ole for noncompact 
geometries---they cannot be discarded. Another difficulty which arises with 
surface terms has to do with spatially homogeneous cosmologies: genuine 
Lagrangian and Hamiltonian formulations are lacking for class B models (see 
\cite{macca=QuaCos} and references therein). This is because the very 
symmetries that are imposed prevent the boundary terms to be vanishing; hence 
the equations derived from symmetry-preserving variations are the wrong 
equations. (We re-examine this issue in greater detail in Subsection~%
\ref{subsec:BiaCosSpaCosHamCos}.) 
\end{rem}

\paragraph*{Canonical quantisation---Wheeler--DeWitt equation.}

From equation \eqref{CanGRmom1} we see that there are primary constraints,
$\Pi \approx 0$ and $\Pi_i \approx 0$; their complete treatment was performed 
by Dirac, DeWitt, and others---this is discussed in several textbooks 
(see, \eg \cite{HELD=GenRel1,ESPOS=QuaGra}). 
\nl
Requiring the preservation in time of the primary constraints one finds as 
secondary constraints the \emph{super-Hamiltonian}, $\EuH \approx 0$, and  
\emph{super-momentum}, $\EuH_i \approx 0$. (The consistency algorithm does not
lead to any new constraints.) For compact geometries the total Hamiltonian is
thus given by
\begin{equation} \label{CanGRTotHam}
   H_{\rmT} = \int_{\Sigma_t}  
              \Bigl( N \EuH + N^i \EuH_i + \mu \Pi + \mu^i \Pi_i \Bigr),
\end{equation}
where $\mu$ and $\mu^i$ are Lagrange multipliers. All the constraints are
first class: The super-Hamiltonian $\EuH$ is responsible for the dynamics
whereas the super-momenta $\EuH_i$ are the generators of spatial coordinate
transformations on $\Sigma_t$. Strictly speaking, the constraints 
$\EuH_i \approx 0$ express that the state of the universe depends only on the
intrinsic three-geometry of the spatial sections and not on the coordinates.
(This leads naturally to the concept of \emph{Wheeler's superspace} 
\cite{wheel=GeoIss}.)
\nl
In conformity with the canonical formulation it is clear that Einstein's 
equations can be split into two sets: the dynamical equations and the 
initial-value equations. The former are the evolution equations for the 
dynamical variables $h_{ij}$ and $\piadm^{ij}$; these cannot be freely 
specified on a given hypersurface $\Sigma_t$ owing to the constraints, 
which constitute the second set of initial-value equations. Both groups of 
equations are intimately connected. Actually, the initial-value equations 
contain all the dynamics of the gravitational field so that one could say that 
Einstein's equations are highly redundant. Solving the constraints and fixing 
the gauge---\ie picking up a specific coordinate system---one removes four 
\dofs out of the six pairs of canonical variables; hence the number of 
physical \dofs is two, which correspond to the two helicity states of the 
spin--$2$ graviton. 
\nl
There are at least three distinct methods to proceed to the canonical 
quantisation of \gr: the \adm procedure in which all the constraints are 
solved and the gauge is fixed before quantising (reduced formalism); the Dirac 
\emph{modus operandi} in which the constraints are imposed as restrictions on 
the quantum state of the universe; and an intermediate approach devised by 
Kucha{\v r} \cite{kucha=BubTim}. In all these approaches the quantum state of 
the system is represented by a wave functional $\Psi [h_{ij}]$---a functional 
on Wheeler's superspace. An important feature of this wave function, which 
appears when one considers closed cosmological models---the basic assumption 
of quantum cosmology; see \cite{halli=IntLec}---, is that it does not depend 
explicitly on the coordinate time label $t$ (as a direct consequence of the 
vanishing of the super-Hamiltonian). 
\nl 
According to the Dirac quantisation method the wave function is annihilated by 
the operator versions of the (first-class) classical constraints: The usual 
substitutions for the momenta having been made, \viz
\begin{equation}
   \piadm^{ij} \rightarrow -i \varD{}{h_{ij}},
\end{equation}
one obtains the following equations for $\Psi$,%
\footnote{We do not address here the operator-ordering issue.} 
\begin{subequations}
   \begin{align}
      \EuH_k \Psi &= 2 i \, \rmD_l \varD{\Psi}{h_{kl}} = 0 \; \; 
                     \text{\footnotesize{\textsf{[momentum constraint]}}}, 
                     \label{QCMomCon} \\
      \EuH \Psi   &= \biggl[ - 16 \pi G_{ijkl} \, \varD{}{h_{ij}} 
                                                  \varD{}{h_{kl}} 
                             - \frac{\sqrt{h}}{16 \pi} \emb{R} \biggr] \Psi
                   = 0 \; \; 
                   \text{\footnotesize{\textsf{[Wheeler--DeWitt equation]}}}. 
                     \label{QCWDW} 
   \end{align}
\end{subequations}       
The momentum constraint \eqref{QCMomCon} is nothing but the quantum mechanical 
expression of the invariance of the theory under three-dimensional 
diffeomorphisms. The Wheeler--DeWitt equation \eqref{QCWDW} is a second-order 
hyperbolic functional differential equation describing the dynamical evolution 
of the wave equation in superspace; it must be supplemented by appropriate 
boundary conditions. 
\begin{rem}
As an alternative to the canonical quantisation procedure one can construct
the wave function using a path-integral approach (see, \eg \cite{halli=IntLec,
hawki=PatInt2} and references therein).
\end{rem}

\subsection{Hamiltonian formulation of nonlinear gravity theories}
\label{subsec:Hamfr}

Consider the Lagrangian density describing nonlinear theories of gravity,
namely
\begin{equation} \label{HamfrLag1}
   \Fl = \sqrt{- g} \, \fR, \qquad  \text{with} \; \fpp \neq 0,  
\end{equation}     
where $\fR$ is a nonlinear arbitrary smooth function of the scalar curvature 
and primes denote differentiation \wrt the scalar curvature. 
\nl
We assume that the space-time is foliated into three-dimensional Cauchy
hypersurfaces $\Sigma_t$ and we adopt the \adm basis (\cfr $3+1$--splitting, 
on page~\pageref{subsec:CanGR}). On account of the explicit form 
\eqref{CanGRScaCur2} of the scalar curvature in terms of the intrinsic 
geometry and extrinsic three-curvature tensor $K_{ij}$ on $\Sigma_t$ and 
according to the definition \eqref{CanGRExtCur1} of $K_{ij}$ in terms of the 
Lie derivative of the three-metric $h_{ij}$ along the normal to $\Sigma_t$, we 
may rewrite the Lagrangian density \eqref{HamfrLag1} as 
\begin{equation} \label{HamfrLag2}
   \Fl = N \sqrt{h} \, f \Bigl( 
                             R \bigl( 
                                  h_{ij}, \Lie h_{ij}, \Lie^2 h_{ij} 
                               \bigr) 
                          \Bigr).
\end{equation}     
Since the lapse function $N$ and shift vector $N^i$ are not dynamical 
variables, we regard them merely as Lagrange multipliers associated with the
gravitational constraints. Furthermore we interpret the Lie derivative along
the normal to $\Sigma_t$ as a generalised notion of time differentiation.%
\footnote{To bear in mind this interpretation and to make the comparison with
          the general formalism easier we sometimes use a dot as an 
          alternative notation for the Lie derivative.}
We follow the generalised constrained Ostrogradsky construction developed in 
Subsection~\ref{subsec:OstroSingBL} in order to cast the theory into canonical 
form.
\begin{rem}
Very recently, Ezawa \etal have performed a similar analysis of the nonlinear
Lagrangian density \eqref{HamfrLag1} \cite{ezawa=CanFor}. They obtain 
comparable results albeit utilising a slightly different Hamiltonian 
formulation: They intimately follow the generalised Ostrogradsky method 
referred to as Buchbinder--Lyakhovich's method. Strictly speaking, they 
introduce velocities with the definition \eqref{OstroBLVelo} (as indicated 
above, \cfr the remark made on page~\pageref{remvelo}, such an adjunction is 
unnecessary); they treat $N$ and $N^i$ as genuine canonical variables; and
they use the partial time derivative $\partial_t$ instead of the generalised 
notion of time differentiation provided by the Lie derivative $\Lie$ along the 
normal to $\Sigma_t$. In contrast with their approach the method presented 
here is more straightforward and clear-cut.
\end{rem}
\nl%
In regard to the second-order Lagrangian density \eqref{HamfrLag2} the 
generalised Ostrogradsky prescriptions \eqref{OstroBLQ1} entail the following 
natural choice for the auxiliary field variables:
\begin{subequations} \label{HamfrOst1}
   \begin{align}
      q_{0,ij}       &:= h_{ij}, \\
      q_{1,ij}       &:= Q_{1, ij} \bigl( h_{ij}, \Lie h_{ij} \bigr) 
                       = K_{ij}, \label{HamfrOst1b} \\
      \dot{q}_{1,ij} &:= \Lie K_{ij}. 
   \end{align}
\end{subequations}
Owing to the definition \eqref{HamfrOst1b}, conditions \eqref{OstroBLDet1} are 
automatically fulfilled since equation \eqref{OstroBLX} becomes
\begin{equation} \label{HamfrLagCon}
   \Lie h_{ij} \equiv \dot{q}_{0,ij} = -2 q_{1,ij}, 
\end{equation}
which actually corresponds to equation \eqref{OstroBLQ2}. 
\nl
The extended Lagrangian density that includes the above Lagrangian constraint 
\eqref{HamfrLagCon} supersedes the original Lagrangian density 
\eqref{HamfrLag2}; it is given by 
\begin{equation}
     \Flb \bigl( q_0, q_1, \dot{q}_0, \dot{q}_1, \lambda \bigr) =  
        N^{-1} \Fl \bigl( q_0, q_1, \dot{q}_1 \bigr) + 
        \bigl( \dot{q}_{0, ij} + 2 q_{1, ij} \bigr) \lambda^{ij},
\end{equation}     
with suitable Lagrange multipliers $\lambda^{ij}$ so as to recover the 
interpretation in terms of the original set of field variables. (Observe that 
an overall $N$ has been factorised.) 
\nl%
The momenta canonically conjugate to the field variables are defined
respectively by
\begin{subequations} \label{HamfrConjMom1}    
   \begin{align}
      p^{ij}_0 &= p^{ij} := \parD{\Flb}{\dot{q}_{0,ij}} = \lambda^{ij}, 
         \label{HamfrConjMom1a} \\  
      p^{ij}_1 &= \cp^{ij} := \parD{\Flb}{\dot{q}_{1,ij}} 
                = N^{-1} \parD{\Fl}{\bigl(\Lie K_{ij}\bigr)}
                = -2 \sqrt{h} \, h^{ij} \fpR, 
         \label{HamfrConjMom1b} \\  
      \Pi^{(\lambda)}_{ij} &:= \parD{\Flb}{\dot{\lambda}^{ij}} = 0.
         \label{HamfrConjMom1c}  
   \end{align}
\end{subequations}
Hence we have the primary constraints
\begin{subequations} \label{HamfrPriCon}    
   \begin{align}
      &\varphi^{ij} = p^{ij} - \lambda^{ij} \approx 0, \\
      &\phi^{ij} \bigl( q_0, q_1, p_1 \bigr) = \cp^{\ttT ij}
       \approx 0, \label{HamfrPriConb} \\
      &\Pi^{(\lambda)}_{ij} \approx 0, 
   \end{align}
\end{subequations}
where $\cp^{\ttT ij}$ is the traceless part of $\cp^{ij}$. The constraint 
\eqref{HamfrPriConb} arises from the indeterminacy of 
$\bigl( \Lie K_{ij} \bigr)^{\ttT}$ in the definition of momenta 
\eqref{HamfrConjMom1b}.
\nl
Now we perform a restricted Legendre transformation. From equations
\eqref{CanGRScaCur2} and \eqref{HamfrConjMom1b} we obtain 
\begin{equation} \label{HamfrResLeg}
   \dot{q}_{1,ij} \, p_1^{ij} 
      = \cp^{ij} \Lie K_{ij} 
      = - \frac{\cp}6 
          \Bigl( 
             R - \emb{R} + 3 K_{ij} K^{ij} - K^2 + 2 N^{-1} h^{ij} N_{\mid ij} 
          \Bigr).
\end{equation}     
The last term in equation \eqref{HamfrResLeg} gives rise to the surface 
integral at spatial infinity%
\footnote{We keep the same notations as those used in the discussion on the
          boundary terms present in the gravitational action; 
          pp.~\pageref{subsub:CanGRADM} ff.}
\begin{equation} \label{HamfrSurTer}
   \frac13 \oint_{\Sigma^{\infty}} \! \rmd^3 x  
      \Bigl( \cp N^{\mid k} - N \cp^{\mid k} \Bigr) r_k,
\end{equation}     
which, in general, does not vanish for noncompact geometries. Henceforth we 
assume that we are dealing with actions that do not produce such nonzero 
surface terms. (We shall come back to this specific issue in the study of
spatially homogeneous cosmologies within the Hamiltonian framework, in 
Subsection~\ref{subsec:BiaCosSpaCosHamCos}.) The surface term 
\eqref{HamfrSurTer} is the only one arising in the theory, in 
contradistinction to canonical \gr, where the additional boundary term
$\partial_{\fn} (\sqrt{h} \, K)$ must be cancelled out by the Gibbons--%
Hawking's compensating boundary term.
\nl
The restricted Hamiltonian density is 
\begin{equation}
   \begin{split}
      \Fh_{\rmr} \bigl(h, K, \cp \bigr) 
         &= \cp^{ij} \Lie K_{ij} - N^{-1} \Fl \bigl( R \bigr) \\
         &= \frac{\cp}6 \Bigl( \emb{R} - 3 K_{ij} K^{ij} + K^2 \Bigr) - 
            \frac13 h^{ij} \cp_{\mid ij} + V(\cp),
   \end{split}
\end{equation}     
where the `potential term'
\begin{equation} \label{HamfrPot}
     V(\cp) := \sqrt{h} \Bigl[ R \fpR - \fR \Bigr]_{R \rightarrow F(\cp)}
\end{equation}     
has been introduced---upon eliminating $R$ for $F(\cp)$, the function that is
obtained by solving the trace of equation \eqref{HamfrConjMom1b} for the
trace of the momentum $\cp^{ij}$. 
\nl
The canonical Hamiltonian density is given by                       
\begin{equation}
   \Fh_{\rmc} \bigl(h, K, p, \cp \bigr) 
      = \Fh_{\rmr} \bigl(h, K, \cp \bigr) - 2 p^{ij} K_{ij}.
\end{equation}     
Therefore, the Hamiltonian form of the gravitational action based on the 
original Lagrangian density \eqref{HamfrLag1} can be written, up to boundary 
terms, as 
\begin{equation} \label{HamfrHamAct}
   S = \int_{\EuM} \! N \Bigl[ 
                           p^{ij} \, \Lie h_{ij} + 
                           \cp^{ij} \, \Lie K_{ij} - 
                           \Fh_{\rmc} \bigl( h, K, p, \cp \bigr) 
                        \Bigr].
\end{equation}
The ensuing Dirac Hamiltonian density is
\begin{equation}
   \Fh_{\EuD} = N \EuH + N^k \EuH_k + \mu_{kl} \varphi^{kl} + 
                \nu^{kl} \Pi^{(\lambda)}_{kl} + \xi_{kl} \phi^{kl}.
\end{equation}     
Preservation in time of the primary constraints \eqref{HamfrPriCon} leads to 
the secondary constraints
\begin{equation} \label{HamfrSecCon}
     \chi^{ij} = 2 p^{\ttT ij} + \frac13 \cp K^{\ttT ij} \approx 0
\end{equation}
and enables to determine the Lagrange multipliers $\mu_{ij}$, $\nu^{ij}$, and 
$\xi_{ij}$. Aside from the usual super-Hamiltonian and super-momentum 
constraints, which are always first class, all other constraints are second 
class by virtue of the Poisson brackets
\begin{subequations}
   \begin{align}  
      \bigl\{ \varphi^{ij} (\bfx), \Pi^{(\lambda)}_{kl} (\bfy) \bigr\} 
        &= - \delta^i_k \delta^j_l \, \delta \bigl( \bfx - \bfy \bigr), \\ 
      \bigl\{ \phi^{ij} (\bfx), \chi^{kl} (\bfy) \bigr\} 
        &= \frac{\cp}3 \bigl( h^{ik} h^{jl} - \frac13 h^{ij} h^{kl} \bigr) 
           \, \delta \bigl( \bfx - \bfy \bigr). 
   \end{align}
\end{subequations}
Therefore, we regard them as strong equations provided we introduce the 
appropriate Dirac bracket. 
\nl
Expressing the Lie derivative in terms of the usual time derivative we recover 
the familiar form of the canonical action, namely
\begin{equation} 
   S = \int_{\EuM} \Bigl( 
                      p^{ij} \, \dot{h}_{ij} + 
                      \cp^{ij} \, \dot{K}_{ij} - 
                      N \EuH - N^k \EuH_k 
                   \Bigr),
\end{equation}
where the super-Hamiltonian and super-momentum constraints are given 
respectively by
\begin{subequations}
   \begin{align}  
      \EuH   &= \frac{\cp}6 \Bigl( \emb{R} - 3 K_{ij} K^{ij} + K^2 \Bigr) + 
                V(\cp) - \frac13 h^{ij} \cp_{\mid ij} - 2 p^{ij} K_{ij}, \\ 
      \EuH_k &= \frac{\cp}3 K_{\mid k} - 
                \frac23 \bigl( \cp K_k^{\ j} \bigr)_{\mid j} - 
                2 h_{ik} \, p^{ij}_{\ \ \mid j}.
   \end{align}
\end{subequations}
\nl
Counting the number of physical \dofs is straightforward: Once the constraints 
have been enforced and some coordinate system has been chosen, there remains 
only three pairs of independent canonical variables. This is in total 
agreement with the particle content of the theory and with the fact that 
nonlinear theories of gravity are dynamically equivalent to scalar-tensor 
theories (\cfr Section~\ref{sec:VarConStr}). We proceed now to a deeper 
analysis on this last property in the light of the Hamiltonian formalism.

\subsubsection{Dynamical equivalence with scalar-tensor theories}
\label{subsub:HamfrDynEquST}

Nothing whatsoever can prevent us from introducing a scalar field, $\Phi$, as 
a new independent \dof, by adopting, instead of the nonlinear Lagrangian 
density \eqref{HamfrLag1}, the equivalent constrained Lagrangian density
\begin{equation} \label{HamfrLag3}
   \Fl = \sqrt{- g} \Bigl[ \fPhi + \fpPhi \bigl( R - \Phi \bigr) \Bigr],  
\end{equation}     
which is sometimes referred to as a \emph{Helmholtz} Lagrangian density.
\nl
According to the Ostrogradsky prescriptions, we define the auxiliary variables
\begin{align*}
   q_{0,ij}       &= h_{ij},                 &q_{0,\Phi} &= \Phi, \\
   q_{1,ij}       &= K_{ij},           &\dot{q}_{0,\Phi} &= \Lie \Phi, \\
   \dot{q}_{1,ij} &= \Lie K_{ij},
\end{align*}
and replace the original Lagrangian density \eqref{HamfrLag3} by the extended
one,
\begin{equation}
     \Flb \bigl( q_0, q_1, \dot{q}_0, \dot{q}_1, \Phi, \lambda \bigr) =  
        N^{-1} \Fl \bigl( q_0, q_1, \dot{q}_1, \Phi \bigr) + 
        \bigl( \dot{q}_{0, ij} + 2 q_{1, ij} \bigr) \lambda^{ij}.
\end{equation}     
The conjugate momenta are given by \eqref{HamfrConjMom1} and $\Pi^{(\Phi)}=0$.
In addition to the primary constraints \eqref{HamfrPriCon} we also obtain  
\begin{align*}
   &\Xi = \cp + 6 \sqrt{h} \fpPhi \approx 0, \\
   &\Pi^{(\Phi)} \approx 0. 
\end{align*}
The Dirac Hamiltonian density is
\begin{equation}
   \Fh_{\EuD} = \Fh_{\rmc} + \mu_{kl} \varphi^{kl} + 
                \nu^{kl} \Pi^{(\lambda)}_{kl} + \xi_{kl} \phi^{kl} +
                \zeta \Xi + \eta \Pi^{(\Phi)},
\end{equation}     
where the canonical Hamiltonian density is given by
\begin{equation}
   \Fh_{\rmc} = \frac{\cp}6 \emb{R} - \frac{\cp}2 K_{ij} K^{ij} + 
                \frac{\cp}6 K^2 - \frac13 h^{ij} \cp_{\mid ij} + V(\Phi) -
                2 p^{ij} K_{ij},
\end{equation}     
with a `potential term'
\begin{equation}
     V(\Phi) := \sqrt{h} \Bigl[ \Phi \fpPhi - \fPhi \Bigr].
\end{equation}     
Time evolution of the primary constraints yields the determination of the
Lagrange multipliers and produces the secondary constraint 
\eqref{HamfrSecCon}. All constraints are second class; we can readily 
eliminate the spurious \dofs associated with $\lambda$ and $\Phi$ from the 
action, bearing in mind that the function $\fpPhi$ has to be inverted. 
The outcome of this procedure is precisely the canonical formalism that was 
developed for the Lagrangian density \eqref{HamfrLag1}, where the variable 
$\cp$ plays the r\^ole of the new independent scalar \dof.

\subsubsection{Extended gravity theories}
\label{subsub:HamfrExtGra}

The Ostrogradsky method is also well suited for building up a canonical 
formalism of theories derived from Lagrangians that are functions not only of 
the scalar curvature $R$ but also $\square^n R$ (such terms can be generated 
by quantum corrections to \gr). The resulting theories, called generically 
\emph{extended gravity theories}, have been studied in the context of 
inflationary cosmology; in particular, Wands has discussed their relationship 
with scalar-tensor theories \cite{wands=ExtGra} (see also Section~%
\ref{sec:VarConStr}). 
\nl
Consider, for instance, the Lagrangian density that describes sixth-order 
gravity, namely
\begin{equation} \label{HamfrLagExt1}
   \Fl = \sqrt{- g} \, \Bigl( R + \gamma R \square R \Bigr),  
\end{equation}     
which is dynamically equivalent to the scalar-tensor Lagrangian density (with 
two scalar fields)
\begin{equation} \label{HamfrLagExt2}
   \Fl = \sqrt{- g} \, 
         \Bigl[ 
            R \bigl( 1 + \gamma \varphi_1 + \gamma \square \varphi_0 \bigr) -
            \gamma \varphi_0 \varphi_1
         \Bigr].  
\end{equation}     
We may define the auxiliary Ostrogradsky variables
\begin{align*}
   q_{0,ij}       &= h_{ij},      &\varphi^{(0)}_0       &= \varphi_0, \\
   q_{1,ij}       &= K_{ij},      &\varphi^{(0)}_1       &= \Lie \varphi_0, \\
   \dot{q}_{1,ij} &= \Lie K_{ij}, &\dot{\varphi}^{(0)}_1 &= \Lie^2 \varphi_0, 
\end{align*}
and replace the original Lagrangian density \eqref{HamfrLagExt2} by the 
extended one
\begin{equation*}
   \begin{split}
      h^{-1/2} \Flb &= 
         R \bigl( q_0, q_1, \dot{q}_1 \bigr)  
            \biggl\{ 
               1 + \gamma \Bigl( 
                             \varphi_1 - \dot{\varphi}^{(0)}_1
                          \Bigr)
                 + \gamma q_0^{kl} \Bigl[ 
                                      \varphi^{(0)}_{0 \mid kl} +
                                      N^{-1} N_{\mid k} 
                                         \varphi^{(0)}_{0 \mid l}
                                   \Bigr]
            \biggr\} \\
                    & \qquad -
         \gamma \varphi^{(0)}_0 \varphi_1 +
         \lambda^{ij} \bigl( \dot{q}_{0,ij} + 2 q_{1,ij} \bigr) +
         \mu \Bigl( \dot{\varphi}^{(0)}_0 - \varphi^{(0)}_1 \Bigr),
   \end{split}
\end{equation*}     
where we have used the useful formula
\begin{equation*}
   \square \phi = h^{ab} \rmD_a \rmD_b \phi - \Lie^2 \phi + 
                  a^c \partial_c \phi.
\end{equation*}     
Then we may proceed in the same way as in the nonlinear $\fR$ case. (We do not 
elaborate further on this example since the procedure is systematic---although 
the ensuing expressions are very awkward.)

\subsection{Link between the Ostrogradsky and \textsc{adm} formulations of
            Einstein's theory}
\label{subsec:HamOstADM}

\subsubsection{Ostrogradsky--\adm equivalence theorem}
\label{subsub:HamOstADMEquThe}

Now that---as a direct application of the generalised constrained Ostrogradsky 
construction---we have achieved a consistent Hamiltonian formulation of 
nonlinear theories of gravity, it is of great interest to examine whether this 
method could likewise be used on the \emph{linear} Lagrangian of \gr. This 
prospect arose in connection with the problem of determining appropriate 
boundary conditions in a theory of gravity with higher derivatives. We have
seen on page~\pageref{HamfrSurTer} that the only boundary term in the 
nonlinear case is a surface integral at spatial infinity. This can be easily 
understood since there is no need whatsoever to discard a total divergence 
that embodies second-order derivatives of the metric: The Ostrogradsky method 
is inherently appropriate to cope with higher derivatives. If one could treat%
---as we surmise---\gr within the Ostrogradsky scheme, then exactly the same 
argument would apply: No boundary terms other than the analogue of the surface 
integral \eqref{HamfrSurTer} would arise. What would then be the relationship 
between the Ostrogradsky formalism and the \adm version of canonical gravity? 
We intend, in this subsection, to unravel that possible connection.
\nl
Our starting point is the Einstein--Hilbert Lagrangian density,
\begin{equation} \label{HamOstADMLag1}
   \Fl = \sqrt{- g} \, R.  
\end{equation}     
We rely on what we did in the previous subsection since the linear Lagrangian
density \eqref{HamOstADMLag1} may be viewed as a special instance of the 
nonlinear theory, provided we relax the condition on the second derivative of 
the function $\fR$, that is $\fpp=0$.
\nl
Instead of utilising expression \eqref{CanGRScaCur4} of the scalar curvature 
as in the \adm method, we exploit the formula \eqref{CanGRScaCur2}. Thus we 
can formally rewrite the Lagrangian density \eqref{HamOstADMLag1} as 
\begin{equation} \label{HamOstADMLag2}
   \Fl = N \sqrt{h} \, R \bigl( h_{ij}, \Lie h_{ij}, \Lie^2 h_{ij} \bigr). 
\end{equation}     
Following the prescriptions of the generalised Ostrogradsky construction we
introduce the auxiliary variables \eqref{HamfrOst1} together with the 
Lagrangian constraint \eqref{HamfrLagCon} and we replace the original 
Lagrangian density \eqref{HamOstADMLag2} by the appropriate extended 
Lagrangian density with Lagrange multipliers $\lambda^{ij}$.
\nl
The momenta canonically conjugate to the field variables are given by 
equations \eqref{HamfrConjMom1}, where $\fpR=1$, that is 
$p^{ij} = \lambda^{ij}$, $\cp^{ij}= -2 \sqrt{h} \, h^{ij}$, and
$\Pi^{(\lambda)}_{ij} = 0$. We thus obtain the primary constraints
\begin{subequations} \label{HamOstADMPriCon}    
   \begin{align}
      &\varphi^{ij} = p^{ij} - \lambda^{ij} \approx 0, \\
      &\phi^{ij} = \cp^{\ttT ij} \approx 0, \\
      &\Xi = \cp + 6 \sqrt{h} \approx 0, \\ 
      &\Pi^{(\lambda)}_{ij} \approx 0. 
   \end{align}
\end{subequations}
(Note that the definition \eqref{HamfrConjMom1b} of the momenta $p_1^{ij}$ is
now a primary constraint since $\fpR=1$.)
\nl 
Performing a Legendre transformation we obtain the canonical Hamiltonian 
density,
\begin{equation} \label{HamOstADMCanHam}
   \Fh_{\rmc} \bigl(h, K, p, \cp \bigr) 
      = \frac{\cp}6 \Bigl( \emb{R} - 3 K_{ij} K^{ij} + K^2 \Bigr) - 
        2 p^{ij} K_{ij},
\end{equation}     
and a surface integral at spatial infinity, 
\begin{equation} 
   - 2 \oint_{\Sigma^{\infty}} \! \rmd \sigma_k N^{\mid k},
\end{equation}     
which can also be derived straightly by setting $\cp=-6\sqrt{h}$ in the 
expression \eqref{HamfrSurTer}. The Dirac Hamiltonian density of the system is 
given by
\begin{equation}
   \Fh_{\EuD} = \Fh_{\rmc} + \mu_{kl} \varphi^{kl} + 
                \nu^{kl} \Pi^{(\lambda)}_{kl} + \xi_{kl} \phi^{kl} +
                \zeta \Xi.
\end{equation}     
Preservation in time of the primary constraints \eqref{HamOstADMPriCon} 
produces the secondary constraints
\begin{equation} \label{HamOstADMSecCon}
     \chi^{ij} = 2 p^{ij} + \frac13 \cp K^{ij} \approx 0
\end{equation}
and enables one to determine all the Lagrange multipliers. All constraints are 
second class; we can impose them as strong equations in the action, provided 
we define the appropriate Dirac bracket. We now demonstrate that the reduction 
process that consists in eliminating all the second-class constraints leads to 
the \adm form of the canonical action. 
\begin{OstADM}
The action of general relativity in the `Ostrogradsky-Hamiltonian' form
\begin{equation} \label{HamOstADMHamAct1}
   \begin{split}
      S_{\text{\textsc{ostro}}} &= 
         \int \! \rmd t 
            \biggl[ 
               \int_{\Sigma_t} \!
                  N \Bigl( 
                       p^{ij} \, \Lie h_{ij} + 
                       \cp^{ij} \, \Lie K_{ij} - 
                       \Fh_{\rmc} (h, K, p, \cp) 
                    \Bigr) \\ 
                                & \qquad \qquad \qquad \qquad \qquad -
                    \frac1{8 \pi} \oint_{S^{\infty}_t} \!
                       \sqrt{\abs{h}} \bigl( r_j N^{\mid j} \bigr)
            \biggr],
   \end{split}
\end{equation}
with the canonical Hamiltonian density given by equation 
\eqref{HamOstADMCanHam}, coincides exactly, after all the second-class 
constraints have been solved, with the \adm action of \gr (without the 
\emph{ad hoc} compensating boundary term $\oint \! K$),
\begin{equation} \label{HamOstADMHamAct2}
   \begin{split}
      S_{\text{\adm}} &= 
         \int \! \rmd t 
            \biggl[ 
               \int_{\Sigma_t} 
                  \Bigl( 
                     \piadm^{ij} \, \partial_t h_{ij} - N \EuH - N^i \EuH_i -
                     2 \partial_{\fn} \bigl( \sqrt{h} \, K \bigr)
                  \Bigr) \\
                      & \qquad \qquad \qquad \qquad -
                  \frac1{8 \pi} \oint_{S^{\infty}_t} 
                     \Bigl( 
                        r_j \bigl( 
                               \sqrt{\abs{h}} \, N^{\mid j} + 
                               N_i \, \piadm^{ij} 
                            \bigr) 
                     \Bigr)
            \biggr], 
   \end{split}
\end{equation}
with the super-Hamiltonian and super-momentum defined by \eqref{CanGRSuper}.
\end{OstADM}
\begin{proof}
Firstly, we expand the `$p \, \dot{q}$' terms in the Ostrogradsky action 
\eqref{HamOstADMHamAct1}, making use of formul{\ae} \eqref{CanGRExtCur3} and 
\eqref{CanGRLieK}, \viz
\begin{equation*}
   \begin{split}
      N \Bigl[ p^{ij} \, \Lie h_{ij} + \cp^{ij} \, \Lie K_{ij} \Bigr] 
      &= p^{ij} \, \partial_t h_{ij} - 2 p^{ij} N_{i \mid j} +
         \cp^{ij} \, \partial_t K_{ij} \\ 
      &\qquad -
       \cp^{ij} \Bigl( N^k K_{ij \mid k} + 2 N^k_{\ \, \mid i} K_{jk} \Bigr).
   \end{split}
\end{equation*}
Solving the second-class constraints amounts to eliminating variables
$\cp^{ij}$ and $K_{ij}$ according to the strong equations
\begin{align*}
   &\cp^{ij} \equiv - 2 \sqrt{h} \, h^{ij},    
   &p^{ij} \equiv  \sqrt{h} \, K^{ij}.
\end{align*}
Hence we obtain successively
\begin{equation*}
   \begin{split}
      p^{ij} \, \partial_t h_{ij} + \cp^{ij} \, \partial_t K_{ij} 
         &= \bigl( - p^{ij} + p h^{ij} \bigr) \partial_t h_{ij} -
            2 \partial_t \bigl( \sqrt{h} \, K \bigr) \\
         &= \piadm^{ij} \, \partial_t h_{ij} - 
            2 \partial_t \bigl( \sqrt{h} \, K \bigr),
   \end{split}
\end{equation*}
and
\begin{equation*}
   \begin{split}
      - 2 p^{ij} N_{i \mid j} 
      - \cp^{ij} \Bigl( N^k K_{ij \mid k} + 2 N^k_{\ \, \mid i} K_{jk} \Bigr)
         &= 2 \Bigl( p^{ij} N_{i \mid j} + N^i p_{\mid i} \Bigr) \\
         &= 2 \bigl( p^{ij} N_i \bigr)_{\mid j} + 
            2 N_i \pi^{ij}_{\text{\normalfont{\adm}} \ \mid j}.  
   \end{split}
\end{equation*}
Summing up the above results we have
\begin{equation*}
   \begin{split}
      N \Bigl[ p^{ij} \, \Lie h_{ij} + \cp^{ij} \, \Lie K_{ij} \Bigr] 
      &= \piadm^{ij} \, \partial_t h_{ij} -
         2 \partial_{\fn} \bigl( \sqrt{h} \, K \bigr) \\
      & \qquad -
         2 \bigl( \piadm^{ij} N_i \bigr)_{\mid j} +   
         2 N_i \pi^{ij}_{\text{\normalfont{\adm}} \ \mid j},  
   \end{split}
\end{equation*}
where we have utilised the identity
\begin{equation*}
   2 \partial_t \bigl( \sqrt{h} \, K \bigr) -
   2 \bigl( p^{ij} N_i \bigr)_{\mid j} \equiv
   2 \partial_{\fn} \bigl( \sqrt{h} \, K \bigr) +
   2 \bigl( \piadm^{ij} N_i \bigr)_{\mid j}.
\end{equation*}
The term $-2 \bigl( \piadm^{ij} N_i \bigr)_{\mid j}$ yields the surface 
integral at spatial infinity in the \adm action \eqref{HamOstADMHamAct2}, \ie
\begin{equation*}
   - \frac1{8 \pi} \oint_{S^{\infty}_t} \Bigl( r_j \piadm^{ij} N_i \Bigr);
\end{equation*}
the term $2 N_i \pi^{ij}_{\text{\normalfont{\adm}} \ \mid j}$ gives the \adm
super-momentum \eqref{CanGRSuperb}.
\nl
Now we examine the Ostrogradsky canonical Hamiltonian density in 
\eqref{HamOstADMHamAct1}. We readily obtain from equation 
\eqref{HamOstADMCanHam} the expression
\begin{equation*}
   \Fh_{\rmc} \bigl(h, K, p, \cp \bigr) 
      \equiv - \sqrt{h} \, \emb{R} 
             + \frac1{\sqrt{h}} 
               \Bigl( 
                  \piadm^{ij} \pi^{\text{\normalfont{\adm}}}_{ij} - 
                  \frac12 \piadm^2 
               \Bigr),
\end{equation*}
which obviously coincides with the \adm super-Hamiltonian \eqref{CanGRSuperb}. 
\end{proof}

\subsubsection{Boundary terms in the light of the Ostrogradsky approach}
\label{subsub:HamOstADMBouCond}

The equivalence of the Ostrogradsky and \adm canonical versions of the 
gravitational action of \gr sheds a new light upon the question of boundary
terms in the gravitational action. We can regard the Einstein--Hilbert action
as a limiting case of the nonlinear gravitational action when $\fR \equiv R$.
In the latter case the only boundary term is the surface integral 
\eqref{HamfrSurTer}. The arguments that lead to the adjunction of the 
compensating term $2 \oint \! K \rmd \Sigma$ in the Einstein--Hilbert action 
come to naught in the higher-order case.%
\footnote{This was firstly argued by Horowitz in the context of Euclidean 
          quantum gravity with a positive-definite action containing quadratic 
          curvature terms \cite{horow=QuaCos} (see also the similar work of 
          Barth, where the Euler characteristic class for the manifold is 
          taken into account \cite{barth=FouGra}). As this applies to the 
          nonlinear case as well, we take up Horowitz's line of thought 
          hereafter. }
\nl
Firstly, there is no need whatsoever to eliminate the second-order derivatives 
of the metric. The Ostrogradsky construction of Subsection~\ref{subsec:Hamfr} 
shows that the configuration space is spanned by the induced metric and the 
trace of the extrinsic curvature. As a consequence, no compensating boundary 
terms are required to cancel surface integrals involving the latter quantity: 
The variation of $K$ vanishes on the boundary $\partial \EuM$.%
\footnote{In \gr imposing that $\delta K$ be zero on the boundary is strictly
          forbidden since this would overdetermine the theory.}
\nl
Secondly, consider a transition from an initial configuration 
$(h^{(1)}_{ab}, K^{(1)})$ on a hypersurface $\Sigma_1$ to a configuration 
$(h^{(2)}_{ab}, K^{(2)})$ on an intermediate hypersurface $\Sigma_2$, then 
followed by a transition to a final configuration $(h^{(3)}_{ab}, K^{(3)})$ on 
a hypersurface $\Sigma_3$. As indicated above (on page~\pageref{path-int}), in 
a consistent path-integral formulation of the theory one expects that the 
amplitude to go from the initial to the final configuration should be obtained 
by integrating over all configurations on the intermediate hypersurface 
$\Sigma_2$. This amounts to require
\begin{equation} \label{HamOstADMPatInt}
   S \bigl[g^{(I)}_{ab} + g^{(II)}_{ab} \bigr] = 
      S \bigl[g^{(I)}_{ab} \bigr] + S \bigl[g^{(II)}_{ab} \bigr],
\end{equation}
where $g^{(I)}_{ab}$ is the metric between $\Sigma_1$ and $\Sigma_2$, which 
induces $(h^{(1)}_{ab}, K^{(1)}_{ab})$ on $\Sigma_1$ and 
$(h^{(2)}_{ab}, K^{(2)})$ on $\Sigma_2$; $g^{(II)}_{ab}$ is the metric between 
$\Sigma_2$ and $\Sigma_3$, which induces $(h^{(2)}_{ab}, K^{(2)})$ on 
$\Sigma_2$ and $(h^{(3)}_{ab}, K^{(3)})$ on $\Sigma_3$; and 
$g^{(I)}_{ab}+g^{(II)}_{ab}$ is the metric obtained by joining together the 
two regions. In \gr the three-metrics induced on the intermediate hypersurface 
$\Sigma_2$ by $g^{(I)}_{ab}$ and $g^{(II)}_{ab}$ agree; but the extrinsic 
curvatures need not: this results in a $\delta$-function in the Ricci 
curvature of $g^{(I)}_{ab}+g^{(II)}_{ab}$ of strength given by the difference 
of the extrinsic curvatures on $\Sigma_2$ \cite{hawki=PatInt1}. In the 
nonlinear case the fact that the respective traces of the extrinsic curvature
tensors are one and the same on $\Sigma_2$ is sufficient to remove the 
difficulty: This is fairly confirmed by the explicit computation of the 
boundary variation of the nonlinear action \cite{madse=DeSitter}, \viz
\begin{equation} \label{HamOstADMMadsen}
   \delta S \bigr|_{\partial \EuM} = 2 \oint_{\partial \EuM} \! \fpr \delta K.
\end{equation}
It is not, in general, possible to write the right-hand side of equation 
\eqref{HamOstADMMadsen} as the variation on the boundary of a functional, as 
is the case in \gr; the only circumstances under which this programme can 
effectively been achieved are when the space-time manifold is maximally 
symmetric \cite{madse=DeSitter}. As a matter of fact, according to the 
discussion above on the crucial r\^ole of $K$, the right-hand side of equation 
\eqref{HamOstADMMadsen} actually vanishes: The possible compensating term is 
reduced to naught. 
\nl
Furthermore, when passing from the nonlinear case to the linear Einstein--%
Hilbert action, one clearly sees how the boundary term 
$2 \oint \! K \rmd \Sigma$ materialises: The trace $K$ becomes a spurious \dof 
and thereby cannot be held fixed any longer on the boundary.

\subsection{Hamiltonian formulation of quadratic theories of gravity}
\label{subsec:HamQua}

In the introduction of Section \ref{sec:TheHig} on page~%
\pageref{subsec:TheHigIntro} we have reviewed the works inspired by the 
Ostrogradsky method, with regard to the canonical formulation of \hotg. 
In addition, there are two major contributions dealing with quadratic theories of 
gravity that do not explicitly resort to the Ostrogradsky construction. Both 
of them require the reduction of the theory to a first-order form. To achieve 
this end Safko and Elston examined the connection between the Lagrange 
multiplier approach, thoroughly analysed in Section~\ref{sec:VarConHOG}, and 
the \adm formalism \cite{safko=LagMul}. They pointed out that there is a sharp 
difference between the constrained Palatini variational method and the 
Lagrange multiplier method that they used to cast higher-order curvature 
invariants into canonical form. Applying an `Ostrogradsky-like' constrained 
formalism they succeeded---at least formally---in formulating a canonical 
version of the pure $R^2$ theory. We are indebted to Boulware for the second 
contribution, in which he has undertaken the quantisation programme of 
quadratic gravity and has addressed a certain number of technical and physical 
issues arising when higher derivatives are present in a (gravitational) field 
theory \cite{boulw=QuaHig}. In contrast with Buchbinder and Lyakhovich's 
systematic approach (\cfr page~\pageref{page:BL}), Boulware's \emph{modus 
operandi} seems quite heuristic at first glance. (However, in our opinion this 
is still the best reference \wrt canonical quadratic gravity.) One advantage 
of Boulware's formalism lies in its extensive use of the Lie derivative---%
instead of time differentiation---which appreciably simplifies 
the technicalities; this contrasts with Buchbinder and Lyakhovich's formalism, 
which is---in that respect---quite difficult to decipher. 
\nl
In this subsection we rely on the generalised constrained Ostrogradsky scheme,
which was thoroughly analysed in Subsection~\ref{subsec:OstroSingBL} and 
already applied successfully to nonlinear gravity theories in Subsection~%
\ref{subsec:Hamfr}: We intend to cast the generic quadratic gravitational 
action into Hamiltonian form.
\begin{rem}
The approach given here differs slightly from Boulware's as well as Buchbinder 
and Lyakhovich's treatments in the way we set it up. It is perhaps more 
transparent in regard to the Ostrogradsky method. The three methods lead to 
the same results, eventually. 
\end{rem}

\paragraph*{The generic quadratic theory.}

We consider the most general quadratic Lagrangian density in a four-%
dimensional space-time $\bigl( \EuM, g_{ab} \bigr)$, 
\begin{equation} \label{HamQuaLag1}
   \Fl = \sqrt{- g} \, \Bigl( \Lambda + \frac1{2 \kappa^2} R -
                              \frac{\alpha}4 C_{abcd} C^{abcd} +
                              \frac{\beta}8 R^2
                       \Bigr),  
\end{equation}     
where $\Lambda$ is the cosmological constant, $\kappa^2 = 8 \pi G_{\msc{n}}$, 
and $\alpha$, $\beta$ are dimensionless coupling constants. 
\begin{rem}
Owing to the Gauss--Bonnet or Lanczos topological invariant in four 
dimensions, we could equally have chosen $\mu R_{ab} R^{ab} + \nu R^2$ as the 
relevant quadratic Lorentz-invariant combination in the Lagrangian density, 
with $\mu=-\alpha/2$ and $\nu=\beta/8+\alpha/6$. However---it is just a 
matter of convenience---,the choice exhibited in equation \eqref{HamQuaLag1} 
is more appropriate for discussing the particular variants of the general 
theory.
\end{rem}
\nl
As usual we assume that space-time is foliated into three-dimensional Cauchy 
hypersurfaces $\Sigma_t$ and we adopt the \adm basis (\cfr the $3+1$--%
splitting on page~\pageref{subsec:CanGR}). We recall here the expression 
\eqref{CanGRScaCur2} of the scalar curvature, \viz
\begin{equation} \label{HamQuaScaCur} 
   R = \emb{R} + K^2 - 3 K_{ij} K^{ij} - 2 N^{-1} h^{ij} N_{\mid ij} 
               - 2 h^{ij} \Lie K_{ij}, 
\end{equation}
and we specialise formul{\ae} \eqref{CanGRCodYorWeyl} to the \adm basis, \viz
\begin{subequations} \label{HamQuaCodYorWeyl}
   \begin{align} 
      C^{\fn}_{\ \, ijk} &= 
         \Bigl[ 
            \delta_i^r \delta_j^s \delta_k^t - 
            \frac12 h^{rt} \bigl( h_{ik} \delta_j^s - h_{ij} \delta_k^s \bigr)
         \Bigr] 
         \bigl( K_{rs\mid t} - K_{rt\mid s} \bigr), 
      \label{HamQuaCodYorWeyla} \\
      C_{\fn i \fn j}    &=
         \frac12 \bigl( \delta_i^k \delta_j^l - \frac13 h_{ij} h^{kl} \bigr)
            \bigl( 
               \Lie K_{kl} + N^{-1} N_{\mid kl} + K K_{kl} + \emb{R}_{kl}
            \bigr). \label{HamQuaCodYorWeylb}
   \end{align}
\end{subequations}       
\nl
Since the Weyl tensor identically vanishes in three dimensions, the quadratic 
Weyl invariant $C_{abcd} C^{abcd}$ reduces to
\begin{equation} \label{HamQuaWeylInv}
   C_{abcd} C^{abcd} = 4 \bigl( 
                            2 C_{\fn i \fn j} C^{\fn i \fn j} + 
                              C_{\fn ijk} C^{\fn ijk} 
                         \bigr),
\end{equation}
where the explicit forms of $C_{\fn ijk}$ and $C_{\fn i \fn j}$ are displayed 
in equations \eqref{HamQuaCodYorWeyl} above. 
\nl
As in the nonlinear case (\cfr Subsection~\ref{subsec:Hamfr}), we regard the 
Lagrangian density \eqref{HamQuaLag1} as a functional of the three-metric and 
its successive Lie derivatives, namely
$\Fl \bigl( h_{ij}, \Lie h_{ij}, \Lie^2 h_{ij} \bigr)$; we then introduce 
Ostrogradsky auxiliary variables according to the same prescriptions as given 
by equations \eqref{HamfrOst1}; we replace the original Lagrangian density 
\eqref{HamQuaLag1} by an extended one, \ie $\Flb$, and define the canonically 
conjugate momenta: 
\begin{subequations} \label{HamQuaConjMom1}    
   \begin{align}
      p^{ij}   &:= \parD{\Flb}{\Lie h_{ij}} = \lambda^{ij}, 
         \label{HamQuaConjMom1a} \\  
      \cp^{ij} &:= \parD{\Flb}{\Lie K_{ij}} 
                 = N^{-1} \parD{\Fl}{\bigl(\Lie K_{ij}\bigr)}
                 = - \sqrt{h} \Bigl[ 
                                 \bigl( \kappa^{-2} + \frac{\beta}2 R \bigr) + 
                                 2 \alpha C^{\fn i \fn j} 
                              \Bigr],
         \label{HamQuaConjMom1b} \\  
      \Pi^{(\lambda)}_{ij} &:= \parD{\Flb}{\Lie \lambda^{ij}} = 0.
         \label{HamQuaConjMom1c}  
   \end{align}
\end{subequations}
At this stage note that the definition of $\cp^{ij}$ differs from Boulware's 
\emph{choice}, which requires that $\cp^{ij}$ be zero at flat space. This 
option is adopted by Boulware because the linear term and the quadratic terms 
in the departing action are not treated on an equal footing in his formalism: 
The former is handled as in the \adm canonical version of \gr whereas the 
latter are reduced to first order by means of a generalised Legendre 
transformation. In the formalism presented here---by contrast---we refrain to 
make such a segregation for we have shown previously that the generalised 
Ostrogradsky construction may be consistently applied to the \adm action (\cfr 
Subsection~\ref{subsec:HamOstADM}). 
\nl
The trace and traceless parts of $\cp^{ij}$ are respectively
\begin{align} \label{HamQuacp}
   &\cp = - 3 \sqrt{h} \bigl( \kappa^{-2} + \frac{\beta}2 R \bigr), 
   &\cp^{\ttT ij} = - 2 \alpha \sqrt{h} \, C^{\fn i \fn j}. 
\end{align}
The sole primary constraints of the general quadratic theory are 
$\varphi^{ij} \approx 0$ and $\Pi^{(\lambda)}_{ij} \approx 0$. Therefore, in
agreement with the particle content of the theory, there are eight physical 
\dofs.%
\footnote{The number of physical \dofs is determined by the equation 
          $N_{\text{phys.}} = K - N_1 - \tfrac12 N_2$, where $K$ is the total 
          number of pairs of canonical variables, $N_1$ and $N_2$ are the 
          number of first-class and second-class constraints respectively; see 
          \cite[p.~29]{HENNE=QuaGau}.}  
\nl
The next step in the Ostrogradsky construction consists in performing a 
generalised Legendre transformation. Here we want to solve the `velocities'
$\Lie K_{ij}$ for the canonical variables and their conjugate momenta. We
achieve this goal by firstly solving equations \eqref{HamQuaScaCur} and 
\eqref{HamQuaCodYorWeylb} for $R$ and $C^{\fn i \fn j}$ respectively, and then 
expressing these latter quantities in terms of $\cp$ and $\cp^{\ttT ij}$
respectively, with the help of formul{\ae} \eqref{HamQuacp}, thereby obtaining
\begin{equation} \label{HamQuaVelo} 
   \begin{split}
      \Lie K_{ij} &= - \frac{\cp^{\ttT ij}}{\alpha \sqrt{h}} 
                     + \frac1{3 \beta} \Bigl( 
                                          \kappa^{-2} + \frac{\cp}{3 \sqrt{h}}
                                       \Bigr) h_{ij} 
                     - \frac{N_{\mid ij}}{N}
                     - \emb{R}_{ij} \\
                  & \qquad  
                     - K K_{ij} \cp^{ij}
                     + \frac{\cp}2 \bigl( 
                                      \emb{R} + K^2 - K_{ij} K^{ij} 
                                   \bigr). 
   \end{split}
\end{equation}
Hence the restricted Hamiltonian density is
\begin{equation}
   \begin{split}
      \Fh_{\rmr} \bigl(h, K, \cp \bigr) 
         &= \cp^{ij} \Lie K_{ij} - N^{-1} \Fl \\
         &= \sqrt{h} \bigl( 
                        \alpha C_{\fn ijk} C^{\fn ijk} + V(\cp) - \Lambda
                     \bigr) +
            \frac{\cp}2 \bigl( \emb{R} + K^2 - K_{ij} K^{ij} \bigr) \\
         &  \qquad -
            \frac{\cp^{\ttT ij} \cp^{\ttT}_{ij}}{2 \alpha \sqrt{h}} -
            \cp^{ij} \bigl(
                        \emb{R}_{ij} + K K_{ij} + N^{-1} N_{\mid ij}
                     \bigr),
   \end{split}
\end{equation}     
where we have defined the `potential term'
\begin{equation}
     V(\cp) := \frac1{2 \beta} \Bigl( 
                                  \kappa^{-2} + \frac{\cp}{3 \sqrt{h}} 
                               \Bigr)^2.
\end{equation}     
The canonical Hamiltonian density is thus given by                       
\begin{equation}
   \Fh_{\rmc} \bigl(h, K, p, \cp \bigr) 
      = \Fh_{\rmr} \bigl(h, K, \cp \bigr) - 2 p^{ij} K_{ij}.
\end{equation}     
Therefore, the Hamiltonian form of the gravitational action based on the 
original Lagrangian density \eqref{HamQuaLag1} may be written as 
\begin{equation} \label{HamQuaHamAct}
   \begin{split}
      S &= \int \! \rmd t 
              \biggl[ 
                 \int_{\Sigma_t} \! N \Bigl( 
                                         p^{ij} \, \Lie h_{ij} + 
                                         \cp^{ij} \, \Lie K_{ij} - 
                                         \Fh_{\rmc} \bigl( h, K, p, \cp \bigr) 
                                      \Bigr) \\
        & \qquad \qquad \qquad -
                 \oint_{S^{\infty}_t}   
                    \Bigl( 
                       N \cp^{jk}_{\ \ \, \mid j} - N_{\mid j} \cp^{jk} 
                    \Bigr) r_k
              \biggr], 
   \end{split}
\end{equation}
where the canonical Hamiltonian density is given by
\begin{equation} \label{HamQuaCanHam}
   \begin{split}
      \Fh_{\rmc} \bigl(h, K, p, \cp \bigr) 
         &= - 2 p^{ij} K_{ij} 
            + \sqrt{h} \bigl( 
                          \alpha C_{\fn ijk} C^{\fn ijk} + V(\cp) - \Lambda
                       \bigr) \\
         &  \qquad 
            - \frac{\cp^{\ttT ij} \cp^{\ttT}_{ij}}{2 \alpha \sqrt{h}} 
            + \frac{\cp}2 \bigl( \emb{R} + K^2 - K_{ij} K^{ij} \bigr) \\
         & \qquad
            - \cp^{ij}_{\ \ \mid ij}
            - \cp^{ij} \bigl( \emb{R}_{ij} + K K_{ij} \bigr). 
   \end{split}
\end{equation}     
Expanding the Lie derivatives in the action \eqref{HamQuaHamAct} and 
integrating by parts we obtain the final form of the quadratic canonical 
action, where we have retained all the surface terms, \viz
\begin{equation} \label{HamQuaCanAct}
   \begin{split}
      S &= \int \! \rmd t 
              \biggl[ 
                 \int_{\Sigma_t} \! N \Bigl( 
                                         p^{ij} \, \partial_t h_{ij} + 
                                         \cp^{ij} \, \partial_t K_{ij} - 
                                         N \EuH - N^k \EuH_k 
                                      \Bigr) \\
        & \qquad \qquad \qquad -
                 \oint_{S^{\infty}_t}   
                    \Bigl[ 
                       N \cp^{jk}_{\ \ \, \mid j} - N_{\mid j} \cp^{jk} +
                       2 N_j \bigl( p^{jk} + \cp^{ik} K^j_{\ i} \bigr)
                    \Bigr] r_k
              \biggr], 
   \end{split}
\end{equation}
where the super-Hamiltonian $\EuH$ constraint coincides with the canonical 
Hamiltonian density \eqref{HamQuaCanHam} and the super-momentum constraint is 
given explicitly by
\begin{equation} \label{HamQuaSupMom}
   \EuH_k = \cp^{ij} K_{ij \mid k} - 
            2 \bigl( \cp^{ij} K_{ik} \bigr)_{\mid j} -
            2 h_{ik} p^{ij}_{\ \ \mid j}.
\end{equation}     
\begin{rem} \label{rem:PBBoulware}
Since the constraints are local functions of the field quantities at a 
particular point of space, the Poisson bracket of two such objects is a sum of 
delta functions and derivatives of delta functions. In many instances it is 
simpler to integrate the constraints over the spatial sections with a test 
function: The constraint $\varphi(\bfx) \approx 0$ becomes
\begin{equation*}
   \varphi \bigl[ \psi \bigr] := 
      \int \! \rmd^3 x \, \psi (\bfx) \varphi (\bfx) \approx 0
\end{equation*}
and it must be functionally differentiated \wrt the canonical variables so 
that the Poisson bracket of two such (integrated) constraints be given by
\begin{equation} \label{HamQuaPBDef}
   \begin{split}
      &\Bigl\{ 
         \varphi_{\rmA} \bigl[ \psi \bigr], 
         \varphi_{\rmB} \bigl[ \psi^{\prime} \bigr]  
       \Bigr\} := \\
   &\qquad \int \! \rmd^3 x 
      \Biggl[ 
         \varD{\varphi_{\rmA} \bigl[ \psi \bigr]}{h_{kl}}
         \varD{\varphi_{\rmB} \bigl[ \psi^{\prime} \bigr]}{p^{kl}} +
         \varD{\varphi_{\rmA} \bigl[ \psi \bigr]}{K_{kl}}
         \varD{\varphi_{\rmB} \bigl[ \psi^{\prime} \bigr]}{\cp^{kl}} -
         \Bigl( \varphi_{\rmA} \longleftrightarrow \varphi_{\rmB} \Bigr)
      \Biggr],
   \end{split}
\end{equation}
where the canonical variables are varied freely, \ie independently of the 
constraints. 
\nl
The Poisson brackets involving the super-momentum constraint are easily 
calculated and understood, for $\EuH_k$ is the generator of spatial 
diffeomorphisms. For instance, a direct application of the definition 
\eqref{HamQuaPBDef} yields
\begin{equation*}
   \Bigl\{ \sfG \bigl[ \boldsymbol{\psi} \bigr], h_{kl} (\bfx) \Bigr\} = 
      - \psi_{(k \mid l)} (\bfx),
\end{equation*}
where we have defined the integrated super-momentum as
\begin{equation*}
   \sfG \bigl[ \boldsymbol{\psi} \bigr] := 
      \int \! \rmd^3 x \, \EuH_m (\bfx) \psi^m (\bfx).
\end{equation*}
On the other hand, it is very difficult to compute the brackets involving the
super-Hamiltonian constraint. (Such a tedious task was carried out by 
Boulware; see \cite[Appendix B]{boulw=QuaHig}.)
\end{rem}
\nl
We now look at the possible simplifications that would arise if we specialised
the generic quadratic action \eqref{HamQuaLag1}. Amongst the five distinct 
variants of the general theory, which were analysed by Buchbinder and 
Lyakhovich \cite{buchb=CanQua}, only one turns out to be relevant for our
purpose here, namely the conformally invariant theory ($\beta=0$, 
$\Lambda=0=\kappa^{-1}$). Strictly speaking, the case $\alpha=0$---and in 
particular the pure $R^2$ theory---reduces to studying gravitational actions 
of the nonlinear type (\cfr Subsection~\ref{subsec:Hamfr}) whereas the cases 
corresponding to the vanishing of the sole coupling constant $\beta$ may be 
set aside if one invokes the local conformal symmetry \emph{ab initio}.

\paragraph*{Conformally invariant theory.}

In the conformally invariant case the Lagrangian density is 
\begin{equation} \label{HamQuaWeylLag}
   \Fl = - \alpha N \sqrt{h} \Bigl( 
                                2 C_{\fn i \fn j} C^{\fn i \fn j} + 
                                  C_{\fn ijk} C^{\fn ijk} 
                             \Bigr)
\end{equation}
owing to the identity \eqref{HamQuaWeylInv}. The only alteration in the 
conjugate momenta \eqref{HamQuaConjMom1} involves $\cp^{ij}$, which becomes
\begin{equation} \label{HamQuaWeylConjMom}
   \cp^{ij} = - 2 \alpha \sqrt{h} \, C^{\fn i \fn j}. 
\end{equation}
Now, in addition to $\varphi^{ij} \approx 0$ and 
$\Pi^{(\lambda)}_{ij} \approx 0$, we obtain the primary constraint
\begin{equation} \label{HamQuaWeylPriCon}
   \phi = h_{ij} \cp^{ij} = \cp \approx 0,
\end{equation}
which arises because only the traceless part of the `velocities', \ie
$\bigl( \Lie K_{ij} \bigr)^{\ttT}$, can be extracted from equation 
\eqref{HamQuaCodYorWeylb}. Performing as usual a Legendre transformation on 
the appropriate extended Lagrangian density we obtain the canonical
Hamiltonian density
\begin{equation} \label{HamQuaWeylCanHam}
   \begin{split}
      \Fh_{\rmc} \bigl(h, K, p, \cp \bigr) 
         = &- 2 p^{ij} K_{ij} + \alpha \sqrt{h} \, C_{\fn ijk} C^{\fn ijk} 
            - \frac{\cp^{\ttT ij} \cp^{\ttT}_{ij}}{2 \alpha \sqrt{h}} \\
           &- \cp^{\ttT ij}_{\ \ \ \mid ij}
            - \cp^{\ttT ij} \bigl( \emb{R}_{ij} + K K^{\ttT}_{ij} \bigr). 
   \end{split}
\end{equation}     
Therefore, we may cast the conformally invariant action into canonical form,
that is, up to boundary terms,
\begin{equation} \label{HamQuaHamWeylAct}
   S = \int_{\EuM} \! N \Bigl[ 
                           p^{ij} \, \Lie h_{ij} + 
                           \cp^{ij} \, \Lie K_{ij} - 
                           \Fh_{\rmc} \bigl( h, K, p, \cp \bigr) 
                        \Bigr].
\end{equation}
The associated Dirac Hamiltonian density is
\begin{equation} \label{HamQuaWeylDirHam}
   \Fh_{\EuD} = \Fh_{\rmc} + \mu_{kl} \varphi^{kl} 
                           + \nu^{kl} \Pi^{(\lambda)}_{kl} 
                           + \xi \phi.
\end{equation}
Consistency of the primary constraints \eqref{HamQuaWeylPriCon} when time 
evolution is considered yields the determination of the multipliers $\mu_{kl}$ 
and $\nu^{kl}$ and gives rise to the secondary constraint
\begin{equation} \label{HamQuaWeylSecCon}
   \chi = 2 p + K^{\ttT}_{kl} \cp^{\ttT kl} \approx 0.
\end{equation}
According to the remark made on page~\pageref{rem:PBBoulware}, we may utilise 
test functions in order to simplify the computation of Poisson brackets.
After tedious calculation we obtain
\begin{align*}
   &\Bigl\{ \sfG \bigl[ \boldsymbol{\psi} \bigr], \chi (\bfx) \Bigr\} \propto
       \bigl( \psi^m \chi \bigr)_{\mid m} (\bfx) \approx 0,
   &\Bigl\{ \sfH \bigl[ \boldsymbol{\zeta} \bigr], \chi (\bfx) \Bigr\} 
       \propto \zeta (\bfx) \, \Fh_{\rmc} (\bfx) \approx 0,
\end{align*}
where we have defined
\begin{equation*}
   \sfH \bigl[ \boldsymbol{\zeta} \bigr] := 
      \int \! \rmd^3 x \, \Fh_{\rmc} (\bfx) \zeta (\bfx).
\end{equation*}
Hence the constraint \eqref{HamQuaWeylSecCon} is the unique secondary 
constraint stemming from the consistency algorithm. Furthermore it is first 
class---it is indeed the generator of conformal transformations---by virtue of
the Poisson bracket      
\begin{equation*}
   \Bigl\{ \phi (\bfx), \chi (\bfy) \Bigr\} = 2 \cp \, \delta (\bfx - \bfy)
      \approx 0.
\end{equation*}
Therefore, the number of physical \dofs is six. This is in total agreement 
with the particle content of the linearised theory.
\nl
To proceed further we need to impose a gauge-fixing condition; we shall see
in Chapter~\ref{chap:BiaCos} how this can be realised, in the specific case of
spatially homogeneous cosmologies, so that the trace of variables $K_{ij}$ and 
$\cp^{ij}$ be removed from the set of canonical variables and the conformal
constraint \eqref{HamQuaWeylSecCon} be automatically satisfied.


\chapter{Higher-order spatially homogeneous cosmologies}
\label{chap:BiaCos}

\begin{minipage}{11cm}
\begin{quote}
\textit{``This cosmos, which is the same for all, has not been made by any 
          god or man, but it always has been, is, and will be, an ever-living
          fire, kindling itself by regular measures and going out by regular
          measures.''} 
\nl
\hfill
--- Heraclitus of Ephesus.
\end{quote}
\end{minipage}

\vspace{1cm}

\dropping{3}{T}\textsc{he} simplest class of space-times that give physically
reasonable cosmological models are the famed Friedmann--Lema\^itre--Robertson%
--Walker (\flrw) spaces, which are isotropic and are homogeneous on spacelike
sections. It is conventional wisdom that \flrw models describe faithfully the 
large-scale properties of our universe. They indeed successfully account for 
many of its observed features and constitute the basic pillar of the standard 
model of cosmology, which is not yet free from certain riddles: the so-called 
horizon, `flatness', and `smoothness' problems. These are addressed, more or 
less adequately, by incorporating the inflationary paradigm into the `bare 
old-fashioned' isotropic cosmology and by invoking the existence of hot and 
cold dark matter blends. Be that as it may, mathematical cosmology cannot be 
satisfied with those models for their degree of simplicity, though it 
guarantees that one could easily find closed-form solutions, also implies that
deeper investigations on specific issues---such as the asymptotic approach 
towards the initial singularity and the related question of the genericness of 
oscillatory, chaotic dynamical regimes---are screened by the stringent 
symmetry requirement that the universe be isotropic \emph{ab initio}. Even 
though the observed universe seems to be highly isotropic, there may have been 
large anisotropies at an earlier epoch; in that respect, it is of great 
interest to understand why the actual universe is so much less anisotropic 
than Einstein's field equations allow it to be in principle. 
\nl
One possible answer is to consider, as Penrose surmises, that the `initial 
conditions' logically entail a vanishing Weyl tensor to ensure an isotropic 
universe from its very inception; another attitude towards that problem, which 
we adopt below, proceeds through generalising the \flrw models by dropping the
isotropy hypothesis while keeping spatial homogeneity.%
\footnote{As isotropy at each point entails spatial homogeneity, anisotropic 
          cosmologies constitute the first class of cosmological models that 
          are (fairly) more general than the \flrw models; they have been 
          extensively studied for a couple of decades. By contrast, it is 
          only very recently that \emph{inhomogeneous} cosmological models 
          have regained a genuine interest amongst researchers; see 
          Krasi\'nsky's `encyclop{\ae}dia' for a recent and thorough account 
          \cite{KRASI=InhCos}.}
In particular, this endows anisotropic cosmologies with a nonzero Weyl tensor, 
the existence of which may alter the character of the singularities that 
typically arise in cosmology (see, \eg \cite{colli=SinBia}). Moreover, the 
field equations remain \odes since only time variations are nontrivial.
\nl
In Section \ref{sec:BiaCosSpaHom} we give a short account of spatially 
homogeneous cosmologies: Firstly, we summarise the geometrical setting that 
enables one to define and classify the spatially homogeneous anisotropic 
Bianchi cosmological models (for a very good synthesis the reader is referred 
to \cite{ellis=GeoCos}); then we discuss briefly the specific problem of
nonvanishing surface terms in the Lagrangian and Hamiltonian formulations of 
Bianchi cosmologies both in \gr and \hotg.
\nl 
In Section \ref{sec:BiaCosMixCos} we analyse the behaviour of the Bianchi-type 
IX model in the pure general quadratic theory of gravity on approach to the 
initial singularity in a four-dimensional space-time. This work was carried 
out in collaboration with S. Cotsakis, J. Demaret, and Y. De Rop 
\cite{cotsa=MixUni}. (We focus on the analytical treatment, leaving aside the 
details of the numerical analysis.) The---empty---Bianchi-type IX or 
\emph{mixmaster} model is of fundamental importance in mathematical cosmology 
for it is the most general spatially homogeneous model and it contains the
closed \flrw model as a special case. Its major feature is its---chaotic%
---oscillatory behaviour when one reaches the initial singularity. The 
question that is addressed in this respect is whether this behaviour is 
altered when one considers more general settings than \gr such as, for 
instance, purely quadratic theories of gravity. In this framework we show that 
the mixmaster universe possesses a \emph{nonchaotic} solution which is stable
and we prove that the so-called \textsc{bkl} approximation scheme is 
structurally unstable.
\nl
In Section \ref{sec:BiaCosHam} we make use of the general Hamiltonian 
formalism that we have developed in Subsection \ref{subsec:HamQua} 
(p.~\pageref{subsec:HamQua} ff.) to study the specific case of spatially 
homogeneous cosmologies. Firstly, we adapt the Hamiltonian formalism to 
Misner's parameterisation; for this purpose we define a canonical 
transformation the virtue of which is to disentangle terms stemming 
respectively from the pure $R$-squared and Weyl-squared variants of the 
general quadratic theory. This is helpful since one is able to treat those 
cases separately, thereby simplifying the analysis. In the first variant 
($R$-squared) we derive---for those Bianchi types admitting a canonical 
formulation---the super-Hamiltonian constraint, the \emph{reduced} Hamiltonian 
density, and the corresponding canonical equations, which constitute an 
autonomous differential system; in particular, we solve the latter 
analytically for Bianchi type I and thereafter compare our results with those 
found in the literature; for the Bianchi-type IX model we reduce the (first-%
order) canonical equations to three coupled second-order differential 
equations for the physical \dofs; finally, we extend the discussion to \flrw 
models. This work was first undertaken during my \MSc \cite{quere=ForHam} and 
was carried out with J. Demaret \cite{demar=HamFor}. In the second variant 
(Weyl-squared), referred to as \emph{conformal gravity}, we consider the 
simplest spatially homogeneous space-time that exhibits nontrivial physical 
\dofs, namely the Bianchi-type I model (the isotropic \flrw space-times are 
indeed conformally flat). We derive the explicit forms of the super-%
Hamiltonian and the constraint expressing the conformal invariance of the 
theory and we write down the system of canonical equations. To seek out exact 
solutions to this system we add extra constraints on the canonical variables 
and we go through a global involution algorithm which eventually leads to the 
closure of the constraint algebra. The Painlev\'e approach provides us with a 
proof of non-integrability, as a consequence of the presence of movable 
logarithms in the general solution of the problem.%
\footnote{I am indebted to C. Scheen for providing me with the material on the
          analytic structure of the Bianchi-type I model in conformal gravity
          (p.~\pageref{subsub:HamWeylAnaStr} ff.), which is partly contained 
          in \cite{demar=HamForWeyl}. For a thorough investigation on methods 
          that probe the analytic structure of differential systems and their 
          application to relativistic cosmology, the interested reader may 
          consult Scheen's \PhD thesis \cite{schee=PhD}.}
We extract all possible particular solutions that may be written in closed 
analytical form. This enables one to demonstrate that the global involution 
algorithm has proven to be exhaustive in the search for exact solutions. 
Finally, we discuss the conformal relationship or absence thereof of our 
solutions with Einstein spaces. More specifically we show that the necessary
condition for a Bianchi-type I space to be conformal to an \einspace, obtained
from purely geometrical considerations, becomes---in conformal gravity---a 
sufficient condition as well; this enables us to determine which solution 
amongst the whole set of exact solutions is \emph{not} conformal to an 
\einspace. This work was carried out in collaboration with J. Demaret and C. 
Scheen \cite{demar=HamForWeyl} (see also \cite{quere=ISMC}).

\section{Spatially homogeneous cosmologies}
\label{sec:BiaCosSpaHom}

\subsection{Geometrical setting}
\label{subsec:BiaCosSpaCosGeoSet}

A four-dimensional space-time $\bigl( \EuM, g_{ab} \bigr)$ is said to be 
\emph{spatially homogeneous} if it possesses an $r$-dimensional isometry group 
$\sfG^{(r)}$ (invariance group of the metric $g_{ab}$) that acts transitively 
on a one-parameter family of spacelike hypersurfaces, the orbits of the group%
\footnote{The orbit of a point $p$ under a group $\sfG^{(r)}$ is the set of 
          points into which $p$ is moved by the action of all elements of the 
          group.}
(so that $r \geqslant 3$), which provides a natural slicing of the space-time. 
At any point $q$ on any such hypersurface $\Sigma$ there are (at least) three 
nonzero linearly independent Killing vector fields tangent to $\Sigma$. Since 
a Killing vector field $\xi^a$ is completely determined by the values of 
$\xi^a$ and $\nabla_a \xi_b$ at any point $q$ of $\Sigma$, there can be at 
most $r_{\text{max}}=\tfrac12 n(n+1)$ linearly independent Killing vector 
fields on a manifold of dimension $n$ (see, \eg \cite[Appendix C.3]%
{WALD=GenRel}); here, this means $3\leqslant r \leqslant 6$. When 
$r=r_{\text{max}}$, the Riemannian space has constant curvature. Furthermore, 
a theorem due to Fubini states that a Riemannian space (for $n>2$) cannot 
admit an isometry group of dimension $r=\tfrac12 n(n+1)-1$ (see, \eg 
\cite{EISEN=RieGeo}); here, this implies that the only possible values of $r$ 
are $3,4$, and $6$. If $r=6$, the space-time is not only spatially homogeneous 
but also spatially isotropic, that is, locally spherically symmetric, and 
belongs to the \flrw class, which features maximally symmetric spacelike 
sections, \ie sections of constant curvature.%
\footnote{Note that in the case of space-times ($n=4$), the ten-parameter 
          Poincar\'e isometry group of flat space is an example of 
          $\sfG^{(10)}$; Minkowski and de Sitter space-times are maximally 
          symmetric.}
If $r=4$, the space-time is locally rotationally symmetric (\textsc{lrs}); 
there exists a three-parameter subgroup $\overline{\sfG}^{(3)}$ that acts 
either \emph{simply} or \emph{multiply} transitively on the spacelike 
hypersurfaces: The latter case includes the Kantowski--Sachs cosmological 
models (the orbits of the group are two-dimensional, maximally symmetric, and 
with positive constant curvature); the former corresponds to \textsc{lrs} 
Bianchi models and thus points to the last possible value of $r$, namely 
$r=3$, which corresponds to the Bianchi class. Thus, if we set aside the 
Kantowski--Sachs models, we are left with isometry groups that act 
\emph{simply} transitively on the spacelike hypersurfaces $\Sigma$, \ie such 
that $\dim \sfG = \dim \Sigma = 3$. Hence, in the light of the preceding 
discussion, we can state a more precise definition of Bianchi cosmologies: 
\textsl{A Bianchi cosmology is a model the metric of which admits a three-%
dimensional group of isometries that acts simply transitively on spacelike 
hypersurfaces $\Sigma$, which are surfaces of homogeneity in space-time}. The 
complete list of Bianchi cosmological models is obtained by classifying the 
three-parameter isometry groups, that is, the three-dimensional real Lie 
algebras generated by the associated Killing vector fields; the outcome is 
nine nonequivalent Bianchi types---or ten equivalence classes---, named 
according to Bianchi's own terminology. Let us briefly sketch the 
classification procedure according to Wahlquist and Behr, Ellis and MacCallum,
and Siklos.%
\footnote{For thorough reviews on the Bianchi--Behr classification and 
          related questions see, \eg \cite{macca=CosMod,macca=AniInh,
          macca=MatAni,jantz=SpaHom,barro=AsySta}.} 
\nl
Suppose $X_i$ for $i=1,2,3$ form a basis of a three-dimensional Lie algebra 
with structure constants $C^k_{\ ij}$, which are defined through the 
commutators $[X_i,X_j]=C^k_{\ ij}X_k$ (they are antisymmetric and satisfy the 
Jacobi identity). Given $C^k_{\ ij}$, we can define a three-vector $a^i$ and a
symmetric $3\times3$ matrix $n^{ij}$ by $a_i:=\tfrac12 C^l_{\ li}$ and 
$n^{ij}:=\epsilon^{ikl} (\tfrac12  C^j_{\ kl} - \delta^j_k a_l)$ respectively, 
where $\epsilon^{ijk}$ is the unique totally antisymmetric tensor satisfying 
$\epsilon^{ijk} \epsilon_{ijk}=3!=6$. It is easy to show that the structure 
constants can be written as 
$C^k_{\ ij} = n^{kl} \epsilon_{lij} + 2 \delta^k_{[i} a_{j]}$. Substitution of 
this expression into the Jacobi identity yields the simple result 
$n^{ij} a_j =0$. According to Ellis and MacCallum, the classification now 
gives two broad classes: `class A' ($a_i=0$) and `class B' ($a_i\neq0$); the 
resulting Lie algebras are divided into several types according to the rank 
and the (modulus of the) signature of $n^{ij}$. In class A there exist 
precisely six distinct Lie algebras whereas in class B there are only four 
possible values for the rank and the signature of $n^{ij}$. Some types (VI and 
VII) can be subclassified with the help of a further invariant $h$, which can 
be determined by the formula 
$a_i a_j = \tfrac12 h \, n^{kr} n^{ls} \epsilon_{rsi} \epsilon_{klj}$. The 
resulting algebraic classification is summarised in Table \ref{tab:BianchiA} 
and Table \ref{tab:BianchiB} ($n_i$ for $i=1,2,3$ denotes the diagonal 
elements of the symmetric matrix $n^{ij}$; $a$ is the nonzero component of the 
vector $a_i$ obtained after a suitable rotation of axis and rescaling; the 
number $p$ refers to the dimension of the orbits of the group---it is related 
to the automorphism \dofs---; and the number $q$ describes the degree of 
generality of the most general vacuum solution of each group type.)
\begin{table} 
\caption{The Bianchi types for class A.}
\label{tab:BianchiA}
\begin{center}
\begin{tabular}{ccccccc}
\hline
\hline
   $a$ & $n_1$ & $n_2$ & $n_3$ & Bianchi type & $p$ & $q$ \\
   $0$ & $0$   & $0$   & $0$   & I            & $0$ & $1$ \\
   $0$ & $+1$  & $0$   & $0$   & II           & $3$ & $2$ \\
   $0$ & $0$   & $+1$  & $-1$  & VI$_0$       & $5$ & $3$ \\
   $0$ & $0$   & $+1$  & $+1$  & VII$_0$      & $5$ & $3$ \\
   $0$ & $+1$  & $+1$  & $-1$  & VIII         & $6$ & $4$ \\
   $0$ & $+1$  & $+1$  & $+1$  & IX           & $6$ & $4$ \\
\hline
\hline
\end{tabular}
\end{center}
\end{table}
\begin{table} 
\caption{The Bianchi types for class B.}
\label{tab:BianchiB}
\begin{center}
\begin{tabular}{ccccccc}
\hline
\hline
   $a$         & $n_1$ & $n_2$ & $n_3$ & Bianchi type     & $p$ & $q$ \\
   $1$         & $0$   & $0$   & $0$   & V                & $3$ & $1$ \\
   $1$         & $0$   & $0$   & $+1$  & IV               & $5$ & $3$ \\
   $1$         & $0$   & $+1$  & $-1$  & III or VI$_{-1}$ & $5$ & $3$ \\
   $\sqrt{-h}$ & $0$   & $+1$  & $-1$  & VI$_h$ ($h<0$)   & $5$ & $3$ \\
   $\sqrt{h}$  & $0$   & $+1$  & $+1$  & VII$_h$ ($h>0$)  & $5$ & $3$ \\
\hline
\hline
\end{tabular}
\end{center}
\end{table}
This completes the Bianchi classification. 
\nl 
Given the classification, one can seek for each Bianchi type the corresponding
three-manifold endowed with a metric and isometry group of that type; this is
achieved by choosing time and spatial congruences, and with the help of 
invariant one-forms (as well as the automorphism \dofs in order to simplify 
the spatial metric \cite{siklo=FieEqu,siklo=EinEqu,roque=AutGro}). We follow 
here the metric approach rather than the orthonormal frame approach (see 
\cite{ellis=GeoCos}). The spatially homogeneous metrics can be written as 
$\rmd s^2 = - \rmd \tau^2 + h_{ij} (\tau) \omega^i \omega^j$ or, in terms of 
an arbitrary time variable, as 
$\rmd s^2 = - N^2(t) \rmd t^2 + h_{ij} (t) \omega^i \omega^j$, where $N$ is
the lapse function and $\omega^i$ for $i=1,2,3$ are the basis dual one-forms, 
which satisfy (owing to Cartan's first structure equations) the relation 
$\rmd \omega^i = \tfrac12 C^i_{\ kj} \, \omega^j \wedge \omega^k$. 
The explicit formul{\ae} for types I, II, V, and IX are summed up in Table
\ref{tab:BiaMet}.
\begin{table} 
\caption{Basis one-forms in Bianchi types I, II, V, and IX.}
\label{tab:BiaMet}
\begin{center}
\begin{tabular}{c|cccc}
\hline
\hline
Type      & I      & II             & V         & IX \\
\hline
          &$\rmd x$&$\rmd x-z\rmd y$&$\rmd x$   &$\cos y\cos z\rmd x-
                                                  \sin z\rmd y$ \\
$\omega^i$&$\rmd y$&$\rmd y$        &$e^x\rmd y$&$\cos y\sin z\rmd x+
                                                  \cos z\rmd y$ \\
          &$\rmd z$&$\rmd z$        &$e^x\rmd z$&$-\sin y\rmd x+\rmd z$ \\
\hline
\hline
\end{tabular}
\end{center}
\end{table}

\subsection{Hamiltonian cosmology}
\label{subsec:BiaCosSpaCosHamCos}

Besides its interest with regard to quantisation, the Hamiltonian formulation
of \gr has also been applied to a large variety of cosmological problems---%
both classical and quantum. One can trace back this kind of investigation to 
DeWitt who specialised the canonical formalism to the closed \flrw model, 
which was viewed as a toy-model for quantum gravity \cite{dewit=QuaThe1}. 
Independently, Misner laid the foundations of \emph{Hamiltonian cosmology} 
\cite{misne=MixUni,misne=QuaCosI}, that is, the study of cosmological models 
by means of equations of motion in Hamiltonian form---to be more specific, 
through the \adm reduction procedure and with the introduction of the fruitful 
concept of \emph{minisuperspace} \cite{misne=Minisu}. 
\nl
Typically, the advantage of the Hamiltonian treatment in \gr is twofold. 
Firstly, it enables one to give an heuristic analysis of the generic 
asymptotic behaviour of Bianchi models near the singularity (and at late 
times) without doing lengthy calculations: The analysis is reduced to the 
qualitative description of the motion of a point in a plane under the 
influence of a time-dependent potential \cite{RYAN=HamCos,uggla=HamCos}. This 
type of investigation constitutes the \emph{qualitative Hamiltonian 
cosmology}. Secondly, the Hamiltonian formulation enables one to produce a 
Hamiltonian in a certain `Lagrangian canonical' form that reveals the 
mathematical similarities amongst different `hypersurface-homogeneous' models 
\cite{uggla=ExaHyp}.
\nl
As regards the scope of this thesis, such qualitative treatment of Bianchi 
dynamics is quite difficult to adopt for we are dealing with higher-order 
theories, having more \dofs than \gr; hence the aforementioned qualitative 
analysis by means of simple potential diagrams breaks down. Notwithstanding 
this hindrance we would like to examine whether the Hamiltonian methods might 
produce some simplification of the intricacies that occur when one deals with 
higher-order field equations. The first aspect we want to discuss has to do 
with the variational principle as applied to spatially homogeneous 
cosmological models and, more specifically, to Bianchi cosmologies.

\subsubsection*{Variational principle and boundary terms.}
\label{subsubsec:VarPriBouTer}
 
In Section \ref{sec:VarConGR} we have given a short account of the metric 
variational principle in \gr, without any reference whatsoever to spatially 
homogeneous cosmologies. Now if one tries to derive the field equations of the 
various Bianchi types by means of the Hilbert variational principle, then in 
general (as was firstly noticed by Hawking \cite{hawki=RotUni}, then confirmed 
by MacCallum and Taub \cite{macca=VarPri}) one will not obtain the correct 
field equations. It was firstly claimed---erroneously---that the origin of the 
trouble laid in the utilisation of non-coordinate frames to perform the 
calculations \cite{ryan=HamCosDea}.%
\footnote{This is still advocated by Ryan \cite{ryan=HamFor}.}
But the true reason was given by Sneddon \cite{snedd=HamCos}: The requirement 
of spatial homogeneity prevents a boundary term being set equal to zero.%
\footnote{See also the remark on page \pageref{rem:SurTer}.} 
However, it turns out that for class A Bianchi models this boundary term 
identically vanishes. More specifically, if the metric is assumed to be 
spatially homogeneous, then the spatial integration in the \adm form of the 
Einstein--Hilbert action \eqref{HamOstADMHamAct2} can be performed to give
\begin{equation} \label{ADMSpaHom}
   S_{\text{\adm}} = \int \! \rmd t 
                        \Bigl[ 
                           \piadm^{ij} \, \partial_t h_{ij} - 
                           N \EuH - N^i \EuH_i -
                           2 \bigl( 
                                \piadm^{ik} N_i -
                                \tfrac12 \piadm N^k 
                             \bigr)_{\mid k} 
                        \Bigr].
\end{equation}
On account of the formula given on page \pageref{conv:Cartan} the spatial 
divergence in the action \eqref{ADMSpaHom} reduces to the expression
\begin{equation}
   \bigl( \piadm^{ik} N_i - \tfrac12 \piadm N^k \bigr)_{\mid k} =
   \bigl( \piadm^{ik} N_i - \tfrac12 \piadm N^k \bigr) C^j_{\ jk}.
\end{equation}
Hence, unless the trace of the structure coefficients vanishes---the condition 
$C^j_{\ jk}=0$ is precisely the characterisation of class A Bianchi models---, 
the variation of the \adm action \wrt the shift vector produces additional 
unwanted terms and results in a wrong super-momentum constraint. Although most
class B models do not admit a Hamiltonian formulation, some particular 
nondiagonal and all diagonal%
\footnote{For example, Sneddon has contrived a method to ``heal'' the 
          variational principle in the specific case of the diagonal Bianchi-%
          type V model \cite{snedd=HamCos}.}
models do: This can be achieved whenever the super-momentum constraint is 
holonomic, \ie can be expressed as the vanishing of a total derivative; then 
it can be integrated and used to reduce the number of gravitational \dofs 
\cite{uggla=ExaHyp}.
\nl
Since at the end of Chapter \ref{chap:HamFor} we have developed a Hamiltonian 
formulation of the generic quadratic gravity theory it is natural to address 
the question of under which circumstances the spatial divergences occurring in 
the action vanish so as to warrant a well-posed variational principle.
Consider the quadratic action \eqref{HamQuaCanAct} in canonical form and focus 
on the surface integral; assuming that the metric be spatially homogeneous and
that spatial integration have been performed we readily identify the 
corresponding spatial divergence, namely  
\begin{equation} \label{HOGSpaDiv}
   \Bigl[ 
      N \cp^{jk}_{\ \ \, \mid j} +
      2 N_j \bigl( p^{jk} + \cp^{ik} K^j_{\ i} \bigr)
   \Bigr]_{\mid k}.
\end{equation}
From mere inspection it is obvious that the variation of this term will not
vanish unless one considers class A models: the expression \eqref{HOGSpaDiv}
is indeed proportional to the trace of the structure coefficients. Therefore,
we end up with the same conclusion as in \gr: All class A models do admit a
Hamiltonian formulation. This result also holds for a nonlinear gravitational
action, for the spatial divergence \eqref{HOGSpaDiv} becomes in that case
\begin{equation} \label{HOGSpaDivfr}
   2 \Bigl[ N_j \bigl( p^{jk} + \cp K^{jk} \bigr) \Bigr]_{\mid k}
\end{equation}
and is proportional to the trace of the structure coefficients as well.
\begin{rem} \label{rem:supmom}
Whereas the super-momentum constraint $\EuH_k$ is automatically satisfied for 
all diagonal class A models in \gr, this is not necessarily the case in 
higher-order gravity, for the expression \eqref{HamQuaSupMom} of $\EuH_k$ is
not as simple as in \gr. As a matter of fact the first term in that expression
is not of the form $\Fa^{ij}_{\ \ \mid j}$, where $\Fa^{ij}$ is a tensor
density; a direct calculation shows it is zero for all diagonal class A 
Bianchi types but type VI$_0$ in the specific conformally invariant case.%
\footnote{Making use of the variables defined on page \pageref{NewVar} we 
          indeed obtain $\EuH_1\equiv0\equiv\EuH_2$ and 
          $\EuH_3 = - \tfrac{2 \sqrt{3}}{3} (\cp_+ K_- + \cp_- K_+) \approx 0$ 
          for type VI$_0$.}
This means that, unless one considers the latter situation, one can always 
safely choose the shift vector to be zero, which is the usual assumption in 
Hamiltonian cosmology. By contrast, in the aforementioned peculiar case one 
must check that this very choice does not break the equivalence with the field 
equations. 
\end{rem}

\section{The mixmaster universe in fourth-order gravity}
\label{sec:BiaCosMixCos}

\subsection{On the mixmaster chaotic dynamics}
\label{subsec:BiaCosMixCosMixCha}

The term `mixmaster' refers to the Bianchi-type IX cosmological model in
vacuum \cite{misne=MixUni}; it suggests the nice features of the Bianchi-type 
IX dynamics, \ie the oscillatory regimes on approach towards the initial 
singularity (see \cite{LANDA=TheCha}). Recently, this terminology has also 
been used to discriminate amongst \emph{asymptotically velocity dominated 
cosmological models} (where the spatial curvature terms in the Hamiltonian 
constraint become negligible as compared to the square of the expansion rate 
as the singularity is approached; this corresponds to a Kasner-like behaviour) 
and oscillatory-like cosmologies the prototype of which is Bianchi type IX 
(and type VIII), which is of fundamental importance with regard to the 
investigations that address the issue of the genericness of chaos in 
relativistic cosmology on approach towards the initial singularity and of the 
nature of that singularity when spatial homogeneity is relaxed---\cfr the 
Belinskii--Khalatnikov--Lifshitz (\textsc{bkl}) conjecture---(see 
\cite{berge=NumApp}, and references therein). Interest in the mixmaster 
dynamics has increased dramatically in the last fifteen years (see the review 
in \cite{cotsa=AdiInv}); controversies have arisen on the chaotic nature of 
the Bianchi-type IX model and in particular on the problem of 
(diffeomorphism-)invariant characterisation of chaos in relativistic systems 
(see \cite{HOBIL=DetCha}). 
\nl
That the mixmaster dynamics be chaotic is now fairly well confirmed---though
it has not yet reached the unambiguous status of a theorem---by means of 
different methodologies: qualitative methods, numerical techniques, and 
analytical tools. \emph{Qualitative methods} embody: the \textsc{bkl} 
piecewise approximation methods \cite{belin=OscApp,khala=ConThe}; Hamiltonian 
methods (Misner, Ryan, and others) \cite{RYAN=HamCos,RYAN=HomRel,
uggla=HamCos}; and dynamical systems methods \cite{WAINW=DynSys}. The 
\textsc{bkl} approach (see, \eg \cite{LANDA=TheCha}) has shown that the 
evolution in time of the mixmaster model can be approximated as a sequence of 
time periods (Kasner epochs and eras) during which certain terms in the field 
equations may be neglected, thereby leading to a description of the dynamics 
in terms of the Bianchi-types I and II models, and `bounce laws' from one 
Kasner era to the other that are sensitive to initial conditions 
\cite{barro=ChaBeh,chern=ChaMix}. Hamiltonian cosmology (see Subsection
\ref{subsec:BiaCosSpaCosHamCos}) has reduced the analysis of the field 
equations to that of a time-dependent Hamiltonian system in two dimensions for 
a particle that bounces on moving `potential walls', which approximate the 
time-dependent potentials characterising the various Bianchi types. The 
dynamical systems approach is based on the fact that Einstein's field 
equations for spatially homogeneous cosmologies can be written as an 
autonomous system of first-order differential equations, thereby defining a 
dynamical system; it was initiated by Collins \cite{colli=QuaCos,colli=MorQua} 
and developed extensively by Bogoyavlensky \cite{BOGOY=MetQua}, Wainwright 
\cite{wainwhsu=DynSys}, and others. Historically, \emph{numerical techniques} 
have proven to be misleading, but the associated controversies have been 
smoothed away. More reliable algorithms are now used, mainly to probe the 
structure of the singularity in inhomogeneous space-times \cite{berge=NumApp}. 
Most recently, Cornish and Levin gave a very strong indication pointing 
towards the chaotic character of the mixmaster model, by exploiting fractal 
methods \cite{corni=MixUniCha,corni=MixUniFar}.%
\footnote{One must bear in mind, however, that the fractal structure is
          obtained numerically: One must still be cautious as regards the 
          conclusions of that analysis, even if the whole construction is
          very interesting conceptually.}
\emph{Analytical tools} tackle the `chaoticity issue' in terms of 
integrability concepts---integrability of differential systems, which is 
inherently encoded into their analytic structure. The Painlev\'e method (see 
p.~\pageref{subsub:HamWeylAnaStr} ff.), which provides such an analytical 
tool, rests on the \emph{Painlev\'e property}---a differential system 
possesses this property if and only if its general solution is uniformisable 
or, equivalently, exhibits no movable critical singularities. Integrable 
systems are defined in the sense of Painlev\'e as those systems that possess 
the Painlev\'e property. The trivial part of the Painlev\'e method, known as 
the Painlev\'e test, produces \emph{necessary} but not sufficient conditions 
for a system to enjoy the Painlev\'e property and requires local single-%
valuedness of the general solution in a vicinity of all possible families of 
movable singularities (see, \eg \cite{raman=PaiPro,conte=SinDif,
conte=PaiApp}). Notwithstanding the fact that it is not conclusive---%
especially for real-time chaos---, the Painlev\'e method has proven to be 
helpful in understanding the intricate structure of systems such as the 
mixmaster model \cite{latif=BiaCos}; even more, the fruitful interplay of the 
Painlev\'e method and complex-time numerical integrations has been exemplified 
very recently \cite{schee=AnaStr,schee=PhD}.
\nl
It has also been demonstrated that the mixmaster chaotic behaviour, which 
appears in a four-dimensional space-time in \gr, is generically sustained as 
one increases the space-time dimensionality up to ten, but disappears in any 
universe with space-time dimension greater than ten \cite{demar=NonBeh,
demar=FatMix,hosoy=CriDim,elske=ChaKal,demar=ChaNon}. So, in the context of 
\gr, the mixmaster evolutionary picture is dimensionally dependent. 
\nl
In the next subsection we address the question of whether this picture is 
sensibly modified when one considers more general frameworks---in particular, 
\hotg. One possible scope of this investigation would be to determine how 
general the features of chaotic evolution met in \gr are in the framework of 
all possible physically interesting theories of gravity. This programme also 
hopes to shed new light into the cosmological structure of gravity theories 
with higher derivatives: One possibility would be, for instance, that nice 
nonchaotic properties emerge near the space-time singularity, as is the case 
for the mixmaster model in a theory described by a Lagrangian of the type 
$L=R+\gamma_1 R^2$ \cite{barro=ChaBehHOG,cotsa=CosMod} or by a scale-invariant 
Lagrangian such as $L=R^2$ \cite{barro=MixCos,spind=AsyBeh}. Here we examine 
the structure of the Bianchi-type IX cosmological model in the pure quadratic 
theory of gravity, \ie the theory that is described by the action 
\eqref{eq:VPHOGMetQuaAct} without the \EH term of \gr.%
\footnote{Discarding that term is consistent with our purpose since we 
          consider asymptotic solutions to the field equations hereafter.}

\subsection{Asymptotic analysis of the field equations}
\label{subsec:BiaCosMixCosAsyAna}

The vacuum field equations derived from the general quadratic action
\begin{equation} \label{BiaQuaActMix}
   S = \int_{\EuM} \rmd^4 x \Bigl[ 
                               \gamma_1 \Fl_1 +
                               \gamma_2 \Fl_2 +
                               \gamma_3 \Fl_3
                            \Bigr],
\end{equation}
where $\gamma_i$ for $i=1,2,3$ denote coupling constants, are obtained by 
gathering the \EL derivatives \eqref{eq:VPHOGELDer} together, \viz
\begin{equation} \label{MixFieEqu}
   \begin{split}
      &\gamma_1 \Bigl(
                   2 \nabla^a \nabla^b R - 
                   2 g^{ab} \square R + 
                   \frac12 g^{ab} R^2 -
                   2 R R^{ab}
                \Bigr) +
       \gamma_2 \Bigl( 
                   \frac12 g^{ab} R_{cd} R^{cd} +
                   \nabla^a \nabla^b R \\
      &\qquad - 
                   2 R^{bcad} R_{cd}  -
                   \square R^{ab} - 
                   \frac12 g^{ab} \square R 
                \Bigr) +
       \gamma_3 \Bigl( 
                   \frac12 R^{cdef} R_{cdef} g^{ab} -
                   2 R^{cdeb} R_{cde}^{\ \ \ \, a} \\
      &\qquad -
                   4 \square R^{ab} + 
                   2 \nabla^a \nabla^b R - 
                   4 R^{bcad} R_{cd} + 
                   4 R^{ca} R^b_{\ c}
                \Bigr) = 0. 
   \end{split}
\end{equation}
\nl
We are interested in the behaviour of the spatially homogeneous Bianchi 
cosmological model of type IX, which is described by a metric of the form
\begin{equation} \label{MixMet1}
   \rmd s^2 = - \rmd t^2 + h_{ij} (t) \omega^i \omega^j, 
\end{equation}
where $\omega^i$ for $i=1,2,3$ are the SO$(3)$-invariant differential
forms that characterise Bianchi type IX (given explicitly in Table
\ref{tab:BiaMet}). The induced three-metric is assumed to be diagonal and of
the form
\begin{equation} \label{MixMet2}
   h_{ij} = \diag [ a^2 (t), b^2 (t), c^2 (t)], 
\end{equation}
where $a, b, c$ are the scale factors. Substituting the Bianchi-type IX metric
\eqref{MixMet1} together with the explicit form \eqref{MixMet2} into the 
vacuum field equations \eqref{MixFieEqu} we obtain the mixmaster field 
equations in terms of the scale factors $a, b,$ and $c$. We write only the 
mixed $(^0_{\ 0})$- and $(^1_{\ 1})$-components (the $(^2_{\ 2})$- and 
$(^3_{\ 3})$-components can be obtained from the $(^1_{\ 1})$-component by 
cyclic permutations). The result is
\begin{subequations} \label{mixeqcomp}
   \begin{align} 
      \bigl[ 
         \gamma_1 \Fl_1 + \gamma_2 \Fl_2 + \gamma_3 \Fl_3 
      \bigr]^0_{\ 0} = 0, \label{00comp} \\
      \bigl[ 
         \gamma_1 \Fl_1 + \gamma_2 \Fl_2 + \gamma_3 \Fl_3 
      \bigr]^1_{\ 1} = 0, \label{11comp} 
   \end{align} 
\end{subequations}
where $\Fl_i^{ab}$ for $i=1,2,3$ are the \EL derivatives associated with 
$\Fl_i$ for $i=1,2,3$ respectively, which are written down extensively in the 
Appendix at the end of this section on page \pageref{mixappix} ff. In 
equations \eqref{mixeqcomp} the contributions from $R^{2}$, $R^{ab} R_{ab}$, 
and $R^{abcd} R_{abcd}$ can be easily identified since they are multiplied by 
$\gamma_1$, $\gamma_2$, and $\gamma_3$ respectively. 
\nl
We now proceed to examine the evolution of the Bianchi IX model on approach to 
the singularity (occurring at $t=0$); this is dictated by the system 
\eqref{mixeqcomp} supplemented by the remaining components of the field 
equations. The basic idea of the asymptotic method is based on the search for 
existence (or nonexistence) of chaotic behaviour in the Bianchi-type IX 
evolution according to the field equations \eqref{MixFieEqu}, which is
intimately connected to the nonexistence (or existence) of power-law 
asymptotic solutions of the system on approach to the singularity as 
$t \rightarrow 0$. This method was first applied successfully by Barrow and 
Cotsakis \cite{barro=ChaBehHOG,cotsa=CosMod} who showed that, in the nonlinear 
theory based on a Lagrangian $L=\fR$ that is a polynomial function of the 
scalar curvature, the vacuum Bianchi-type IX model is nonchaotic and possesses 
monotonic, power-law asymptotes on approach to the singularity. Accordingly, 
we look for power-law asymptotes that, to lower order, have the form
\begin{equation} \label{Mixps}
   (a,b,c)= \bigl( t^{p_1}, t^{p_2}, t^{p_3} \bigr)
\end{equation}
as $t \rightarrow 0$. For simplicity we define constants $q^{s}$ by
\begin{equation} \label{Mixqs}
   q^{s} = \sum_{i=1}^{3} p_i^s.
\end{equation}
{\allowdisplaybreaks
Substituting the \emph{ansatz} \eqref{Mixps} into the Bianchi-type IX field 
equations \eqref{00comp} and \eqref{11comp} (similar terms appear in the 
$(^2_{\ 2})$- and $(^3_{\ 3})$-components) we obtain, for the $(^0_{\ 0})$-%
component,
\begin{subequations} \label{MixEL00}
   \begin{align}
      \bigl( \Fl_1 \bigr)^0_{\ 0} &=
         \frac12 t^{-4} \bigl[ q^2 + (q^1)^2 - 2 q^1 \bigr] 
                        \bigl[ 3 q^2 - (q^1)^2 + 6 q^1 \bigr] + 
         [\star \star \star], \label{MixEL00a} \\ 
      \bigl( \Fl_2 \bigr)^0_{\ 0} &=
         \frac12 t^{-4} \bigl[ 
                           3 (q^2)^2 - q^2 (q^1)^2 + 2 (q^1)^3 
         \notag \\
         & \qquad \qquad
                   + 2 q^2 q^1 - 3 q^2 - 3 (q^1)^2 
                \bigr] + [\star \star \star], \label{MixEL00b} \\ 
      \bigl( \Fl_3 \bigr)^0_{\ 0} &=
         t^{-4} \bigl[ 
                   3 q^4 + 3 (q^2)^2 - 4 q^3 q^1 + 4 q^2 q^1 - 6 q^2
                \bigr] + [\star \star \star], \label{MixEL00c} 
   \end{align}
\end{subequations}
and, for the $(^1_{\ 1})$-component,
\begin{subequations} \label{MixEL11}
   \begin{align}
      \bigl( \Fl_1 \bigr)^1_{\ 1} &=
         \frac12 t^{-4} \bigl[ 
                           4 p_1 (q^1 - 3) - q^2 - (q^1)^2 + 10 q^1 - 24
                        \bigr] \notag \\
         & \qquad \qquad 
                     \times \bigl[ q^2 + (q^1)^2 - 2 q^1 \bigr] + 
                     [\star \star \star], \label{MixEL11a} \\ 
      \bigl( \Fl_2 \bigr)^1_{\ 1} &=
         \frac12 t^{-4} \bigl[ 
                           4 p_1 (q^1 - 3)(q^2-1) - (q^2)^2 - q^2 (q^1)^2 + 
                           2 (q^1)^3 \notag \\
         & \qquad \qquad 
                         + 6 q^2 q^1 - 7 q^2 - 11 (q^1)^2 + 12 q^1
                        \bigr] + [\star \star \star], \label{MixEL11b} \\ 
      \bigl( \Fl_3 \bigr)^1_{\ 1} &=
         \frac12 t^{-4} \bigl[ 
                           8 p_1 (q^1 - 3)(p_1^2 - p_1 q^1 + q^1 + q^2 - 2) 
                           \notag \\
         & \qquad \qquad 
                         - 2 q^4 - 2 (q^2)^2 + 8 q^3 - 4 q^2
                        \bigr] + [\star \star \star], \label{MixEL11c}  
   \end{align}
\end{subequations}
}%
where the explicitly written terms correspond to Bianchi type I and 
$[\star \star \star]$ are additional terms generated by the Bianchi-type IX 
potential. We know that the only power-law solutions to the Bianchi-type I 
field equations derived from equations \eqref{MixFieEqu} in a four-dimensional 
space-time are \cite{derue=AppCos,capra=PowTyp}:
\begin{enumerate}
   \item the well-known Kasner solution
         \begin{equation} \label{MixKas}
            \rmd s^2 = - \rmd t^2 
                       + \sum_{i=1}^{3} t^{2 p_i} (\rmd x^i)^2,
            \qquad \text{with} \; q^1 = q^2 = 1,
         \end{equation}
         where the $p$'s can be represented in the parametric form
         \begin{align} \label{MixKasPar}
            p_1(s)  &= \frac{-s}{1+s+s^2},
            &p_2(s) &= \frac{s(1+s)}{1+s+s^2}, 
            &p_3(s) &= \frac{1+s}{1+s+s^2}, 
         \end{align}
         where the \emph{Kasner parameter} $s$ varies in the range 
         $s\geqslant1$;
   \item the isotropic solution
         \begin{equation} \label{MixIso}
            \rmd s^2 = - \rmd t^2 
                       + \sum_{i=1}^{3} t^{2 p_i} (\rmd x^i)^2,
           \qquad \text{with} \; p_1 = p_2 = p_3 = \frac12.
         \end{equation}
\end{enumerate}
This plays a key r\^ole in determining the evolution of the Bianchi-type IX 
model near the singularity in the fourth-order theory of gravity described by 
the action \eqref{BiaQuaActMix}: Looking for power asymptotes to the Bianchi 
type IX field equations amounts to looking for Kasner or isotropic asymptotes 
to these equations. 
\nl
We consider the algebraic system $\{$\eqref{MixEL00b}, \eqref{MixEL00c},
\eqref{MixEL11b}, \eqref{MixEL11c}$\}$. Firstly, we seek for Kasner asymptotic 
solutions of the form \eqref{Mixps}, \eqref{Mixqs}, and \eqref{MixKas}. Most 
of the additional terms appearing in the equations are ``nondangerous'' in the 
sense that they grow slower than the $t^{-4}$ contributions (present in the 
Bianchi-type I field equations): they are unimportant as we approach the 
singularity. However, the following terms, which appear in equations 
\eqref{MixEL11b} and \eqref{MixEL11c}, namely 
\begin{subequations}
   \begin{align}
      &   \frac{15 a^4}{8 b^4 c^4} 
        - \frac{2 a^2}{b^2 c^2} \biggl( 
                                   \frac{\ddot{b}}{b} + \frac{\ddot{c}}{c}
                                \biggr)
        + \frac{a^2}{b^2 c^2} \biggl[ 
                                 \Bigl( \frac{\dot{b}}{b} \Bigr)^2 +
                                 \Bigl( \frac{\dot{c}}{c} \Bigr)^2
                              \biggr] \notag \\
      & \qquad \qquad
        - 2 a \dot{a} \biggl(
                         \frac{\dot{c}}{b^2 c^3} + \frac{\dot{b}}{b^3 c^2}
                      \biggr) 
        - \frac{2 a^2 \dot{b} \dot{c}}{b^3 c^3} 
        + \frac{\dot{a}^2}{b^2 c^2} 
        + \frac{2 a \ddot{a}}{b^2 c^2}, \label{MixDang1} \\
\intertext{and}
      &   \frac{55 a^4}{8 b^4 c^4}
        - \frac{7 a^2}{b^2 c^2} \biggl(
                                   \frac{\ddot{b}}{b} + \frac{\ddot{c}}{c}
                                \biggr)
        + \frac{8 a^2}{b^2 c^2} \biggl[
                                   \Bigl( \frac{\dot{b}}{b} \Bigr)^2 +
                                   \Bigl( \frac{\dot{c}}{c} \Bigr)^2
                                \biggr] \notag \\
      & \qquad \qquad
        - 12 a \dot{a} \biggl( 
                          \frac{\dot{c}}{b^2 c^3} + \frac{\dot{b}}{b^3 c^2}
                        \biggr) 
        - \frac{a^2 \dot{b} \dot{c}}{b^3 c^3} 
        + \frac{6 \dot{a}^2}{b^2 c^2} 
        + \frac{12 a \ddot{a}}{b^2 c^2} \label{MixDang2}
   \end{align}
\end{subequations}
respectively, are of the form $t^{-4+8p_1}$  (the very first terms in 
expressions \eqref{MixDang1} and \eqref{MixDang2}) and $t^{-4+4p_1}$ (the 
remaining nine terms in expressions \eqref{MixDang1} and \eqref{MixDang2}). 
Since $p_1<0$ in the anisotropic case all these terms grow faster than 
$t^{-4}$ as $t \rightarrow 0$: They will be the dominant ones in the field 
equations on approach to the singularity. Thus in this case ``dangerous 
terms'' appear in the Ricci-squared and Riemann-squared \EL expressions. This 
is completely analogous to what happens in \gr as demonstrated by the 
\textsc{bkl} approximation method: At a certain stage the description of the 
asymptotic dynamics in terms of the Bianchi-type I Kasner solution breaks down 
and one must take into account the aforementioned ``dangerous terms'' in the 
field equations; the dynamics is then described by the Taub solution of the 
Bianchi-type II model and this corresponds to a transition from one ``Kasner 
epoch'' to another. Hence, in our case, one should retrieve the same kind of 
oscillatory behaviour as that found in \gr by \textsc{bkl} on approach to the 
singularity as $t \rightarrow 0$. One can indeed explicitly check that the 
\textsc{bkl} solution 
\cite{belin=OscApp}
\begin{equation}
   \begin{split}
      a^2 &= \frac{- 2 p_1 \Lambda}{\cosh (2 p_1 \Lambda \tau)}, \\
      b^2 &= b(0)^2 \exp \bigl[ 2 \Lambda (p_1 + p_2) \tau \bigr] 
                    \cosh (2 p_1 \Lambda \tau), \\
      c^2 &= c(0)^2 \exp \bigl[ 2 \Lambda (p_1 + p_3) \tau \bigr] 
                    \cosh (2 p_1 \Lambda \tau),
   \end{split}
\end{equation}
where $\rmd t = abc \, \rmd \tau$ and $\Lambda$ is a constant, satisfies the 
field equations \eqref{00comp} and \eqref{11comp} for Bianchi type IX to 
leading order on approach to the singularity, wherein only the ``dangerous'' 
and the ``Kasnerian'' terms---those containing four dots---have been retained.
\nl
On the other hand, if we introduce the asymptotic isotropic solution
\eqref{MixIso} in the algebraic system $\{$\eqref{MixEL00b}, \eqref{MixEL00c},
\eqref{MixEL11b}, \eqref{MixEL11c}$\}$, it appears that all the terms in the 
equations are nondangerous. This proves the possibility of reaching the 
cosmological singularity in a monotonic, nonchaotic way. 
\nl
The conclusion from this analysis is that the fourth-order Bianchi-type IX
equations admit no anisotropic monotonic power-law solutions all the way to 
the singularity, but only an isotropic one, given by \eqref{MixIso}.
\nl
Now it is known that each vacuum solution of \gr also satisfies the equations 
\eqref{MixFieEqu} in space-time dimension $n\leqslant4$ since the $R$-squared 
equation (\cfr \eqref{ELQuaVar1}) and the Ricci-squared equation (\cfr 
\eqref{ELQuaVar2}) are obviously satisfied if $R_{ab}=R=0$ and because the 
variation of the quadratic Lovelock Lagrangian identically vanishes in space-%
time dimension $n<5$. However, the solution space of higher-order gravity is 
larger than in \gr; in particular, the Kasner solution \eqref{MixKas} is not 
the general solution of the Bianchi-type I model in the fourth-order theory 
based on the action \eqref{BiaQuaActMix}. Therefore, it is natural to address 
the question of the stability of the exact polynomial solutions \eqref{MixKas} 
and \eqref{MixIso}. This is achieved through a perturbation analysis.
\nl
Consider for instance the pure Riemann-squared theory and choose the 
parameterisation
\begin{align} \label{MixPar}
   T &= \ln(t), &\alpha &= \ln(a), &\beta &= \ln(b), &\gamma &= \ln(c).
\end{align}
The linearised field equations for small perturbations $\epsilon_1$,
$\epsilon_2$, and $\epsilon_3$ of $\alpha=p_1 T$, $\beta=p_2 T$, and 
$\gamma=p_3T$ respectively are
\begin{align} \label{MixPerSys}
   \dot{x_1} &= x_2, \notag \\
   \dot{x_2} &= x_3, \notag \\
   \dot{x_3} &= \sum_{i=1}^{9} f_i (p_1, p_2, p_3) x_i, \notag \\
   \dot{x_4} &= x_5, \notag \\
   \dot{x_5} &= x_6, \\
   \dot{x_6} &= \sum_{i=1}^{3} f_{i + 6} (p_2, p_3, p_1) x_i +
                \sum_{i=4}^{9} f_{i - 3} (p_2, p_3, p_1) x_i, \notag \\
   \dot{x_7} &= x_8, \notag \\
   \dot{x_8} &= x_9, \notag \\
   \dot{x_9} &= \sum_{i=1}^{6} f_{i + 3} (p_3, p_1, p_2) x_i +
                \sum_{i=7}^{9} f_{i - 6} (p_3, p_1, p_2) x_i, \notag 
\end{align}
where we have defined 
\begin{align*} 
   x_1  &:= \dot{\epsilon_1}, 
   &x_4 &:= \dot{\epsilon_2}, 
   &x_7 &:= \dot{\epsilon_3},
\end{align*}
and
\begin{align*} 
   f_1 (p_1, p_2, p_3) &= \bigl( 
                             2 p_1^2 - 2 p_1 p_2 - 2 p_1 p_3 + 3 p_1 + p_2^2 +
                             p_2 + p_3^2 + p_3 - 2 
                          \bigr) \\
                       &  \qquad \qquad \times 
                          \bigl( p_1 + p_2 + p_3 - 3 \bigr), \\
   f_2 (p_1, p_2, p_3) &= - p_1 \bigl( 4 p_2 + 4 p_3 - 9 \bigr) 
                          - p_2 \bigl( 2 p_3 - 7 \bigr) 
                          + p_1^2 + 7 p_3 - 11, \\
   f_3 (p_1, p_2, p_3) &= - \bigl( 2 p_1 + 2 p_2 + 2 p_3 - 6 \bigr), \\
   f_4 (p_1, p_2, p_3) &= \bigl[
                             2 p_2 (p_3-2) + 3 p_2^2 + p_3^2 + 2 p_3 - 5
                          \bigr] p_1 -
                          p_1^2 \bigl( p_2 + 2 p_3 - 5 \bigr) \\
                       &  \qquad \qquad -
                          p_2 \bigl(p_3^2 + 1 \bigr) - 2 p_2^3 + 3 p_2^2, \\
   f_5 (p_1, p_2, p_3) &= p_1 \bigl( 2 p_2 + 1 \bigr) - p_1^2 - p_2^2 + p_2, 
                          \\
   f_6 (p_1, p_2, p_3) &= 0, \\
   f_7 (p_1, p_2, p_3) &= p_1 \bigl[ 
                                 2 p_2 (p_3+1) + p_2^2 + 3 p_3^2 - 4 p_3 - 5
                              \bigr] -
                          p_1^2 \bigl( 2 p_2 + p_3 - 5 \bigr) \\
                       &  \qquad \qquad -
                          p_2^2 p_3 - 2 p_3^3 + 3 p_3^2 - p_3, \\
   f_8 (p_1, p_2, p_3) &= p_1 \bigl( 2 p_3 +1 \bigr) - p_1^2 - p_3^2 + p_3, \\
   f_9 (p_1, p_2, p_3) &= 0. 
\end{align*}
The characteristic polynomial related to the differential system 
\eqref{MixPerSys} takes the forms
\begin{equation*}
   - \big( \lambda + 1 \bigr) 
     \bigl( \lambda - 2 \bigr)^5
     \bigl( \lambda - 3 \bigr) \lambda^2
\end{equation*}
for the Kasner solution \eqref{MixKas} and
\begin{equation*}
   - \frac{1}{64} \bigl( 2 \lambda - 1 \bigr)^2
                  \bigl( 2 \lambda - 3 \bigr)^3
                  \bigl( 2 \lambda - 5 \bigr)
                  \bigl( \lambda + 1 \bigr)
                  \bigl( \lambda - 1 \bigr)^2
\end{equation*}
for the isotropic solution \eqref{MixIso}. Hence the corresponding evolution 
laws for the perturbations are
\begin{subequations} \label{MixPerEvo}
   \begin{align} 
      \epsilon_i (T) & \simeq c_{i0} + c_{i1} \exp(-T) + c_{i2} T + 
                              c_{i3} \exp(2T) + c_{i4} \exp(3T), \\
      \epsilon_i (T) & \simeq c'_{i0} + c'_{i1} \exp(-T) + c'_{i2} \exp(T/2) +
                              c'_{i3} \exp(T) \notag \\
                     & \qquad \qquad 
                            + c'_{i4} \exp(3T/2) + c'_{i5} \exp(5T/2),
   \end{align}
\end{subequations}
for $i=1,2,3$, with the conditions
\begin{align*}
   &c_{11} = k p_1, \quad c_{21} = k p_2, \quad c_{31} = k p_3, \\
   &p_1 c_{12} + p_2 c_{22} + p_3 c_{32} = 0, \\
   &c_{14} = c_{24} = c_{34} = 0, \\
   &c'_{11} = c'_{21} = c'_{31} = \tfrac{k}2, \\
   &c'_{12} + c'_{22} + c'_{32} = 0, \\
   &c'_{13} + c'_{23} + c'_{33} = 0, \\
   &c'_{15} = c'_{25} = c'_{35} = 0,
\end{align*}
where $k$ is an arbitrary but---as it stems from the perturbation analysis---%
in\-fin\-i\-tes\-i\-mal constant. Both perturbations given in equations
\eqref{MixPerEvo} grow exponentially for $T \rightarrow + \infty$: The 
corresponding solutions are unstable. The same conclusion seems to apply in 
the neighbourhood of the initial singularity, \ie as $T \rightarrow - \infty$, 
due to the presence of the terms $c_{i1}\exp(-T)$, $c_{i2}T$ and 
$c'_{i1}\exp(-T)$. However, the mere r\^ole of the terms involving $\exp(-T)$ 
is to shift the position of the initial singularity on the time axis. More 
explicitly, looking at the scale factor $\alpha$ and assuming that 
$k\exp(-T)\ll1$ we obtain
\begin{subequations}
   \begin{align}
      \alpha (T) + \epsilon_1 (T) 
         & \simeq \alpha_0 + p_1 T + c_{11} \exp(-T) + c_{12} T \notag \\
         & \simeq \alpha_0 + p_1 \ln \Bigl[ 
                                        \exp(T) \bigl( 1 + k \exp(-T) \bigr)
                                     \Bigr] + c_{12} T \notag \\
         & \simeq \alpha_0 + p_1 \ln (t+k) + c_{12} \ln(t) 
           \label{MixPerAlpa} \\
\intertext{in the case of the Kasner metric and}
      \alpha (T) + \epsilon_1 (T) 
         & \simeq \alpha_0 + \frac12 T + c'_{11} \exp(-T) \notag \\
         & \simeq \alpha_0 + \frac12 \ln \Bigl[
                                            \exp(T) \bigl( 
                                                       1 + k \exp(-T) 
                                                    \bigr)
                                         \Bigr] \notag \\
         & \simeq \alpha_0 + \frac12 \ln (t+k) 
           \label{MixPerAlpb}
   \end{align}
\end{subequations}
for the isotropic solution. Similar relations hold for the scale factors 
$\beta$ and $\gamma$. 
\nl
In conclusion, due to the presence of the divergent logarithmic term in 
\eqref{MixPerAlpa} the Kasner metric \eqref{MixKas} is unstable on approach to 
the singularity, but the isotropic metric \eqref{MixIso} is stable. To resume, 
the important facts of our analysis are:
\begin{itemize}
   \item The quadratic theory based on the action \eqref{BiaQuaActMix} admits 
         no anisotropic polynomial solutions other than the Kasner solution
         \eqref{MixKas}. In \gr this is the general--thus stable---solution to 
         the Bianchi-type I model; this leads in the Bianchi-type IX case to a 
         shift from a Kasner solution to another with a different set of 
         Kasner exponents, thereby giving rise to the characteristic 
         oscillatory behaviour of the mixmaster universe. Here we have shown
         that the Kasner polynomial solution is itself an unstable solution of 
         the Bianchi-type I fourth-order field equations. This implies that an 
         oscillatory behaviour based on Kasner asymptotes cannot be generic in 
         the Bianchi-type IX fourth-order dynamics. The typical mixmaster 
         oscillatory behaviour is of zero measure amongst all possible 
         behaviours since it is unstable \wrt small perturbations.%
         \footnote{This is supported by the numerical analysis of the 
                   Bianchi-type IX field equations in fourth-order gravity 
                   \cite{cotsa=MixUni}.}
 
   \item There exists one stable, isotropic and monotonic solution, given by 
         the metric \eqref{MixIso}, which attracts sufficiently close 
         trajectories in the phase space. Such a situation is not met in \gr,
         for the isotropic metric \eqref{MixIso} is not a solution of 
         Einstein's vacuum field equations.
\end{itemize}
\nl
Inclusion of matter fields does not alter the conclusions since they become 
dynamically negligible near the singularity: In \gr matter fields are 
negligible with respect to the dominant metric terms, which grow as $t^{-2}$ 
as one approaches the singularity; in the fourth-order dynamics the metric 
terms grow typically as $t^{-4}$ as $t \rightarrow 0$; hence matter terms do 
not influence the Bianchi-type IX evolution.
\begin{rem}
When the space-time dimension is greater than four, the Kasner metric is no
more a Bianchi-type I solution to the fourth-order theory derived from the 
action \eqref{BiaQuaActMix} \cite{capra=PowTyp}. Therefore, it seems 
reasonable to conjecture that, in contrast to the general nondiagonal case of
multidimensional spatially homogeneous cosmology in \gr, it is impossible to 
build a mixmaster universe in the fourth-order Kaluza--Klein theory based on 
the \textsc{bkl} approximation scheme.
\end{rem}

\subsection*{Appendix}
\label{mixappix}

We give here the mixed components of the \EL derivatives 
\eqref{eq:VPHOGELDer}, specialised to Bianchi type IX. 
\begin{align*}
   &\gamma_1 \bigl( \Fl_1 \bigr)^0_{\ 0} = \\
   &\qquad - \gamma_1 \bigl( 
                           4 \dddot{a} \dot{a} a^{-2} 
                         + 4 \dddot{a} \dot{b} a^{-1} b^{-1} 
                         + 4 \dddot{a} \dot{c} a^{-1} c^{-1}  
                         - 2 \ddot{a}^2 a^{-2} 
                         - 4 \ddot{a} \dot{a}^2 a^{-3} 
                         - 4 \ddot{a} \ddot{b} a^{-1} b^{-1} \\
   & \qquad \qquad \quad
                         + 4 \ddot{a} \dot{b}^2 a^{-1} b^{-2}  
                         + 8 \ddot{a} \dot{b} \dot{c} a^{-1} b^{-1} c^{-1} 
                         - 4 \ddot{a} \ddot{c} a^{-1} c^{-1} 
                         - 4 \dot{a}^3 \dot{b} a^{-3} b^{-1} \\
   & \qquad \qquad \quad
                         + 4 \ddot{a} \dot{c}^2 a^{-1} c^{-2} 
                         - 4 \dot{a}^3 \dot{c} a^{-3} c^{-1} 
                         + 4 \dot{a}^2 \ddot{b} a^{-2} b^{-1} 
                         - 6 \dot{a}^2 \dot{b}^2 a^{-2} b^{-2} \\ 
   & \qquad \qquad \quad
                         - 4 \dot{a}^2 \dot{b} \dot{c} a^{-2} b^{-1} c^{-1} 
                         + 4 \dot{a}^2 \ddot{c} a^{-2} c^{-1}
                         - 6 \dot{a}^2 \dot{c}^2 a^{-2} c^{-2} 
                         - 4 \dot{a}^2 a^{-4} \\
   & \qquad \qquad \quad
                         + 2 \dot{a}^2 a^{-4} b^2 c^{-2} 
                         + 2 \dot{a}^2 a^{-4} b^{-2} c^2 
                         - 2 \dot{a}^2 b^{-2} c^{-2} 
                         + 4 \dot{a} \dddot{b} a^{-1} b^{-1} \\
   & \qquad \qquad \quad
                         + 8 \dot{a} \ddot{b} \dot{c} a^{-1} b^{-1} c^{-1} 
                         - 4 \dot{a} \dot{b}^3 a^{-1} b^{-3}
                         - 4 \dot{a} \dot{b}^2 \dot{c} a^{-1} b^{-2} c^{-1} 
                         - 2 \dot{a} \dot{b} a^{-1} b^{-3} \\
   & \qquad \qquad \quad
                         + 8 \dot{a} \dot{b} \ddot{c} a^{-1} b^{-1} c^{-1} 
                         - 4 \dot{a} \dot{b} \dot{c}^2 a^{-1} b^{-1} c^{-2} 
                         -   \dot{a} \dot{b} a b^{-3} c^{-2}
                         -   \dot{a} \dot{b} a^{-3} b c^{-2} \\
   & \qquad \qquad \quad
                         + 2 \dot{a} \dot{b} a^{-1} b^{-1} c^{-2} 
                         -   \dot{a} \dot{b} a^{-3} b c^{-2} 
                         - 2 \dot{a} \dot{b} a^{-3} b^{-1}
                         - 4 \dot{a} \dot{c}^3 a^{-1} c^{-3} \\
   & \qquad \qquad \quad
                         + 3 \dot{a} \dot{b} a^{-3} b^{-3} c^2 
                         + 4 \dot{a} \dddot{c} a^{-1} c^{-1} 
                         -   \dot{a} \dot{c} a b^{-2} c^{-3}
                         -   \dot{a} \dot{c} a^{-3} b^{-2} c \\
   & \qquad \qquad \quad
                         - 2 \dot{a} \dot{c} a^{-1} c^{-3} 
                         + 2 \dot{a} \dot{c} a^{-1} b^{-2} c^{-1} 
                         + 3 \dot{a} \dot{c} a^{-3} b^2 c^{-3} 
                         - 2 \dot{a} \dot{c} a^{-3} c^{-1} 
                         - 4 \ddot{b} \dot{b}^2 b^{-3} \\
   & \qquad \qquad \quad
                         + 4 \dddot{b} \dot{b} b^{-2} 
                         + 4 \dddot{b} \dot{c} b^{-1} c^{-1} 
                         - 2 \ddot{b}^2 b^{-2} 
                         + 4 \ddot{b} \dot{c}^2 b^{-1} c^{-2} 
                         - 4 \ddot{b} \ddot{c} b^{-1} c^{-1} \\
   & \qquad \qquad \quad
                         - 4 \dot{b}^3 \dot{c} b^{-3} c^{-1} 
                         + 4 \dot{b}^2 \ddot{c} b^{-2} c^{-1} 
                         + 2 \dot{b}^2 a^2 b^{-4} c^{-2} 
                         - 6 \dot{b}^2 \dot{c}^2 b^{-2} c^{-2} \\
   & \qquad \qquad \quad
                         + 2 \dot{b}^2 a^{-2} b^{-4} c^2 
                         - 2 \dot{b}^2 a^{-2} c^{-2} 
                         - 4 \dot{b} \dot{c}^3 b^{-1} c^{-3} 
                         - 4 \dot{b}^2 b^{-4} 
                         + 4 \dot{b} \dddot{c} b^{-1} c^{-1} \\
   & \qquad \qquad \quad
                         + 3 \dot{b} \dot{c} a^2 b^{-3} c^{-3} 
                         -   \dot{b} \dot{c} a^{-2} b c^{-3} 
                         -   \dot{b} \dot{c} a^{-2} b^{-3} c 
                         + 2 \dot{b} \dot{c} a^{-2} b^{-1} c^{-1} \\
   & \qquad \qquad \quad
                         -   \dot{b} \dot{c} a^{-2} b^{-3} c 
                         - 2 \dot{b} \dot{c} b^{-1} c^{-3} 
                         - 2 \dot{b} \dot{c} b^{-3} c^{-1} 
                         + 4 \dddot{c} \dot{c} c^{-2} 
                         - 2 \ddot{c}^2 c^{-2} 
                         - 4 \ddot{c} \dot{c}^2 c^{-3} \\
   & \qquad \qquad \quad
                         + 2 \dot{c}^2 a^2 b^{-2} c^{-4} 
                         + 2 \dot{c}^2 a^{-2} b^2 c^{-4} 
                         - 2 \dot{c}^2 a^{-2} b^{-2} 
                         - 4 \dot{c}^2 c^{-4} 
                         + \tfrac18 a^4 b^{-4} c^{-4} \\
   & \qquad \qquad \quad
                         - \tfrac12 a^2 b^{-2} c^{-4} 
                         - \tfrac12 a^2 b^{-4} c^{-2} 
                         - \tfrac12 a^{-2} b^2 c^{-4} 
                         + \tfrac12 a^{-2} b^{-2} 
                         - \tfrac12 a^{-2} b^{-4} c^2 \\
   & \qquad \qquad \quad
                         + \tfrac12 a^{-2} c^{-2} 
                         + \tfrac18 a^{-4} b^4 c^{-4} 
                         - \tfrac12 a^{-4} b^2 c^{-2} 
                         - \tfrac12 a^{-4} b^{-2} c^2 
                         + \tfrac18 a^{-4} b^{-4} c^4 \\
   & \qquad \qquad \quad
                         + \tfrac34 a^{-4} 
                         + \tfrac12 b^{-2} c^{-2} 
                         + \tfrac34 b^{-4} 
                         + \tfrac34 c^{-4}
                      \bigr);
\end{align*}
\begin{align*}
   &\gamma_2 \bigl( \Fl_2 \bigr)^0_{\ 0} = \\
   &\qquad - \gamma_2 \bigl( 
                           2 \dddot{a} \dot{a} a^{-2}
                         +   \dddot{a} \dot{b} a^{-1} b^{-1}
                         +   \dddot{a} \dot{c} a^{-1} c^{-1}
                         -   \ddot{a}^2 a^{-2}
                         - 2 \ddot{a} \dot{a}^2 a^{-3}
                         +   \ddot{a} \dot{a} \dot{b} a^{-2} b^{-1} \\
   & \qquad \qquad \quad
                         +   \ddot{a} \dot{a} \dot{c} a^{-2} c^{-1}
                         -   \ddot{a} \ddot{b} a^{-1} b^{-1}
                         +   \ddot{a} \dot{b}^2 a^{-1} b^{-2}
                         + 2 \ddot{a} \dot{b} \dot{c} a^{-1} b^{-1} c^{-1} \\
   & \qquad \qquad \quad
                         -   \ddot{a} \ddot{c} a^{-1} c^{-1}
                         +   \ddot{a} \dot{c}^2 a^{-1} c^{-2}
                         -   \dot{a}^3 \dot{b} a^{-3} b^{-1}
                         +   \dot{a}^2 \ddot{b} a^{-2} b^{-1}
                         -   \dot{a}^3 \dot{c} a^{-3} c^{-1} \\
   & \qquad \qquad \quad
                         - 3 \dot{a}^2 \dot{b}^2 a^{-2} b^{-2}
                         -   \dot{a}^2 \dot{b} \dot{c} a^{-2} b^{-1} c^{-1}
                         - 3 \dot{a}^2 \dot{c}^2 a^{-2} c^{-2}
                         +   \dot{a}^2 \ddot{c} a^{-2} c^{-1} \\
   & \qquad \qquad \quad
                         +   \dot{a}^2 a^{-4} b^2 c^{-2}
                         +   \dot{a}^2 a^{-4} b^{-2} c^2
                         +   \dot{a}^2 b^{-2} c^{-2}
                         +   \dot{a} \dddot{b} a^{-1} b^{-1}
                         - 2 \dot{a}^2 a^{-4} \\
   & \qquad \qquad \quad
                         +   \dot{a} \ddot{b} \dot{b} a^{-1} b^{-2}
                         -   \dot{a} \dot{b}^3 a^{-1} b^{-3}
                         -   \dot{a} \dot{b}^2 \dot{c} a^{-1} b^{-2} c^{-1}
                         + 2 \dot{a} \ddot{b} \dot{c} a^{-1} b^{-1} c^{-1} \\
   & \qquad \qquad \quad
                         + 2 \dot{a} \dot{b} \ddot{c} a^{-1} b^{-1} c^{-1}
                         - 2 \dot{a} \dot{b} a b^{-3} c^{-2}
                         - 2 \dot{a} \dot{b} a^{-3} b c^{-2}
                         -   \dot{a} \dot{b} \dot{c}^2 a^{-1} b^{-1} c^{-2} \\
   & \qquad \qquad \quad
                         + 2 \dot{a} \dot{b} a^{-3} b^{-3} c^2
                         +   \dot{a} \ddot{c} \dot{c} a^{-1} c^{-2}
                         -   \dot{a} \dot{c}^3 a^{-1} c^{-3}
                         +   \dot{a} \dddot{c} a^{-1} c^{-1} \\
   & \qquad \qquad \quad
                         - 2 \dot{a} \dot{c} a b^{-2} c^{-3}
                         - 2 \dot{a} \dot{c} a^{-3} b^{-2} c
                         + 2 \dddot{b} \dot{b} b^{-2}
                         + 2 \dot{a} \dot{c} a^{-3} b^2 c^{-3} \\
   & \qquad \qquad \quad
                         +   \dddot{b} \dot{c} b^{-1} c^{-1}
                         - 2 \ddot{b} \dot{b}^2 b^{-3}
                         +   \ddot{b} \dot{b} \dot{c} b^{-2} c^{-1}
                         -   \ddot{b}^2 b^{-2} 
                         -   \ddot{b} \ddot{c} b^{-1} c^{-1} \\
   & \qquad \qquad \quad
                         -   \dot{b}^3 \dot{c} b^{-3} c^{-1}
                         +   \dot{b}^2 \ddot{c} b^{-2} c^{-1}
                         - 3 \dot{b}^2 \dot{c}^2 b^{-2} c^{-2}
                         +   \ddot{b} \dot{c}^2 b^{-1} c^{-2} 
                         +   \dot{b}^2 a^{-2} b^{-4} c^2 \\
   & \qquad \qquad \quad
                         +   \dot{b}^2 a^2 b^{-4} c^{-2} 
                         +   \dot{b}^2 a^{-2} c^{-2}
                         - 2 \dot{b}^2 b^{-4}
                         -   \dot{b} \dot{c}^3 b^{-1} c^{-3}
                         +   \dot{b} \ddot{c} \dot{c} b^{-1} c^{-2} \\
   & \qquad \qquad \quad
                         +   \dot{b} \dddot{c} b^{-1} c^{-1} 
                         + 2 \dot{b} \dot{c} a^2 b^{-3} c^{-3}
                         + 2 \dddot{c} \dot{c} c^{-2}
                         - 2 \dot{b} \dot{c} a^{-2} b c^{-3}
                         - 2 \dot{b} \dot{c} a^{-2} b^{-3} c \\
   & \qquad \qquad \quad
                         -   \ddot{c}^2 c^{-2}
                         +   \dot{c}^2 a^2 b^{-2} c^{-4}
                         +   \dot{c}^2 a^{-2} b^2 c^{-4}
                         +   \dot{c}^2 a^{-2} b^{-2}
                         - 2 \ddot{c} \dot{c}^2 c^{-3} \\
   & \qquad \qquad \quad
                         - 2 \dot{c}^2 c^{-4} 
                         + \tfrac38 a^4 b^{-4} c^{-4}
                         - \tfrac12 a^2 b^{-2} c^{-4}
                         - \tfrac12 a^2 b^{-4} c^{-2}
                         + \tfrac12 a^{-2} b^{-2} \\
   & \qquad \qquad \quad
                         - \tfrac12 a^{-2} b^2 c^{-4}
                         - \tfrac12 a^{-2} b^{-4} c^2
                         + \tfrac12 a^{-2} c^{-2}
                         + \tfrac38 a^{-4} b^4 c^{-4}
                         - \tfrac12 a^{-4} b^{-2} c^2 \\
   & \qquad \qquad \quad
                         - \tfrac12 a^{-4} b^2 c^{-2}
                         + \tfrac38 a^{-4} b^{-4} c^4
                         + \tfrac14 a^{-4}
                         + \tfrac12 b^{-2} c^{-2}
                         + \tfrac14 b^{-4}
                         + \tfrac14 c^{-4}
                      \bigr);
\end{align*}
\begin{align*}
   &\gamma_3 \bigl( \Fl_3 \bigr)^0_{\ 0} = \\
   &\qquad - \gamma_3 \bigl( 
                           4 \dddot{a} \dot{a} a^{-2}
                         - 2 \ddot{a}^2 a^{-2}
                         - 4 \ddot{a} \dot{a}^2 a^{-3}
                         + 4 \ddot{a} \dot{a} \dot{b} a^{-2} b^{-1}
                         + 4 \ddot{a} \dot{a} \dot{c} a^{-2} c^{-1} \\
   & \qquad \qquad \quad
                         - 6 \dot{a}^2 \dot{b}^2 a^{-2} b^{-2} 
                         - 6 \dot{a}^2 \dot{c}^2 a^{-2} c^{-2}
                         + 2 \dot{a}^2 a^{-4} b^2 c^{-2}
                         + 2 \dot{a}^2 a^{-4} b^{-2} c^2 \\
   & \qquad \qquad \quad
                         + 6 \dot{a}^2 b^{-2} c^{-2}
                         - 4 \dot{a}^2 a^{-4} 
                         + 4 \dot{a} \ddot{b} \dot{b} a^{-1} b^{-2}
                         - 7 \dot{a} \dot{b} a b^{-3} c^{-2} \\
   & \qquad \qquad \quad
                         + 2 \dot{a} \dot{b} a^{-1} b^{-3}
                         - 7 \dot{a} \dot{b} a^{-3} b c^{-2}
                         - 2 \dot{a} \dot{b} a^{-1} b^{-1} c^{-2} 
                         + 2 \dot{a} \dot{b} a^{-3} b^{-1} \\
   & \qquad \qquad \quad
                         + 4 \dot{a} \ddot{c} \dot{c} a^{-1} c^{-2}
                         - 7 \dot{a} \dot{c} a b^{-2} c^{-3}
                         - 2 \dot{a} \dot{c} a^{-1} b^{-2} c^{-1}
                         + 5 \dot{a} \dot{b} a^{-3} b^{-3} c^2 \\
   & \qquad \qquad \quad
                         + 2 \dot{a} \dot{c} a^{-1} c^{-3} 
                         + 5 \dot{a} \dot{c} a^{-3} b^2 c^{-3}
                         - 7 \dot{a} \dot{c} a^{-3} b^{-2} c
                         + 2 \dot{a} \dot{c} a^{-3} c^{-1}
                         + 4 \dddot{b} \dot{b} b^{-2} \\
   & \qquad \qquad \quad
                         - 2 \ddot{b}^2 b^{-2}
                         - 4 \ddot{b} \dot{b}^2 b^{-3}
                         + 2 \dot{b}^2 a^2 b^{-4} c^{-2}
                         + 4 \ddot{b} \dot{b} \dot{c} b^{-2} c^{-1}
                         - 6 \dot{b}^2 \dot{c}^2 b^{-2} c^{-2} \\
   & \qquad \qquad \quad
                         + 2 \dot{b}^2 a^{-2} b^{-4} c^2
                         + 6 \dot{b}^2 a^{-2} c^{-2}
                         + 4 \dot{b} \ddot{c} \dot{c} b^{-1} c^{-2}
                         + 5 \dot{b} \dot{c} a^2 b^{-3} c^{-3}
                         - 4 \dot{b}^2 b^{-4} \\
   & \qquad \qquad \quad
                         - 7 \dot{b} \dot{c} a^{-2} b c^{-3}
                         - 7 \dot{b} \dot{c} a^{-2} b^{-3} c
                         + 2 \dot{b} \dot{c} b^{-1} c^{-3}
                         - 2 \dot{b} \dot{c} a^{-2} b^{-1} c^{-1} \\
   & \qquad \qquad \quad
                         + 2 \dot{b} \dot{c} b^{-3} c^{-1}
                         - 2 \ddot{c}^2 c^{-2}
                         - 4 \ddot{c} \dot{c}^2 c^{-3}
                         + 2 \dot{c}^2 a^2 b^{-2} c^{-4}
                         + 4 \dddot{c} \dot{c} c^{-2} \\
   & \qquad \qquad \quad
                         + 2 \dot{c}^2 a^{-2} b^2 c^{-4} 
                         + 6 \dot{c}^2 a^{-2} b^{-2}
                         - 4 \dot{c}^2 c^{-4}
                         + \tfrac{11}{8} a^4 b^{-4} c^{-4}
                         - \tfrac32 a^2 b^{-4} c^{-2} \\
   & \qquad \qquad \quad
                         - \tfrac32 a^2 b^{-2} c^{-4}
                         - \tfrac32 a^{-2} b^2 c^{-4}
                         + \tfrac32 a^{-2} b^{-2}
                         - \tfrac32 a^{-2} b^{-4} c^2
                         + \tfrac{11}{8} a^{-4} b^4 c^{-4} \\
   & \qquad \qquad \quad
                         + \tfrac32 a^{-2} c^{-2}
                         - \tfrac32 a^{-4} b^2 c^{-2}
                         - \tfrac32 a^{-4} b^{-2} c^2
                         + \tfrac14 a^{-4}
                         + \tfrac32 b^{-2} c^{-2} \\
   & \qquad \qquad \quad
                         + \tfrac{11}{8} a^{-4} b^{-4} c^4
                         + \tfrac14 b^{-4}
                         + \tfrac14 c^{-4}
                      \bigr);
\end{align*}
\begin{align*}
   &\gamma_1 \bigl( \Fl_1 \bigr)^1_{\ 1} = \\
   &\qquad - \gamma_1 \bigl( 
                           4 \ddddot{a} a^{-1}
                         - 8 \dddot{a} \dot{a} a^{-2}
                         + 8 \dddot{a} \dot{b} a^{-1} b^{-1}
                         + 8 \dddot{a} \dot{c} a^{-1} c^{-1}
                         + 8 \ddot{a} \dot{a}^2 a^{-3}
                         - 6 \ddot{a}^2 a^{-2} \\
   & \qquad \qquad \quad
                         - 20 \ddot{a} \dot{a} \dot{b} a^{-2} b^{-1}
                         - 20 \ddot{a} \dot{a} \dot{c} a^{-2} c^{-1}
                         - 4 \ddot{a} \dot{b}^2 a^{-1} b^{-2}
                         + 8 \ddot{a} \ddot{b} a^{-1} b^{-1} \\
   & \qquad \qquad \quad
                         + 8 \ddot{a} \dot{b} \dot{c} a^{-1} b^{-1} c^{-1}
                         + 8 \ddot{a} \ddot{c} a^{-1} c^{-1}
                         - 4 \ddot{a} a b^{-2} c^{-2}
                         - 4 \ddot{a} \dot{c}^2 a^{-1} c^{-2} \\
   & \qquad \qquad \quad
                         + 4 \ddot{a} a^{-3} b^2 c^{-2}
                         + 4 \ddot{a} a^{-3} b^{-2} c^2
                         - 8 \ddot{a} a^{-3}
                         + 8 \dot{a}^3 \dot{c} a^{-3} c^{-1}
                         + 8 \dot{a}^3 \dot{b} a^{-3} b^{-1} \\
   & \qquad \qquad \quad
                         - 8 \dot{a}^2 \ddot{b} a^{-2} b^{-1}
                         + 2 \dot{a}^2 \dot{b}^2 a^{-2} b^{-2}
                         - 8 \dot{a}^2 \ddot{c} a^{-2} c^{-1}
                         - 12 \dot{a}^2 \dot{b} \dot{c} a^{-2} b^{-1} c^{-1} \\
   & \qquad \qquad \quad
                         + 2 \dot{a}^2 \dot{c}^2 a^{-2} c^{-2}
                         - 6 \dot{a}^2 a^{-4} b^2 c^{-2}
                         + 4 \dot{a} \dddot{b} a^{-1} b^{-1}
                         - 6 \dot{a}^2 a^{-4} b^{-2} c^2 \\
   & \qquad \qquad \quad
                         + 12 \dot{a}^2 a^{-4}
                         - 2 \dot{a}^2 b^{-2} c^{-2}
                         + 4 \dot{a} \ddot{b} \dot{c} a^{-1} b^{-1} c^{-1}
                         + 4 \dot{a} \dot{b}^3 a^{-1} b^{-3}
                         - 8 \dot{a} \ddot{b} \dot{b} a^{-1} b^{-2} \\
   & \qquad \qquad \quad
                         - 4 \dot{a} \dot{b}^2 \dot{c} a^{-1} b^{-2} c^{-1}
                         - 4 \dot{a} \dot{b} \dot{c}^2 a^{-1} b^{-1} c^{-2}
                         + 4 \dot{a} \dot{b} a b^{-3} c^{-2}
                         + 4 \dot{a} \dot{b} \ddot{c} a^{-1} b^{-1} c^{-1} \\
   & \qquad \qquad \quad
                         + 12 \dot{a} \dot{b} a^{-3} b c^{-2}
                         - 4 \dot{a} \dot{b} a^{-3} b^{-3} c^2
                         + 4 \dot{a} \dddot{c} a^{-1} c^{-1}
                         - 8 \dot{a} \dot{b} a^{-3} b^{-1} \\
   & \qquad \qquad \quad
                         - 8 \dot{a} \ddot{c} \dot{c} a^{-1} c^{-2}
                         + 4 \dot{a} \dot{c} a b^{-2} c^{-3}
                         - 4 \dot{a} \dot{c} a^{-3} b^2 c^{-3}
                         + 4 \dot{a} \dot{c}^3 a^{-1} c^{-3} \\
   & \qquad \qquad \quad
                         + 12 \dot{a} \dot{c} a^{-3} b^{-2} c
                         + 4 \ddddot{b} b^{-1}
                         - 4 \dddot{b} \dot{b} b^{-2}
                         + 8 \dddot{b} \dot{c} b^{-1} c^{-1}
                         - 8 \dot{a} \dot{c} a^{-3} c^{-1} \\
   & \qquad \qquad \quad
                         - 2 \ddot{b}^2 b^{-2}
                         - 8 \ddot{b} \dot{b} \dot{c} b^{-2} c^{-1}
                         + 12 \ddot{b} \ddot{c} b^{-1} c^{-1}
                         - 4 \ddot{b} \dot{c}^2 b^{-1} c^{-2}
                         + 4 \ddot{b} \dot{b}^2 b^{-3} \\
   & \qquad \qquad \quad
                         - \ddot{b} a^{-2} b c^{-2}
                         - 2 \ddot{b} a^{-2} b^{-1}
                         + 3 \ddot{b} a^{-2} b^{-3} c^2
                         + 2 \ddot{b} b^{-1} c^{-2}
                         - \ddot{b} a^2 b^{-3} c^{-2} \\
   & \qquad \qquad \quad
                         + 4 \dot{b}^3 \dot{c} b^{-3} c^{-1}
                         - 4 \dot{b}^2 \ddot{c} b^{-2} c^{-1}
                         + 2 \dot{b}^2 \dot{c}^2 b^{-2} c^{-2}
                         - 4 \dot{b}^2 a^{-2} b^{-4} c^2
                         - 2 \ddot{b} b^{-3} \\
   & \qquad \qquad \quad
                         - 4 \dot{b}^2 a^{-2} c^{-2}
                         - 4 \dot{b}^2 a^2 b^{-4} c^{-2} 
                         + 8 \dot{b}^2 b^{-4}
                         + 8 \dot{b} \dddot{c} b^{-1} c^{-1}
                         + 4 \dot{b} \dot{c}^3 b^{-1} c^{-3} \\
   & \qquad \qquad \quad
                         - 7 \dot{b} \dot{c} a^2 b^{-3} c^{-3}
                         - 8 \dot{b} \ddot{c} \dot{c} b^{-1} c^{-2} 
                         + 9 \dot{b} \dot{c} a^{-2} b c^{-3}
                         + 9 \dot{b} \dot{c} a^{-2} b^{-3} c \\
   & \qquad \qquad \quad
                         - 2 \dot{b} \dot{c} b^{-1} c^{-3}
                         - 2 \dot{b} \dot{c} b^{-3} c^{-1}
                         - 4 \dddot{c} \dot{c} c^{-2}
                         - 2 \ddot{c}^2 c^{-2}
                         - 2 \dot{b} \dot{c} a^{-2} b^{-1} c^{-1} \\
   & \qquad \qquad \quad
                         + 4 \ddot{c} \dot{c}^2 c^{-3}
                         -   \ddot{c} a^2 b^{-2} c^{-3}
                         -   \ddot{c} a^{-2} b^{-2} c
                         - 2 \ddot{c} a^{-2} c^{-1}
                         + 2 \ddot{c} b^{-2} c^{-1}
                         + 4 \ddddot{c} c^{-1} \\
   & \qquad \qquad \quad
                         + 3 \ddot{c} a^{-2} b^2 c^{-3} 
                         - 2 \ddot{c} c^{-3}
                         - 4 \dot{c}^2 a^{-2} b^{-2}
                         + 8 \dot{c}^2 c^{-4}
                         - 4 \dot{c}^2 a^{-2} b^2 c^{-4} \\
   & \qquad \qquad \quad
                         - 4 \dot{c}^2 a^2 b^{-2} c^{-4} 
                         + \tfrac58 a^4 b^{-4} c^{-4}
                         - \tfrac32 a^2 b^{-4} c^{-2}
                         + \tfrac12 a^{-2} b^2 c^{-4}
                         - \tfrac32 a^2 b^{-2} c^{-4} \\ 
   & \qquad \qquad \quad
                         + \tfrac12 a^{-2} b^{-4} c^2
                         - \tfrac12 a^{-2} c^{-2} 
                         - \tfrac38 a^{-4} b^4 c^{-4}
                         + \tfrac32 a^{-4} b^2 c^{-2}
                         - \tfrac12 a^{-2} b^{-2} \\
   & \qquad \qquad \quad
                         - \tfrac38 a^{-4} b^{-4} c^4
                         - \tfrac94 a^{-4} 
                         + \tfrac32 a^{-4} b^{-2} c^2 
                         + \tfrac12 b^{-2} c^{-2}
                         + \tfrac34 b^{-4}
                         + \tfrac34 c^{-4}
                      \bigr);
\end{align*}
\begin{align*}
   &\gamma_2 \bigl( \Fl_2 \bigr)^1_{\ 1} = \\
   &\qquad - \gamma_2 \bigl( 
                           2 \ddddot{a} a^{-1}
                         - 4 \dddot{a} \dot{a} a^{-2}
                         + 4 \dddot{a} \dot{b} a^{-1} b^{-1}
                         + 4 \dddot{a} \dot{c} a^{-1} c^{-1}
                         - 3 \ddot{a}^2 a^{-2}
                         + 4 \ddot{a} \dot{a}^2 a^{-3} \\
   & \qquad \qquad \quad
                         - 7 \ddot{a} \dot{a} \dot{b} a^{-2} b^{-1}
                         - 7 \ddot{a} \dot{a} \dot{c} a^{-2} c^{-1}
                         + 3 \ddot{a} \ddot{b} a^{-1} b^{-1}
                         - 2 \ddot{a} \dot{b}^2 a^{-1} b^{-2} \\
   & \qquad \qquad \quad
                         + 4 \ddot{a} \dot{b} \dot{c} a^{-1} b^{-1} c^{-1}
                         + 3 \ddot{a} \ddot{c} a^{-1} c^{-1}
                         - 2 \ddot{a} \dot{c}^2 a^{-1} c^{-2}
                         + 2 \ddot{a} a b^{-2} c^{-2} \\
   & \qquad \qquad \quad
                         + 2 \ddot{a} a^{-3} b^2 c^{-2}
                         + 2 \ddot{a} a^{-3} b^{-2} c^2
                         - 4 \ddot{a} a^{-3}
                         + 2 \dot{a}^3 \dot{b} a^{-3} b^{-1}
                         + 2 \dot{a}^3 \dot{c} a^{-3} c^{-1} \\
   & \qquad \qquad \quad
                         - 2 \dot{a}^2 \ddot{b} a^{-2} b^{-1}
                         +   \dot{a}^2 \dot{b}^2 a^{-2} b^{-2}
                         - 3 \dot{a}^2 \dot{b} \dot{c} a^{-2} b^{-1} c^{-1}
                         - 2 \dot{a}^2 \ddot{c} a^{-2} c^{-1} \\
   & \qquad \qquad \quad
                         +   \dot{a}^2 \dot{c}^2 a^{-2} c^{-2}
                         - 3 \dot{a}^2 a^{-4} b^2 c^{-2}
                         - 3 \dot{a}^2 a^{-4} b^{-2} c^2
                         + 6 \dot{a}^2 a^{-4}
                         +   \dot{a}^2 b^{-2} c^{-2} \\
   & \qquad \qquad \quad
                         +   \dot{a} \dddot{b} a^{-1} b^{-1}
                         - 4 \dot{a} \ddot{b} \dot{b} a^{-1} b^{-2}
                         +   \dot{a} \ddot{b} \dot{c} a^{-1} b^{-1} c^{-1}
                         + 2 \dot{a} \dot{b}^3 a^{-1} b^{-3} \\
   & \qquad \qquad \quad
                         - 2 \dot{a} \dot{b}^2 \dot{c} a^{-1} b^{-2} c^{-1}
                         +   \dot{a} \dot{b} \ddot{c} a^{-1} b^{-1} c^{-1}
                         - 2 \dot{a} \dot{b} \dot{c}^2 a^{-1} b^{-1} c^{-2}
                         - 2 \dot{a} \dot{b} a b^{-3} c^{-2} \\
   & \qquad \qquad \quad
                         + 6 \dot{a} \dot{b} a^{-3} b c^{-2}
                         - 4 \dot{a} \dot{b} a^{-3} b^{-1}
                         - 2 \dot{a} \dot{b} a^{-3} b^{-3} c^2
                         +   \dot{a} \dddot{c} a^{-1} c^{-1} \\
   & \qquad \qquad \quad
                         - 4 \dot{a} \ddot{c} \dot{c} a^{-1} c^{-2}
                         + 2 \dot{a} \dot{c}^3 a^{-1} c^{-3}
                         - 2 \dot{a} \dot{c} a b^{-2} c^{-3}
                         - 2 \dot{a} \dot{c} a^{-3} b^2 c^{-3} \\
   & \qquad \qquad \quad
                         + 6 \dot{a} \dot{c} a^{-3} b^{-2} c
                         - 4 \dot{a} \dot{c} a^{-3} c^{-1}
                         +   \ddddot{b} b^{-1}
                         -   \dddot{b} \dot{b} b^{-2}
                         + 2 \dddot{b} \dot{c} b^{-1} c^{-1} \\
   & \qquad \qquad \quad
                         +   \ddot{b} \dot{b}^2 b^{-3}
                         - 2 \ddot{b} \dot{b} \dot{c} b^{-2} c^{-1}
                         + 3 \ddot{b} \ddot{c} b^{-1} c^{-1}
                         -   \ddot{b} \dot{c}^2 b^{-1} c^{-2}
                         - 2 \ddot{b} a^2 b^{-3} c^{-2} \\
   & \qquad \qquad \quad
                         - 2 \ddot{b} a^{-2} b c^{-2}
                         + 2 \ddot{b} a^{-2} b^{-3} c^2
                         +   \dot{b}^3 \dot{c} b^{-3} c^{-1}
                         -   \dot{b}^2 \ddot{c} b^{-2} c^{-1}
                         +   \dot{b}^2 \dot{c}^2 b^{-2} c^{-2} \\
   & \qquad \qquad \quad
                         +   \dot{b}^2 a^2 b^{-4} c^{-2}
                         - 3 \dot{b}^2 a^{-2} b^{-4} c^2
                         - 3 \dot{b}^2 a^{-2} c^{-2}
                         + 2 \dot{b}^2 b^{-4}
                         + 2 \dot{b} \dddot{c} b^{-1} c^{-1} \\
   & \qquad \qquad \quad
                         - 2 \dot{b} \ddot{c} \dot{c} b^{-1} c^{-2}
                         +   \dot{b} \dot{c}^3 b^{-1} c^{-3}
                         - 2 \dot{b} \dot{c} a^2 b^{-3} c^{-3}
                         + 6 \dot{b} \dot{c} a^{-2} b c^{-3} \\
   & \qquad \qquad \quad
                         + 6 \dot{b} \dot{c} a^{-2} b^{-3} c
                         +   \ddddot{c} c^{-1}
                         -   \dddot{c} \dot{c} c^{-2}
                         +   \ddot{c} \dot{c}^2 c^{-3}
                         + 2 \ddot{c} a^{-2} b^2 c^{-3}
                         - 2 \ddot{c} a^2 b^{-2} c^{-3} \\
   & \qquad \qquad \quad
                         - 2 \ddot{c} a^{-2} b^{-2} c
                         +   \dot{c}^2 a^2 b^{-2} c^{-4}
                         - 3 \dot{c}^2 a^{-2} b^2 c^{-4}
                         + 2 \dot{c}^2 c^{-4}
                         - 3 \dot{c}^2 a^{-2} b^{-2} \\
   & \qquad \qquad \quad
                         + \tfrac{15}{8} a^4 b^{-4} c^{-4}
                         - \tfrac32 a^2 b^{-2} c^{-4}
                         - \tfrac32 a^2 b^{-4} c^{-2}
                         - \tfrac12 a^{-2} b^{-2}
                         + \tfrac12 a^{-2} b^2 c^{-4} \\
   & \qquad \qquad \quad
                         + \tfrac12 a^{-2} b^{-4} c^2
                         - \tfrac12 a^{-2} c^{-2}
                         - \tfrac98 a^{-4} b^4 c^{-4}
                         + \tfrac32 a^{-4} b^{-2} c^2
                         + \tfrac32 a^{-4} b^2 c^{-2} \\
   & \qquad \qquad \quad
                         - \tfrac98 a^{-4} b^{-4} c^4
                         - \tfrac34 a^{-4}
                         + \tfrac12 b^{-2} c^{-2}
                         + \tfrac14 c^{-4}
                         + \tfrac14 b^{-4} 
                      \bigr);
\end{align*}
\begin{align*}
   &\gamma_3 \bigl( \Fl_3 \bigr)^1_{\ 1} = \\
   &\qquad - \gamma_3 \bigl( 
                           4 \ddddot{a} a^{-1}
                         - 8 \dddot{a} \dot{a} a^{-2}
                         + 8 \dddot{a} \dot{b} a^{-1} b^{-1}
                         + 8 \dddot{a} \dot{c} a^{-1} c^{-1}
                         - 6 \ddot{a}^2 a^{-2}
                         + 8 \ddot{a} \dot{a}^2 a^{-3} \\
   & \qquad \qquad \quad
                         - 8 \ddot{a} \dot{a} \dot{b} a^{-2} b^{-1}
                         - 8 \ddot{a} \dot{a} \dot{c} a^{-2} c^{-1}
                         + 4 \ddot{a} \ddot{b} a^{-1} b^{-1}
                         - 4 \ddot{a} \dot{b}^2 a^{-1} b^{-2} \\
   & \qquad \qquad \quad
                         + 8 \ddot{a} \dot{b} \dot{c} a^{-1} b^{-1} c^{-1}
                         + 4 \ddot{a} \ddot{c} a^{-1} c^{-1}
                         - 4 \ddot{a} \dot{c}^2 a^{-1} c^{-2}
                         + 12 \ddot{a} a b^{-2} c^{-2} \\
   & \qquad \qquad \quad
                         + 4 \ddot{a} a^{-3} b^2 c^{-2}
                         + 4 \ddot{a} a^{-3} b^{-2} c^2
                         - 8 \ddot{a} a^{-3}
                         + 2 \dot{a}^2 \dot{b}^2 a^{-2} b^{-2}
                         + 2 \dot{a}^2 \dot{c}^2 a^{-2} c^{-2} \\
   & \qquad \qquad \quad
                         - 6 \dot{a}^2 a^{-4} b^2 c^{-2}
                         - 6 \dot{a}^2 a^{-4} b^{-2} c^2
                         + 12 \dot{a}^2 a^{-4}
                         + 6 \dot{a}^2 b^{-2} c^{-2}
                         - 8 \dot{a} \ddot{b} \dot{b} a^{-1} b^{-2} \\
   & \qquad \qquad \quad
                         + 4 \dot{a} \dot{b}^3 a^{-1} b^{-3}
                         - 4 \dot{a} \dot{b}^2 \dot{c} a^{-1} b^{-2} c^{-1}
                         - 4 \dot{a} \dot{b} \dot{c}^2 a^{-1} b^{-1} c^{-2}
                         - 12 \dot{a} \dot{b} a b^{-3} c^{-2} \\
   & \qquad \qquad \quad
                         + 12 \dot{a} \dot{b} a^{-3} b c^{-2}
                         - 8 \dot{a} \dot{b} a^{-3} b^{-1}
                         - 4 \dot{a} \dot{b} a^{-3} b^{-3} c^2
                         - 8 \dot{a} \ddot{c} \dot{c} a^{-1} c^{-2} \\
   & \qquad \qquad \quad
                         + 4 \dot{a} \dot{c}^3 a^{-1} c^{-3}
                         - 12 \dot{a} \dot{c} a b^{-2} c^{-3}
                         - 4 \dot{a} \dot{c} a^{-3} b^2 c^{-3}
                         + 12 \dot{a} \dot{c} a^{-3} b^{-2} c \\
   & \qquad \qquad \quad
                         - 8 \dot{a} \dot{c} a^{-3} c^{-1}
                         + 2 \ddot{b}^2 b^{-2}
                         - 7 \ddot{b} a^2 b^{-3} c^{-2}
                         - 7 \ddot{b} a^{-2} b c^{-2}
                         + 2 \ddot{b} a^{-2} b^{-1} \\
   & \qquad \qquad \quad
                         + 5 \ddot{b} a^{-2} b^{-3} c^2
                         - 2 \ddot{b} b^{-1} c^{-2}
                         + 2 \ddot{b} b^{-3}
                         + 2 \dot{b}^2 \dot{c}^2 b^{-2} c^{-2}
                         + 8 \dot{b}^2 a^2 b^{-4} c^{-2} \\
   & \qquad \qquad \quad
                         - 8 \dot{b}^2 a^{-2} b^{-4} c^2
                         - 8 \dot{b}^2 a^{-2} c^{-2}
                         -   \dot{b} \dot{c} a^2 b^{-3} c^{-3}
                         + 15 \dot{b} \dot{c} a^{-2} b c^{-3} \\
   & \qquad \qquad \quad
                         + 2 \dot{b} \dot{c} a^{-2} b^{-1} c^{-1}
                         + 15 \dot{b} \dot{c} a^{-2} b^{-3} c
                         + 2 \dot{b} \dot{c} b^{-1} c^{-3}
                         + 2 \dot{b} \dot{c} b^{-3} c^{-1}
                         + 2 \ddot{c}^2 c^{-2} \\
   & \qquad \qquad \quad
                         - 7 \ddot{c} a^2 b^{-2} c^{-3}
                         + 5 \ddot{c} a^{-2} b^2 c^{-3}
                         - 7 \ddot{c} a^{-2} b^{-2} c
                         + 2 \ddot{c} a^{-2} c^{-1}
                         - 2 \ddot{c} b^{-2} c^{-1} \\
   & \qquad \qquad \quad
                         + 2 \ddot{c} c^{-3} 
                         + 8 \dot{c}^2 a^2 b^{-2} c^{-4}
                         - 8 \dot{c}^2 a^{-2} b^2 c^{-4}
                         - 8 \dot{c}^2 a^{-2} b^{-2}
                         + \tfrac{55}{8} a^4 b^{-4} c^{-4} \\
   & \qquad \qquad \quad
                         - \tfrac92 a^2 b^{-2} c^{-4} 
                         - \tfrac92 a^2 b^{-4} c^{-2}
                         + \tfrac32 a^{-2} b^2 c^{-4}
                         - \tfrac32 a^{-2} b^{-2}
                         + \tfrac32 a^{-2} b^{-4} c^2 \\
   & \qquad \qquad \quad
                         - \tfrac32 a^{-2} c^{-2} 
                         - \tfrac{33}{8} a^{-4} b^4 c^{-4}
                         + \tfrac92 a^{-4} b^2 c^{-2}
                         + \tfrac92 a^{-4} b^{-2} c^2
                         - \tfrac{33}{8} a^{-4} b^{-4} c^4 \\
   & \qquad \qquad \quad
                         - \tfrac34 a^{-4} 
                         + \tfrac32 b^{-2} c^{-2}
                         + \tfrac14 b^{-4}
                         + \tfrac14 c^{-4}
                      \bigr).
\end{align*}

\section{Quadratic Hamiltonian cosmologies}
\label{sec:BiaCosHam}

\subsection{Canonical formalism and Misner's parameterisation}
\label{subsec:BiaCosHamMisPar}

We henceforth assume that the Bianchi-type metrics be diagonal so as to 
simplify the investigation.%
\footnote{This assumption might be wrong for some Bianchi types; the question 
          of whether the Bianchi metrics are diagonalisable or not in 
          generalised theories of gravity remains still unanswered.}
According to the possible forms of the Bianchi metrics, we start from the 
spatial element $\rmd l^2 = h_{ij} (t) \omega^i \omega^j$ for which several 
parameterisations are equally acceptable; one usually adopts either 
$h_{ij}=\diag[a^2(t),b^2(t),c^2(t)]$ or
$h_{ij}=\diag[e^{2\alpha(t)},e^{2\beta(t)},e^{2\gamma(t)}]$ as possible 
\emph{ansatz}, with proper time or logarithmic time. It is useful to represent 
the anisotropic scale factors in terms of the logarithmic volume $\mu$ and 
orthogonal anisotropic `shears' $\beta_{\pm}$. These prescriptions, first
introduced by Misner, are simply%
\footnote{See, \eg \cite{misne=MixCosMet}.}
\begin{align} \label{MisPar}
   \alpha  &= \mu + \beta_+ + \sqrt{3} \beta_-, 
   &\beta  &= \mu + \beta_+ - \sqrt{3} \beta_-, 
   &\gamma &= \mu - 2 \beta_+, 
\end{align}
and imply that the matrix formed with the $\beta$'s be traceless. Adopting
Misner's parameterisation we may write the three-metrics of Bianchi models in 
the form
\begin{equation} \label{BiaCosMisParMet}    
   \rmd l^2 = e^{2 \mu} 
                 \Bigl[ 
                    e^{2(\beta_+ + \sqrt{3} \beta_-)} 
                       \bigl( \omega^1 \bigr)^2 +
                    e^{2(\beta_+ - \sqrt{3} \beta_-)}
                       \bigl( \omega^2 \bigr)^2 + 
                    e^{-4 \beta_+} \bigl( \omega^3 \bigr)^2 
                 \Bigr],
\end{equation}
where $\mu$, $\beta_+$, and $\beta_-$ are functions of time only and 
$\omega^i$ for $i=1,2,3$ are the invariant differential forms that 
characterise the Bianchi type under study.
\nl
We rely on the Ostrogradsky canonical formalism of quadratic gravity theories 
developed in Subsection \ref{subsec:HamQua}, which is completely generic, that 
is, applicable to space-times without isometries. Henceforth, we intend to 
specify this general Hamiltonian formalism to spatially homogeneous 
cosmologies. To this aim we perform an appropriate canonical transformation 
that renders Misner's parameterisation manifest: It maps the original set of 
Ostrogradsky canonical variables $\{h_{ij},K_{ij};p^{ij},\cp^{ij}\}$ onto the 
set%
\footnote{The symbol $S_{\pm}$ stands for $\{S_+,S_-\}$ and the notation
          $A_{\pm} B_{\pm}$ must be understood as $A_+ B_+ + A_- B_-$.} 
$\bigl\{\mu,\beta_{\pm},K_{\circ},K_{\pm}$;$
   \Pi_{\mu},\Pi_{\pm},\cp_{\circ},\cp_{\pm}\bigr\}$
and is explicitly defined according to the following prescriptions: 
\label{NewVar} 
\begin{align*} 
   &h_{11} = e^{2 \mu} e^{2 (\beta_+ + \sqrt{3} \beta_-)}, 
   &&h_{22} = e^{2 \mu} e^{2 (\beta_+ - \sqrt{3} \beta_-)},
   &&h_{33} = e^{2 \mu} e^{- 4 \beta_+},
\end{align*}
and 
\begin{equation*}
   p^{ij} = \Pi^{ij} + K^{\ttT i}_{\ \ k} \cp^{\ttT jk} 
                     + \frac{\cp_{\circ}}{\sqrt{3}} K^{\ttT ij} 
                     + \frac{K_{\circ}}{\sqrt{3}} \cp^{\ttT ij} 
                     + \frac{K_{\circ} \cp_{\circ}}{3} h^{ij}, 
\end{equation*}
where the new momenta $\Pi^{ij}$ are given by 
\begin{align*} 
   &\Pi^1_{\ 1} = \frac{1}{12} \bigl( 
                                   2 \Pi_{\mu} + \Pi_+ + \sqrt{3} \Pi_-
                                \bigr), \\
   &\Pi^2_{\ 2} = \frac{1}{12} \bigl( 
                                   2 \Pi_{\mu} + \Pi_+ - \sqrt{3} \Pi_-
                                \bigr), \\
   &\Pi^3_{\ 3} = \frac16 \bigl( \Pi_{\mu} - \Pi_+ \bigr),  
\end{align*}
and the new $K$'s and $\cp$'s by 
\begin{alignat*}{2} 
   K_{ij} &= K^{\ttT}_{ij} + \frac{K_{\circ}}{\sqrt{3}} h_{ij},
   &\qquad
   &K^{\ttT i}_{\ \ j} = \frac{1}{\sqrt{6}} 
                            \diag \bigl(
                                     K_+ + \sqrt{3} K_-, 
                                     K_+ - \sqrt{3} K_-, 
                                     - 2 K_+
                                   \bigr), \\ 
   \cp^{ij} &= \cp^{\ttT ij} + \frac{\cp_{\circ}}{\sqrt{3}} h^{ij},
   &\qquad
   &\cp^{\ttT i}_{\ \ j} = \frac{1}{\sqrt{6}} 
                              \diag \bigl(
                                       \cp_+ + \sqrt{3} \cp_-, 
                                       \cp_+ - \sqrt{3} \cp_-, 
                                       - 2 \cp_+
                                    \bigr). 
\end{alignat*}
\nl
In terms of these new variables the original action in Hamiltonian form 
\eqref{HamQuaHamAct} becomes 
\begin{equation} \label{HamActQuaLQ}
   S = \int \! \rmd t 
       \Bigl[
          \Pi_{\mu} \dot{\mu} + 
          \Pi_{\pm} {\dot{\beta}_{\pm}} + 
          \cp_{\circ} \dot{K}_{\circ} + 
          \cp_{\pm} \dot{K}_{\pm} - 
          N \EuH - N_i \EuH^i
      \Bigr], 
\end{equation}
where the spatial integration 
$\int \! \omega^1 \wedge \omega^2 \wedge \omega^3$ has been performed. The 
super-Hamiltonian constraint is, in general, a quite complicated expression 
that is obtained by specifying the canonical Hamiltonian density 
\eqref{HamQuaCanHam} in terms of Bianchi isometries. For instance, the 
Bianchi-type IX super-Hamiltonian in the pure qua\-drat\-ic gravity theory is
given explicitly by the expression%
\footnote{The ensuing canonical equations are far more simpler to write down
          than the intricate \EL expressions given in the appendix of 
          Section \ref{sec:BiaCosMixCos}.} 
\begin{equation} \label{BiaCosSupHamIX}
   \begin{split} 
      \EuH_{\text{IX}} &= \alpha e^{\mu} \bf{C}\cdot\bf{C} -
                          \bf{P}\cdot\bf{R} \\
                       &\quad +
                           \frac{\sqrt{6}}3 \cp_+ 
                              \bigl( K_-^2 - K_+^2 \bigr) +
                           \frac{\sqrt{3}}3 \Vo{K} 
                              \bigl( K_\pm \cp_\pm - \Pi_{\mu} \bigr) \\
                       &\quad -
                           \frac{\sqrt{6}}6 K_\pm \Pi_\pm +
                           \frac{2\sqrt{6}}3 K_+ K_- \cp_- +
                           \frac{\sqrt{3}}6 \Vo{\cp} 
                              \bigl( K_+^2 + K_-^2 \bigr) \\
                       &\quad +
                           \frac{2\sqrt{3}}3 \Vo{K}^2 \Vo{\cp} +
                           \frac1{6\beta} e^{-3\mu} \Vo{\cp}^2 -
                           \frac1{2\alpha} e^{-3\mu} 
                              \bigl( \cp_+^2 + \cp_-^2 \bigr),
   \end{split}
\end{equation}
where we have denoted 
\begin{equation} 
   \begin{split} 
      \bf{C}\cdot\bf{C} 
         := &\, e^{2\mu} C_{\fn ijk} C^{\fn ijk} \\
          = &\, \frac32 K_+^2 \Bigl[
                              2 e^{4\beta_+} \bigl( 
                                                1 - \cosh (4\sqrt{3}\beta_-) 
                                             \bigr) -
                              3 e^{-8\beta_+}
                           \Bigr] \\
          &\quad + 
             K_-^2 \Bigl[
                      4 e^{-2\beta_+} \cosh(2\sqrt{3}\beta_-) -
                      e^{4\beta_+} \bigl( 
                                      1 + 7 \cosh(4\sqrt{3}\beta_-) 
                                   \bigr) -
                      \frac12 e^{-8\beta_+}
                   \Bigr] \\
          &\quad - 
             4 \sqrt{3} K_+ K_- \Bigl[
                                   e^{-2\beta_+} \sinh(2\sqrt{3}\beta_-) +
                                   e^{4\beta_+} \sinh(4\sqrt{3}\beta_-)
                                \Bigr] 
   \end{split}
\end{equation}
and 
\begin{equation} 
   \begin{split} 
      \bf{P}\cdot\bf{R} 
         := &\, \cp^{\ttT ij} \, \emb{R}_{ij} \\
          = &\, e^{-2\mu} 
                \Bigl[
                   \frac{\sqrt{6}}3 \cp_+ \parD{\cv}{\beta_+} + 
                   \sqrt{2} \cp_- \parD{\cv}{\beta_-} + 
                   \frac{\sqrt{3}}{12} \Vo{\cp} 
                      \cv \bigl( \beta_+, \beta_- \bigr)
                \Bigr],
   \end{split}
\end{equation}
and where the function $\cv(\beta_+, \beta_-)$ is given explicitly in Table
\ref{tab:BianchiPots} at the type IX entry.
\nl
As regards the super-momentum constraint, we reiterate that one may 
consistently choose the shift vector to be zero in all cases but type VI$_{0}$ 
(\cfr the remark on page \pageref{rem:supmom}). To simplify the analysis we 
shall consider those canonical systems that correspond respectively to the 
pure $R$-squared and Weyl-squared variants of the general quadratic theory.

\subsection[Pure $R^2$ Bianchi cosmologies]%
           {Pure $\boldsymbol{R^2}$ Bianchi cosmologies}
\label{subsec:BiaCosHamR2}

\subsubsection{Reduced Hamiltonians and exact solutions}

We first study the pure $R$-squared Bianchi cosmologies, the field equations
of which were analysed by several authors with the aim of examining the 
asymptotic behaviour of the mixmaster model in that specific quadratic 
variant; see, \eg \cite{barro=MixCos,barro=ChaBehHOG,spind=AsyBeh,
spind=CosModR2}. In contradistinction to these works, we lay our analysis on 
the canonical system rather than the field equations. Our goal is to 
appraise the achievements of the Hamiltonian formalism; in particular, to
determine whether it simplifies the investigation, and to compare our results 
with those that are published.
\nl                 
The canonical action \eqref{HamActQuaLQ} reduces to
\begin{equation} \label{HamActR2}
   S = \int \! \rmd t 
       \Bigl[
          \Pi_{\mu} \dot{\mu} + 
          \Pi_{\pm} {\dot{\beta}_{\pm}} + 
          \Vo{\cp} \dot{K}_{\circ} - 
          N \EuH 
      \Bigr], 
\end{equation}
due to the existence of the primary constraint \eqref{HamfrPriConb}, which
translates into $\cp_{\pm} \approx 0$ in terms of the new variables introduced
in the preceding subsection. The super-Hamiltonian is 
\begin{equation} \label{BiaCosSupHamR2a}
   \begin{split}
      \EuH &= \frac{\sqrt{3}}3 
                 \Bigl[
                    2 \Vo{\cp} \Vo{K}^2 +
                    \frac12 \Vo{\cp} \bigl( K_+^2 + K_-^2 \bigr) -
                    \Vo{K} \Pi_{\mu} -
                    \frac{\sqrt{2}}2 K_{\pm} \Pi_{\pm}
                 \Bigr] \\
           & \qquad \qquad +
             \frac16 e^{-3\mu} \Vo{\cp}^2 -
             \frac{\sqrt{3}}{12} e^{-2\mu} 
                \Vo{\cp} \, \cv (\beta_+, \beta_-),              
   \end{split}
\end{equation}
where $\cv (\beta_+,\beta_-)$ denotes the potential of the Bianchi type 
considered (\cfr Table \ref{tab:BianchiPots}), which stems from the three-%
dimensional scalar curvature, and the coupling constant $\beta$ of the $R$-%
squared term in the action \eqref{HamQuaLag1} has been set equal to one.  
\begin{table} 
\caption{Potentials for types I, II, VII$_0$, VIII, and IX.}
\label{tab:BianchiPots}
\begin{center}
\begin{tabular}{c|l}
\hline
\hline
Type    & $\cv (\beta_+, \beta_-)$ \\
\hline
I       &$0$ \\
II      &$\exp (-8\beta_+)$ \\
VII$_0$ &$2 \exp (4\beta_+) [\cosh (4\sqrt{3} \beta_-) - 1]$ \\
VIII    &$\exp (-8\beta_+) +  
         2 \exp (4\beta_+) [\cosh (4\sqrt{3} \beta_-) - 1]$ \\
        &$\qquad + 4 \exp (-2\beta_+) \cosh (2\sqrt{3} \beta_-)$ \\
IX      &$\exp (-8\beta_+) +
         2 \exp (4\beta_+) [\cosh (4\sqrt{3} \beta_-) - 1]$ \\
        &$\qquad - 4 \exp (-2\beta_+) \cosh (2\sqrt{3} \beta_-)$ \\
\hline
\hline
\end{tabular}
\end{center}
\end{table}
\nl
We must apply the Dirac--Bergmann consistency algorithm to this constrained 
system, the Poisson bracket being of course defined \wrt the new variables; 
or, equivalently, we can translate the results of Subsection
\ref{subsec:HamQua} in terms of those variables. The Dirac Hamiltonian density 
is $\Fh_{\EuD}=N \EuH+\lambda_{\pm} \cp_{\pm}$, with Lagrange multipliers 
$\lambda_{\pm}$. Conservation of the primary constraints $\cp_{\pm} \approx 0$ 
brings forth secondary constraints $\chi_{\pm} \approx 0$, \viz
\begin{equation*}
   \cp_{\pm} \approx 0 \tcons \chi_{\pm} = 
   \Vo{\cp} K_{\pm} - \frac{\sqrt{2}}2 \Pi_{\pm} \approx 0,
\end{equation*}
the time evolution of which determines the Lagrange multipliers 
$\lambda_{\pm}$. Since we have the nonzero Poisson bracket
$\{\cp_{\pm},\chi_{\pm}\}\approx -\Vo{\cp}$ the above primary and secondary
constraints are second class; hence the associated \dofs, \ie $\cp_{\pm}$ and
$K_{\pm}$, are unphysical and can be removed. The super-Hamiltonian 
\eqref{BiaCosSupHamR2a} becomes simply
\begin{equation} \label{BiaCosSupHamR2b}
   \begin{split}
      \EuH &= \frac{\sqrt{3}}3 
                 \biggl[
                    2 \Vo{\cp} \Vo{K}^2 -
                    \Vo{K} \Pi_{\mu} -
                    \frac{\Pi_+^2 + \Pi_-^2}{4 \Vo{\cp}}
                 \biggr] +
              \frac16 e^{-3\mu} \Vo{\cp}^2 \\
           &  \qquad \qquad \qquad -
              \frac{\sqrt{3}}{12} e^{-2\mu} 
                 \Vo{\cp} \, \cv (\beta_+, \beta_-),              
   \end{split}
\end{equation}
and the ensuing canonical equations are
\begin{subequations} \label{BiaCosR2CanEqu1}
   \begin{align}
      \dot{\mu}         &= - \frac{\sqrt{3}}3 N \Vo{K}, \\
      \dot{\beta}_{\pm} &= - \frac{\sqrt{3}}6 N 
                             \frac{\Pi_{\pm}}{\Vo{\cp}}, \\
      \dot{K}_{\circ}   &= N \biggl[ 
                                \frac{\sqrt{3}}3 \biggl(
                                                    2 K^2_{\circ} +
                                                    \frac{\Pi_+^2 + \Pi_-^2}
                                                         {4 \Vo{\cp}^2}
                                                 \biggr) +
                                    \frac13 \Vo{\cp} e^{-3\mu} -
                                    \frac{\sqrt{3}}{12} e^{-2 \mu} 
                                       \cv (\beta_+, \beta_-)
                             \biggr], \\
      \dot{\Pi}_{\mu}   &= N \Bigl[ 
                                \frac12 \cp^2_{\circ} e^{-3\mu} -
                                \frac{\sqrt{3}}6 \Vo{\cp} e^{-2\mu}
                                   \cv (\beta_+, \beta_-)
                             \Bigr], \\
      \dot{\Pi}_{\pm}   &= \frac3{12} N \Vo{\cp} e^{-2\mu}
                           \parD{\cv}{\beta_{\pm}}, \\ 
      \dot{\cp}_{\circ} &= - \frac{\sqrt{3}}3 N \bigl( 
                                                   4 \Vo{K} \Vo{\cp} -
                                                   \Pi_{\mu}
                                                \bigr). 
                           \label{BiaCosR2CanEqu1f}
   \end{align}
\end{subequations}
This canonical system possesses the first integral 
\begin{equation} \label{BiaCosR2FirInt}
   \Pi_{\mu} - \Vo{K} \Vo{\cp} = \Vo{k}, 
\end{equation}
where $\Vo{k}$ is an arbitrary constant.
\nl
In the usual \emph{minisuperspace} approaches of Hamiltonian cosmology one 
goes further by firstly fixing the lapse function, then solving the super-%
Hamiltonian for an adequately chosen canonical variable. The outcome is a 
\emph{nonvanishing} `reduced' Hamiltonian density. We proceed likewise and 
firstly consider space-times of nonconstant scalar curvature.

\paragraph*{Nonconstant scalar curvature.}
First of all, we fix the time gauge by setting
\begin{equation} \label{BiaCosR2Time}
   R = \rho e^t,
\end{equation}
where $\rho=\pm1$ (this allows for a separate discussion on space-times with 
positive or negative scalar curvature). According to the definition 
\eqref{HamQuacp} the variable $\Vo{\cp}$ becomes
\begin{equation}
   \Vo{\cp} = - \frac{\sqrt{3}}2 \rho e^{3\mu+t}.
\end{equation}
Differentiating this expression and making use of the canonical equations 
\eqref{BiaCosR2CanEqu1} and the first integral \eqref{BiaCosR2FirInt} we 
obtain the form of the lapse function that corresponds to the gauge choice 
\eqref{BiaCosR2Time}, namely
\begin{equation}
   N(t) = \frac{\sqrt{3}}{\Vo{k}} \Vo{\cp}
        = - \frac{3 \rho}{2 \Vo{k}} e^{3\mu+t}. 
\end{equation}
Solving the super-Hamiltonian \eqref{BiaCosSupHamR2b} for $\Vo{K}$ and
performing the canonical transformation
\begin{align*}
    \mu       &\longrightarrow \frac12 (\mu-t)
   &\Pi_{\mu} &\longrightarrow 2 \Pi_{\mu} + 3 \Vo{K} \Vo{\cp}     
\end{align*}
we obtain, after little algebraic manipulation, the reduced form of the 
Bianchi $R$-squared canonical action \eqref{HamActR2}, namely
\begin{equation} \label{HamActR2ADM}
   S = \int \! \rmd t 
            \Bigl[ 
               \Pi_{\mu} \dot{\mu} + 
               \Pi_{\pm} {\dot{\beta}_{\pm}} - 
               H 
            \Bigr], 
\end{equation}
where the reduced Hamiltonian density is
\begin{equation} \label{BiaCosADMHamR2}
   H = \frac1{4 \Vo{k}} \biggl[
                           4 \Pi_{\mu}^2 - \Pi_+^2 - \Pi_-^2 -
                           \frac{3\rho^2}4 e^{2\mu} 
                              \Bigl(
                                 \rho e^{\mu} + \cv (\beta_+, \beta_-)              
                              \Bigr) +
                           \Vo{k}^2
                        \biggr]. 
\end{equation}
The canonical equations are readily derived from the action 
\eqref{HamActR2ADM}, \viz
\begin{subequations} \label{BiaCosR2CanEqu2}
   \begin{align}
      \dot{\mu}         &= \frac2{\Vo{k}} \Pi_{\mu}, 
                           \label{BiaCosR2CanEqu2a} \\
      \dot{\beta}_{\pm} &= - \frac1{2\Vo{k}} \Pi_{\pm}, 
                           \label{BiaCosR2CanEqu2b} \\
      \dot{\Pi}_{\mu}   &= \frac{3\rho^2}{16\Vo{k}}
                              \bigl[ 
                                 3 \rho e^{\mu} +
                                 2 \cv (\beta_+, \beta_-)
                              \bigr] e^{2\mu}, 
                           \label{BiaCosR2CanEqu2c} \\
      \dot{\Pi}_{\pm}   &= \frac{3\rho^2}{16\Vo{k}} e^{2\mu}
                              \parD{\cv}{\beta_{\pm}}. 
                           \label{BiaCosR2CanEqu2d}
   \end{align}
\end{subequations}
They are equivalent to the autonomous system of second-order coupled \odes
\begin{subequations} \label{BiaCosR2CanEqu3}
   \begin{align}
      \ddot{\mu}         &= \frac{9\rho^3}{8\Vo{k}^2} e^{3\mu} +
                            \frac{3\rho^2}{4\Vo{k}^2} e^{2\mu}
                               \cv (\beta_+, \beta_-), 
                            \label{BiaCosR2CanEqu3a} \\
      \ddot{\beta}_{\pm} &= - \frac{3\rho^2}{32\Vo{k}^2} e^{2\mu}
                               \parD{\cv}{\beta_{\pm}}.
   \end{align}
\end{subequations}
By mere inspection, we see that the system \eqref{BiaCosR2CanEqu3} decouples 
when the potential term vanishes, \ie for type I; hence we carry on to 
analysing the corresponding type I system.
\nl
From the canonical equation \eqref{BiaCosR2CanEqu2d} with zero potential, we 
conclude that the momenta $\Pi_{\pm}$ are constants of motion. Hence we can 
readily integrate the equation \eqref{BiaCosR2CanEqu2b}: the variables 
$\beta_{\pm}$ are simply linear functions of $t$, namely
\begin{equation} \label{BiaCosR2BIbeta}
   \beta_{\pm} (t) = - \frac{\Pi_{\pm}}{2\Vo{k}} t + b_{\pm}, 
\end{equation}
where the constants of integration $b_{\pm}$ can be removed by a change of 
scale of the spatial coordinates. Furthermore, the equations 
\eqref{BiaCosR2CanEqu2a} and \eqref{BiaCosR2CanEqu2c} reduce to
\begin{subequations} \label{BiaCosR2CanEqu4}
   \begin{align}
      \dot{\mu}         &= \frac2{\Vo{k}} \Pi_{\mu}, 
                           \label{BiaCosR2CanEqu4a} \\
      \dot{\Pi}_{\mu}   &= \frac{9\rho^3}{16\Vo{k}} e^{3\mu}. 
                           \label{BiaCosR2CanEqu4b}
   \end{align}
\end{subequations}
This is easy to solve and we obtain 
\begin{subequations} \label{BiaCosR2CanSol4}
   \begin{align}
      e^{3\mu(t)}  &= \frac{64\Vo{k}^2\Vo{\omega}^2}{27\rho^3}
                      \frac{\Vo{n}e^{2\Vo{\omega} t}}
                           {\bigl( 1 - \Vo{n} e^{2\Vo{\omega} t} \bigr)^2}, 
                      \label{BiaCosR2CanSol4a} \\
      \Pi_{\mu}(t) &= \frac{\Vo{k} \Vo{\omega}}{3} 
                      \frac{1 + \Vo{n} e^{2\Vo{\omega} t}}
                           {1 - \Vo{n} e^{2\Vo{\omega} t}}, 
                      \label{BiaCosR2CanSol4b}
   \end{align}
\end{subequations}
where $\Vo{n}$ is a new constant of integration and 
$\Vo{\omega}^2:=\tfrac94 [1+(\Pi_+^2+\Pi_-^2)/\Vo{k}^2]$. Therefore, the 
general Bianchi-type I metric is given by
\begin{equation} \label{BiaCosR2BISolMet}
   \rmd s^2 = - \frac{9 \rho^2}{4 \Vo{k}^2} e^{3\mu(t)} e^{-t} \rmd t^2 
              + e^{\mu(t)} e^{-t} \sum_{i=1}^3 e^{2\nu_i t} 
                                               \bigl( \rmd x^i \bigr)^2,               
\end{equation}
where the constants $\nu_i$ for $i=1,2,3$ satisfy
\begin{align*}
   &\sum_{i=1}^3 \nu_i = 0,
   &\sum_{i=1}^3 \nu_i^2 &= \frac3{2 \Vo{k}^2} \Bigl( 
                                                     \Pi_+^2 + \Pi_-^2 
                                                  \Bigr) 
                          = \frac23 \bigl( \Vo{\omega}^2 - \frac94 \bigr).
\end{align*}
\nl
Now, it is interesting to compare the form \eqref{BiaCosR2BISolMet} of the 
general solution with those representations that were obtained by Buchdahl 
\cite{buchd=FieEqu} and independently by Spindel and Zinque 
\cite{spind=AsyBeh} from a direct analysis of the Bianchi-type I higher-order 
field equations. In the former article the time coordinate is defined as 
$t_{\text{[\textsc{b}]}}=e^t$ and the logarithmic volume as 
$\mu_{\text{[\textsc{b}]}}=\tfrac12 (\mu-t)$ (with these relations one easily 
passes from the expression \eqref{BiaCosR2BISolMet} to Buchdahl's solution); 
in the latter the time coordinate is given by $t_{\text{[\textsc{sz}]}}=t$ 
whereas the variable $\Delta_{\text{[\textsc{sz}]}}$ coincides with $3\mu$. 
Upon a mere redefinition of the constants of integration in the general 
solution \eqref{BiaCosR2CanSol4a} we recover Spindel and Zinque's solution for 
$\Delta_{\text{[\textsc{sz}]}}(t)$, namely 
\begin{equation} \label{BiaCosR2BISpin}
   e^{\Delta_{\text{[\textsc{sz}]}}(t)} = 
   e^{3\mu(t)}  = \frac43 Q^2 a^2 \frac{e^{Q t}}
                                       {\bigl( 
                                           1 - \rho \lambda^2 a^2 e^{Q t} 
                                        \bigr)^2}, 
\end{equation}
where $Q^2$, $\lambda^2$, and $a^2$ are positive constants defined in terms of
$\rho$, $\Vo{k}$, $\Vo{\omega}$, and $\Vo{n}$ by
\begin{align*}
    Q^2       &:= 4 \Vo{\omega}^2,
   &\lambda^2 &:= \frac{9\rho^2}{4\Vo{k}^2},
   &a^2       &:= \frac{4\Vo{n} \Vo{k}^2}{9\rho^3},
\end{align*}
respectively. The proper time $\tau$ is obtained, up to constant factors, by 
integrating 
\begin{equation} \label{BiaCosR2Tau}
   \frac{\rmd \tau}{\rmd t} = \pm \frac{e^{\frac12 (Q-1)t}}
                                       {1 - \rho \lambda^2 a^2 e^{Q t}}. 
\end{equation}
We thus recover the two distinct behaviours of the general solution 
\eqref{BiaCosR2BISpin} found by Spindel and Zinque \cite{spind=AsyBeh} whose 
analysis of the Bianchi-type I field equations generalises Buchdahl's work
\cite{buchd=FieEqu}. Strictly speaking, if $\rho=+1$, \ie positive curvature,
then the function in the \rhs of equation \eqref{BiaCosR2Tau} exhibits a 
vertical asymptote at $e^{Q t_{\infty}}=\lambda^{-2} a^{-2}$ corresponding to 
$\tau \rightarrow + \infty$; this behaviour features an asymptotic de Sitter 
space-time. On the other hand, if $\rho=-1$, \ie negative curvature,%
\footnote{This is the sole case considered by Buchdahl \cite{buchd=FieEqu}.}
then integration of equation \eqref{BiaCosR2Tau} for $t \in \mathbb{R}$ yields 
a finite proper time after which the Bianchi-type I universe recollapses.
\begin{rem}
The metric \eqref{BiaCosR2BISolMet} becomes isotropic if and only if 
$\Pi_{\pm}=0$, \ie $\Vo{\omega}=\tfrac32$; that is, up to rescaled 
coordinates,
\begin{equation} \label{BiaCosR2BISolMetIso}
   \rmd s^2 = - \frac{e^{2t}}{\bigl( 1 - \Vo{n} e^{3t} \bigr)^2} \rmd t^2 
              + \frac{\rmd l^2}{\bigl( 1 - \Vo{n} e^{3t} \bigr)^{\frac23}}.
\end{equation}
\end{rem}

\paragraph*{Constant scalar curvature.}
When the scalar curvature is constant the field equations of the $R$-squared
variant are satisfied by an arbitrary \einspace, \ie 
$R_{ab}=\tfrac14Rg_{ab}=\Lambda g_{ab}$ with $\Lambda$ constant. The problem 
of determining the general form of the metrics corresponding to \einspaces was 
undertaken by Petrov who proved, in particular, the following result
\cite[p.~84]{PETRO=EinSpa}.%
\footnote{We have modified the original statement in accordance with our 
          conventions and corrected a misprint.}
\begin{thm} \label{thm:Petrov}
The metric 
\begin{equation*}
   \rmd s^2 = - \bigl( \rmd x^n \bigr)^2
              + \sum_{i=1}^{n-1} f_i (x^n) \bigl( \rmd x^i \bigr)^2               
\end{equation*}
defines a $n$-dimensional \einspace $R_{ab}=\tfrac1{n}Rg_{ab}=\Lambda g_{ab}$ 
if the functions $f_i$ for $\range{i}{1}{n-1}$ are chosen as follows:
\begin{equation} \label{petrovsol}
   f_i (x^n) = \begin{cases}
                  \sin^{\frac2{n-1}}(\alpha x^n) 
                  \tan^{\frac{2\alpha_i}{\alpha}}(\frac12 \alpha x^n)
                  &\text{if $\Lambda<0$}, \\
                  \sinh^{\frac2{n-1}}(\beta x^n) 
                  \tanh^{\frac{2\beta_i}{\beta}}(\frac12 \beta x^n)
                  &\text{if $\Lambda>0$},
               \end{cases}
\end{equation}
where $\alpha$, $\beta$, $\alpha_i$, and $\beta_i$ are constants; and if 
$\Lambda=0$ one obtains either an $S^n$, \ie a space of constant curvature, or 
the generalised Kasner metric%
\footnote{If $\Lambda=0$ and $n=4$, then the Kasner metric is the general
          solution.} 
\begin{align*}
   f_i (x^n) &= \bigl( x^n \bigr)^{2p_i},
   &\text{with}&\; \sum_{i=1}^{n-1} p_i=1=\sum_{i=1}^{n-1} p_i^2. 
\end{align*}
\end{thm}
We demonstrate how the closed-form solution \eqref{petrovsol} can be derived 
from the canonical formalism, in the case $n=4$ and for Bianchi type I.%
\footnote{The simple procedure given here yields the same results as 
          Buchdahl's \emph{modus operandi} \cite{buchd=FieEqu}.}
First of all, observe that a constant scalar curvature, $R=4\Lambda$, entails 
the vanishing of the constant $\Vo{k}$ in the first integral 
\eqref{BiaCosR2FirInt} owing to the fact that the momentum $\Vo{\cp}$ reduces 
to $\Vo{\cp}=-2\sqrt{3}\Lambda e^{3\mu}$; hence we have 
$\Pi_{\mu}=\Vo{K}\Vo{\cp}$. We are free to fix a time gauge by choosing
a specific lapse function. 
To recover Petrov's representation of Theorem \ref{thm:Petrov} we could either
start anew from the canonical equations \eqref{BiaCosR2CanEqu1} with zero 
potential and $N=1$ or, equivalently, choose $N=\sqrt{3} \Vo{\cp}$ and perform 
the transformation from the coordinate time $t$ to the proper time $\tau$ on 
the result. In both cases we obtain the differential equation 
\begin{equation} \label{BiaCosR2RcstPodetau}
   \Vo{\cp}^{''} -  \Vo{\omega}^2 \Vo{\cp} = 0,
\end{equation}
where a prime denotes differentiation \wrt $\tau$ and 
$\Vo{\omega}^2:=3\Lambda$. According to the sign of $\Lambda$ there are two 
distinct solutions, which involve either trigonometric or hyperbolic 
functions, as this is expected from Theorem \ref{thm:Petrov}. For instance, if 
$\Lambda>0$, then equation \eqref{BiaCosR2RcstPodetau} integrates and gives 
$\Vo{\cp}(\tau)=\Vo{p} \sinh(\Vo{\omega} \tau)$, where 
$\Vo{p}^2=\tfrac34 (\Pi_+^2+\Pi_-^2)/\Vo{\omega}^2$. 
The anisotropic shears are obtained upon integrating 
\begin{equation} \label{BiaCosR2Rcstbetaodetau}
   \beta_{\pm}^{'} +  
   \frac{\sqrt{3}\Pi_{\pm}}{6\Vo{p}} \sinh^{-1} (\Vo{\omega} \tau) = 0;
\end{equation}
that is,
\begin{equation} \label{BiaCosR2Rcstbetatau}
   \beta_{\pm} (\tau) =   
   - \frac{\sqrt{3}\Pi_{\pm}}{6\Vo{\omega} \Vo{p}} 
     \ln \Bigl\lvert \tanh \Bigl( \frac{\Vo{\omega} \tau}2 \Bigr) \Bigr\rvert.
\end{equation}
Therefore, the general form of the metric can be written, up to rescaled 
constant factors, as
\begin{equation} \label{BiaCosR2RcstSol}
   \rmd s^2 = - \rmd \tau^2
              + \sinh^{\frac23} (\Vo{\omega} \tau) 
                \sum_{i=1}^{3} \Bigl[ 
                                  \tanh \Bigl( \frac{\Vo{\omega} \tau}2 \Bigr)
                               \Bigr]^{\frac{2 \nu_i}{\Vo{\omega}}} 
                \bigl( \rmd x^i \bigr)^2,               
\end{equation}
where the parameters $\nu_i$ satisfy the identities 
\begin{align*} 
   &\sum_{i=1}^3 \nu_i = 0,
   &\sum_{i=1}^3 \nu_i^2 &= \frac23 \Vo{\omega}^2 = 2 \Lambda.
\end{align*}
We have thus recovered Petrov's representation \eqref{petrovsol} for the 
positive-curvature case.%
\footnote{One may proceed likewise for $\Lambda<0$: the hyperbolic functions
          in the solution \eqref{BiaCosR2RcstSol} are then replaced by the
          corresponding trigonometric ones (\cfr \cite{buchd=FieEqu}).}
\begin{rem}
As this was first noticed by Buchdahl for a negative scalar curvature 
\cite{buchd=FieEqu}, if one defines the new parameters 
$\tau^{\star}:=\tfrac2{\Vo{\omega}}\tanh(\tfrac12\Vo{\omega}\tau)$ and
$p_i:=\nu_i/\Vo{\omega}+\frac13$ for $i=1,2,3$, then one may take the limit of
the spatial element in the metric \eqref{BiaCosR2RcstSol} as $\Lambda$ tends
to zero: 
\begin{equation*}
   \lim_{\Vo{\omega} \rightarrow 0} 
      \biggl[ 
         \cosh^{\frac23} \bigl( \Vo{\omega} \tau \bigr)
         \sum_{i=1}^{3} \bigl( \tau^{\star} \bigr)^{2 p_i} 
         \bigl( \rmd x^i \bigr)^2 
      \biggl] =
      \sum_{i=1}^{3} \tau^{2 p_i} \bigl( \rmd x^i \bigr)^2,               
\end{equation*}
where the parameters $p_i$ satisfy the relations
\begin{align*} 
   &\sum_{i=1}^3 p_i = 1,
   &\sum_{i=1}^3 p_i^2 &= 1.
\end{align*}
This is precisely the Kasner metric. 
\end{rem}

\subsubsection{Isotropic models}

Consider the closed \flrw cosmological model, which is contained in Bianchi
type IX, the three-metric of which is
\begin{equation} \label{BiaCosR2IsoCloMet}
   \rmd l^2 = A^2(t) \biggl[ 
                        \frac{\rmd r^2}{1-r^2} + r^2 \rmd \Omega^2 
                     \biggr],                
\end{equation}
where $\rmd \Omega^2 = \rmd \vartheta^2 + \sin^2(\vartheta) \rmd \varphi$; $r$, 
$\vartheta$, $\varphi$ are the spherical coordinates; and $A(t)$ is the scale
factor.
\nl
We take $A$ and $\Vo{K}:=\tfrac{\sqrt{3}}3 A^2 K$ as configuration variables; 
their respective conjugate momenta are $\Pi_A:=6A p^{ij}$ (for $i=j$) and
$\Vo{\cp}:=\tfrac{\sqrt{3}}3 A^{-2} \cp$. For an arbitrary nonlinear 
Lagrangian the latter momentum becomes $\Vo{\cp}:=-2\sqrt{3} A \fpr(R)$ 
according to the definition \eqref{HamfrConjMom1b}. The canonical action is
simply
\begin{equation} \label{HamActR2IsoClo}
   S = \int \! \rmd t 
       \Bigl[
          \Pi_A \dot{A} + \Vo{\cp} \dot{K}_{\circ} - N \EuH 
       \Bigr], 
\end{equation}
where the super-Hamiltonian is
\begin{equation} \label{R2IsoClosuphamfr}
   \EuH = \sqrt{3} \Vo{\cp} - 
          \frac{\sqrt{3}}3 \Vo{K} \frac{\Pi_A}{A} +
          V \bigl( \Vo{\cp} \bigr),
\end{equation}
with the potential $V$ defined by the expression \eqref{HamfrPot}. 
\nl
For instance, if one considers the Einstein--Hilbert Lagrangian with an 
additional $R$-squared term (\cfr the Lagrangian density \eqref{HamQuaLag1} 
with $\Lambda=0=\alpha$), then the above potential reduces to 
$V(\Vo{\cp})=\tfrac{1}{2\beta}A(\tfrac{\sqrt{3}}3 \Vo{\cp} + 
             \kappa^{-2} A)^2$.
The quantum cosmology based on such a variant of the generic quadratic theory
was investigated by Hawking and Luttrell \cite{hawki=HigDer}. They showed that
the wave function of the closed \flrw minisuperspace universe could be
interpreted as corresponding in the classical limit to a family of---%
particular---solutions that feature a period of inflation followed by a 
matter-dominated era. This era exhibits a rapidly oscillating scale factor 
superimposed on an overall expansion, after which the universe recollapses.
A direct comparison with Hawking and Luttrell's canonical formalism shows that
their canonical Hamiltonian coincides exactly with the super-Hamiltonian 
\eqref{R2IsoClosuphamfr} (with the specific aforementioned potential) if the
lapse function is taken as $N(t):=A(t)$ and once the trivial canonical 
transformation $\bigl\{ Q_{\text{[\textsc{hl}]}} \rightarrow \Vo{\cp}$;
$\Pi_{Q_{\text{[\textsc{hl}]}}} \rightarrow -\Vo{K} \bigr\}$ from their 
variables to ours has been performed. 
\nl
We henceforth investigate the pure $R$-squared variant of the generic 
nonlinear case. The super-Hamiltonian is given by
\begin{equation} \label{R2IsoClosupham}
   \EuH = \sqrt{3} \Vo{\cp} - 
          \frac{\sqrt{3}}3 \Vo{K} \frac{\Pi_A}{A} +
          \frac16 A \Vo{\cp}^2
\end{equation}
and the ensuing canonical system, 
\begin{subequations} \label{R2IsoCloCanEqu1}
   \begin{align}
      \dot{A}          &= - \frac{\sqrt{3}}3 N \frac{\Vo{K}}{A}, \\
      \dot{K}_{\circ}  &= N \bigl( \frac13 A \Vo{\cp} + \sqrt{3} \bigr), \\
      \dot{\Pi}_A      &= - N \Bigl(
                                 \frac16 \Vo{\cp}^2 + 
                                 \frac{\sqrt{3}}3 \Vo{K} \frac{\Pi_A}{A^2}
                              \Bigr), \\
      \dot{\cp}_{\circ} &= \frac{\sqrt{3}}3 N \frac{\Pi_A}{A},
   \end{align}
\end{subequations}
possesses the first integral 
\begin{equation} \label{R2IsoCloFirInt}
   A \Pi_A + \Vo{K} \Vo{\cp} = \Vo{k}, 
\end{equation}
where $\Vo{k}$ is an arbitrary constant. Firstly, we consider space-times of 
nonconstant scalar curvature.

\paragraph*{Nonconstant scalar curvature.}
First of all, we fix the time gauge by setting
\begin{equation} \label{R2IsoCloTime}
   R = \rho e^t,
\end{equation}
where $\rho=\pm1$. According to its definition the variable $\Vo{\cp}$ becomes
\begin{equation}
   \Vo{\cp} = - \frac{\sqrt{3}}2 \rho A e^t.
\end{equation}
Differentiating this expression and making use of the canonical equations 
\eqref{R2IsoCloCanEqu1} and the first integral \eqref{R2IsoCloFirInt} we 
obtain the form of the lapse function that corresponds to the gauge choice 
\eqref{R2IsoCloTime}, namely
\begin{equation}
   N(t) = - \frac{3 \rho}{2 \Vo{k}} A^3 e^t. 
\end{equation}
Solving the super-Hamiltonian \eqref{R2IsoClosupham} for $\Vo{K}$ and
performing the canonical transformation
\begin{align*}
    A       &\longrightarrow e^{\frac12(\alpha-t)}
   &\Pi_A   &\longrightarrow 2 \Bigl( \Pi_{\alpha} + \frac{\Vo{k}}2 \Bigr)
                             e^{-\frac12(\alpha-t)}
\end{align*} 
we obtain the reduced form of the departing canonical action, namely
\begin{equation} \label{R2IsoCloADM}
   S = \int \! \rmd t 
            \Bigl[ \Pi_{\alpha} \dot{\alpha} - H \Bigr], 
\end{equation}
where the reduced Hamiltonian density is
\begin{equation} \label{IsoCloADMHamR2}
   H = \frac1{4 \Vo{k}} \Bigl[
                           4 \Pi_{\alpha}^2 -
                           \frac{3\rho^3}4 e^{3\alpha} +
                           9\rho^2 e^{2\alpha} +
                           \Vo{k}^2
                        \Bigr]. 
\end{equation}
The resulting canonical system, 
\begin{subequations} \label{R2IsoCloCanEqu2}
   \begin{align}
      \dot{\alpha}         &= \frac2{\Vo{k}} \Pi_{\alpha}, \\
      \dot{\Pi}_{\alpha}   &= \frac{9\rho^3}{16\Vo{k}} e^{3\alpha} -
                              \frac{9\rho^2}{2\Vo{k}} e^{2\alpha}, 
   \end{align}
\end{subequations}
is equivalent to the autonomous second-order differential equation%
\footnote{Observe that this equation is nothing else than equation 
          \eqref{BiaCosR2CanEqu3a} particularised to the isotropic case.} 
\begin{equation} \label{R2IsoCloODE}
   \ddot{\alpha} = \frac{9\rho^3}{8\Vo{k}^2} e^{3\alpha} -
                   \frac{9\rho^2}{\Vo{k}^2} e^{2\alpha},
\end{equation}
which can be formally solved: its general solution is given in terms of 
elliptic integrals.%
\footnote{We prefer not to write down the details of the resolution since this
          is not a very exciting task.}

\paragraph*{Constant scalar curvature.}
When the scalar curvature is constant, \ie $R=4\Lambda$, we obtain from the 
canonical system \eqref{R2IsoCloCanEqu1} (with $N=1$) the elementary second-%
order \ode
\begin{equation}
   \ddot{A} - \omega^2 A = 0,
\end{equation}
where we have defined $\omega^2:=\tfrac{\Lambda}3$, and the first integral
\begin{equation}
   \dot{A}^2 - \omega^2 A^2 + 1 = 0.
\end{equation}
Once again there are two distinct cases depending on whether the scalar
curvature be positive or negative: 
\begin{equation}
   A(t)=\begin{cases}
           \Lambda^{-\frac12} \cosh (\sqrt{\Lambda} t)  
           &\text{if $\Lambda>0$}, \\
           \abs{\Lambda}^{-\frac12} \cos (\sqrt{\abs{\Lambda}} t)  
           &\text{if $\Lambda<0$},
        \end{cases}
\end{equation}
where we have rescaled the coordinates to eliminate irrelevant numerical 
factors. The former corresponds asymptotically to the de Sitter space-time.
\begin{rem}
When the scalar curvature is zero, the general solution to the canonical 
system \eqref{R2IsoCloCanEqu1} (with $N=1$) is readily found; it is explicitly
$A(t) = \sqrt{c_1 + c_2 t - t^2}$, where $c_1$ and $c_2$ are constants of 
integration, and contains the special solution that corresponds to an 
\einspace filled with incoherent radiation (\cfr \cite{schmi=InfClo}). 
\end{rem}

\paragraph*{Einstein--de Sitter minisuperspace.}

The three-metric of the Einstein-de Sitter model is
\begin{equation} \label{BiaCosR2IsoEdSMet}
   \rmd l^2 = A^2(t) \bigl( \rmd x^2 + \rmd y^2 + \rmd z^2 \bigr).  
\end{equation}
We proceed as in the case of the closed \flrw model. The super-Hamiltonian is
\begin{equation} \label{R2IsoEdSsupham}
   \EuH = \frac16 A \Vo{\cp}^2 - \frac{\sqrt{3}}3 \Vo{K} \frac{\Pi_A}{A}
\end{equation}
and the ensuing canonical system, 
\begin{subequations} \label{R2IsoEdSCanEqu1}
   \begin{align}
      \dot{A}          &= - \frac{\sqrt{3}}3 N \frac{\Vo{K}}{A}, \\
      \dot{K}_{\circ}  &= \frac13 N A \Vo{\cp}, \\
      \dot{\Pi}_A      &= - N \Bigl(
                                 \frac16 \Vo{\cp}^2 + 
                                 \frac{\sqrt{3}}3 \Vo{K} \frac{\Pi_A}{A^2}
                              \Bigr), \\
      \dot{\cp}_{\circ} &= \frac{\sqrt{3}}3 N \frac{\Pi_A}{A},
   \end{align}
\end{subequations}
possesses two first integrals: $\Pi_A=aA^2$ and $A\Pi_A+\Vo{K}\Vo{\cp}=b$, 
with $a$ and $b$ constant. 
\nl
When the scalar curvature is not constant we choose $R=\rho e^t$ and 
straightly integrate the canonical system to obtain the general solution in 
the form
\begin{equation} \label{BiaCosR2IsoEdSSol}
   \rmd s^2 = - \frac{e^{2t}}{\bigl( 1 - \Vo{n} e^{3t} \bigr)^2} \rmd t^2 
              + \frac{\rmd l^2}{\bigl( 1 - \Vo{n} e^{3t} \bigr)^{\frac23}},
\end{equation}
where $\Vo{n}$ is a constant of integration. Therefore we see that this
representation coincides with the isotropic limit \eqref{BiaCosR2BISolMetIso}
of the Bianchi-type I solution. 
\nl
On the other hand, if we consider spaces of constant curvature, \ie 
$R=4\Lambda$, then we obtain from the canonical system \eqref{R2IsoEdSCanEqu1} 
and the super-Hamiltonian constraint \eqref{R2IsoEdSsupham} the elementary 
differential equation $\dot{A}/A=\pm \omega$, where 
$\omega^2:=\tfrac13 \Lambda$, whence we derive two distinct solutions (up to 
rescaled coordinates), 
\begin{equation}
   A(t)=\begin{cases}
           \Lambda^{-\frac12} e^{\pm \sqrt{\Lambda} t} 
           &\text{if $\Lambda>0$}, \\
           \abs{\Lambda}^{-\frac12} \cos (\sqrt{\abs{\Lambda}} t) 
           &\text{if $\Lambda<0$},
        \end{cases}
\end{equation}
which correspond respectively to a de Sitter space and an anti-de Sitter 
space. This exemplifies the relevance of the $R$-squared theory to the 
inflationary scenario: In the field equations the terms that come specifically 
from the quadratic part of the Lagrangian density can play the r\^ole of a 
cosmological constant; moreover it is well known that the addition of the 
$R^2$ term in the gravitational action introduces a new spin-$0$ scalar field 
which may act as a natural inflaton in the early universe (\cfr the 
Introduction). 
\nl
For a zero scalar curvature the canonical equations are easily integrated and 
we obtain the solution $\rmd s^2=-\rmd t^2 + t \rmd l^2$, which was found 
independently by Caprasse \etal from a search of Groebner bases of the system 
of algebraic equations associated with power-law type solutions of the 
fourth-order field equations \cite{capra=PowTyp}.

\subsection{Conformal Bianchi-type I model}
\label{subsec:BiaCosHamWeyl}

\subsubsection{Canonical equations and super-Hamiltonian}

In conformal gravity the canonical action \eqref{HamActQuaLQ} reduces to
\begin{equation} \label{HamWeylCanAct1}
   S = \int \! \rmd t 
       \Bigl[ 
          \Pi_{\mu} \dot{\mu} + 
          \Pi_{\pm} \dot{\beta}_{\pm} + 
          \cp_{\pm} \dot{K}_{\pm} - 
          N \EuH
       \Bigr],     
\end{equation}   
owing to the existence of the primary constraint \eqref{HamQuaWeylPriCon}, 
which translates into $\Vo{\cp} \approx 0$ in terms of the new variables 
introduced on page \pageref{NewVar}. The explicit form of the super-%
Hamiltonian constraint depends on the Bianchi type considered; for type IX, 
for instance, this is given by the expression \eqref{BiaCosSupHamIX} without 
those few terms involving $\Vo{\cp}$. In the action \eqref{HamWeylCanAct1} we 
have not written the super-momentum constraints since they identically vanish 
for all diagonal class A models but type VI$_0$; we also reiterate that the 
surface terms at spatial infinity arising in the canonical formalism come to 
naught when one considers space-time isometries corresponding to spatially 
homogeneous Bianchi cosmologies (\cfr the discussion on boundary terms on 
pages \pageref{subsubsec:VarPriBouTer}--\pageref{rem:supmom}).
\nl
We focus on the simplest cosmological model that exhibits nontrivial \dofs in
conformal gravity, \ie Bianchi type I---the isotropic \flrw space-times are 
conformally flat. In this case the super-Hamiltonian is
\begin{equation} \label{HamWeylSupHam1}
   \begin{split}
      \EuH &= \frac{\sqrt{6}}6 
                 \Bigl[ 
                    4 \cp_- K_+ K_- -
                    2 \cp_+ \bigl( K^2_+ - K^2_- \bigr) -
                    K_{\pm} \Pi_{\pm}
                 \Bigr] \\
           &\qquad +
              \frac{\sqrt{3}}3 \Vo{K} 
                 \bigl( K_{\pm} \cp_{\pm} - \Pi_{\mu} \bigr) -
              \frac12 e^{-3\mu} 
                 \bigl( \cp^2_+ + \cp^2_- \bigr). 
   \end{split}
\end{equation}
(The coupling constant $\alpha$ of the Weyl-squared term has been set equal to
one.) 
\nl
We apply the Dirac--Bergmann consistency algorithm to this constrained system. 
The Dirac Hamiltonian density is $\Fh_{\EuD}=N \EuH+\Vo{\lambda}\Vo{\cp}$, 
with a Lagrange multiplier $\Vo{\lambda}$. Conservation of the primary 
constraint $\Vo{\cp}\approx0$ brings forth the secondary constraint 
$\Vo{\chi}\approx0$, \viz
\begin{equation*}
   \Vo{\cp} \approx 0 \tcons \Vo{\chi} = 
   \Pi_{\mu} - K_{\pm} \cp_{\pm} \approx 0.
\end{equation*}
Since we have the zero Poisson bracket $\{\Vo{\cp},\Vo{\chi}\}\approx0$ the 
above primary and secondary constraints are first class (the secondary one is
the generator of conformal transformations). To proceed we must impose a 
gauge-fixing condition associated with the `conformal constraint' 
$\Vo{\chi}\approx0$. We introduce the coordinate condition $\mu\approx0$ as an 
additional (primary) constraint---this amounts to turning the conformal 
constraint into second class. Conservation of this additional constraint 
yields the secondary constraint $\Vo{K}\approx0$. Therefore we end up with a 
set of four constraints, which are obviously second class. This enables one to
eliminate the pairs of canonical variables $\{\mu,\Pi_{\mu}\}$ and 
$\{\Vo{K},\Vo{\cp}\}$. The canonical action \eqref{HamWeylCanAct1} reduces to
\begin{equation} \label{HamWeylCanAct2}
   S = \int \! \rmd t  
       \Bigl[ 
          \Pi_{\pm} \dot{\beta}_{\pm} + \cp_{\pm} \dot{K}_{\pm} - \EuH
       \Bigr],     
\end{equation}   
where, in accordance with the conformal invariance of the theory and the form
of the Bianchi metrics \eqref{BiaCosMisParMet}, we have parameterised the 
lapse function as $N=e^{\mu}$; hence the gauge fixing described above amounts 
to choosing a specific conformal factor for the type I metric. The super-%
Hamiltonian constraint \eqref{HamWeylSupHam1} now reduces to 
\begin{equation} \label{HamWeylSupHam2}
   \EuH = \frac{\sqrt{6}}6 
             \Bigl[ 
                4 \cp_- K_+ K_- -  
                2 \cp_+ \bigl( K^2_+ - K^2_- \bigr) -
                K_{\pm} \Pi_{\pm} 
             \Bigr] - 
          \frac12 \bigl( \cp^2_+ + \cp^2_- \bigr).
\end{equation}
Varying the action \eqref{HamWeylCanAct2} \wrt the canonical variables and 
conjugate momenta we obtain the canonical equations of Bianchi type I in
conformal gravity: 
\begin{subequations} \label{HamWeylCanEq1}
   \begin{align}
      \dot{\beta}_{\pm} &= - \frac{\sqrt{6}}6 K_{\pm}, 
                             \label{HamWeylCanEq1a} \\ 
      \dot{\Pi}_{\pm}   &= 0, \label{HamWeylCanEq1b} \\
      \dot{K}_+         &= \frac{\sqrt{6}}3 \bigl( K^2_- - K^2_+ \bigr) - 
                           \cp_+, \label{HamWeylCanEq1c} \\         
      \dot{K}_-         &= \frac{2\sqrt{6}}3 K_+ K_-  - \cp_-, 
                           \label{HamWeylCanEq1d} \\        
      \dot{\cp}_+       &= \frac{\sqrt{6}}6 
                              \bigl( 
                                 \Pi_+ - 4 K_- \cp_- + 4 K_+ \cp_+ 
                              \bigr), 
                           \label{HamWeylCanEq1e} \\
      \dot{\cp}_-       &= \frac{\sqrt{6}}6 
                              \bigl( 
                                 \Pi_- - 4 K_- \cp_+ - 4 K_+ \cp_- 
                              \bigr). 
                           \label{HamWeylCanEq1f}
   \end{align} 
\end{subequations}
This system supplemented with the super-Hamiltonian constraint 
\eqref{HamWeylSupHam2} is equivalent to the Bach equations $B_{ab}=0$. (The
components of the Bach tensor $B_{ab}$ are written down in the Appendix at the 
end of this section on page \pageref{bachappix} ff.) We have indeed explicitly 
checked that: \textit{(i)} the canonical equations \eqref{HamWeylCanEq1b} are 
in fact the (spatial) \EL equations, \ie $B_{ii}=0$ for $i=1,2,3$, derived 
from the conformal quadratic action, owing to the recursion relations for the 
momenta $\Pi_{\pm}$ that are obtained from equations \eqref{HamWeylCanEq1e} 
and \eqref{HamWeylCanEq1f};%
\footnote{This equivalence is best understood with a glance at the box on
          page \pageref{box:GenOstThe}.}
\textit{(ii)} the super-Hamiltonian constraint \eqref{HamWeylSupHam2} 
coincides exactly with the `time-time' component of the Bach tensor, \ie 
$B_{00}=0$.
\nl
Having at our disposal---instead of the Bach equations---the nice differential 
system \eqref{HamWeylCanEq1} and the algebraic constraint 
\eqref{HamWeylSupHam2} we intend to seek out exact solutions by performing a 
global involution algorithm on appropriate extra constraints that yield closed 
constraint algebras. 

\subsubsection{Global involution algorithm}
\label{subsubsec:HamWeylGloInvAlg}

The involution method consists in applying the Dirac--Bergmann consistency
algorithm to our classical system, with the Poisson bracket defined \wrt the 
canonical variables $\beta_{\pm}$, $\Pi_{\pm}$, $K_{\pm}$, $\cp_{\pm}$, and 
after suitable conditions, \ie additional \emph{ad hoc} constraints, have been 
imposed.%
\footnote{For a rigorous treatment on the concept of involution as applied to 
          constrained systems, we refer the interested reader to Seiler and 
          Tucker's article \cite{seile=InvConI} (see also 
          \cite{gerdt=InvMet}).} 
Strictly speaking, the steps of the global involution algorithm are:
\begin{enumerate}
   \item Impose an appropriate extra constraint on the canonical variables.
         \label{step1}
   \item Require that this constraint be preserved when time evolution is
         considered. This gives rise to secondary constraints and possibly to
         the determination of the Lagrange multiplier associated with the
         extra constraint. \label{step2}
   \item Repeat Step (\ref{step2}) until no new information comes out. 
         \label{step3}
\end{enumerate}
Once the involution algorithm has been performed we may classify all the
constraints into first class and second class and proceed further to the
analysis of the particular system. 
 
\paragraph*{Constraint $\varphi_k \approx 0$.}
\label{par:PhiK}

The first extra constraint we consider expresses that the ratio of the 
variables $K_{\pm}$ be constant, namely
\begin{equation} \label{HamWeylPhiK0} 
   \varphi_k^{(0)} = K_- - k K_+ \approx 0,  \qquad k \in \mathbb{R}. 
\end{equation}   
The Poisson bracket of $\varphi_k^{(0)}$ and $\EuH$ yields the secondary
constraint
\begin{equation} \label{HamWeylPhiK1}
   \varphi_k^{(1)} = \frac{\sqrt{6}}3 k \bigl( k^2 - 3 \bigr) K^2_+ +
                     \bigl( \cp_- - k \cp_+ \bigr) \approx 0. 
\end{equation}   
The Poisson bracket of $\varphi_k^{(1)}$ and the Dirac Hamiltonian density
$\EuH_{\EuD} := \EuH + \lambda_k \varphi_k^{(0)}$ leads to the determination 
of the Lagrange multiplier, \viz
$\lambda_k \approx \sqrt{6}(\Pi_- - k \Pi_+)/[6(1+k^2)]$,  
which turns out to be constant, and the algorithm stops. Both constraints 
\eqref{HamWeylPhiK0} and \eqref{HamWeylPhiK1} are second class since their 
Poisson bracket is equal to $1+k^2$. Thus we can eliminate the corresponding 
spurious \dofs. To this end we perform the canonical transformation 
\begin{align} \label{HamWeylCanTra}
    &\cp_+ \longrightarrow \frac{\cp_+ - k \cp_-}{1+k^2}, 
   &&\cp_- \longrightarrow \frac{k \cp_+ + \cp_-}{1+k^2}. 
\end{align}   
The action \eqref{HamWeylCanAct2} then reduces to 
\begin{equation} \label{HamWeylCanAct3}
   S = \int \! \rmd t  
          \Bigl[ 
             \Pi_{\pm} \dot{\beta}_{\pm} + 
             \cp_+ \dot{K}_+ - 
             \EuH^{\prime} 
          \Bigr],     
\end{equation}   
where the super-Hamiltonian \eqref{HamWeylSupHam2} is now given by
\begin{equation} \label{HamWeylSupHam3}
   \begin{split}
      \EuH^{\prime} &= \frac{k^2 (k^2-3)^2}{3(1+k^2)} K^4_+ +
                       \frac{\sqrt{6}(3k^2-1)}{3(1+k^2)} K^2_+ \cp_+ \\
                    &\qquad - 
                       \frac{\sqrt{6}}6 \bigl( \Pi_+ + k \Pi_- \bigr) K_+ -
                       \frac{\cp^2_+}{2(1+k^2)}.
   \end{split}
\end{equation}
Varying the action \eqref{HamWeylCanAct3} \wrt the pair of canonical variables 
$K_+$ and $\cp_+$ we obtain the canonical equations
\begin{subequations} \label{HamWeylCanEq2}
   \begin{align}
      \dot{K}_+   &= \frac{\sqrt{6}(3k^2-1)}{3(1+k^2)} K^2_+ -
                     \frac{\cp_+}{1+k^2}, \label{HamWeylCanEq2a} \\         
      \dot{\cp}_+ &= \frac{4k^2(k^2-3)^2}{3(1+k^2)^2} K^3_+ + 
                     \frac{2\sqrt{6}(3k^2-1)}{3(1+k^2)} K_+ \cp_+ -
                     \frac{\sqrt{6}}6 \bigl( \Pi_+ + k \Pi_- \bigr).
                     \label{HamWeylCanEq2b} 
   \end{align} 
\end{subequations}
Taking into account the explicit form of the super-Hamiltonian 
\eqref{HamWeylSupHam3} we obtain a nonlinear first-order differential equation 
for $K_+(t)$,
\begin{equation} \label{HamWeylBriBou0}
   \dot{K}^2_+ = \frac23 (1+k^2) K^4_+ -
                 \frac{\sqrt{6} \bigl( \Pi_+ + k \Pi_- \bigr)}{3(1+k^2)} K_+,  
\end{equation}   
which can be written as one binomial equation of Briot and Bouquet, 
\begin{equation}
   \dot{u}^2 = u^4 + \gamma^3 u,
\end{equation}   
in terms of a new variable $u$ defined by the homography
$u=\pm\tfrac{\sqrt{6}}3K_+\sqrt{1+k^2}$, and with 
$\gamma^3=\mp\tfrac23\bigl(\Pi_+ + k\Pi_-\bigr)(1+k^2)^{-1/2}$. Direct 
calculation shows that its general solution is given by 
\begin{equation} \label{HamWeylBriBou1}
   u(t) = \frac{\gamma^3}{4 \cw}, 
\end{equation}   
where $\cw$ stands for the Weierstrass elliptic function $\wp(t-t_0;g_2,g_3)$, 
with invariants $g_2=0$ and 
$g_3=-\tfrac1{36}\bigl(\Pi_+ + k\Pi_-\bigr)^2/(1+k^2)$, and with one arbitrary 
constant $t_0$ (\cfr \cite[p.~627 ff.]{ABRAM=HanMat}). In terms of the 
canonical pair $(K_+,\cp_+)$ the representation \eqref{HamWeylBriBou1} 
corresponds to the expressions 
\begin{subequations} \label{HamWeylBBSol}
   \begin{align}
      K_+(t)   &= - \frac{\sqrt{6}\bigl(\Pi_+ + k\Pi_-\bigr)}{12(1+k^2)} 
                    \frac1{\cw}, \label{HamWeylBBSola} \\ 
      \cp_+(t) &= - \frac{\sqrt{6}\bigl(\Pi_+ + k\Pi_- \bigr)}{72\cw^2} 
                       \Biggl[ 
                          \frac{1-3k^2}{(1+k^2)^2} 
                             \bigl( \Pi_+ + k \Pi_- \bigr) + 
                          6 \cw \frac{\rmd \ln \cw}{\rmd t}
                       \Biggr],   
   \end{align}
\end{subequations}   
where                           
\begin{equation*}                          
   \cw \frac{\rmd \ln \cw}{\rmd t} = 
       \pm \frac16 \sqrt{144 \cw^3 + 
                         \frac{\bigl( \Pi_+ + k \Pi_- \bigr)^2}{1 + k^2}}. 
\end{equation*}   
To obtain the analytic form of the anisotropic metric functions 
$\beta_{\pm}(t)$ we must integrate the \rhs of equation \eqref{HamWeylBBSola}. 
We apply the following result \cite[p.~641]{ABRAM=HanMat}.%
\footnote{A prime denotes differentiation \wrt $z$, the argument of the 
          Weierstrassian functions considered hereafter.}
\begin{prop}
If $\wp^{\prime}(z_0)\neq 0$, then 
\begin{equation*}
   \wp^{\prime}(z_0) \int \frac{\rmd z}{\wp(z) - \wp(z_0)} 
      = 2 z \, \zeta(z_0) + \ln \sigma(z-z_0) - \ln \sigma(z+z_0),    
\end{equation*}   
with Weierstrassian functions $\zeta(z)$ and $\sigma(z)$ defined by 
$\zeta^{\prime}(z)+\wp(z)=0$ and $\sigma^{\prime}(z)-\sigma(z)\zeta(z)=0$ 
respectively. 
\end{prop}
To achieve our aim we take $z=t$ and $z_0=t_{\rmz}$, where $t_{\rmz}$ is a 
zero of the Weierstrass elliptic function, \ie $\wp(t_{\rmz})=0$. Indeed, we 
obtain
\begin{equation} \label{HamWeylIntWei}
   \int \frac{\rmd t}{\cw(t)} 
      = \pm \frac{6 \sqrt{1+k^2}}{\Pi_+ + k \Pi_-}
        \bigl[ 
           2 t \, \zeta (t_{\rmz}) + 
           \ln \sigma (t-t_{\rmz}) - \ln \sigma (t+t_{\rmz})
        \bigr]    
\end{equation}   
and we can write the corresponding homogeneous metrics of type I under the 
form 
\begin{equation} \label{HamWeylMetPhiK}
   \begin{split}
      \rmd s^2 =  - \rmd t^2  
                 &+ \exp \Biggl[ 
                            \pm \frac{2(1+\sqrt{3}k)}
                                     {\sqrt{1+k^2}} t \, \zeta(t_{\rmz})
                         \Biggr]
                    \Biggl[ 
                       \frac{\sigma (t-t_{\rmz})}{\sigma (t+t_{\rmz})}
                    \Biggr]^{\pm \frac{1+\sqrt{3}k}{\sqrt{1+k^2}}} \rmd x^2 \\
                 &+ \exp \Biggl[ 
                            \pm \frac{2(1-\sqrt{3}k)}
                                     {\sqrt{1+k^2}} t \, \zeta(t_{\rmz})
                         \Biggr]
                    \Biggl[ 
                       \frac{\sigma (t-t_{\rmz})}{\sigma (t+t_{\rmz})}
                    \Biggr]^{\pm \frac{1-\sqrt{3}k}{\sqrt{1+k^2}}} \rmd y^2 \\
                 &+ \exp \Biggl[ 
                            \mp \frac4{\sqrt{1+k^2}} t \, \zeta(t_{\rmz})
                         \Biggr]
                    \Biggl[ 
                       \frac{\sigma (t-t_{\rmz})}{\sigma (t+t_{\rmz})}
                    \Biggr]^{\mp \frac2{\sqrt{1+k^2}}} \rmd z^2.
   \end{split} 
\end{equation}
\nl
As a particular case of the present analysis we can specialise the above
solution \eqref{HamWeylBriBou1} to the axisymmetric case, for which the secondary 
constraint \eqref{HamWeylPhiK1} reduces to the conditions
\begin{equation} \label{HamWeylAxiCon}
   \frac{K_-}{K_+} \approx k \approx \frac{\cp_-}{\cp_+},
   \qquad k \in \bigl\{ 0, \pm \sqrt{3} \bigr\}.
\end{equation}   
According to the canonical transformation \eqref{HamWeylCanTra} the new 
variable $\cp_-$ vanishes automatically. The super-Hamiltonian 
\eqref{HamWeylSupHam3} becomes 
\begin{equation} \label{HamWeylSupHam4}
   \EuH_{\text{axi}}^{\prime} = 
      \frac{\sqrt{6}(3k^2-1)}{3(1+k^2)} K^2_+ \cp_+ - 
      \frac{\sqrt{6}}6 \bigl( \Pi_+ + k \Pi_- \bigr) K_+ -
      \frac{\cp^2_+}{2(1+k^2)}.
\end{equation}
and the corresponding canonical equations are
\begin{subequations} \label{HamWeylCanEq3}
   \begin{align}
      \dot{K}_+   &= \frac{\sqrt{6}(3k^2-1)}{3(1+k^2)} K^2_+ -
                     \frac{\cp_+}{1+k^2}, \label{HamWeylCanEq3a} \\         
      \dot{\cp}_+ &= \frac{2\sqrt{6}(3k^2-1)}{3(1+k^2)} K_+ \cp_+ -
                     \frac{\sqrt{6}}6 \bigl( \Pi_+ + k \Pi_- \bigr).
                     \label{HamWeylCanEq3b} 
   \end{align} 
\end{subequations}
As in the general case we find out the binomial equation of Briot and Bouquet 
\eqref{HamWeylBriBou0}, the solution \eqref{HamWeylBriBou1} of which is valid 
for any value of the parameter $k$; hence the axisymmetric solution is 
obtained by setting $k$ to $0$ or $\pm \sqrt{3}$ in expressions 
\eqref{HamWeylBBSol} and finally in the metric \eqref{HamWeylMetPhiK}. 

\paragraph*{Constraint $\varphi_p \approx 0$.}
\label{par:HamWeylPhiP}

Consider the extra constraint expressing that the ratio of the variables 
$\cp_{\pm}$ be constant, namely
\begin{equation} \label{HamWeylPhiP0}
   \varphi_p^{(0)} = \cp_- - p \cp_+ \approx 0,  \qquad p \in \mathbb{R}. 
\end{equation}   
The Poisson bracket of $\varphi_p^{(0)}$ and 
$\EuH_{\EuD}:=\EuH+\lambda_p\varphi_p^{(0)}$ yields the secondary constraint
\begin{equation} \label{HamWeylPhiP1}
   \varphi_p^{(1)} =  4 \cp_+ \bigl[ K_- (p^2-1) - 2 p K_+ \bigr] +
                      \bigl( \Pi_- - p \Pi_+ \bigr) \approx 0. 
\end{equation}   
To check the consistency of the involution algorithm we consider two distinct 
cases: 
\begin{enumerate}
   \item The secondary constraint \eqref{HamWeylPhiP1} is identically 
         satisfied if $\Pi_- - p \Pi_+ \approx 0$ and 
         $K_-(p^2-1)-2pK_+\approx 0$. Again, in order to proceed, we split up 
         the analysis into two subdivisions: 
         \begin{enumerate}
            \item If $p \neq \sigma$, with $\sigma=\pm 1$, the following weak 
                  equality holds: $K_- \approx 2 p K_+ /(p^2-1)$. The Poisson 
                  bracket of $\varphi_p^{(1)}$ and $\EuH_{\EuD}$ yields the 
                  constraint
                  \begin{equation} 
                     \begin{split}
                        \varphi_p^{(2)} &= (1-3p^2) \lambda_p -
                                           p(p^2-3) \cp_+ \\
                                        &\qquad -
                                           \frac{2\sqrt{6}p(1-3p^2)(p^2-3)}
                                                {3(p^2-1)^2}
                                         K^2_+ \approx 0, 
                     \end{split}
                  \end{equation}   
                  where $p \neq \tfrac{\sigma}{\sqrt{3}}$ since we consider 
                  nonzero canonical variables $\cp_{\pm}$. Consistency then 
                  leads to the determination of the Lagrange multiplier 
                  $\lambda_p$, \viz
                  \begin{equation*}
                     \lambda_p \approx 
                        \frac{2\sqrt{6}p(p^2-3)}{3(p^2-1)^2} K^2_+ -
                        \frac{p(p^2-3)}{3p^2-1} \cp_+.
                  \end{equation*}   
                  When $p \in \bigl\{ 0, \sigma \sqrt{3} \bigr\}$ the Poisson 
                  bracket of $\varphi_p^{(2)}$ and $\EuH_{\EuD}$ vanishes 
                  identically and the involution algorithm does not generate 
                  any new constraints. This subcase corresponds to a 
                  particular case of the axisymmetric solution with $k=p$ and
                  $\Pi_- \approx p \Pi_+$ (\cfr equation 
                  \eqref{HamWeylAxiCon}). For the nonaxisymmetric case, on the 
                  other hand, the next step in the involution algorithm yields 
                  \begin{equation*}
                     \Pi_+ \approx 
                        \Biggl[ 
                           \frac{8\sqrt{6}(3p^2-1)(p^2+1)}
                                {3(p+1)^3(p-1)^3} K^2_+ -
                           \frac{4(p^2+1)}{p^2-1} \cp_+ 
                        \Biggr] K_+
                  \end{equation*}   
                  and the last step gives
                  \begin{equation*}
                     \cp_+ \approx \frac{\sqrt{6}(1-3p^2)(\sigma-3)}
                                        {6(p^2-1)^2} K^2_+. 
                  \end{equation*}   
                  At this stage no more constraints arise. Taking into account 
                  that the super-Hamiltonian \eqref{HamWeylSupHam2} is weakly
                  vanishing we find out that $\Pi_+$ must be equal to zero. 
                  This subcase will be discussed below, as part of our 
                  discussion on the generic case with $\Pi_\pm \approx 0$.
            \item If $p = \sigma$, then $K_+$ is weakly vanishing. The 
                  Poisson bracket of $\varphi_p^{(1)}$ and $\EuH_{\EuD}$ 
                  yields the constraint 
                  \begin{equation} 
                     \varphi_p^{(2)} = \lambda_p - \sigma \cp_+ + 
                                       \frac{\sqrt{6}\sigma}3 K^2_- 
                                       \approx 0 
                  \end{equation}   
                  and consistency determines the Lagrange multiplier 
                  $\lambda_p$. The Poisson bracket of $\varphi_p^{(2)}$ and 
                  $\EuH_{\EuD}$ gives the weak equality
                  \begin{equation*}
                     \Pi_+ \approx 
                        \frac4{\sigma} 
                           \bigl( \frac{\sqrt{6}}3 K^2_- - \cp_+ \bigr) K_-.
                  \end{equation*}   
                  The last step in the involution algorithm yields
                  \begin{equation*}
                     \cp_+ \approx \frac{\sqrt{6}}{12} (3+\sigma) K^2_-. 
                  \end{equation*}   
                  At this stage no more constraints arise. The vanishing of 
                  the super-Hamiltonian \eqref{HamWeylSupHam2} restricts 
                  $\Pi_+$ to be zero. As in the previous non\-axi\-symmetric 
                  case with $p \neq \sigma$, this subcase will be discussed 
                  below.
         \end{enumerate}
   \item When $\Pi_- - p \Pi_+ \neq 0$, the Poisson bracket of 
         $\varphi_p^{(1)}$ and $\EuH_{\EuD}$ yields the constraint 
         \begin{align*} 
            \varphi_p^{(2)} &= 
               \sqrt{6} p (p^2-3) \cp^2_+ -  
               \sqrt{6} (1-3p^2) \lambda_p \cp_+ \\
            &\quad \qquad - 
               4 \cp_+ \bigl( K_+ + p K_- \bigr) \bigl( p K_+ - K_- \bigr) \\
            &\quad \qquad +
               \Pi_- \bigl( K_+ - p K_- \bigr) +
               \Pi_+ \bigl( p K_+ + K_- \bigr) 
            \approx 0. 
         \end{align*}   
         If $p = \tfrac{\sigma}{\sqrt{3}}$, the involution algorithm closes 
         when $\varphi_p^{(5)}$ is computed, but it does not lead to any exact 
         solutions since the super-Hamiltonian is not compatible with the 
         constraints $\varphi_p^{(j)}$ for $\range{j}{1}{5}$. On the other 
         hand, if $p \neq \tfrac{\sigma}{\sqrt{3}}$, we have not been able to 
         close the algorithm. It turns out, however, that the involution 
         algorithm in this case is useless since the system under 
         consideration does not produce an integration case, as it can be 
         shown by a local study of its analytic structure 
         \cite{demar=HamForWeyl}.%
         \footnote{Strictly speaking, the system with the constraint 
                   \eqref{HamWeylPhiP0} does not possess the Painlev\'e
                   property.}
\end{enumerate}

\paragraph*{Constraints $\Pi_{\pm} \approx 0$.}
\label{par:HamWeylPiCons}

Consider now the extra constraints expressing that the canonical variables
$\Pi_{\pm}$ be equal to zero. Contrary to the involution of the previous 
constraints $\varphi_k$ and $\varphi_p$, the consistency algorithm here is 
trivial: it is unable to produce any new information since there are no 
secondary constraints. The constraints $\Pi_{\pm} \approx 0$ are first class; 
we can choose their corresponding Lagrange multipliers $\lambda_{\pm}$ to be 
zero. In that case the constraints remain weak equations to be imposed on the
physical states of the system. Hence, at the classical level, the appropriate 
system of canonical equations is the system \eqref{HamWeylCanEq1}, where we 
set $\Pi_{\pm}$ to zero. Besides equation \eqref{HamWeylCanEq1a}, the relevant 
equations are thus 
\begin{subequations} \label{HamWeylPiEq}
   \begin{align}
      \dot{K}_+   &= \frac{\sqrt{6}}3 \bigl( K^2_- - K^2_+ \bigr) - \cp_+, 
                     \label{HamWeylPiEqa} \\         
      \dot{K}_-   &= \frac{2\sqrt{6}}3 K_+ K_-  - \cp_-, 
                     \label{HamWeylPiEqb} \\        
      \dot{\cp}_+ &= \frac{2\sqrt{6}}3 \bigl( K_+ \cp_+ - K_- \cp_- \bigr), 
                     \label{HamWeylPiEqc} \\
      \dot{\cp}_- &= - \frac{2\sqrt{6}}3 \bigl( K_- \cp_+ + K_+ \cp_- \bigr). 
                     \label{HamWeylPiEqd} 
   \end{align} 
\end{subequations}
The super-Hamiltonian \eqref{HamWeylSupHam2} is now given by
\begin{equation} \label{HamWeylSupHam5}
   \EuH = \frac{\sqrt{6}}3 
             \Bigl[ 
                2 \cp_- K_+ K_- -  
                \cp_+ \bigl( K^2_+ - K^2_- \bigr) 
             \Bigr] - 
          \frac12 \bigl( \cp^2_+ + \cp^2_- \bigr).
\end{equation}
\nl
The general solution of equations \eqref{HamWeylPiEq} is easy to produce under 
closed analytic form. We first solve equations \eqref{HamWeylPiEqa} and 
\eqref{HamWeylPiEqb} for $\cp_+$ and $\cp_-$ respectively and write down 
second-order equations for $K_+$ and $K_-$, namely 
\begin{align} \label{HamWeylPiEq2}
   &3 \ddot{K}_{\pm} = 4 \Sigma \, K_{\pm}, 
   &\text{with}\;& \Sigma := K_+^2 + K_-^2.
\end{align}
Aside from the super-Hamiltonian constraint, the canonical system 
\eqref{HamWeylPiEq} possesses the first integral 
\begin{equation} \label{HamWeylPiEqFirInt}
   K_+ \dot{K}_- - K_- \dot{K}_+ = \delta,
\end{equation}
where $\delta$ is an arbitrary constant. This first integral has been found
independently from a direct analysis of the Bianchi-type I field equations in
conformal gravity \cite{remy=EtuSol}. Making use of the super-Hamiltonian 
constraint \eqref{HamWeylSupHam5} we produce a scalar second-order equation 
for $\Sigma$, namely $\ddot{\Sigma} = 4 \Sigma^2$. We integrate this equation and obtain the 
solution $2 \Sigma = 3 \wp (t-t_0;0,g_3)$, with arbitrary constants $t_0$ and 
$g_3 = \tfrac{16}9 \delta^2$. This reduces the system \eqref{HamWeylPiEq2} to 
linear differential equations of the Lam\'e type, \ie
\begin{equation*}
   \ddot{K}_{\pm} = 2 \wp (t-t_0;0,g_3) K_{\pm}, 
\end{equation*}
which are studied exhaustively in \cite{INCE=OrdDif,WHITT=CouMod}. If we adopt 
Ince's notations, our particular case of the Lam\'e equation is specified by 
$h=0$ and $n=1$.%
\footnote{The general solution to the system \eqref{HamWeylPiEq} is uniform; 
          this confirms the single-valuedness of a possible integrability case 
          detected by the Painlev\'e test: the complete system with 
          $\Pi_{\pm}=0$ has the Painlev\'e property \cite{demar=HamForWeyl}.}
If we introduce $t_{\rmz}$ such that the transcendental equation 
$\wp (t_{\rmz}) = 0$ is satisfied, \ie $t_{\rmz}$ is a zero of the
Weierstrass elliptic function, then the general solution to the system under
study is given by the fundamental set
\begin{subequations} \label{HamWeylLamGen}
   \begin{align}
      K_{\pm, 1} &= \exp \bigl[ - t \, \zeta (t_{\rmz}) \bigr]
                    \frac{\sigma (t+t_{\rmz})}{\sigma (t)}, \\
      K_{\pm, 2} &= \exp \bigl[ + t \, \zeta (t_{\rmz}) \bigr]
                    \frac{\sigma (t-t_{\rmz})}{\sigma (t)},
   \end{align} 
\end{subequations}
with Weierstrassian functions $\zeta(t)$ and $\sigma(t)$ defined by 
$\dot{\zeta}(t)+\wp(t)=0$ and $\dot{\sigma}(t)-\sigma(t)\zeta(t)=0$ 
respectively. The solutions given by the fundamental set \eqref{HamWeylLamGen} 
are distinct, provided that the parameters $e_i$ for $i=1,2,3$, which are 
defined by
\begin{align*}
   &e_1 + e_2 + e_3 = 0, 
   &&4 ( e_2 e_3 + e_3 e_1 + e_1 e_2 ) = - g_2 \equiv 0, 
   &&4 e_1 e_2 e_3 = g_3,
\end{align*}
are not equal to zero; this is indeed the case here whenever $g_3 \neq 0$. If 
$g_3=0=\delta$, then $e_1=e_2=e_3=0$ and the solutions in the fundamental set 
are one and the same:%
\footnote{This case is better understood in the light of its analytic 
          structure: it turns out that one of the relevant singularity 
          families of our general canonical system becomes an exact two-%
          parameter particular solution when $\Pi_{\pm}$ vanish 
          \cite{demar=HamForWeyl}.}
\begin{subequations} \label{HamWeylLamF3}
   \begin{align}
      K_+   &= \pm \frac{\sqrt{6}}2 \frac{\sin 4K}{t - t_0}, \\
      K_-   &= \frac{\sqrt{6}}2 \frac{\cos 4K}{t - t_0}, \\
      \cp_+ &= \frac{\sqrt{6}}2 \frac{1 \pm \sin 4K - 2 \sin^2 4K}
                                     {(t - t_0)^2}, \\
      \cp_- &= \frac{\sqrt{6}}2 \frac{\cos 4K \bigl( 1 \pm 2 \sin 4K \bigr)}
                                     {(t - t_0)^2},
   \end{align} 
\end{subequations}
with $K$ constant. Denoting $\chi:=t-t_0$ and integrating equations 
\eqref{HamWeylCanEq1a} we obtain type I homogeneous metrics under the form
\begin{equation} \label{HamWeylMetF3}
   \rmd s^2 = - \rmd \chi^2 
              + \sum_{i=1}^3 \chi^{2 \nu_i} \bigl( \rmd x^i \bigr)^2,
\end{equation}
where the parameters $\nu_i$ for $i=1,2,3$ satisfy the relations
\begin{align*} 
   &\sum_{i=1}^3 \nu_i = 0,
   &\sum_{i=1}^3 \nu_i^2 &= \frac32.
\end{align*}

\subsubsection{Analytic structure of the Bianchi-type {\normalfont{I}} system}
\label{subsub:HamWeylAnaStr}

\paragraph*{The Painlev\'e strategy.}
The existence of the above solutions, whether particular solutions of the
general differential system or general solutions of specialised systems, tells
nothing about the integrability or non-integrability of the complete system 
and gives no information whatsoever about the mere accessibility of an exact 
and closed-form analytic expression of its general solution. This is due to 
the fact that the global involution algorithm of the extra constraints, as 
operated above, is not related with integrability and may even prove to be 
nonexhaustive. The integrability issue is tackled through an invariant 
investigation method of intrinsic properties of the general solution. In 
particular, the result does not depend on specific choices of the metric,
within some well-defined equivalence class.
\nl
The approach advocated by Painlev\'e proceeds from the main observation,
nowadays frequently exemplified in various areas of theoretical physics, that 
all analytic solutions encountered are single-valued or multiple-valued
\emph{finite} expressions (possibly intricate) depending on a finite number of
\emph{functions}---solutions of linear equations, elliptic functions, and the 
six transcendental functions systematically extracted by Painlev\'e and 
Gambier. Therefore, the building blocks of this process are the functions, 
which are explicitly defined through their single-valuedness. This emphasises 
the relevance of the analytic structure of the solution, as uniformisability 
ensures adaptability to all possible sets of initial conditions. The next step
of this approach deals with integrability in some fundamental sense. Indeed, 
probing the analytic structure of a system requires a global, as opposed to 
local, integrability-related property, namely the \emph{Painlev\'e property}.%
\footnote{A differential system possesses this property if, and only if, its 
          general solution is uniformisable or, equivalently, exhibits no 
          movable critical singularity.} 
At this level, integrable systems are defined in the sense of Painlev\'e as 
those systems that possess the Painlev\'e property.%
\footnote{The Painlev\'e property is invariant under arbitrary holomorphic
          transformations of the independent variable and arbitrary 
          homographic transformations of dependent variables. In particular, 
          the results given in greater detail in \cite{demar=HamForWeyl} are
          invariant under arbitrary analytic reparameterisations of time.} 
Moreover, in keeping with the above observation, the requirement for global
single-valuedness ought to be relaxed to accommodate this definition to 
integrability in the practical sense. Upon using broader classes of
transformations, the (unavoidable) analytic structure of the solution is
easily kept under control and either allows or rules out the possibility to 
produce a mode of representation of the general integral.
\nl
Such an approach requires the analytic continuation of the solution in the
complex domain of the independent variable and computes generic behaviours of 
the solution in a vicinity of each movable singularity. The trivial part of 
the Painlev\'e method, known as the Painlev\'e test, produces necessary but 
not sufficient conditions for a system to enjoy the Painlev\'e property and 
requires local single-valuedness of the general solution in a vicinity of all 
possible families of movable singularities.%
\footnote{Movable essential singularities are difficult to handle, since one 
          lacks methods to write down conditions under which they are indeed 
          noncritical. For a tutorial presentation of basic ideas and 
          constructive algorithms aimed at the generation of integrability 
          conditions, we refer the interested reader to \cite{conte=PaiApp,
          raman=PaiPro}.}

\paragraph*{Analytic proof of non-integrability.}
In \cite{demar=HamForWeyl} we have shown that the complete differential system 
\eqref{HamWeylCanEq1} admits three distinct local leading behaviours in some 
vicinity of movable singularities, that is, three distinct singularity 
families. The first leading behaviour, denoted $f_{(1)}$, is valid in a 
vicinity of the movable point $t=t_1$ and exists only when 
$\Pi_+ \Pi_- \neq 0$. The existence of the second singularity family, referred 
to as $f_{(2), \pm}$, valid in some vicinity of the movable point $t=t_2$, 
requires $\Pi_- (2 \sqrt{3} \Pi_- \pm 3 \Pi_+) \neq 0$. The third and last 
leading behaviour, denoted $f_{(3), \pm}$, is valid in some vicinity of the 
movable point $t=t_3$ and is associated with another arbitrary constant 
parameter, denoted $K$ in \cite{demar=HamForWeyl}; moreover it exists only 
when 
$4K\not\in 
\{0,\pi,-\pi,\tfrac{\pi}2,-\tfrac{\pi}2,\mp\tfrac{\pi}6,\mp\tfrac{5\pi}6\}$. 
The super-Hamiltonian \eqref{HamWeylSupHam2} introduces no new restrictions 
since, at leading order in $\chi:= t - t_i$ for all $i=1,2,3$, it always 
vanishes in some vicinity of movable singularities around which leading 
behaviours $f_{(1)}$, $f_{(2), \pm}$, and $f_{(3), \pm}$ hold. 
\nl
Around each species of movable singularities one must inquire whether it is 
possible to generate \emph{single-valued generic} local representations of the 
general solution. The Fuchsian indices have been computed for each leading 
behaviour and all preliminary conditions have proved to be fulfilled. However, 
near both movable points $t_2$ and $t_3$ in $\mathbb{C}$, it is not possible 
to build single-valued generic local expansions, unless some constraints are 
imposed. Upon introducing movable logarithmic terms in these local series one 
may regain genericness---at the expense of losing the Painlev\'e property.%
\footnote{For a thorough account and a complete proof of these claims, \cfr 
          \cite[Appendix B]{demar=HamForWeyl}.} 
\nl
Such multiple-valuedness, betrayed by local investigations, also pertains to
the general solution itself. This analytically proves that the system under
consideration is not integrable: Its general solution exhibits, in complex
time, an infinite number of logarithmic transcendental essential movable 
singularities; in other words, an analytic structure not compatible with
integrability in the practical sense---the quest for generic, exact and 
closed-form analytic expressions of the solution is hopeless. This result 
holds under space-time transformations within the equivalence class of the
Painlev\'e property; see \cite{conte=PaiApp}. Yet, local information produced
by the Painlev\'e test may still be used in order to extract all particular
systems that may prove to be integrable. In the case under study, it turns out 
that all particular solutions are meromorphic, but this is only a result of 
the application of the method.

\paragraph*{All integrable particular cases.}
Even though this does not provide a proof but merely an indication of
uniformisability, the Painlev\'e test requires local single-valuedness around 
all possible species of movable singularities. Utilising the results obtained 
from the Painlev\'e analysis \cite[Appendix B]{demar=HamForWeyl} and 
comparing them with those derived from the global involution algorithm above
we draw the following conclusions:
\begin{itemize}
   \item Near the movable point $t=t_2$, the complete set of integrability 
         conditions is satisfied if one imposes $\Pi_-=\pm\sqrt{3}\Pi_+$ and 
         simultaneously requires the vanishing of the arbitrary constant 
         parameter associated with the index $j=-3$. On the other hand, if one 
         sets $2\sqrt{3}\Pi_-=\mp3\Pi_+$ or $\Pi_-=0$, the family 
         $f_{(2), \pm}$ dies out and \emph{ipso facto} no restriction needs 
         fulfilment.
   \item Near the movable point $t=t_3$, single-valuedness may only be 
         recovered in two particular cases. The first case deals with a local 
         representation of the general solution of that specialised system 
         obtained with $\Pi_\pm = 0$. In this case local leading behaviours 
         $f_{(1)}$ and $f_{(2), \pm}$ die out whereas, near $t = t_3$, a 
         meromorphic local generic expansion is produced. This corresponds to
         the closed-form solution \eqref{HamWeylLamGen}. The second case deals 
         with a local representation of some particular solution of the 
         complete differential system and requires that the arbitrary 
         parameter $K$ be such that $\sin4K=\Pi_+(\Pi_+^2 + \Pi_-^2)^{-1/2}$
         and $\cos4K=\pm\Pi_-(\Pi_+^2 + \Pi_-^2)^{-1/2}$.
   \item The closed-form solution \eqref{HamWeylMetPhiK} obtained by imposing 
         the extra constraint $\varphi_k \approx 0$ on the system corresponds 
         to one particular integrable case, for it turns out that:
         \begin{enumerate}
            \item all indices that characterise the $f_{(3), \pm}$ singularity
                  family are compatible;
            \item the $f_{(2), \pm}$ singularity family dies out unless 
                  $k \in \{0,\pm \sqrt{3}\}$, but in this latter instance the
                  indices are compatible.
         \end{enumerate}
   \item The particular axisymmetric solution with $\cp_-=\pm\sqrt{3}\cp_+$
         corresponds to the integrable case of the $f_{(2), \pm}$ singularity
         family.
   \item The closed-form solution \eqref{HamWeylMetF3} corresponds to the case
         when the singularity family $f_{(3), \pm}$ becomes an exact two-%
         parameter solution whenever $\Pi_{\pm}=0$.
\end{itemize}
These results clearly indicate that the global involution algorithm of extra
constraints has proven to be exhaustive in the search for exact solutions that 
may be written in closed analytical form.

\subsubsection{Conformal relationship with Einstein spaces}
\label{subsubsec:einspace}

The problem of conformal relationship between Riemannian spaces and \einspaces
was first addressed circa 1920 by Brinkmann who studied the necessary and 
sufficient conditions for $n$-dimensional spaces to be conformally related to 
\einspaces \cite{brink=RieSpa,brink=EinSpa}.%
\footnote{Standard results can be found in Schouten's and Eisenhart's 
          textbooks \cite{SCHOU=RicCal,EISEN=RieGeo}.}
Kozameh, Newman, and Tod reexamined this question for four-dimensional 
manifolds and obtained nice results by addressing the problem at one and the
same time from the tensor and spinor points of view \cite{kozam=ConEin}.%
\footnote{The very first who undertook this problem within the framework of 
          the spinor formalism was Szekeres \cite{szeke=SpaCon}; however, the
          necessary and sufficient conditions he found are extremely difficult
          to translate into tensorial expressions. By contrast, Kozameh, 
          Newman, and Tod's analysis yields simpler results.}
\nl
In the context of conformal gravity it is crucial to determine whether 
closed-form solutions are conformally related to \einspaces or not: Since 
every \einspace or every space conformal to an \einspace 
fulfils the vacuum Bach equations automatically, any closed-form solution that
can be mapped onto an \einspace can be thought of---from the point of view of
generalised gravity theories---as a minor solution; as a matter of fact the 
interesting solutions will be those that are not conformal to an \einspace.%
\footnote{These are also called `nontrivial' solutions (according to Schmidt's 
          terminology \cite{schmi=NonTri}).}
\nl
In the case under study, \ie Bianchi-type I cosmology, despite the fact that
the general system is not integrable, as was proven by the Painlev\'e 
analysis, we have obtained all the particular exact solutions that may be 
written in closed analytical form. In keeping with the remark above, it is 
natural to discuss the conformal relationship of those solutions with
\einspaces \cite{quere=ISMC}. 
\nl
A four-dimensional space-time $(\EuM,g)$ can be mapped onto an \einspace 
under a conformal transformation $\tilde{g}_{ab}=e^{2\sigma(\bfx)} g_{ab}$ if 
and only if there exists a smooth function $\sigma(\bfx)$ that satisfies 
\begin{equation} \label{HamWeylEisen}
   L_{ab} = \nabla_a \sigma \nabla_b \sigma - \nabla_a \nabla_b \sigma 
            - \frac12 g_{ab} g^{cd} \nabla_c \sigma \nabla_d \sigma
            - \frac{\Lambda}{6} e^{2 \sigma} g_{ab},
\end{equation}
where the tensor $L_{ab}$ is defined by $L_{ab}:=\tfrac1{12}(Rg_{ab}-6R_{ab})$ 
and $\Lambda:=\tfrac14\tilde{R}$ denotes the cosmological constant that
characterises the conformal \einspace $(\EuM,\tilde{g})$ \cite{EISEN=RieGeo}. 
The first integrability conditions of equation \eqref{HamWeylEisen} are given 
by
\begin{equation} \label{HamWeylIntCon1}
   \nabla_d C^d_{\ abc} + C^d_{\ abc} \nabla_d \sigma = 0.
\end{equation}
These are the necessary and sufficient conditions for a space to be 
conformally related to a `$C$-space', \ie $\nabla_{[c} L_{b]a}=0$,%
\footnote{There is a natural hierarchy of classes of Riemannian spaces of 
          which the most general class consists of those spaces in which the
          Bianchi identities take the form $\nabla_d C^d_{\ abc}=0$ or,
          equivalently, $\nabla_{[c} L_{b]a}=0$; these are called 
          `$C$-spaces' and contain the \einspaces as a subclass 
          \cite{szeke=SpaCon}.} 
\cite{szeke=SpaCon}. The second integrability conditions of equation 
\eqref{HamWeylEisen} are 
\begin{equation} \label{HamWeylIntCon2}
   B_{ab}:= 2 \nabla^c \nabla^d C_{cabd} + C_{cabd} R^{cd} = 0.
\end{equation}
Therefore, fulfilment of the Bach equations is a necessary condition for a 
space to be conformally related to an \einspace. If considered separately the 
above integrability conditions of equation \eqref{HamWeylEisen} are merely 
necessary conditions with regard to the conformal relationship with 
\einspaces. However, Kozameh, Newman, and Tod have proven that they constitute 
a set of sufficient conditions as well \cite{kozam=ConEin}:
\begin{thm}
A space-time $(\EuM,g)$ is conformally related to an \einspace 
$(\EuM,\tilde{g})$ if and only if equations \eqref{HamWeylIntCon1} and 
\eqref{HamWeylIntCon2} are fulfilled.
\end{thm}
\nl
Specifying equation \eqref{HamWeylEisen} for Bianchi type I we obtain the 
differential system
\begin{subequations}
   \begin{align}
      &2 \ddot{\sigma} - \dot{\sigma}^2 
         + 5 \bigl( \dot{\beta}_+^2 + \dot{\beta}_-^2 \bigr)
         - \frac{\Lambda}3 e^{2 \sigma} = 0, \\
      &\dot{\sigma}^2 
         + 2 \dot{\sigma} \bigl( \dot{\beta}_+ + \sqrt{3} \dot{\beta}_- \bigr)
         + \ddot{\beta}_+ + \sqrt{3} \ddot{\beta}_-
         - \dot{\beta}_+^2 - \dot{\beta}_-^2
         - \frac{\Lambda}3 e^{2 \sigma} = 0, \\
      &\dot{\sigma}^2 
         + 2 \dot{\sigma} \bigl( \dot{\beta}_+ - \sqrt{3} \dot{\beta}_- \bigr)
         - \ddot{\beta}_+ - \sqrt{3} \ddot{\beta}_- 
         - \dot{\beta}_+^2 - \dot{\beta}_-^2
         - \frac{\Lambda}3 e^{2 \sigma} = 0, \\
      &\dot{\sigma}^2 
         - 4 \dot{\sigma} \dot{\beta}_+
         - 2 \ddot{\beta}_+
         - \dot{\beta}_+^2 - \dot{\beta}_-^2
         - \frac{\Lambda}3 e^{2 \sigma} = 0,
   \end{align}
\end{subequations}
which readily integrates to yield the necessary and sufficient condition for a
Bianchi-type I space to be conformally related to an \einspace, namely
\begin{equation} \label{HamWeylEinSpa1}
   K_{\pm} = - \sqrt{6} \dot{\beta}_{\pm} 
             = - \sqrt{6} k_{\pm} e^{- 2 \sigma}, \qquad k_{\pm} \neq 0,
\end{equation}
and provides one with a differential equation that enables one to determine 
explicitly the conformal factor $x:=e^{2 \sigma}$, \viz
\begin{equation} \label{HamWeylEinSpa2}
   \dot{x}^2 = \frac43 \Lambda x^3 + 4 \bigl( k_+^2 + k_-^2 \bigr).
\end{equation}
This last equation can be written as the Weierstrass \ode in terms of the
Weierstrass elliptic function $\cw:=\wp (t-t_0;g_2,g_3)$, with invariants
$g_2=0$ and $g_3=-\tfrac49 \Lambda^2 (k_+^2 + k_-^2)$. Therefore, the 
conformal factor that brings a Bianchi-type I space onto an \einspace is given 
by
\begin{equation} \label{HamWeylConFac1}
   e^{2 \sigma (t)} = 3 \Lambda^{-1} \cw.
\end{equation}
Taking this last result into account we get from equation 
\eqref{HamWeylEinSpa1} the explicit form of the functions $K_\pm (t)$, that is
\begin{equation} \label{HamWeylEinSpa3}
   K_{\pm} (t) = - \frac{\sqrt{6}}3 \Lambda k_{\pm} \cw^{-1}.
\end{equation}
So far we have not used the Bach equations nor the equivalent canonical system 
\eqref{HamWeylCanEq1}. If we do so, we see that the expression 
\eqref{HamWeylEinSpa3} does actually coincide with the solution 
\eqref{HamWeylBBSola} upon identifying 
\begin{align} \label{HamWeylConst1}
   &k \equiv \frac{k_-}{k_+}, 
   &&\Lambda \equiv \frac{k_+ \Pi_+ + k_- \Pi_-}{4 (k_+^2 + k_-^2)}.
\end{align}
Thus we can rewrite type I homogeneous metrics \eqref{HamWeylMetPhiK} under 
the equivalent form 
\begin{equation} \label{HamWeylMetPhip2}
   \begin{split}
      \rmd s^2 = - \rmd t^2  
                 &+ \exp \Biggl[ 
                            \pm \frac{2(k_+ + \sqrt{3} k_-)}
                                     {\sqrt{k_+^2 + k_-^2}} 
                            t \, \zeta(t_{\rmz})
                         \Biggr]
                    \Biggl[ 
                       \frac{\sigma (t-t_{\rmz})}{\sigma (t+t_{\rmz})}
                    \Biggr]^{\pm \frac{k_+ + \sqrt{3} k_-}
                            {\sqrt{k_+^2 + k_-^2}}} \rmd x^2 \\
                 &+ \exp \Biggl[ 
                            \pm \frac{2(k_+ - \sqrt{3} k_-)}
                                     {\sqrt{k_+^2 + k_-^2}} 
                            t \, \zeta(t_{\rmz})
                         \Biggr]
                    \Biggl[ 
                       \frac{\sigma (t-t_{\rmz})}{\sigma (t+t_{\rmz})}
                    \Biggr]^{\pm \frac{k_+ - \sqrt{3} k_-}
                            {\sqrt{k_+^2 + k_-^2}}} \rmd y^2 \\
                 &+ \exp \Biggl[ 
                            \mp \frac{4 k_+}{\sqrt{k_+^2 + k_-^2}} 
                            t \, \zeta(t_{\rmz})
                         \Biggr]
                    \Biggl[ 
                       \frac{\sigma (t-t_{\rmz})}{\sigma (t+t_{\rmz})}
                    \Biggr]^{\mp \frac{2 k_+}{\sqrt{k_+^2 + k_-^2}}} \rmd z^2. 
   \end{split}
\end{equation}
This is in agreement with the fact that imposing the constraint 
\eqref{HamWeylPhiK0}---here, a direct consequence of equation 
\eqref{HamWeylEinSpa1}---on the Bach equations is equivalent to requiring the
condition \eqref{HamWeylIntCon1} to be fulfilled, as it can be proved with the 
help of \textsc{Reduce}. In other words, assuming that the ratio of variables 
$K_{\pm}$ be constant is a necessary condition for a Bianchi-type I space to 
be conformally related to an \einspace; in conformal gravity, this becomes a 
sufficient condition as well. Thus the only way to find out a solution to the 
Bach equations that is not conformally related to an \einspace is to relax the 
constraint \eqref{HamWeylPhiK0}. In accordance with our analysis of the 
preceding sections, we have indeed obtained the only closed-form analytical 
solution that cannot be mapped onto an \einspace, namely the general solution 
\eqref{HamWeylLamGen} to the canonical system \eqref{HamWeylPiEq}, thereby 
confirming Schmidt's conjecture on the existence of such solutions 
\cite{schmi=NonTri,quere=ISMC}.
\nl
Now examine the particular case of a vanishing cosmological constant. Equation 
\eqref{HamWeylEinSpa2} becomes trivial and yields (up to an irrelevant 
constant of integration) 
\begin{equation} \label{HamWeylConFac2}
   e^{2 \sigma (t)} = t - t_0 =: \chi.
\end{equation}
Inserting this result into equation \eqref{HamWeylEinSpa1} and integrating we 
obtain type I homogeneous metrics under the form
\begin{equation} \label{HamWeylLamNul1}
   \rmd s^2 = - \rmd \chi^2 
              + \chi^{2 (k_+ + \sqrt{3} k_-)} \rmd x^2
              + \chi^{2 (k_+ - \sqrt{3} k_-)} \rmd y^2
              + \chi^{- 4 k_+} \rmd z^2, 
\end{equation} 
which coincides with the exact two-parameter particular solution 
\eqref{HamWeylMetF3} in the case of zero $\Pi_{\pm}$, upon identifying 
$k_+ \equiv \mp \tfrac12 \sin 4 K$ and $k_- \equiv - \tfrac12 \cos 4 K$.
Performing a conformal transformation with conformal factor 
\eqref{HamWeylConFac2} and introducing the proper time $\tau:=\chi^{3/2}$ we
derive the metric \eqref{HamWeylLamNul1} that characterises the conformal 
\einspace with zero cosmological constant, namely
\begin{equation} \label{HamWeylLamNul2}
   \rmd s^2 = - \rmd \tau^2 
              + \sum_{i=1}^3 \tau^{2 p_i} \bigl( \rmd x^i \bigr)^2, 
\end{equation} 
where the parameters $p_i$ for $i=1,2,3$ satisfy the relations
\begin{align} \label{HamWeylLamNul3}
   &p_1 + p_2 + p_3 = 1, 
   &&p_1^2 + p_2^2 + p_3^2 = \frac13 \bigl[ 1 + 8 (k_+^2 + k_-^2) \bigr] = 1.
\end{align} 
Hence, as expected, we recover the Bianchi-type I Kasner solution of vacuum 
\gr.

\subsection*{Appendix: Bach tensor}
\label{bachappix}

We have computed the components of the Bach tensor \eqref{eq:BachTens} for 
the Bianchi-type I metric with Misner's parameterisation, utilising the 
\textsc{Excalc} package in \textsc{Reduce}; they are explicitly:
\allowdisplaybreaks{
\begin{align*}
   B_{00} &=  2 \dddot{\beta}_- \dot{\beta}_- -
                \ddot{\beta}_-^2 -
             12 \dot{\beta}_-^4 -
             24 \dot{\beta}_-^2 \dot{\beta}_+^2 +
              2 \dddot{\beta}_+ \dot{\beta}_+ -
                \ddot{\beta}_+^2 -
             12 \dot{\beta}_+^4, \\
   B_{11} &= \sqrt{3} \bigl(    
                            \ddddot{\beta}_- -
                         24 \ddot{\beta}_- \dot{\beta}_-^2 -
                          8 \ddot{\beta}_- \dot{\beta}_+^2 -
                         16 \dot{\beta}_- \ddot{\beta}_+ \dot{\beta}_+ 
                      \bigr) -  
              2 \dddot{\beta}_- \dot{\beta}_- +
                \ddot{\beta}_-^2 \notag \\ 
          &\qquad - 
             16 \ddot{\beta}_- \dot{\beta}_- \dot{\beta}_+ + 
             12 \dot{\beta}_-^4 -  
              8 \dot{\beta}_-^2 \ddot{\beta}_+ +
             24 \dot{\beta}_-^2 \dot{\beta}_+^2 +
                \ddddot{\beta}_+ \notag \\
          &\qquad - 
              2 \dddot{\beta}_+ \dot{\beta}_+ +
                \ddot{\beta}_+^2 -
             24 \ddot{\beta}_+ \dot{\beta}_+^2 +
             12 \dot{\beta}_+^4, \\
   B_{22} &= \sqrt{3} \bigl(    
                            \ddddot{\beta}_- -
                         24 \ddot{\beta}_- \dot{\beta}_-^2 -
                          8 \ddot{\beta}_- \dot{\beta}_+^2 -
                         16 \dot{\beta}_- \ddot{\beta}_+ \dot{\beta}_+ 
                      \bigr) +  
              2 \dddot{\beta}_- \dot{\beta}_- -
                \ddot{\beta}_-^2 \notag \\
          &\qquad + 
             16 \ddot{\beta}_- \dot{\beta}_- \dot{\beta}_+ -
             12 \dot{\beta}_-^4 +  
              8 \dot{\beta}_-^2 \ddot{\beta}_+ -
             24 \dot{\beta}_-^2 \dot{\beta}_+^2 -
                \ddddot{\beta}_+ \notag \\
          &\qquad + 
              2 \dddot{\beta}_+ \dot{\beta}_+ - 
                \ddot{\beta}_+^2 +
             24 \ddot{\beta}_+ \dot{\beta}_+^2 -
             12 \dot{\beta}_+^4, \\
   B_{33} &=  2 \dddot{\beta}_- \dot{\beta}_- -
                \ddot{\beta}_-^2 + 
             32 \ddot{\beta}_- \dot{\beta}_- \dot{\beta}_+ - 
             12 \dot{\beta}_-^4 - 
             16 \dot{\beta}_-^2 \ddot{\beta}_+ - 
             24 \dot{\beta}_-^2 \dot{\beta}_+^2 \notag \\
          &\qquad + 
              2 \ddddot{\beta}_+ +
              2 \dddot{\beta}_+ \dot{\beta}_+ -    
                \ddot{\beta}_+^2 -
             48 \ddot{\beta}_+ \dot{\beta}_+^2 - 
             12 \dot{\beta}_+^4. 
\end{align*}
}

\backmatter


\thispagestyle{empty}

\vspace*{\stretch{1}}

\hfill
\begin{minipage}{9cm}
\large
\begin{quote}
\textit{``When you have read these hastily scrawled pages 
          you may guess, though never fully realize, why 
          it is that I must have forgetfulness or death.''} 
\nl
\hfill
--- H. P. Lovecraft, ``Dagon''
\end{quote}
\normalsize
\end{minipage}

\vspace*{\stretch{2}}

\cleardoublepage


\end{document}